\documentclass[dvips]{article}
\usepackage{supertabular,lscape,epsfig}
\usepackage{amssymb}
\usepackage{amsmath}

\usepackage[polish]{babel}
\usepackage[T1]{fontenc}
\usepackage[latin2]{inputenc}
\usepackage{pslatex}
\usepackage{graphicx}

\DeclareSymbolFont{ppa}{OT1}{ppl}{m}{it}
\DeclareMathSymbol{\vv}{\mathalpha}{ppa}{'166}
\setkeys{Gin}{bb=18 144 592 718}

\begin{document}

\newcommand{\dd}{\,{\rm d}}
\newcommand{\ie}{{\it i.e.},\,}
\newcommand{\etal}{{\it et al.\ }}
\newcommand{\eg}{{\it e.g.},\,}
\newcommand{\cf}{{\it cf.\ }}
\newcommand{\vs}{{\it vs.\ }}
\newcommand{\zdot}{\makebox[0pt][l]{.}}
\newcommand{\up}[1]{\ifmmode^{\rm #1}\else$^{\rm #1}$\fi}
\newcommand{\dn}[1]{\ifmmode_{\rm #1}\else$_{\rm #1}$\fi}
\newcommand{\upd}{\up{d}}
\newcommand{\uph}{\up{h}}
\newcommand{\upm}{\up{m}}  
\newcommand{\ups}{\up{s}}
\newcommand{\arcd}{\ifmmode^{\circ}\else$^{\circ}$\fi}
\newcommand{\arcm}{\ifmmode{'}\else$'$\fi}
\newcommand{\arcs}{\ifmmode{''}\else$''$\fi}
\newcommand{\MS}{{\rm M}\ifmmode_{\odot}\else$_{\odot}$\fi}
\newcommand{\RS}{{\rm R}\ifmmode_{\odot}\else$_{\odot}$\fi}
\newcommand{\LS}{{\rm L}\ifmmode_{\odot}\else$_{\odot}$\fi}

\newcommand{\Abstract}[2]{{\footnotesize\begin{center}ABSTRACT\end{center}
\vspace{1mm}\par#1\par   
\noindent
{~}{\it #2}}}

\newcommand{\TabCap}[2]{\begin{center}\parbox[t]{#1}{\begin{center}
  \small {\spaceskip 2pt plus 1pt minus 1pt T a b l e}
  \refstepcounter{table}\thetable \\[2mm]
  \footnotesize #2 \end{center}}\end{center}}

\newcommand{\TableSep}[2]{\begin{table}[p]\vspace{#1}
\TabCap{#2}\end{table}}

\newcommand{\FigCap}[1]{\footnotesize\par\noindent Fig.\  %
  \refstepcounter{figure}\thefigure. #1\par}

\newcommand{\TableFont}{\footnotesize}
\newcommand{\TableFontIt}{\ttit}
\newcommand{\SetTableFont}[1]{\renewcommand{\TableFont}{#1}}

\newcommand{\MakeTable}[4]{\begin{table}[htb]\TabCap{#2}{#3}
  \begin{center} \TableFont \begin{tabular}{#1} #4
  \end{tabular}\end{center}\end{table}}

\newcommand{\MakeTableSep}[4]{\begin{table}[p]\TabCap{#2}{#3}
  \begin{center} \TableFont \begin{tabular}{#1} #4
  \end{tabular}\end{center}\end{table}}

\newenvironment{references}%
{
\footnotesize \frenchspacing
\renewcommand{\thesection}{}
\renewcommand{\in}{{\rm in }}
\renewcommand{\AA}{Astron.\ Astrophys.}
\newcommand{\AAS}{Astron.~Astrophys.~Suppl.~Ser.}
\newcommand{\ApJ}{Astrophys.\ J.}
\newcommand{\ApJS}{Astrophys.\ J.~Suppl.~Ser.}
\newcommand{\ApJL}{Astrophys.\ J.~Letters}
\newcommand{\AJ}{Astron.\ J.}
\newcommand{\IBVS}{IBVS}
\newcommand{\PASP}{P.A.S.P.}
\newcommand{\Acta}{Acta Astron.}
\newcommand{\MNRAS}{MNRAS}
\renewcommand{\and}{{\rm and }}
\section{{\rm REFERENCES}}
\sloppy \hyphenpenalty10000
\begin{list}{}{\leftmargin1cm\listparindent-1cm
\itemindent\listparindent\parsep0pt\itemsep0pt}}%
{\end{list}\vspace{2mm}}
 
\def\TYLDA{~}
\newlength{\DW}
\settowidth{\DW}{0}
\newcommand{\dw}{\hspace{\DW}}

\newcommand{\refitem}[5]{\item[]{#1} #2%
\def\REFARG{#3}\ifx\REFARG\TYLDA\else, {\it#3}\fi
\def\REFARG{#4}\ifx\REFARG\TYLDA\else, {\bf#4}\fi
\def\REFARG{#5}\ifx\REFARG\TYLDA\else, {#5}\fi.}

\newcommand{\Section}[1]{\section{#1}}
\newcommand{\Subsection}[1]{\subsection{#1}}
\newcommand{\Acknow}[1]{\par\vspace{5mm}{\bf Acknowledgements.} #1}
\pagestyle{myheadings}

\newfont{\bb}{ptmbi8t at 12pt}
\newcommand{\xrule}{\rule{0pt}{2.5ex}}  
\newcommand{\xxrule}{\rule[-1.8ex]{0pt}{4.5ex}}  
\def\thefootnote{\fnsymbol{footnote}}

\begin{center}

{\Large\bf
Binary Lenses in OGLE-III EWS Database. Seasons 2006--2008}

\vskip 1.7cm
{\bf M.~~ J~a~r~o~s~z~y~\'n~s~k~i$^1$,~~ J.~~ S~k~o~w~r~o~n$^{2,1}$,~~ 
A.~~ U~d~a~l~s~k~i$^1$,~~ M.~~ K~u~b~i~a~k$^1$, M.~~ S~z~y~m~a~\'n~s~k~i$^1$,~~ 
G.~~ P~i~e~t~r~z~y~\'n~s~k~i$^1$,~~ I.~~ S~o~s~z~y~\'n~s~k~i$^1$,~~ 
\L.~~ W~y~r~z~y~k~o~w~s~k~i$^3$,
K.~~ U~l~a~c~z~y~k$^1$,~~ and~~ R.~~ P~o~l~e~s~k~i$^1$}

{$^1$Warsaw University Observatory, Al.~Ujazdowskie~4, 00-478~Warszawa, Poland\\
e-mail:(mj,jskowron,udalski,mk,msz,pietrzyn,soszynsk,kulaczyk,rpoleski)\\
@astrouw.edu.pl\\
 $^2$Department of Astronomy, The Ohio State University, 140~W. 18th Ave.,\\
Columbus, OH~43210, USA\\
 $^3$Institute of Astronomy, Cambridge University, Madingley Road,\\
Cambridge CB3 OHA, UK\\
e-mail:wyrzykow@ast.cam.ac.uk}
\end{center}
\vskip1.5cm

\Abstract{We present 27 binary lens candidates from OGLE-III Early
Warning System database for the seasons 2006--2008. The candidates have
been selected by visual light curves inspection. Our sample of binary lens
events consists now of 78 stellar systems and 7 extrasolar planets of
OGLE-III published elsewhere. Examining the distribution of stellar
binaries we find that the number of systems per logarithmic mass ratio
interval increases with mass ratio $q$, in contradiction with our previous
findings. Stellar binaries belong to the region $0.03<q<1$ and there is a
gap between them and a separate population of planets.} {Gravitational
lensing -- Galaxy: center -- binaries: general}

\Section{Introduction}
In this paper we present the results of the search for binary lens events
among microlensing phenomena discovered by the Early Warning System (EWS --
Udalski \etal 1994, Udalski 2003) of the third phase of the Optical
Gravitational Lensing Experiment (OGLE-III) in the seasons
2006--2008. Since OGLE-III Phase ended early in 2009, and we do not find
any binary lens candidates with complete light curves discovered by EWS in
this season, this paper finishes our presentation of binary lens candidates
among OGLE-III events. Our study is the continuation of work on binary
lenses in OGLE-II (Jaroszyński 2002, hereafter Paper I) and OGLE-III
databases (Jaroszyński \etal 2004, hereafter Paper~II Jaroszyński \etal
2006, hereafter Paper~III, and Skowron \etal 2007, hereafter Paper~IV).
The results of the similar search for binary lens events in MACHO data have
been presented by Alcock \etal (2000).

The motivation of the study remains the same -- we are going to obtain a
uniform sample of binary lens events, selected and modeled with the same
methods for all seasons. The sample may be used to study the population of
binary systems in the Galaxy. The method of observation of the binaries
(gravitational lensing) allows to study their mass ratios distribution,
since they are directly given by the models. The binary separations are
more difficult, because only their projection into the sky expressed in
Einstein radius units enters the models. In small number of cases the
estimation of the masses and distances to the lenses may be possible.

Cases of extremely low binary mass ratios ($q\le0.01$) are usually
considered as planetary lensing. Models of six microlensing planetary events 
from the OGLE-III database have been published 
(OGLE-2003-BLG-235 / MOA-2003-BLG-53 Bond \etal, 2004; 
OGLE-2005-BLG-071 Udalski \etal 2005;
OGLE-2005-BLG-169 Gould \etal 2006;
OGLE-2005-BLG-390 Beaulieu \etal 2006;
OGLE-2006-BLG-109 Gaudi \etal 2008;
OGLE-2007-BLG-368 Sumi \etal 2010).
Since the event OGLE-2006-BLG-109 is modeled as caused by a star with
two planets, there are seven planets with known mass ratios.

The adequate modeling of a planetary event requires frequent round the
clock observations of the source, which is achieved by cooperation of
observers at different longitudes on Earth. In cases of less extreme
lenses, the observations of single telescope may be sufficient to obtain
well constrained models of the systems. The present analysis is based on
the OGLE-III data alone.

Our approach follows that of Papers I, II, III and IV, where the references
to earlier work on the subject are given. Some basic ideas for binary lens
analysis can be found in the review article by Paczyński (1996). Paper~I
presents the analysis of 18 binary lens events found in OGLE-II data with
10 safe caustic crossing cases. There are 15 binary lens events reported in
Paper II, 19 in Paper III, and 9 in Paper IV. The event OGLE-2003-BLG-235
included in Paper~II is a planetary lens, and the binary lens
interpretation of the event OGLE-2005-BLG-226 of Paper~IV is less
convincing than for other cases, so these two events are not included in
our sample of stellar binary lenses.

In Section~2 we describe the selection of binary lens candidates. In
Section~3 we describe the improvements of the fitting procedure as compared
to our previous articles on the subject. The results are described in
Section~4, and the discussion follows in Section~5. The extensive graphical
material is shown in Appendix.

\Section{Choice of Candidates} 
The OGLE-III data is routinely reduced with difference photometry ({\sc
dia}, Alard and Lupton 1998, Alard 2000) which gives high quality light
curves of variable objects. The EWS system of OGLE-III (Udalski 2003)
automatically picks up candidate objects with microlensing-like
variability.

There are 581 microlensing event candidates selected by EWS in the season
of 2006, 610 in 2007, 654 in 2008, and 156 in 2009. We visually inspect all
candidate light curves looking for features characteristic for binary
lenses (multiple peaks, U-shapes, asymmetry). Light curves showing
excessive noise are omitted. We select 31 candidate binary events from the
data for further study. For these candidate events we apply our standard
procedure of finding binary lens models (compare Papers~I--IV and
Section~3).

\Section{Fitting Binary Lens Models}
The models of the two point mass lens were investigated by many authors
(Schneider and Weiss 1986, Mao and Paczyński 1991, Mao and DiStefano 1995,
DiStefano and Mao 1996, Dominik 1998, to mention only a few). The effective
methods applicable for extended sources were described by Mao and Loeb
(2001). We follow their approach and use image finding algorithms based on
the Newton method. Our approach has been described in Papers~I--IV. We have
technically improved the search for models in cases of caustic crossing
events, using different, more {\it natural} parametrization of the models
in the calculations, following Cassan (2008) and Skowron (2009). In
reporting our results we use the notation of Papers~I--IV; below we
describe our convention for completeness.

We fit binary lens models using the $\chi^2$ minimization  method for the 
light curves. It is convenient to model the flux at the time $t_i$ as: 
$$F_i=F(t_i)=A(\mathbf{p};t_i)\times F_s+F_b\equiv\left(A(\mathbf{p};t_i)-1\right)\times F_s+F_0\eqno(1)$$
where $A(\mathbf{p};t_i)$ is the binary lens amplification at instant $t_i$
calculated with the model defined by a set of parameters $\mathbf{p}$,
$F_s$ is the flux of the source being lensed, $F_b$ the blended flux (from
the source close neighbors and possibly the lens), and the combination
$F_b+F_s=F_0$ is the total flux, measured long before or long after the
event. The baseline flux ($F_0$) can be reasonably well estimated with
observations performed in seasons when the source was not showing
variability (see below). The binary lens parameters ($\mathbf{p}$) are: the
mass ratio $q$, the binary separation $d$, the source trajectory direction
$\beta$, the impact parameter $b$ relative to the lens center of mass, the
time of the closest center of mass approach $t_0$, the Einstein time
$t_{\rm E}$, the size of the source $r_s$. The separation ($d$), impact
parameter ($b$) and source radius ($r_s$) are all expressed in Einstein
units.

In fitting the models we use rescaled errors (compare Papers~I--IV).
More detailed analysis (\eg Wyrzykowski 2005, Skowron 2009, Skowron
\etal 2009) shows that the OGLE photometric errors are overestimated 
for very faint sources and underestimated for bright ones. Error scaling
used here, based on the scatter of the observed flux in seasons when the
source is supposedly invariable, is the simplest approach. It gives the
estimate of the combined effect of the observational errors and possibly
undetectable, low amplitude internal source variability. We require that
constant flux source model fits well the other seasons data after
introducing error scaling factor $s$:
$$\chi_{\rm other}^2=\sum\limits_{i=1}^{N^\prime}
\frac{(F_i-F_0)^2}{(s\sigma_i)^2}=N^\prime-1\eqno(2)$$
where $F_0$ is the optimized value. The summation is over $N^\prime$
data points, which do not belong to the event's season.
Our analysis of the models, their fit quality etc is based on the
$\chi_1^2$ calculated with the rescaled errors:
$$\chi_1^2\equiv\frac{\chi^2}{s^2}\eqno(3)$$
which is displayed in the tables and plots. For events with multiple models
(representing different local minima of $\chi^2$), we assess the relevance
of each model with the relative weight $w\sim\exp(-\chi_1^2/2)$.

Only the events with characteristics of caustic crossing (apparent
discontinuities in observed light curves, U-shapes) can be treated as safe
binary lens cases. The double peak events may result from cusp approaches,
but may also be produced by double sources (\eg Gaudi and Han 2004). In
such cases we check also the double source fit of the event postulating:
$$F(t)=A_s(u_1(t))\times F_{\rm s1}+A_s(u_2(t))\times F_{\rm s2}+F_{\rm b}\eqno(4)$$
where $F_{\rm s1}$, $F_{\rm s2}$ are the fluxes of the source components,
$F_{\rm b}$ is the blended flux, $u_i$ are the distances between the source
components and the lens expressed in Einstein units, and $A_s(u)$ is the
single lens amplification (Paczyński, 1986).

\Section{Results}
Our fitting procedures applied to 31 candidate events selected give the
results summarized in Table~1. Models of the two events (2006-BLG-038 and
2006-BLG-460) have already been published by Skowron \etal (2009). Our
calculations give very similar sets of parameters for these two events.
Some events have two or three substantially
different models of similar fit quality and in such cases we present all of
them. We do not include models of planetary events published elsewhere.
\MakeTable{ccrccccccccl}{12.7cm}{Binary lenses parameters, excluding
known planetary cases}
{\hline
\noalign{\vskip 3pt}
Event & &  $\chi_1^2/{\rm dof}$ & $s$ & $q$ & $d$ & $\beta$ & $b$ & $t_0$ & $t_E$ & $f$ & $r_s$ \\
\noalign{\vskip3pt}
\hline
06-028  & b &    919.6/770 &  1.89 & 0.195 & 1.086 &    1.73 &  -0.10 &  3799.6 &   41.6 &  0.39 &  \\
06-038  & b &   1097.8/732 &  1.79 & 0.618 & 3.116 &  153.67 &   0.64 &  3810.6 &   12.4 &  1.00 &  \\
06-076  & b &   1357.1/667 &  1.93 & 0.455 & 1.732 &    2.51 &  -0.02 &  3811.6 &    9.7 &  0.93 & 0.0104 \\
06-119  & u &   1260.1/759 &  1.58 & 0.066 & 0.770 &  136.42 &   0.05 &  3842.0 &   63.1 &  0.19 &  \\
06-215  & b &   1131.4/804 &  4.08 & 0.959 & 0.991 &  152.12 &  -0.49 &  3855.2 &   28.9 &  0.69 & 0.0081 \\
06-277  & b &    634.1/441 &  2.52 & 0.593 & 1.368 &   97.80 &   0.12 &  3941.3 &   38.1 &  0.99 & 0.0056 \\
06-284  & b &   1437.2/601 &  1.81 & 0.242 & 0.820 &   23.56 &   0.10 &  3898.9 &   37.2 &  0.99 & 0.0008 \\
06-304  & b &   1022.0/723 &  1.61 & 0.248 & 2.630 &  128.33 &   1.38 &  3844.9 &   49.8 &  0.53 & 0.0009 \\
06-304  & b &   1023.7/723 &  1.61 & 0.308 & 0.628 &   54.93 &   0.06 &  3900.4 &   24.0 &  0.34 & 0.0010 \\
06-335  & b &    884.6/648 &  1.82 & 0.923 & 2.168 &  127.01 &  -0.54 &  3934.4 &   33.9 &  0.98 & 0.0032 \\
06-375  & b &    603.4/568 &  1.70 & 0.501 & 0.891 &  326.83 &   0.67 &  3944.4 &   29.0 &  0.93 &  \\
06-450  & b &    609.1/555 &  1.48 & 0.449 & 0.934 &  113.54 &   0.18 &  3963.2 &    9.2 &  0.88 &  \\
06-460  & b &   1115.7/973 &  1.47 & 0.230 & 3.352 &  181.36 &  -0.11 &  3981.2 &   20.1 &  0.67 &  \\
07-006  & b &    801.6/737 &  1.71 & 0.069 & 0.906 &  177.76 &   0.00 &  4169.9 &   49.9 &  0.67 &  \\
07-069  & b &    792.9/966 &  1.83 & 0.570 & 3.006 &  105.40 &   1.61 &  4159.7 &   61.3 &  0.55 & 0.0016 \\
07-149  & b &    923.8/973 &  1.17 & 0.825 & 2.519 &   40.72 &   0.68 &  4302.4 &  105.4 &  0.25 &  \\
07-149  & b &    927.0/973 &  1.17 & 0.660 & 0.623 &  105.01 &   0.07 &  4217.6 &   49.1 &  0.26 &  \\
07-149  & b &    927.2/973 &  1.17 & 0.108 & 1.007 &  130.24 &   0.25 &  4209.5 &   46.0 &  0.55 &  \\
07-237  & b &    453.4/614 &  2.16 & 0.327 & 1.844 &  100.51 &   0.88 &  4227.3 &   48.3 &  0.05 & 0.0007 \\
07-237  & b &    454.0/614 &  2.16 & 0.577 & 0.965 &  142.41 &   0.12 &  4235.5 &   20.5 &  0.04 & 0.0020 \\
07-237  & b &    458.0/614 &  2.16 & 0.804 & 0.486 &   43.85 &   1.18 &  3893.9 &  248.6 &  0.07 & 0.0001 \\
07-327  & u &   2489.8/652 &  1.87 & 0.012 & 1.723 &  136.06 &   0.74 &  3640.7 &  817.5 &  0.50 &  \\
07-363  & d &  1148.5/1195 &  1.32 & 0.164 & 2.211 &  151.68 &  -0.07 &  4295.6 &   21.5 &  0.13 &  \\
07-373  & b &    855.0/988 &  1.79 & 0.208 & 2.047 &  353.78 &  -0.08 &  4317.5 &   17.5 &  0.40 & 0.0023 \\
07-399  & b &    409.7/426 &  1.24 & 0.041 & 1.271 &   73.09 &   0.47 &  4305.9 &   18.4 &  0.62 &  \\
08-118  & b &  1179.2/1229 &  1.53 & 0.673 & 0.831 &   43.05 &   0.07 &  4604.9 &   81.3 &  0.32 &  \\
08-118  & b &  1184.0/1229 &  1.53 & 0.003 & 2.389 &  117.28 &   1.75 &  3368.3 & 1374.7 &  0.33 &  \\
08-125  & b &    830.7/748 &  1.65 & 0.449 & 0.508 &  174.86 &  -1.65 &  4763.2 &  449.8 &  0.80 &  \\
08-243  & b &    828.6/816 &  1.30 & 0.372 & 1.397 &  171.93 &  -0.11 &  4611.0 &   36.8 &  0.71 & 0.0016 \\
08-263  & b &     93.3/150 &  1.00 & 0.039 & 1.922 &   38.43 &   0.79 &  4740.2 &  131.3 &  0.59 &  \\
08-330  & b &  1526.9/1533 &  1.50 & 0.343 & 0.817 &  155.48 &  -0.48 &  4679.7 &   89.2 &  0.90 &  \\
08-355  & b &  1301.8/1309 &  1.88 & 0.106 & 1.326 &   79.11 &   0.42 &  4642.6 &   33.2 &  0.44 &  \\
08-493  & b &  1426.5/1528 &  1.71 & 0.726 & 2.097 &  187.30 &  -0.24 &  4554.7 &  129.0 &  0.24 &  \\
08-513  & b &  1377.4/1380 &  1.93 & 0.215 & 0.901 &  138.32 &  -0.06 &  4708.2 &   41.9 &  0.11 &  \\
08-559  & b &  1608.8/1671 &  1.29 & 0.227 & 0.872 &  149.65 &   0.00 &  4715.4 &   47.2 &  0.54 &  \\
08-559  & b &  1608.8/1671 &  1.29 & 0.624 & 2.277 &  116.60 &  -0.58 &  4742.9 &   75.6 &  0.31 &  \\
08-584  & b &  1249.3/1304 &  1.58 & 0.355 & 0.572 &   63.76 &   0.05 &  4691.6 &  136.0 &  0.03 &  \\
08-584  & b &  1251.6/1304 &  1.58 & 0.333 & 2.076 &  339.84 &   0.17 &  4610.2 &  239.6 &  0.03 &  \\
08-584  & b &  1251.9/1304 &  1.58 & 0.069 & 1.013 &   51.55 &   0.22 &  4702.5 &   88.4 &  0.10 &  \\
08-592  & d &    828.1/863 &  1.71 & 0.907 & 0.631 &   87.73 &   0.00 &  4705.1 &   21.0 &  0.10 &  \\
08-592  & d &    829.7/863 &  1.71 & 0.093 & 0.799 &   31.61 &   0.12 &  4706.2 &   28.6 &  0.15 &  \\
08-592  & d &    830.2/863 &  1.71 & 0.642 & 1.041 &   20.02 &   0.03 &  4706.4 &   20.0 &  0.31 &  \\
\noalign{\vskip3pt}
\hline
\noalign{\vskip3pt}
\multicolumn{12}{p{12.7cm}}{Note: 
The table contains all 2006--2008 seasons events, which have been modeled
as binary lenses. The columns show: two last digits of the year and the EWS
number, the event classification (``b'' for binary lens, ``d'' for double
source, ``u'' for unsatisfactory), the rescaled $\chi^2_1$ / degree of
freedom number, the scaling factor $s$
($\chi^2_1=\chi^2_{\rm raw}/s^2$), the binary lens model parameters (see
Section~3), and the blending parameter $f\equiv F_{\rm s}/F_0$. For
events with at least one resolved caustic crossing the size of the source
$r_s$ is given; otherwise it is omitted.}}

In the second column of the table we assess the character of the events.
In 27 cases (of 31 investigated) the events are safe binary lens phenomena
in our opinion (designated ``b'' in Table~1). There are two cases
classified as double source events (``d'' in Table~1) and two events with a
low quality, unsatisfactory fit (``u'' in Table~1). The source paths and
model light curves are shown in the first part of Appendix.

Some events have double peaks, but do not show any sharp changes in their
light curves. In such cases we try double source models. If double source
models are acceptable, we skip binary lens modeling. In some cases we
present models of both kinds. Results of double source modeling are
presented in Table~2. The model light curves are shown in Appendix.

\MakeTable{lrccccrcc}{12.7cm}{Parameters of double source modeling}
{\hline
\noalign{\vskip3pt}
Event & $\chi_1^2/$DOF & $b_1$ & $b_2$ & $t_{01}$ & $t_{02}$ & $t_{\rm E}$ &  $f_1$ & $f_2$ \\
\noalign{\vskip3pt}
\hline
06-023 &    842./506 & 0.4066 & 0.0274 & 3801.59 & 3857.87 &    60.2 & 0.637 & 0.363 \\
06-038 &   2352./735 & 0.0582 & 0.0409 & 3827.33 & 3830.68 &    15.4 & 0.154 & 0.175 \\
06-061 &    647./574 & 0.0023 & 0.3595 & 3717.61 & 3827.98 &    45.5 & 0.758 & 0.242 \\
06-238 &   1215./699 & 0.0458 & 0.0000 & 3878.53 & 3879.81 &    23.0 & 0.501 & 0.016 \\
06-393 &    604./647 & 0.5400 & 0.6874 & 4073.06 & 4287.97 &   135.5 & 0.221 & 0.267 \\
06-398 &    904./738 & 0.0047 & 0.0004 & 3931.52 & 3934.78 &   356.7 & 0.001 & 0.000 \\
06-441 &    365./490 & 0.0385 & 0.0508 & 4009.75 & 4333.22 &   962.2 & 0.005 & 0.008 \\
06-444 &    722./646 & 0.1325 & 0.0186 & 3960.27 & 3982.20 &    23.9 & 0.250 & 0.142 \\
06-450 &   3434./554 & 0.1767 & 0.0119 & 3961.86 & 3964.70 &     8.4 & 0.931 & 0.069 \\
06-460 &   1131./974 & 0.0958 & 0.0435 & 3969.47 & 4030.73 &    23.6 & 0.455 & 0.162 \\
06-504 &    466./531 & 0.0415 & 0.0647 & 4010.02 & 4337.93 &   845.8 & 0.005 & 0.009 \\
07-006 &   3825./736 & 0.0013 & 0.0000 & 4168.02 & 4183.50 &   161.5 & 0.173 & 0.019 \\
07-040 &    655./661 & 1.6218 & 0.2823 & 4161.42 & 4319.90 &    76.4 & 0.928 & 0.071 \\
07-080 &    637./624 & 0.2360 & 0.0110 & 4211.02 & 4380.11 &    49.7 & 0.795 & 0.008 \\
07-091 &    912./1037 & 1.2175 & 1.3241 & 4188.87 & 4290.17 &    17.2 & 0.582 & 0.417 \\
07-159 &   1302./1002 & 0.0247 & 0.7771 & 4228.01 & 4230.65 &    19.0 & 0.016 & 0.984 \\
07-236 &    947./1033 & 0.0383 & 0.0429 & 4233.90 & 4239.38 &    20.1 & 0.038 & 0.066 \\
07-237 &    870./613 & 0.0059 & 0.4840 & 4208.92 & 4240.76 &     0.1 & 0.000 & 1.000 \\
07-355 &    583./583 & 0.4516 & 0.0257 & 4301.36 & 4306.21 &    13.1 & 0.861 & 0.047 \\
07-363 &   1231./1194 & 0.0178 & 0.0592 & 4289.86 & 4292.68 &    11.9 & 0.082 & 0.127 \\
07-399 &    493./425 & 0.0018 & 0.0069 & 4301.98 & 4305.34 &   173.9 & 0.003 & 0.011 \\
07-491 &    738./744 & 0.0000 & 0.1769 & 4352.25 & 4356.28 &    51.8 & 0.019 & 0.183 \\
07-594 &    488./608 & 0.0023 & 0.0520 & 4378.67 & 4382.26 &    61.8 & 0.011 & 0.060 \\
08-078 &    660./601 & 0.4376 & 0.0039 & 4545.08 & 4546.68 &    33.6 & 0.659 & 0.040 \\
08-084 &    599./588 & 0.3480 & 0.0101 & 4538.52 & 4547.57 &    27.2 & 0.877 & 0.123 \\
08-092 &    567./619 & 0.0001 & 1.5498 & 4541.25 & 4726.12 &    34.0 & 0.003 & 0.997 \\
08-098 &    848./854 & 0.7560 & 0.3552 & 4616.70 & 4774.99 &   120.7 & 0.648 & 0.352 \\
08-110 &    643./623 & 0.0175 & 0.3668 & 4537.30 & 4541.81 &    32.2 & 0.076 & 0.924 \\
08-143 &   1853./1310 & 1.0437 & 0.4371 & 4579.51 & 4726.86 &    40.3 & 0.276 & 0.724 \\
08-146 &   2259./1361 & 0.0000 & 0.0000 & 4566.83 & 4584.53 & 30628.0 & 0.000 & 0.000 \\
08-210 &   1298./1175 & 0.0433 & 0.0113 & 4606.25 & 4607.07 &    19.5 & 0.639 & 0.219 \\
08-353 &    857./882 & 0.1527 & 0.0478 & 4620.69 & 4652.47 &    20.2 & 0.213 & 0.032 \\
08-513 &   7419./1379 & 0.0087 & 0.0184 & 4706.21 & 4716.68 &    36.4 & 0.072 & 0.014 \\
08-559 &   1842./1670 & 0.0894 & 0.0024 & 4715.51 & 4724.62 &    45.1 & 0.429 & 0.011 \\
08-592 &    836./862 & 0.0178 & 0.1071 & 4701.01 & 4710.59 &    22.2 & 0.040 & 0.226 \\
\noalign{\vskip3pt}
\hline
\multicolumn{9}{p{12.7cm}}{Note: The table contains 
two last digits of the year and the EWS number, the rescaled $\chi^2$ value
and the DOF number, the impact parameters $b_1$ and $b_2$ for the two
source components, times of the closest approaches $t_{01}$ and $t_{02}$,
the Einstein time $t_{\rm E}$, and the blending parameters $f_1=F_{\rm
s1}/(F_{\rm s1}+F_{\rm s2}+F_{\rm b})$ and $f_2=F_{\rm s2}/(F_{\rm
s1}+F_{\rm s2}+F_{\rm b})$.}}

While our double source models usually give formally good fits of the light
curves, not all of them are realistic. There are cases degenerated, where
one or both source components are very weak, but visible due to very small
impact parameters and very large amplifications. The Einstein times for
such models are also unrealistically long. Since double sources modeling
are not our primary goal, we do not further pursue this topic.

Our sample of binary lenses consists now of 78 events of Papers~I--IV, and
the present work, some of them with multiple models, plus 6 published
OGLE-III planetary events, which we also include.  Using the sample we may
study the distributions of various binary lens parameters. In Fig.1 we show
the distribution of the mass ratio.  In the left panel the observed
distributions of planetary and binary events are shown on the same
plot. The two kinds of events are not overlapping: the binary stars with
mass ratio $q<0.03$ are not present in our sample.
\begin{figure}[htb]
\centerline{\includegraphics[height=61.0mm,width=60.0mm]{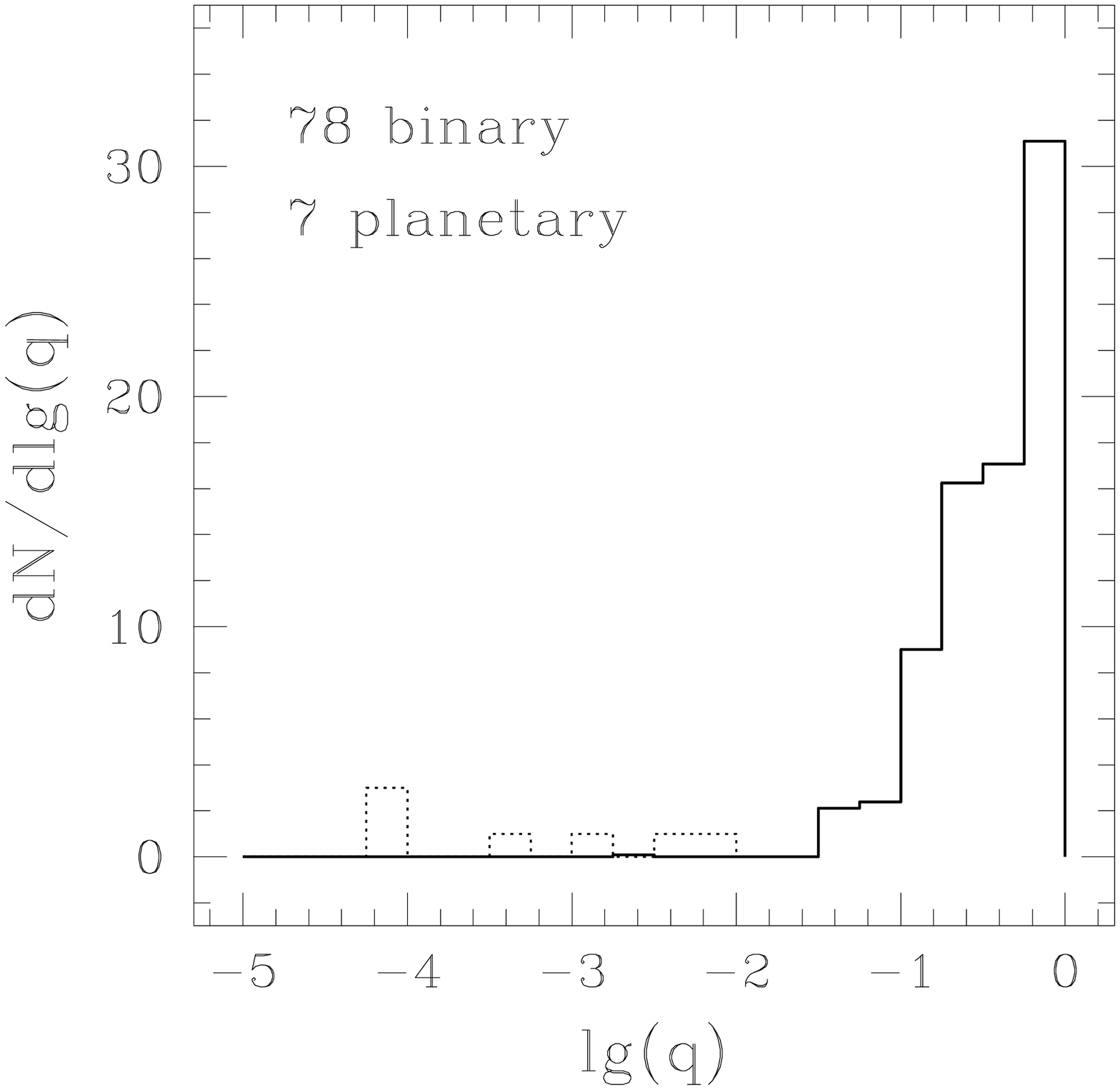}\hfill
\includegraphics[height=61.0mm,width=60.0mm]{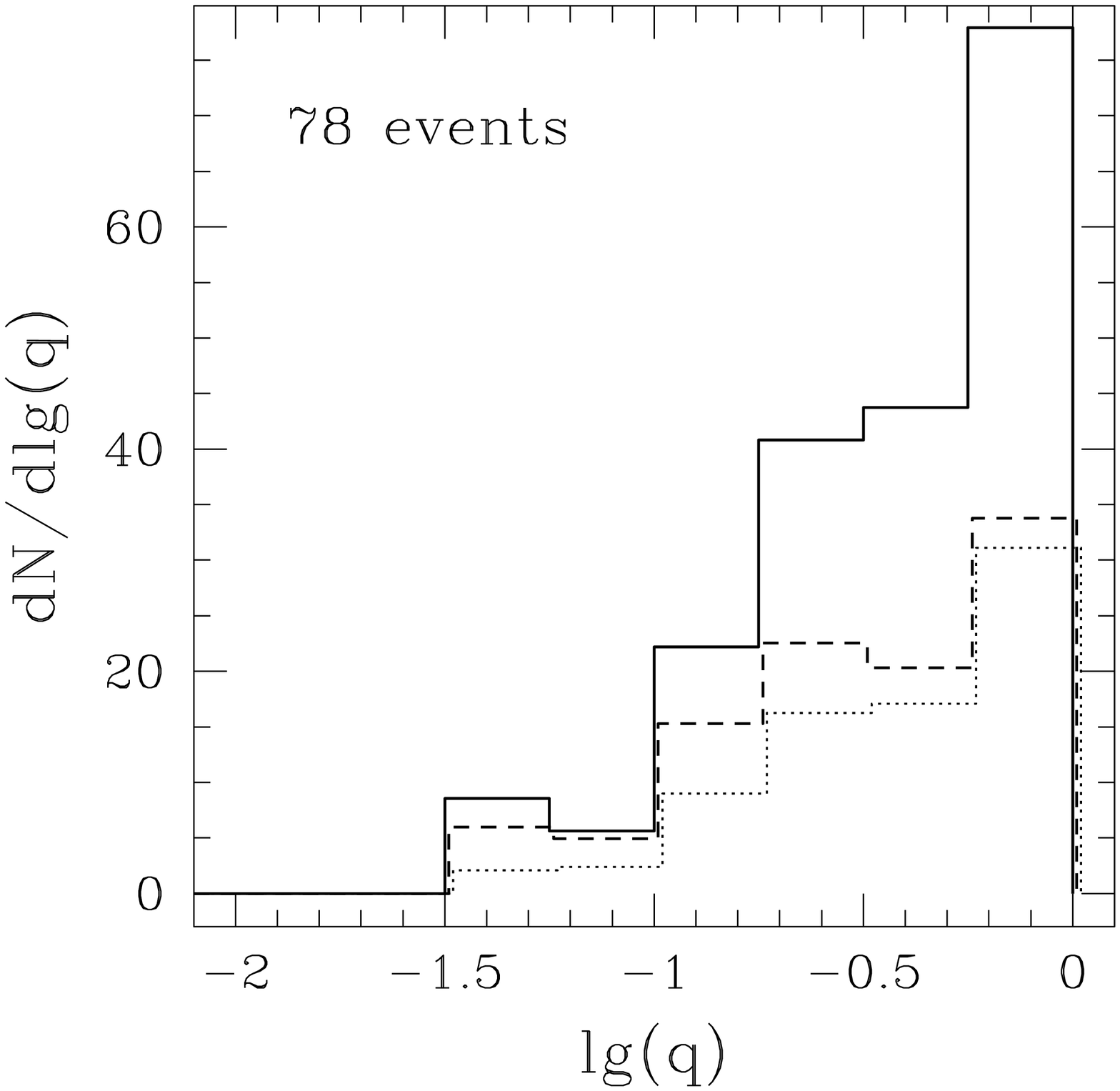}}
\FigCap{The histograms for the mass ratios. In the {\it left panel} we show
the distribution of binary lens ratios (solid line) found in Papers~I--IV
and in this work with superimposed data for OGLE-III planetary events
(dotted) published elsewhere. In the {\it right panel} we compare the raw
distribution for binaries (dotted) with the distribution corrected for
selection effects with the help of method I (dashed) and II
(solid). Corrected histograms are multiplied by artificial factors to avoid
crossings of the plots.}
\end{figure}

The observed distribution of the mass ratio (or any other binary lens
parameter) depends on its intrinsic distribution and on the observational
selection. The contribution of the binary lenses to the rate of
microlensing events depends on many parameters. Modifying formulae
applicable to single lens (Griest 1991, Kiraga and Paczyński 1994, Wood
and Mao 2005) and neglecting the part describing the source, one has for a
binary lens:
$$\Delta\Gamma\propto n(m,q,d,R_{\rm OL})S(q,d)~r_E~f(\vv)~\vv\eqno(5)$$
where $n$ is the number density of binary systems of total mass $m$, mass
ratio $q$, separation $d$, and the distance from the observer
$R_{\rm OL}$; $r_{\rm E}$ is the Einstein radius, $\vv$ is the
relative source--lens velocity component perpendicular to the line of
sight, and $f(\vv)$ is its distribution function. $S(q,d)$ is the angle
averaged width of the caustic, expressed in Einstein units, which gives the
relative probability of a caustic crossing event caused by a binary of
given caustic size and topology. The trajectory of a source has to pass
sufficiently close to a caustic to cause a non crossing event
distinguishable from point lens microlensing, so the relative probability
is roughly given by $S(q,d)$ also in this case. In the case of planetary
events, where the duration of the caustic region crossing plays a role in
detection efficiency, the length of the source trajectory inside caustic
may also be a factor worth consideration making the rate of the detection
proportional to the caustic surface area in the simplest approach, but
according to Sumi \etal, (2010) the dependence on the event duration is
much weaker. We neglect the possible dependence of the detection efficiency
on the duration of the caustic region crossing / approach in the case of
stellar binary lenses.

The full analysis of the {\it a priori} probability of detecting a binary
event, based on stellar mass and velocity distributions goes beyond 
the scope of this paper. We limit our considerations to the dependence of
detection probability on the mass ratio and the binary separation. 
For everything else equal, the observed and intrinsic distributions
of these parameters are related by:
$$N_{\rm obs}(\lg{q},\lg{d})\propto f_q(\lg{q})~f_d(\lg{d})~S(q,d)\eqno(6)$$
where $N_{\rm obs}$ gives the number of observed events in equal
logarithmic bins of $q$ and $d$, and $f_q$, $f_d$ are the intrinsic
distribution functions of these parameters. We have assumed that the
intrinsic mass ratio and separation distributions are independent. For any
assumed separation distribution one can average both sides of the equation
over $\lg{d}$ obtaining (method~I):
$$f_q(\lg{q})\propto\frac{\langle N_{\rm obs}(\lg{q},\lg{d})\rangle}{\langle f_d(\lg{d})~S(q,d)\rangle}\eqno(7)$$

Another method (II) of finding mass ratio distribution reads:
$$f_q(\lg{q})\propto\langle\frac{N_{\rm obs}(\lg{q},\lg{d})}{f_d(\lg{d})~S(q,d)}\rangle>.\eqno(8)$$
In the right panel of Fig.~1 we show the observed and corrected for
detection efficiency distribution of the mass ratios for stellar binary
lenses. The corrections by methods I and II have been made assuming that
the separation distribution is uniform in logarithm ($f_d(\lg{d})\propto
{\rm const}$). All three histograms show that the number of binary systems
per logarithmic interval of the mass ratio increases with $q$.

We also show the observed distributions of the blending factor ($f=
F_s/F_0$) and the distribution of the Einstein times ($t_{\rm E}$) among
the binary lens models belonging to our sample.

\begin{figure}[htb]
\centerline{
\includegraphics[height=61.0mm,width=60.0mm]{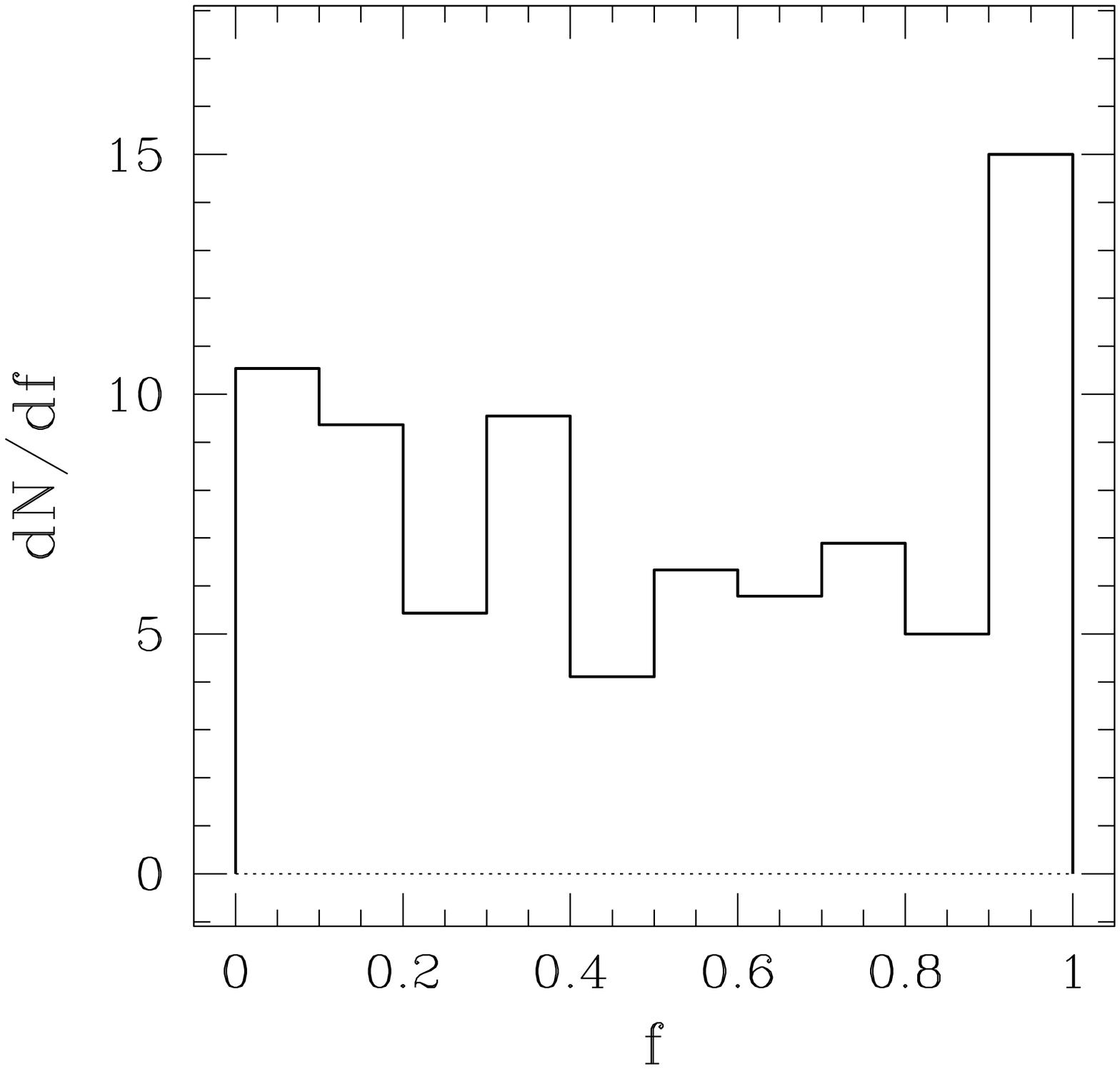}\hfill
\includegraphics[height=61.0mm,width=60.0mm]{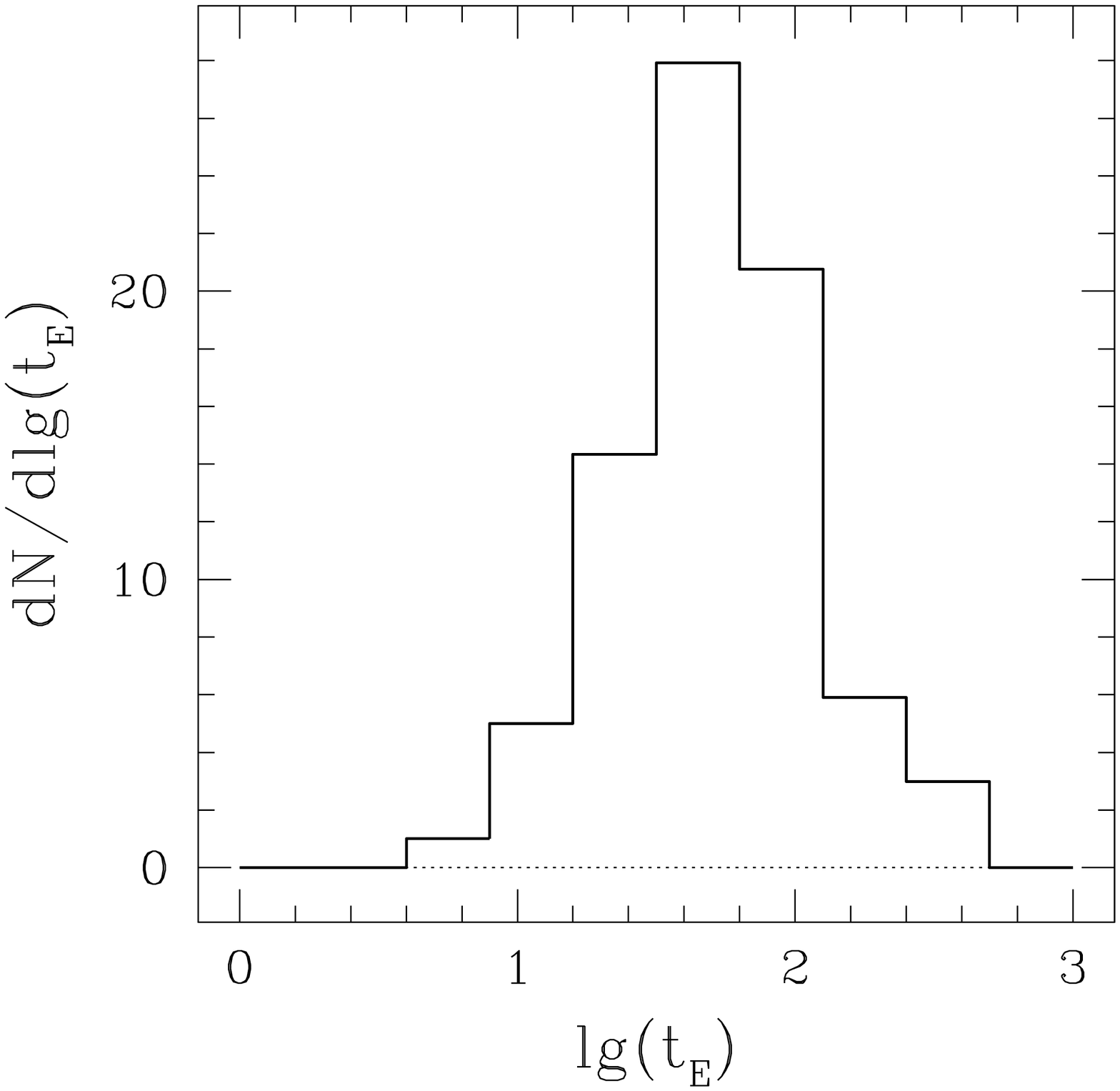}}
\FigCap{The histograms for the blending parameter ({\it left}) and the
logarithm of the Einstein time. The plots are not corrected for any
selection effects.}
\end{figure}

\Section{Discussion}
Our sample of OGLE-II--OGLE-III stellar binary lens events contains now 78
cases and we present the distribution of the binary lens mass ratio in
Fig.~1. The planetary and stellar binary lenses are clearly separate
populations of objects, which do not overlap on our plot. This is not
surprising, since the formation scenarios are different for them. It is
also impossible to draw any conclusions about the relative rate of
planetary and stellar binary events, since the former attract much more
attention from astronomers and their discoveries with the microlensing
methods result from much larger observational and theoretical effort. 
Effectively the two samples are based on different databases. The planetary
mass function is presented by Sumi \etal (2010). We limit our analysis to
the stellar binaries.

Our sample of stellar binary lenses covers now the range $0.03<q<1$. The
observed distribution of number of events per logarithmic mass ratio
interval ($\dd N/\dd\lg{q}$) is an increasing function of $q$. The simple
methods of correction for the observational selection effects do not change
this conclusion: both raw and corrected histograms show the increase of the
number of objects with increasing mass ratio. This conclusion contradicts
our earlier view, that the distribution of mass ratios is uniform in
logarithm ($\dd N/\dd\lg{q}\propto{\rm const}$). According to Trimble
(1990) the uniform distribution is appropriate to spectroscopic binaries
with orbital periods between 10 and 1000~d, but other kinds of binaries can
have different forms of distribution.

Skowron \etal (2009) find, that the uniform distribution of mass ratios is
consistent with their small sample of binaries causing repeating
microlensing events. Simultaneously, under the assumptions of uniform
distributions of binaries in $\lg{q}$ and $\lg{d}$ they find that all stars
should be in binary systems (in contradiction with popular wisdom) to
explain the relatively high rate of repeating microlensing events
classified as caused by binary lenses. One of the ways to circumvent the
contradiction is to change the distribution of mass ratios allowing for
relatively higher number of $q\approx1$ systems, which in turn increases
the probability of detection of repeating events caused by binary lenses.

Jaroszyński and Skowron (2008) investigate the role of double sources in
explaining the rate of repeating events. Similarly as Skowron \etal (2009),
and using similar assumptions they find that all sources should be binary
systems. Since the luminosity ratio is some positive power of mass ratio
(the paper assumes $\propto q^{3.2}$), the sources with small $q$ do not
count: the weaker component is practically undetectable. Again, postulating
relatively higher number of similar mass systems, one gets relatively
higher probability of detection of a repeating event caused by a double
source, thus decreasing the required number of binaries among stars.

Our analysis of the mass ratio distribution is still preliminary. We treat
all events modeled as caused by binary lenses as equally important,
ignoring the differences in light curve coverage, fit quality, and a priori
probability of various physical lens parameters. These topics require
further study.

The blending parameter distribution presented in Fig.~2 agrees with the
result presented in Paper~II. It may also be compared with the distribution
obtained by Smith \etal (2007) for single microlensing events. In our
approach we treat models with blending parameter $f>1$ as unphysical, since
they require negative blended flux. Smith \etal (2007) allow larger values
of this parameter ($f\le2$), interpreting negative fluxes as possible
errors of photometric pipeline. In the range $0\le f\le1$ there is no
contradiction between our distribution and their findings.

\Acknow{The systematic investigation of binary gravitational lenses in OGLE
databases was suggested to us by late Professor Bohdan Paczyński, whose
ideas and enthusiasm were an invaluable help in our studies. We thank Shude
Mao for the permission of using his binary lens modeling software. This
work was supported in part by the Polish Ministry of Science and Higher
Education grant N203 008 32/0709 and by Space Exploration Research Fund
of The Ohio State University.}

%%%%\end{document}
\vfill
\eject

\section{Appendix}

\subsection{Binary lens models}

Below we present the plots for the 31 events for which 
the binary lens modeling has been applied. 
The events are ordered and named according to their position 
in the OGLE EWS database. 
Some of the events,
especially cases without apparent
caustic crossing, may have alternative double source models. In such
cases we show the comparison of the binary lens and double source fits
to the data in the next subsection. 
For events which have multiple, substantially different
models of similar quality, we present up to 3 of them.

Each case is illustrated with two panels. The most interesting part of
the source trajectory, the binary and its caustic structure are shown in
the left panel for the case considered. The labels give the $q$ and $d$
values. On the right the part of the best fit light curve 
is compared with observations. The labels give 
the rescaled $\chi_1^2$ /DOF values and Einstein times resulting from
fits.
Multiple models of the same event usually have similar model light
curves, but have different trajectories, different caustic structure and
different values of the Einstein time.
Below the light curves we show the differences between the observed and
modeled flux in units of rescaled errors. The dotted lines show the
rescaled $\pm 3\sigma$ band.

\vfill

\noindent\parbox{12.7cm}{
\leftline {\bf OGLE 2006-BLG-028} 

\includegraphics[height=62mm,width=63mm]{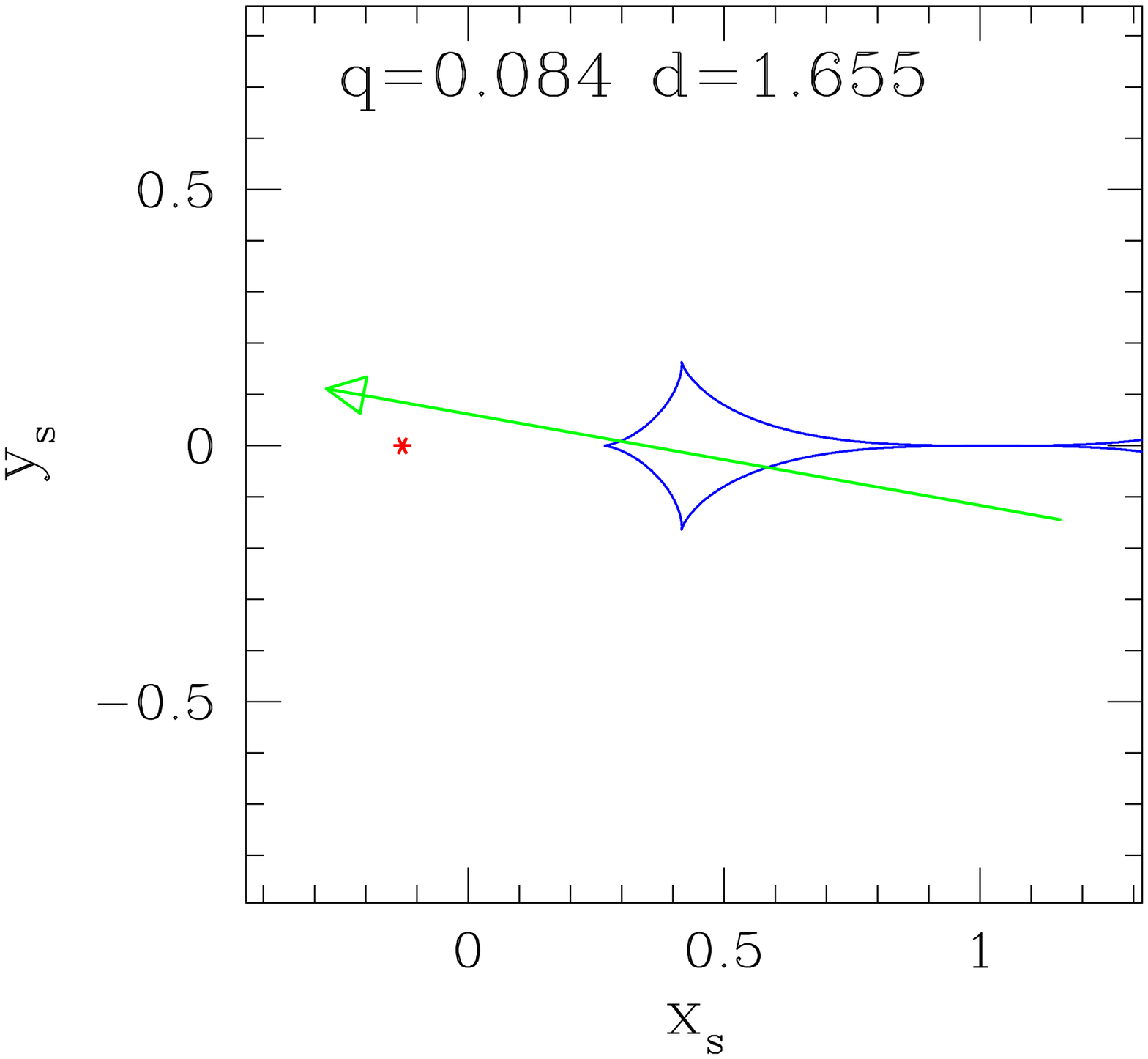} \hfill
\includegraphics[height=62mm,width=63mm]{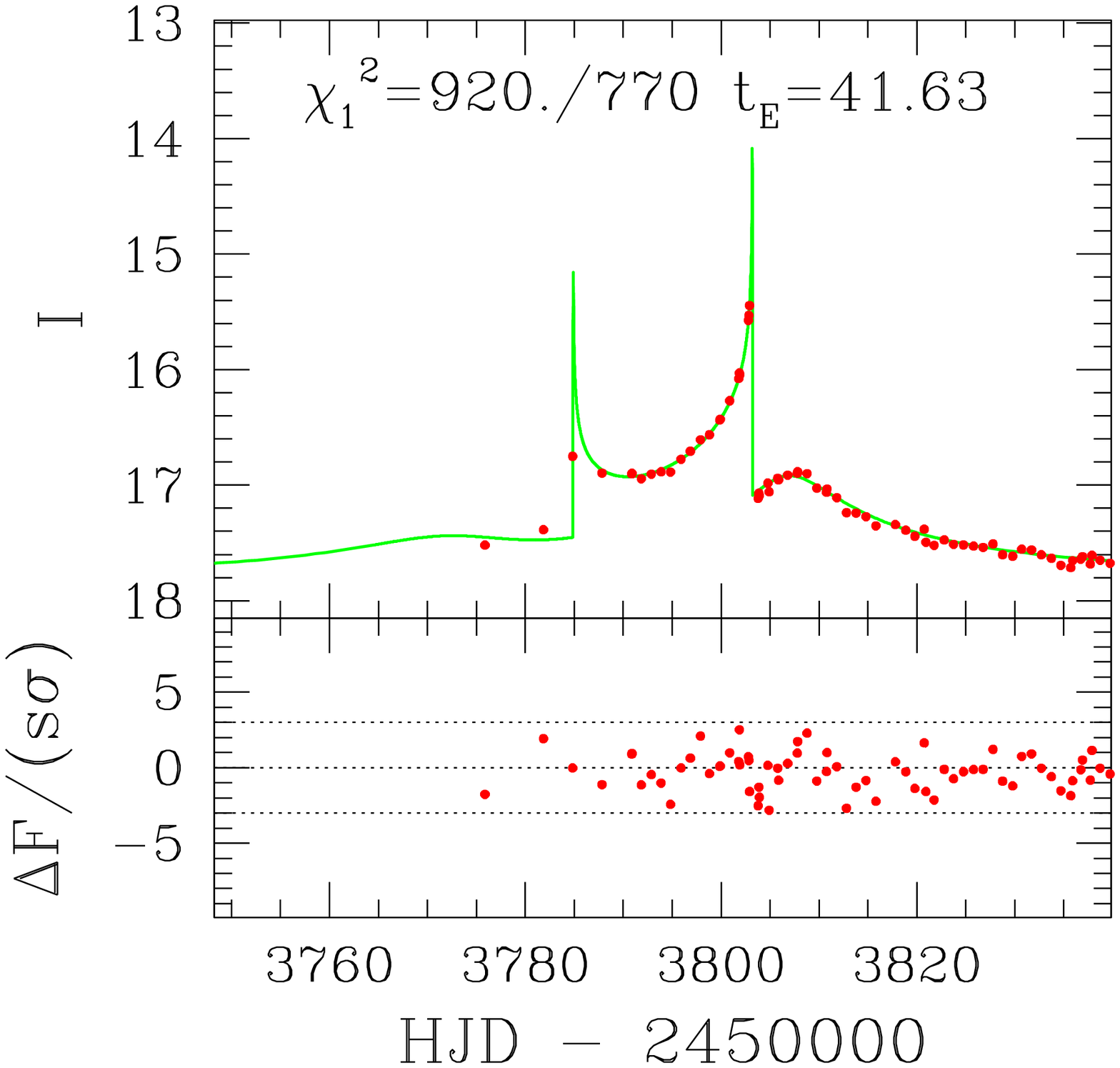}%

}

\noindent\parbox{12.7cm}{
\leftline {\bf OGLE 2006-BLG-038} 

\includegraphics[height=62mm,width=63mm]{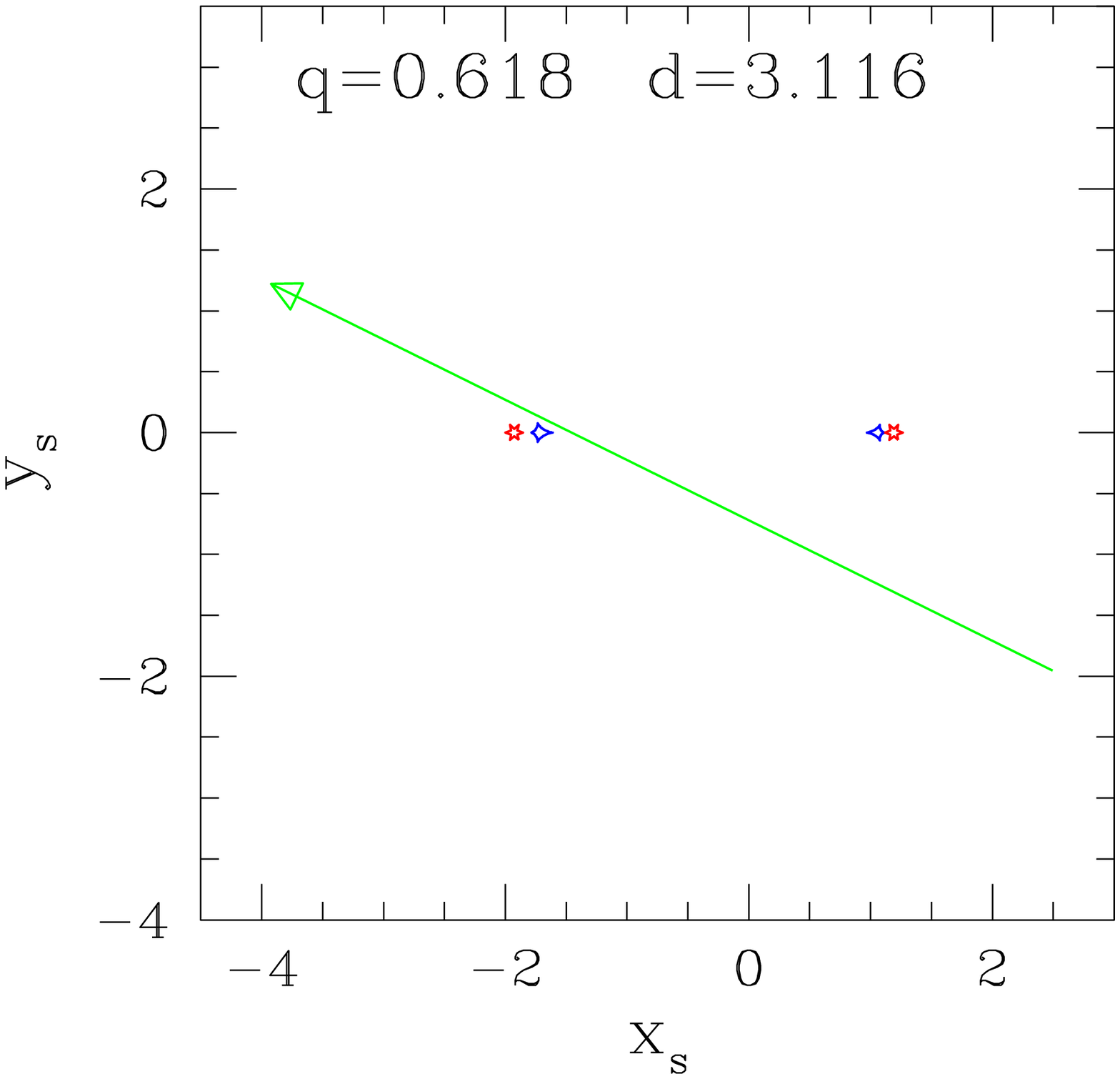} \hfill
\includegraphics[height=62mm,width=63mm]{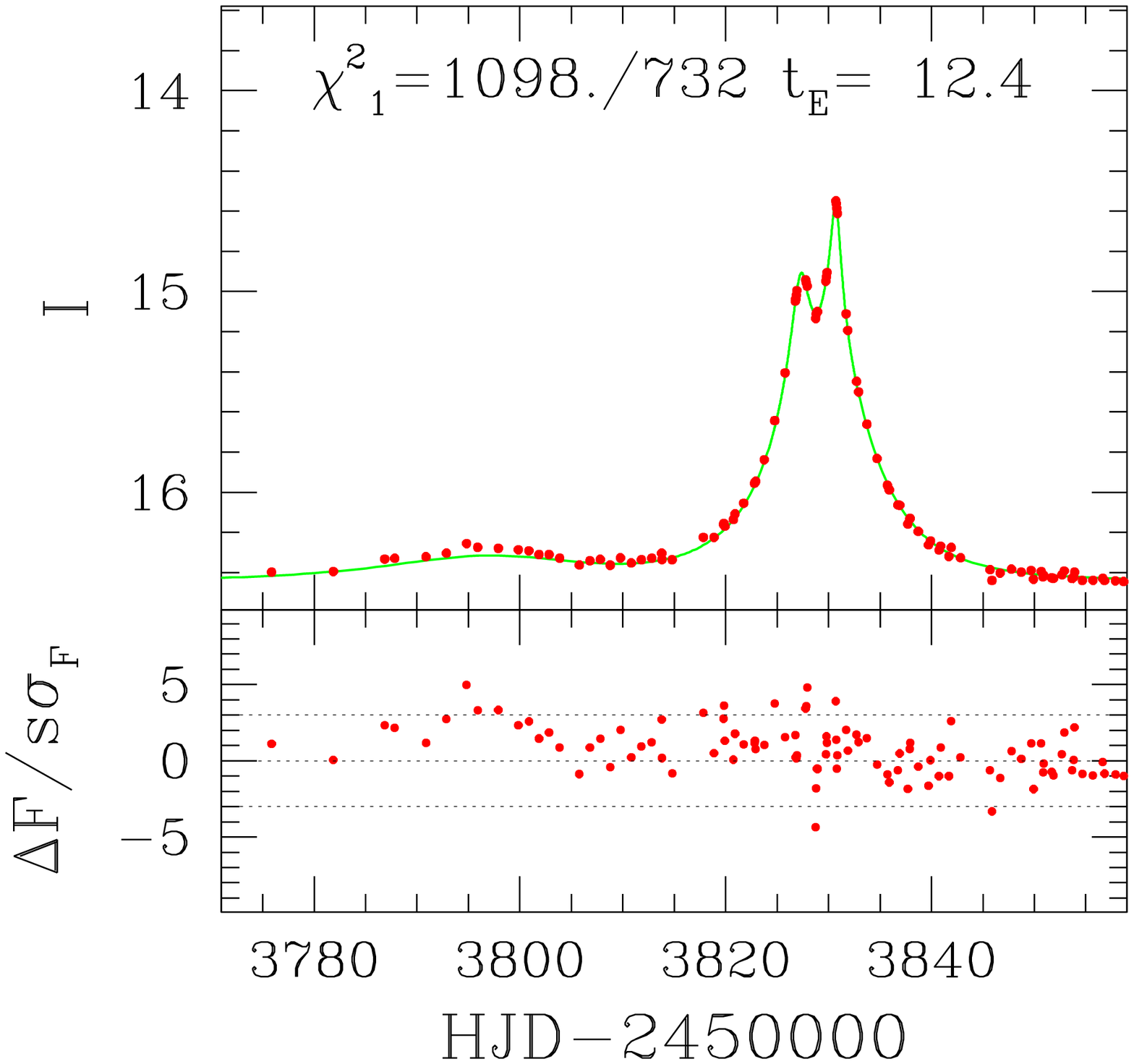}%

}

\noindent\parbox{12.7cm}{
\leftline {\bf OGLE 2006-BLG-076}

\includegraphics[height=62mm,width=63mm]{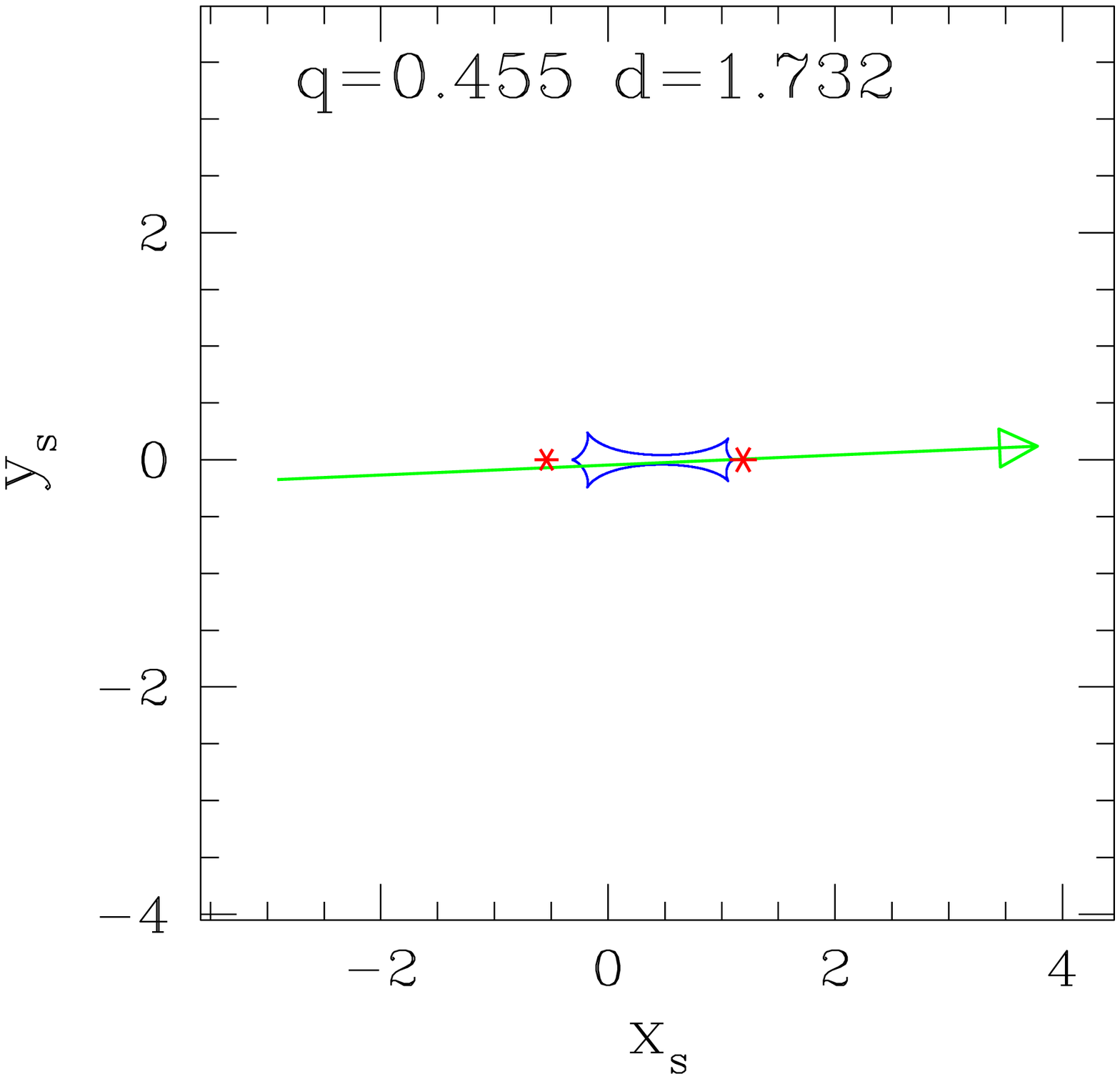} \hfill
\includegraphics[height=62mm,width=63mm]{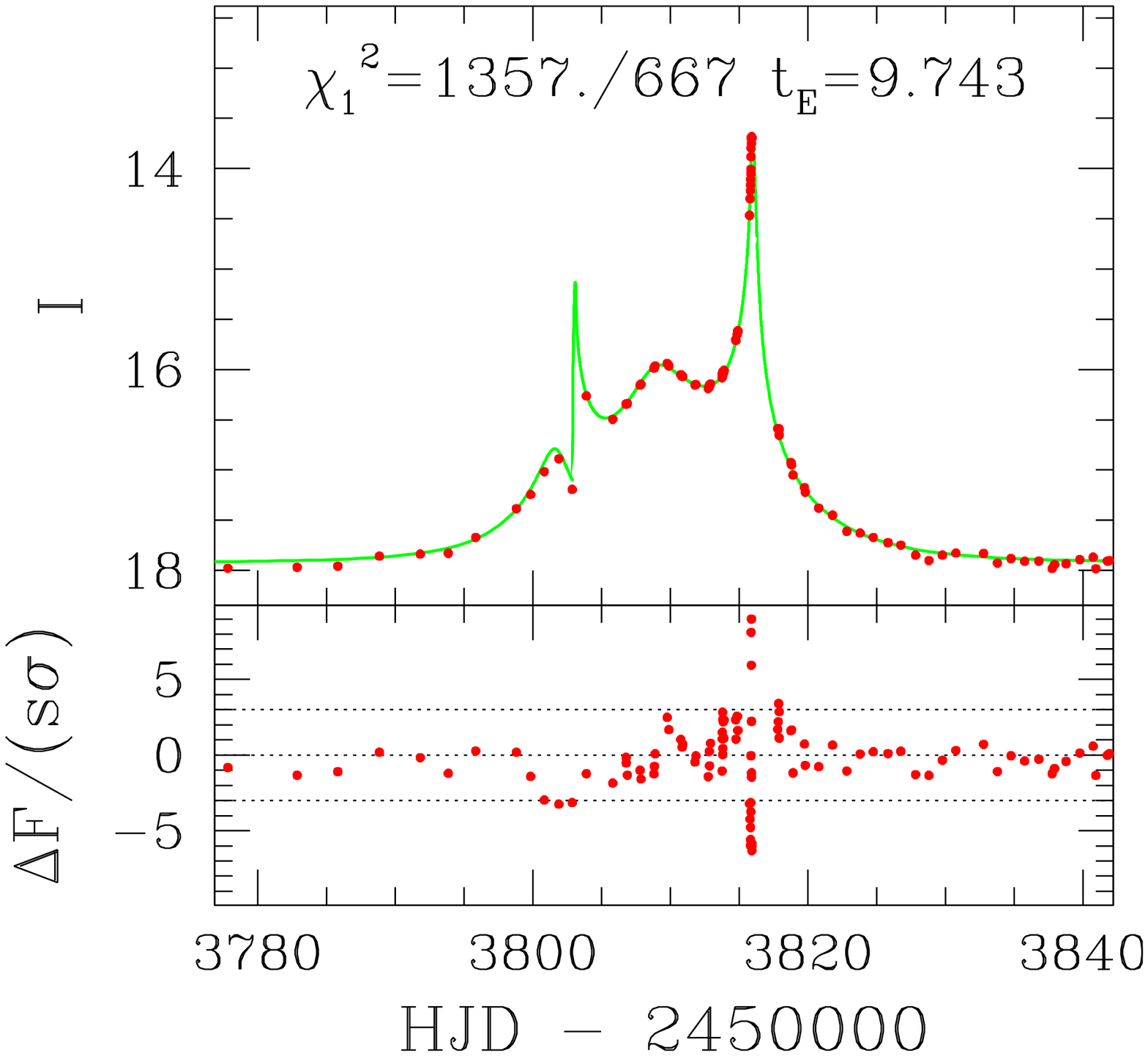}%

}

\noindent\parbox{12.75cm}{
\leftline {\bf OGLE 2006-BLG-119} 

\includegraphics[height=62mm]{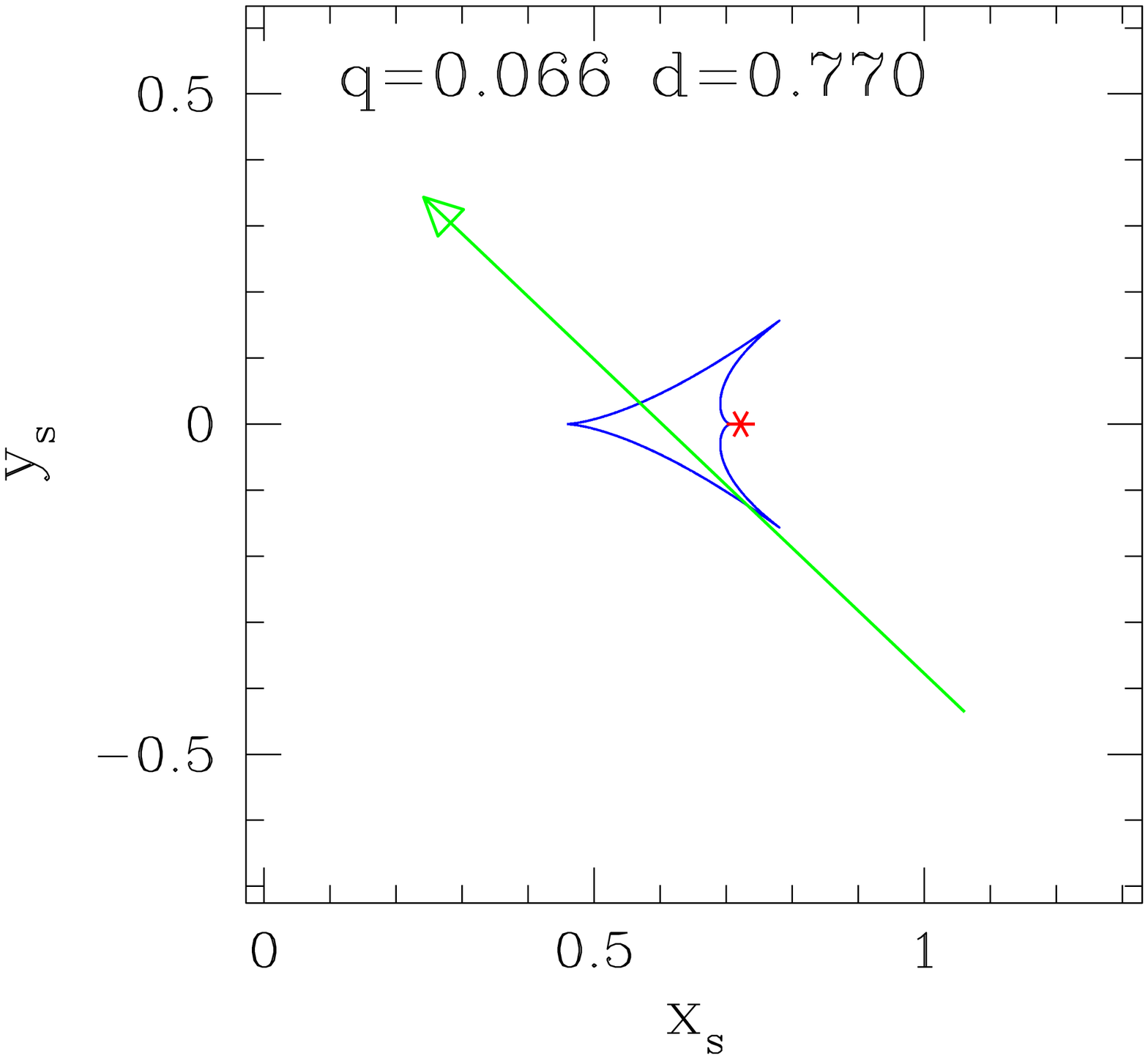} \hfill
\includegraphics[height=62mm]{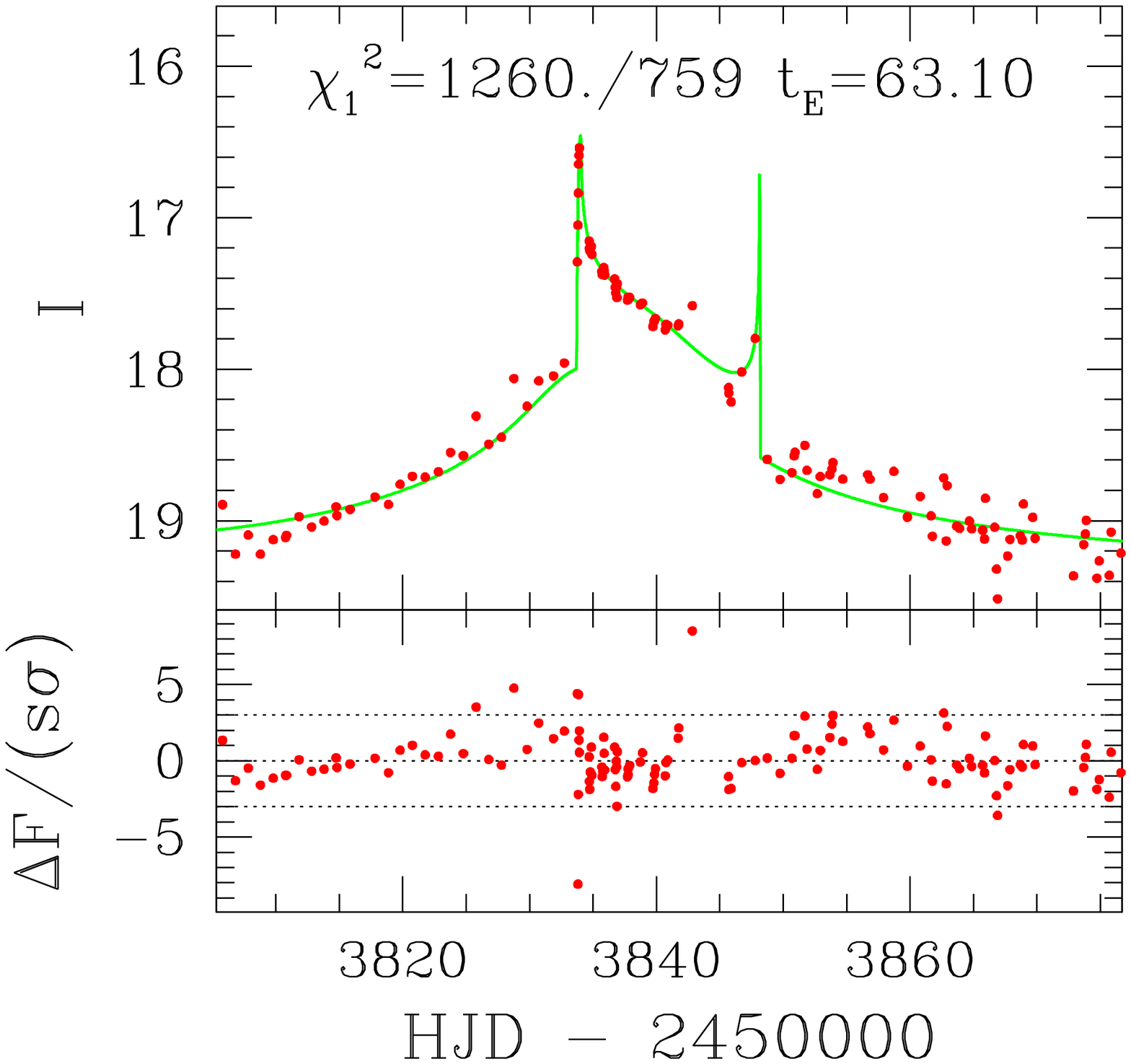}%

}

\noindent\parbox{12.75cm}{
\leftline {\bf OGLE 2006-BLG-215} 

\includegraphics[height=62mm,width=63mm]{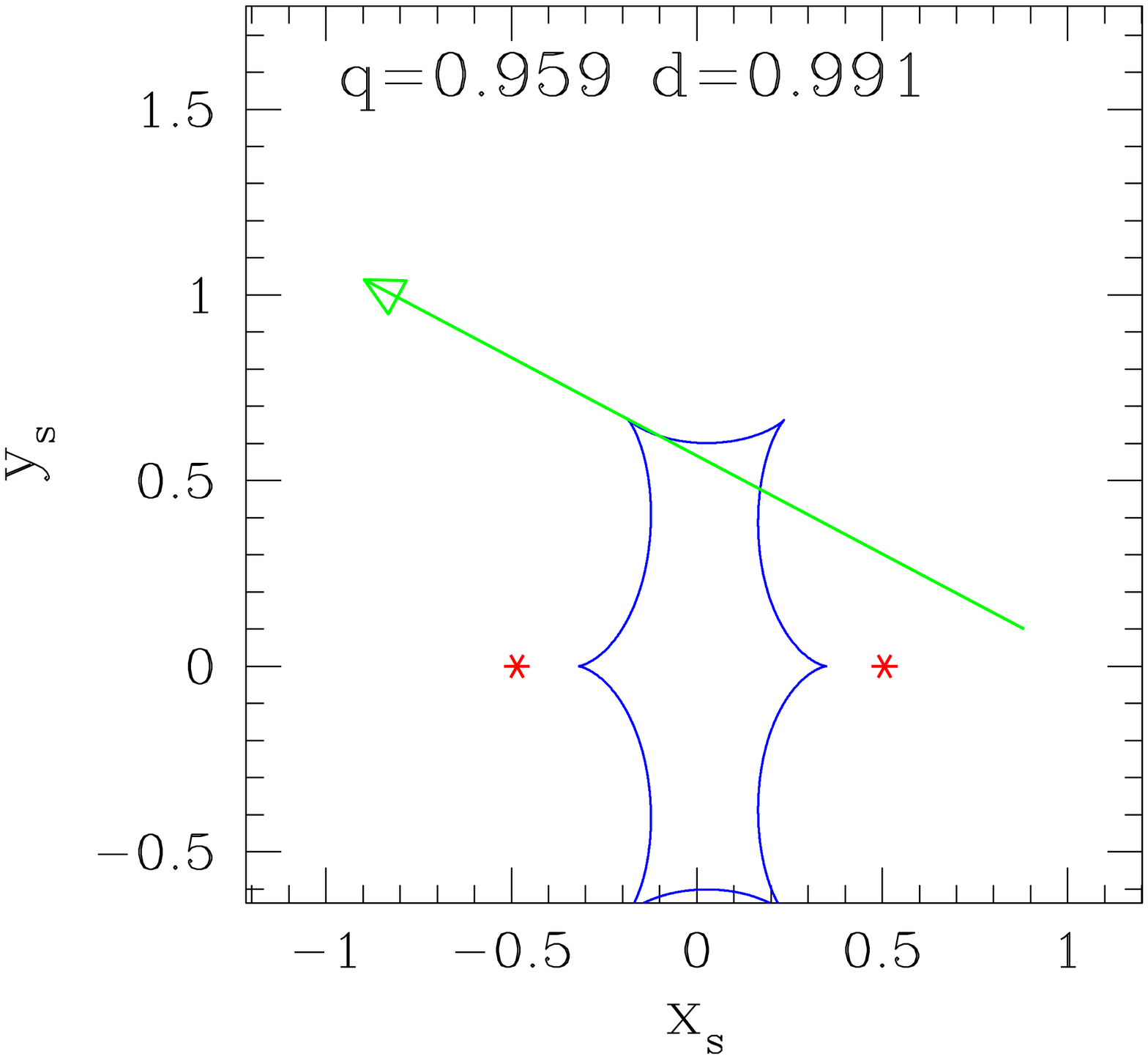} \hfill
\includegraphics[height=62mm,width=63mm]{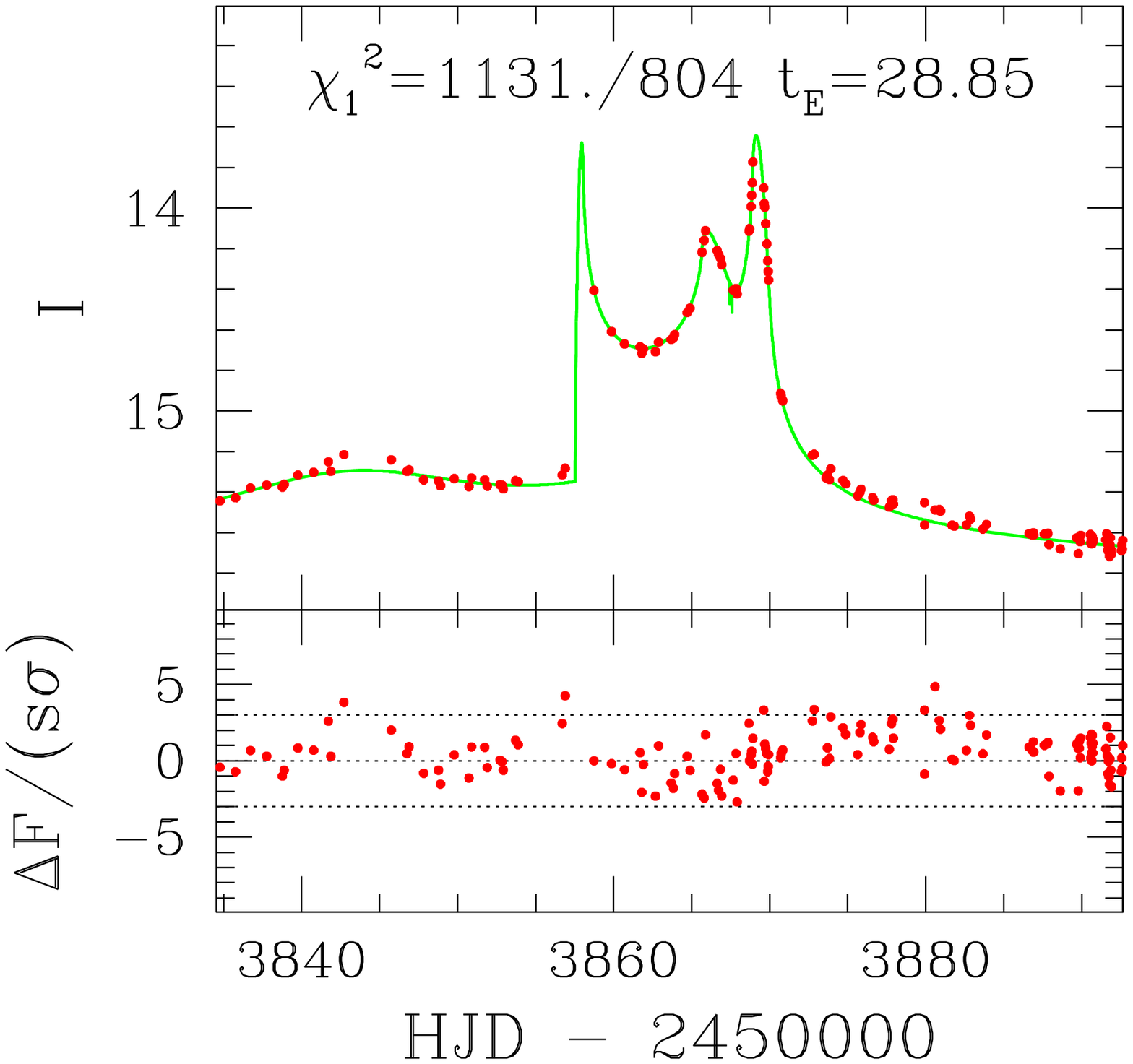}%

}

\noindent\parbox{12.75cm}{
\leftline {\bf OGLE 2006-BLG-277} 

 \includegraphics[height=62mm,width=63mm]{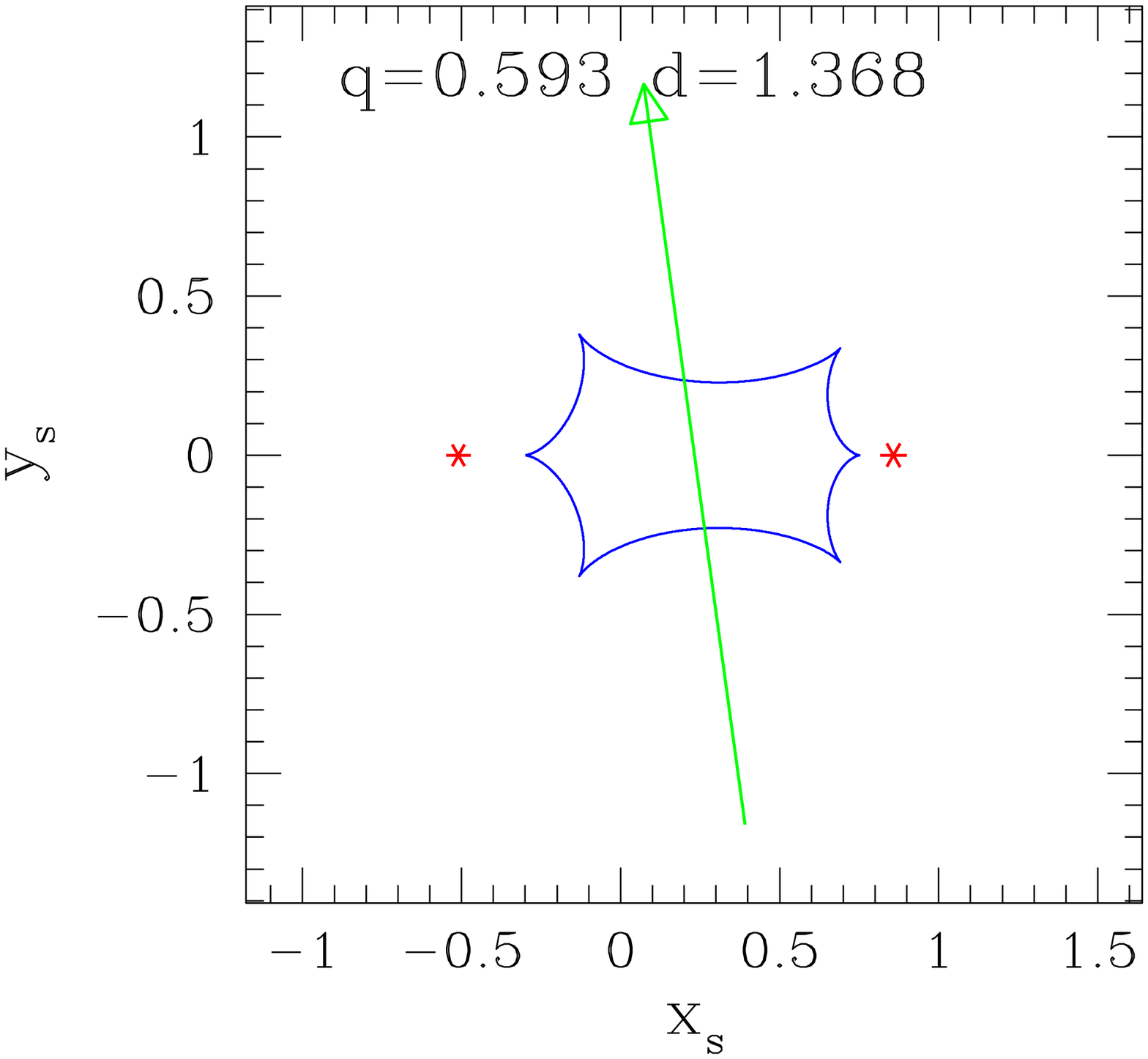} \hfill
 \includegraphics[height=62mm,width=63mm]{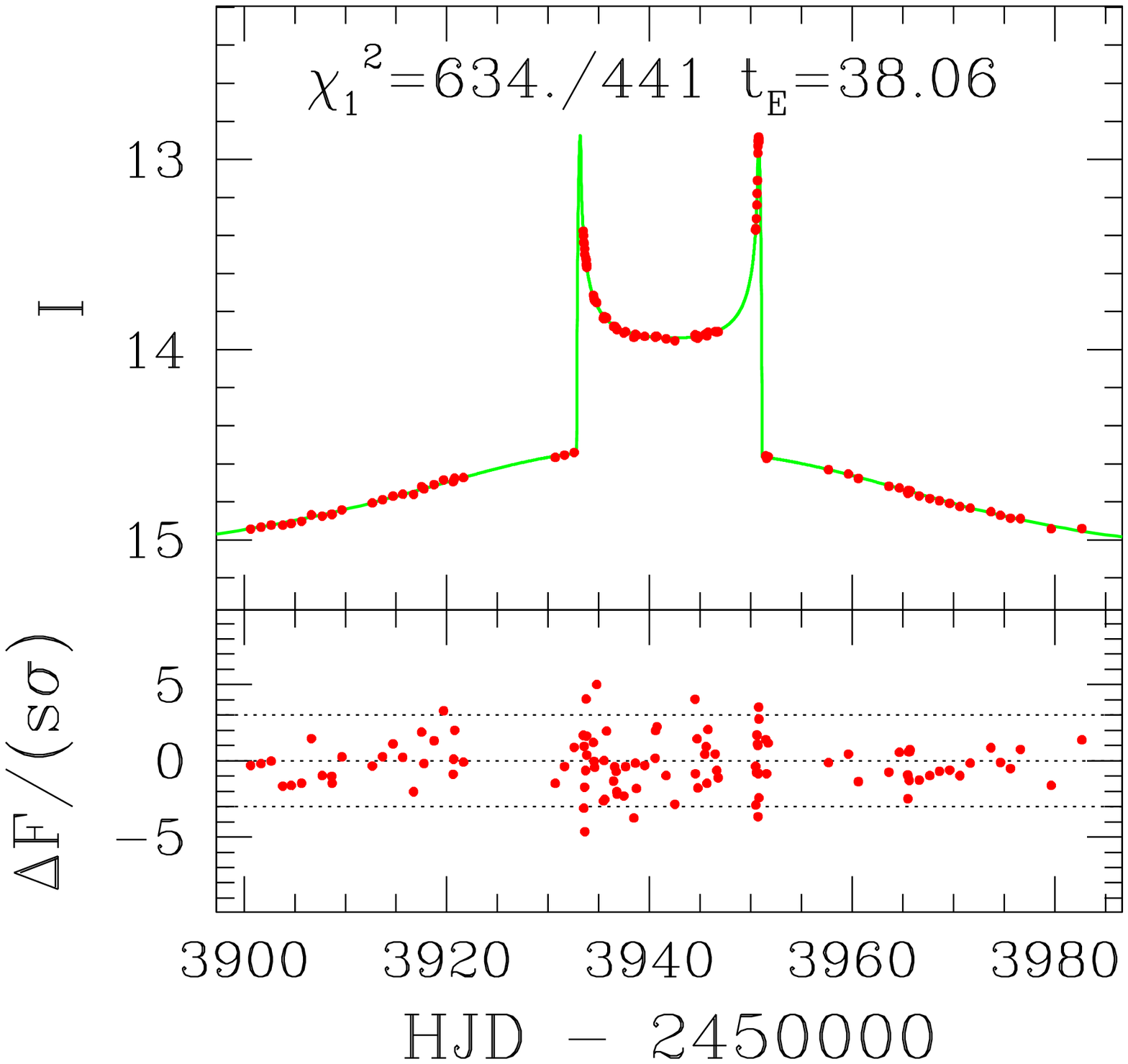}%

}

\noindent\parbox{12.75cm}{
\leftline {\bf OGLE 2006-BLG-284} 

 \includegraphics[height=62mm,width=63mm]{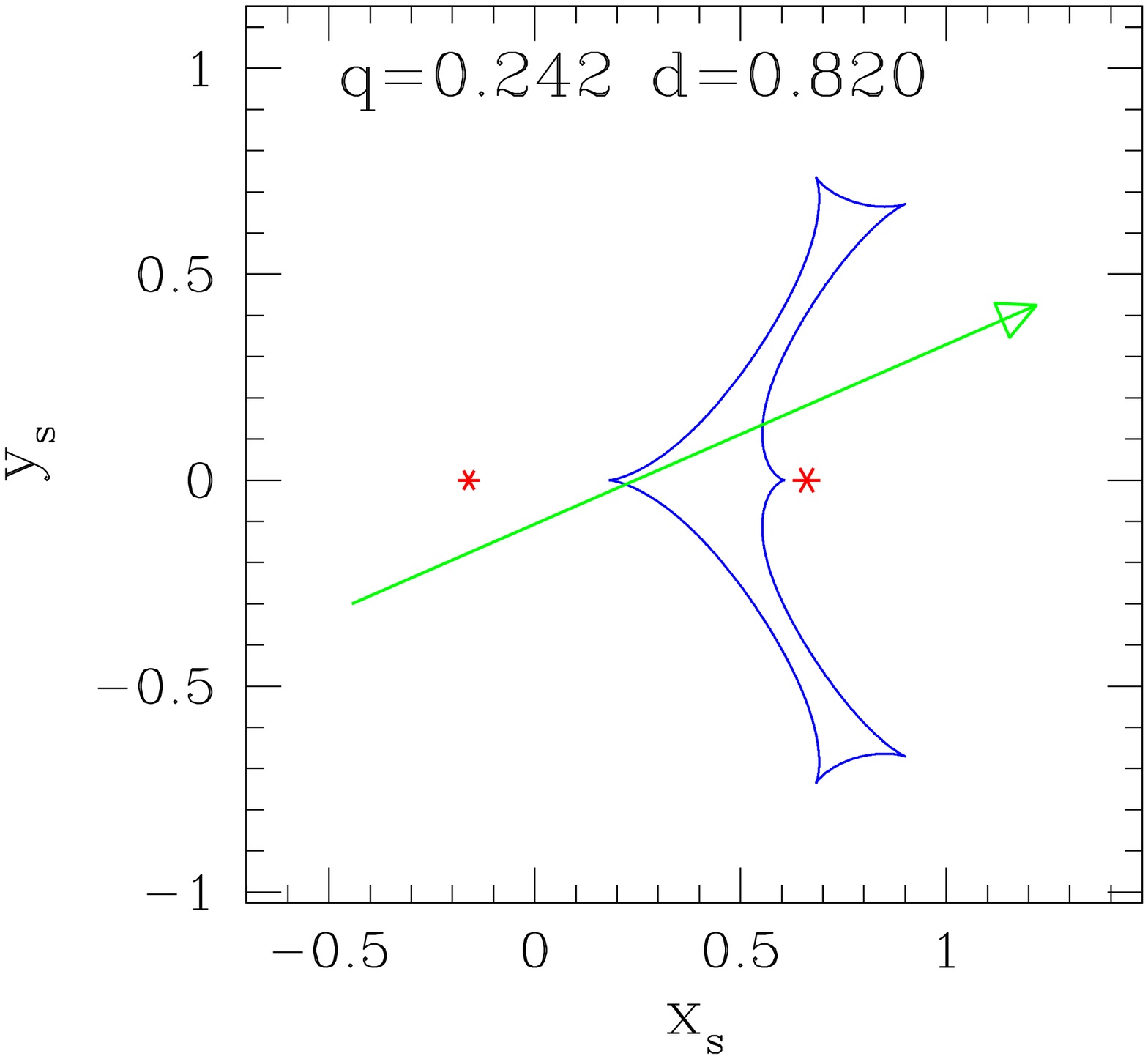} \hfill
 \includegraphics[height=62mm,width=63mm]{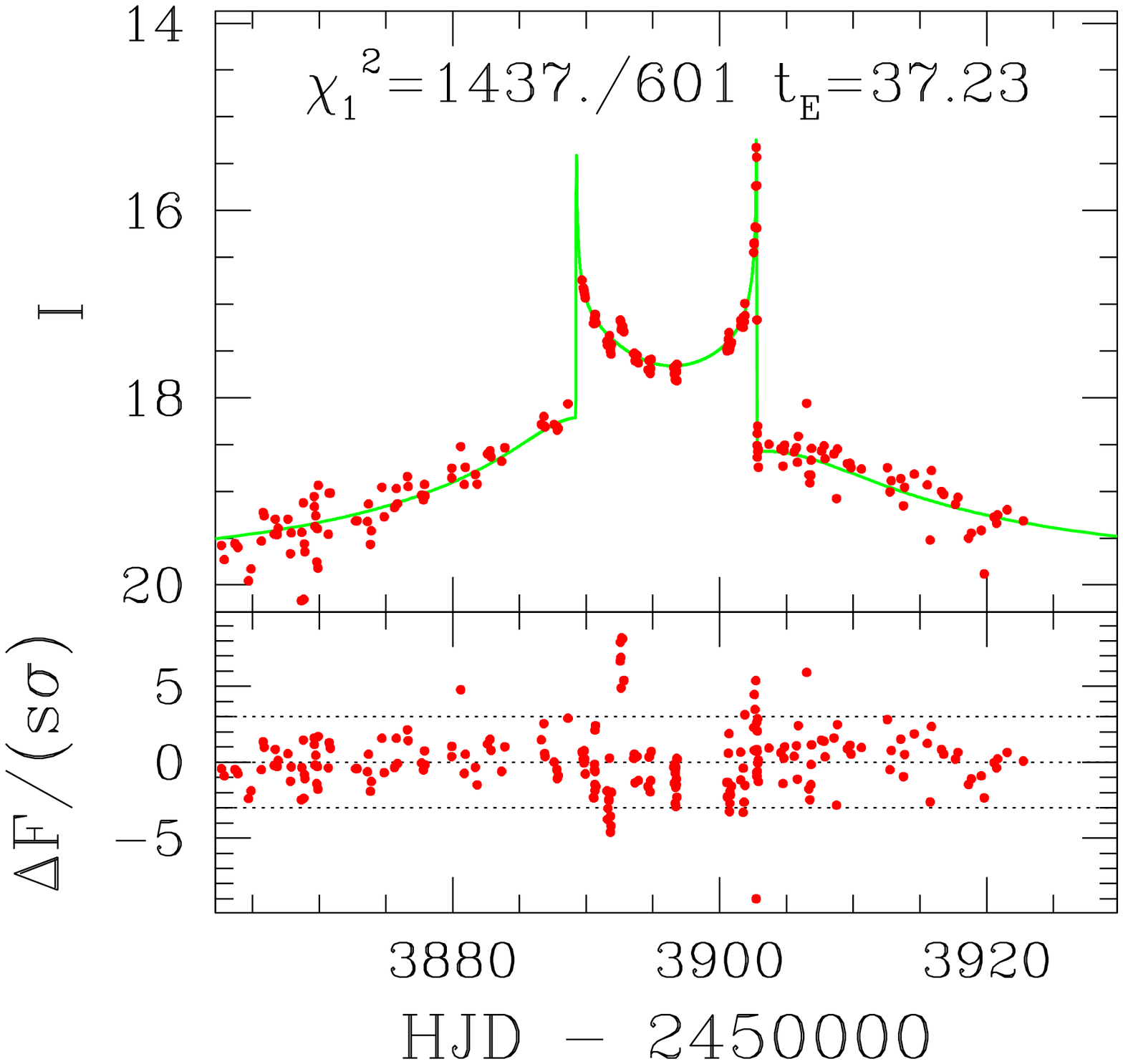}%

}

\noindent\parbox{12.75cm}{
\leftline {{\bf OGLE 2006-BLG-304} (1st model)}

\includegraphics[height=62mm,width=63mm]{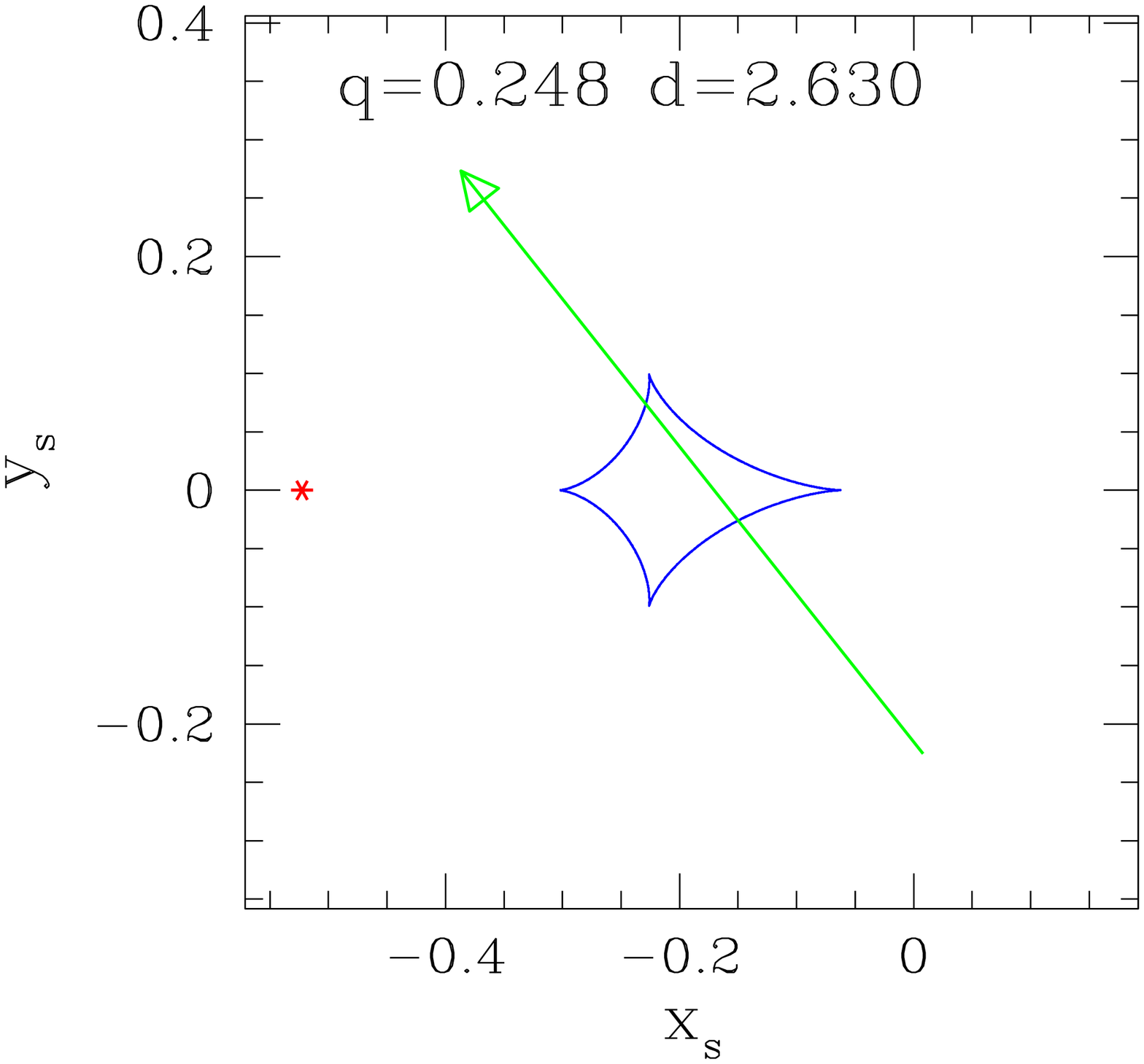} \hfill
\includegraphics[height=62mm,width=63mm]{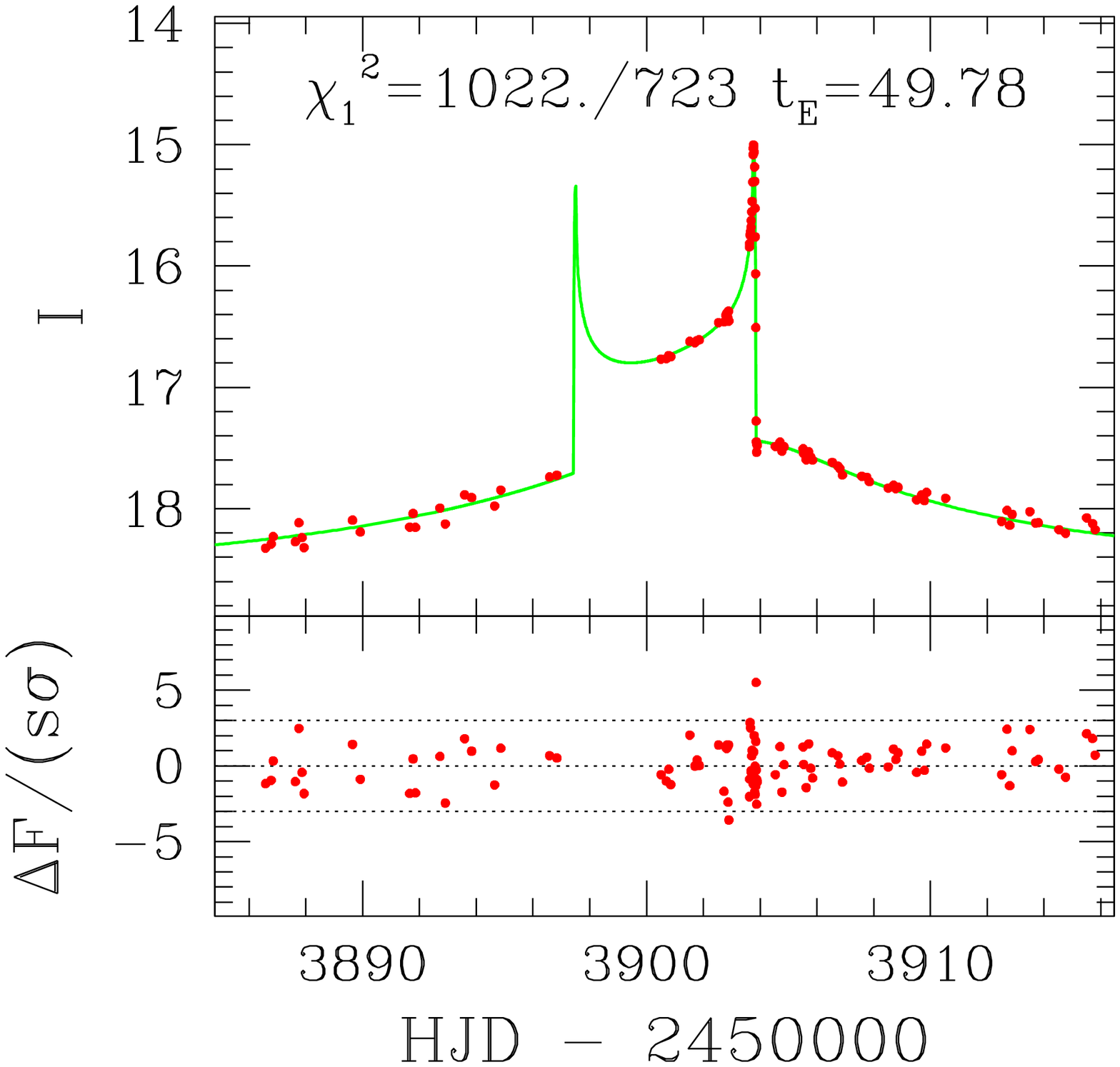}%

}

\noindent\parbox{12.75cm}{
\leftline {{\bf OGLE 2006-BLG-304} (2nd model)}

\includegraphics[height=62mm,width=63mm]{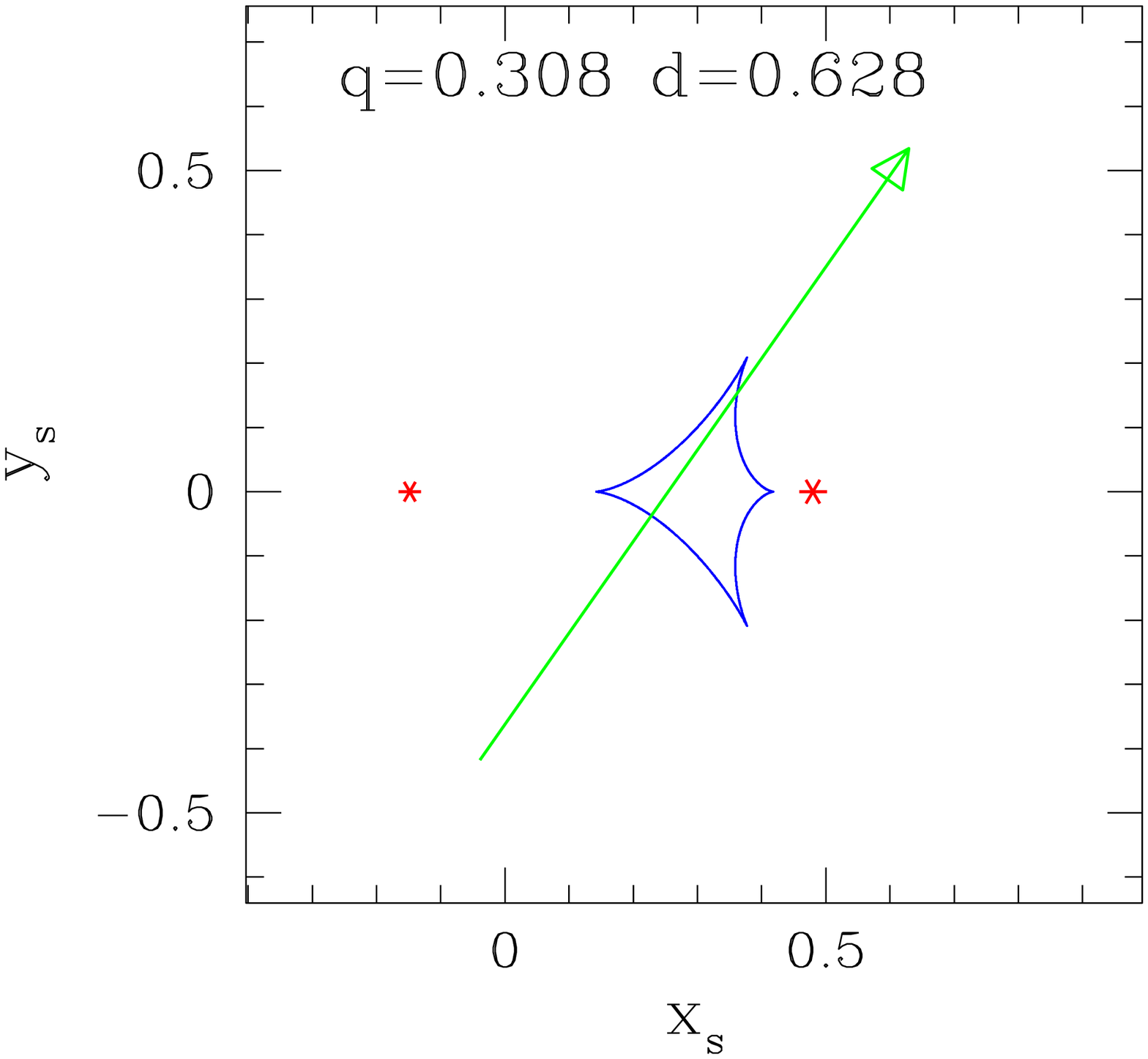} \hfill
\includegraphics[height=62mm,width=63mm]{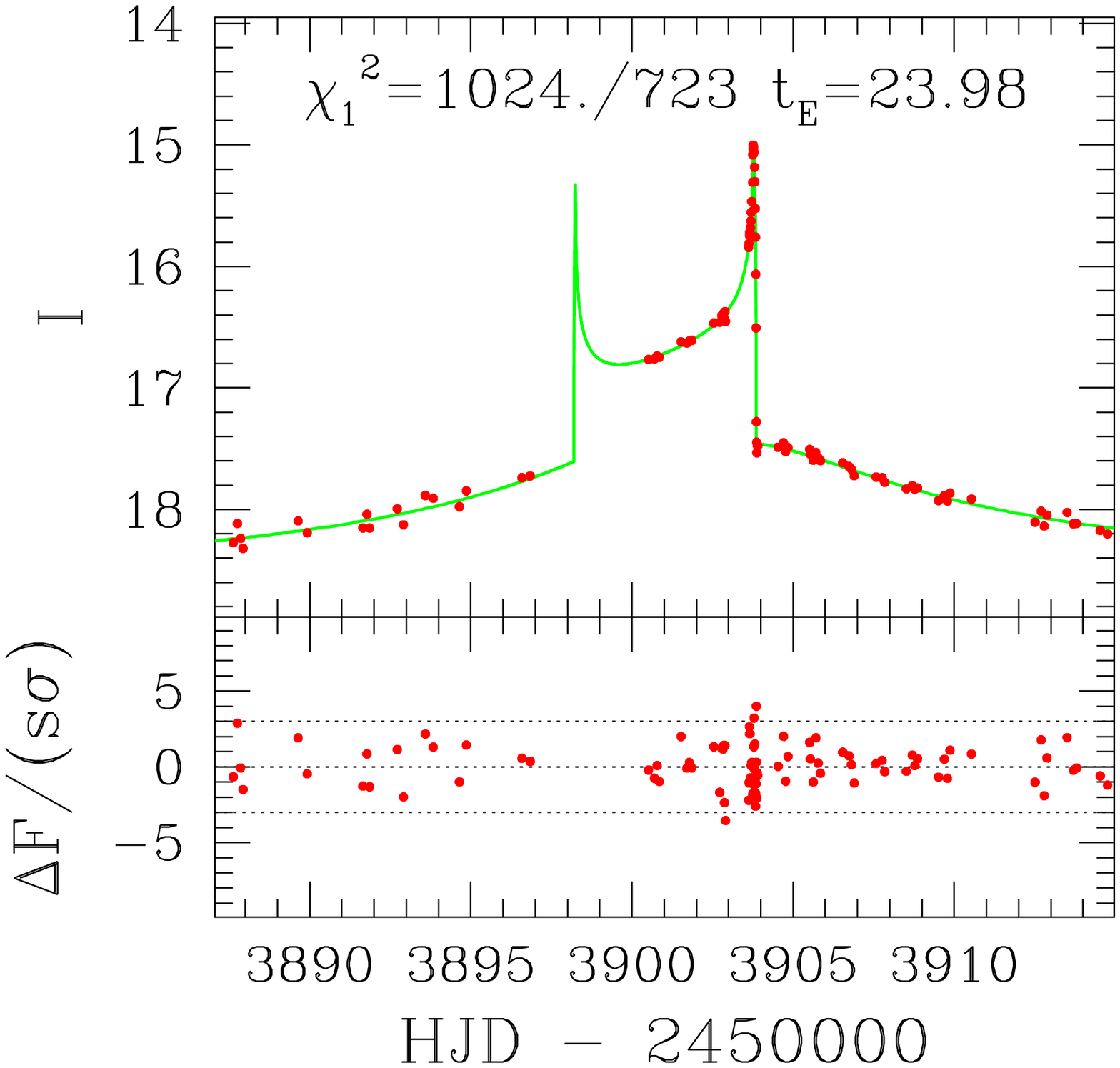}%

}

\noindent\parbox{12.75cm}{
\leftline {\bf OGLE 2006-BLG-335} 

\includegraphics[height=62mm,width=63mm]{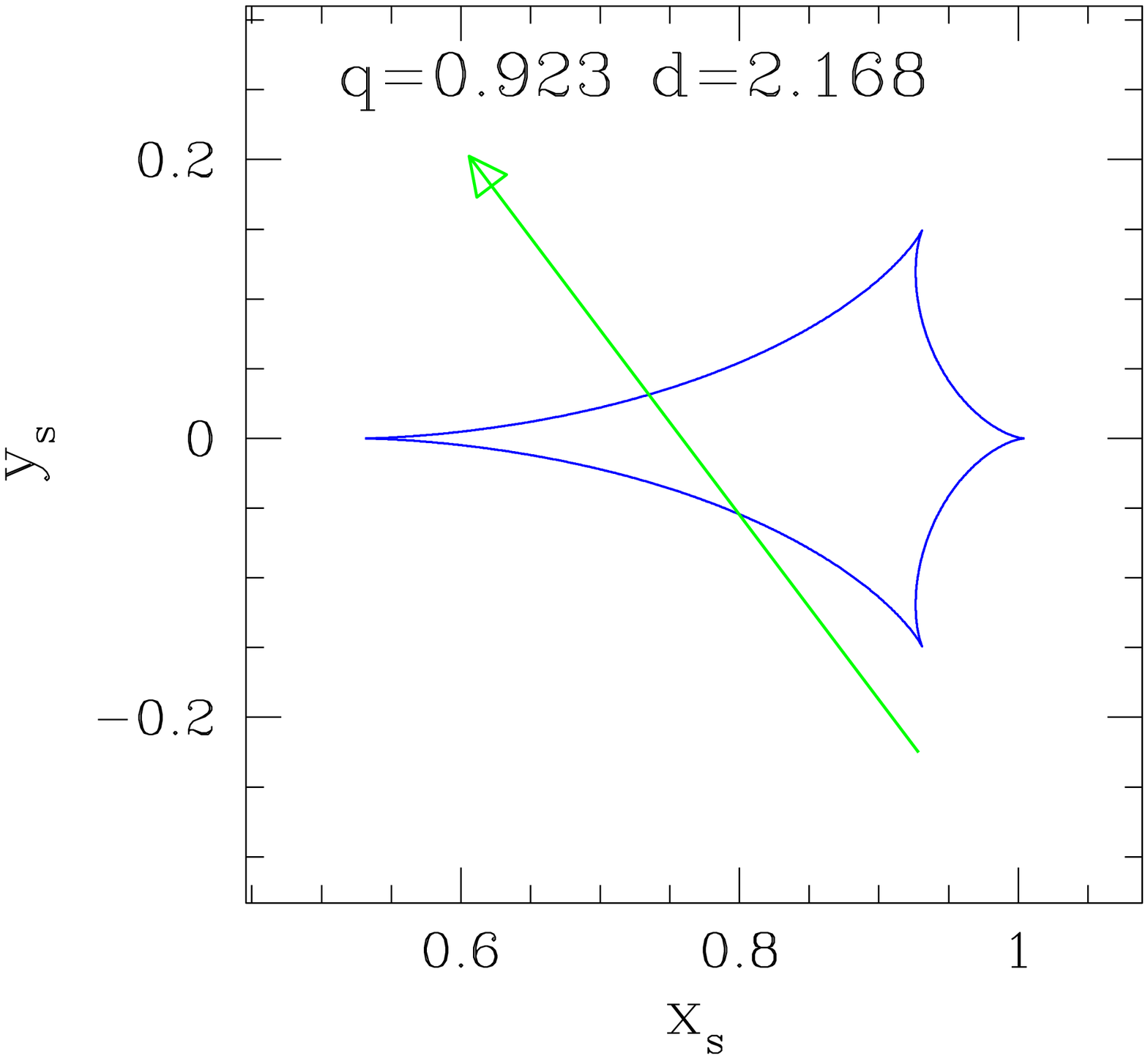} \hfill
\includegraphics[height=62mm,width=63mm]{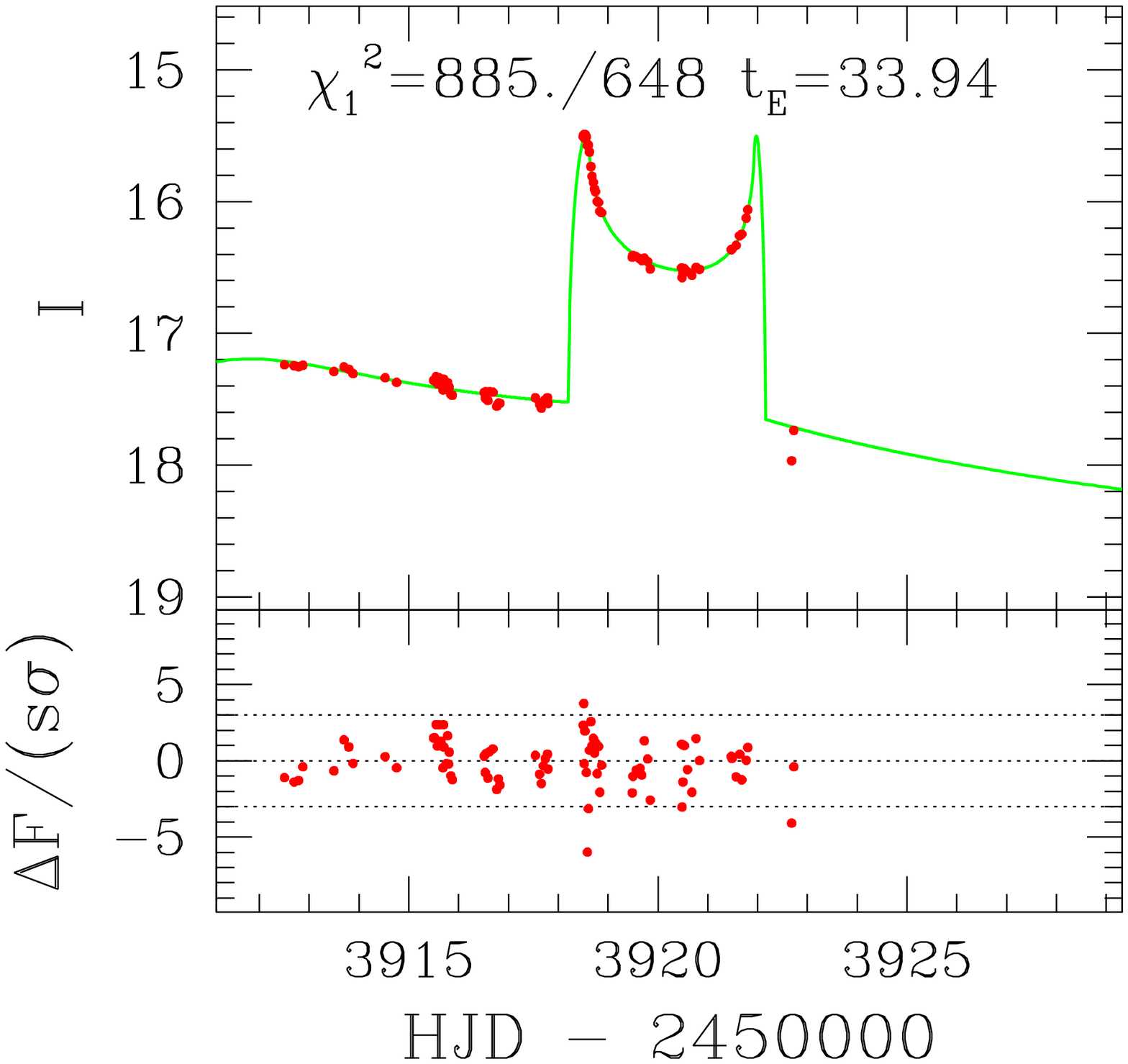}%

}

\noindent\parbox{12.75cm}{
\leftline {\bf OGLE 2006-BLG-375} 

\includegraphics[height=62mm,width=63mm]{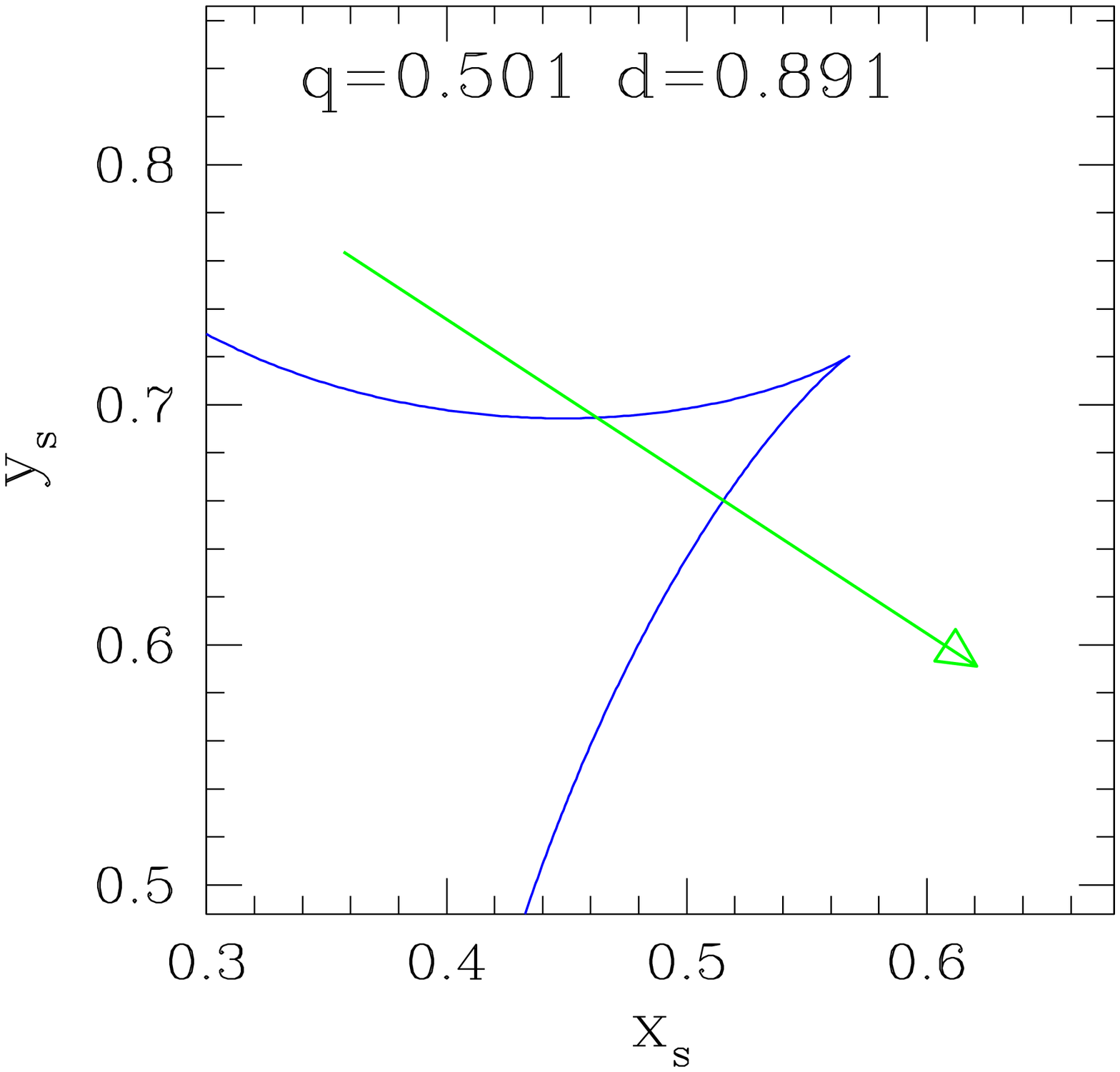} \hfill
\includegraphics[height=62mm,width=63mm]{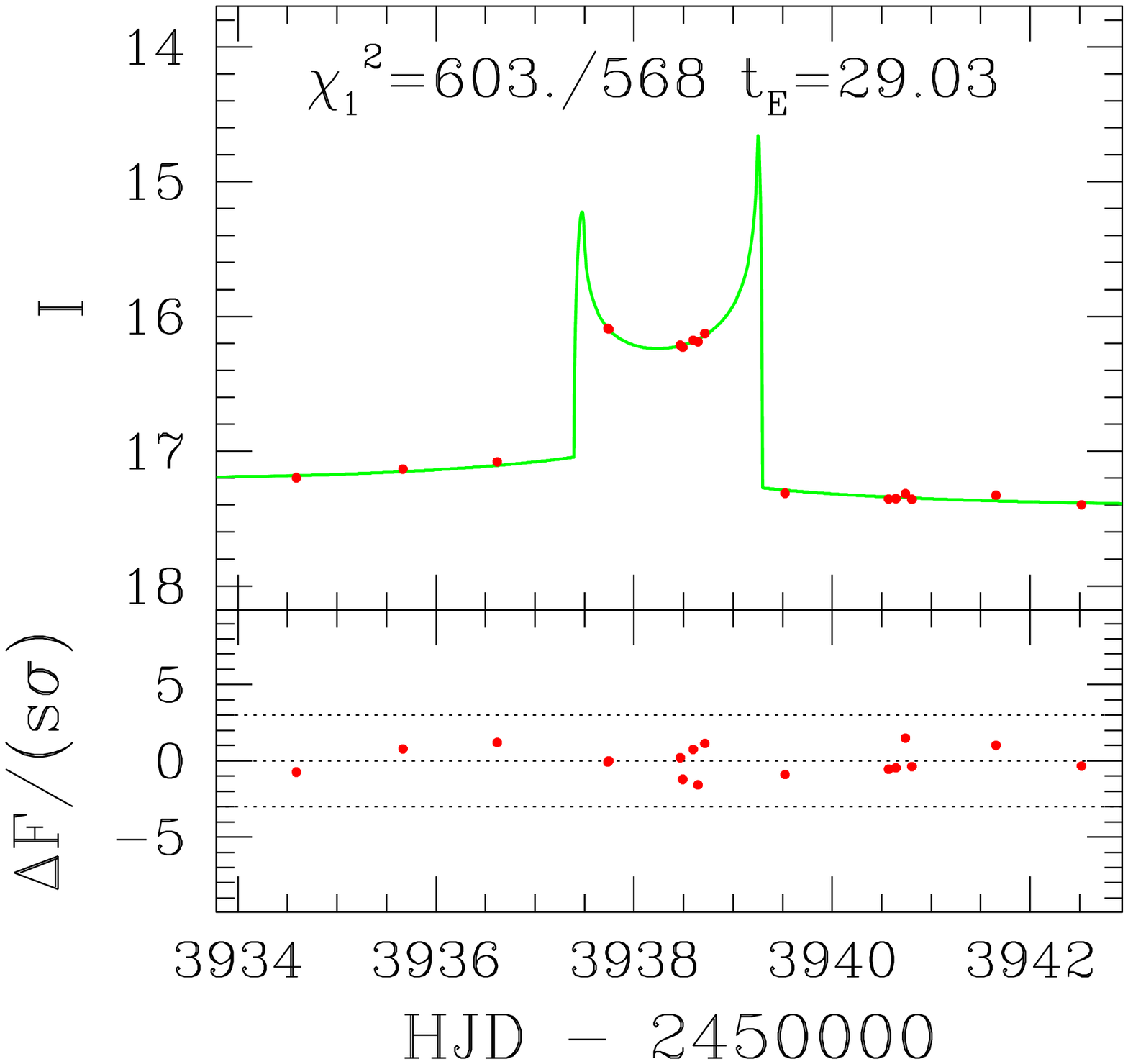}%

}

\noindent\parbox{12.75cm}{
\leftline {\bf OGLE 2006-BLG-450} 

\includegraphics[height=62mm,width=63mm]{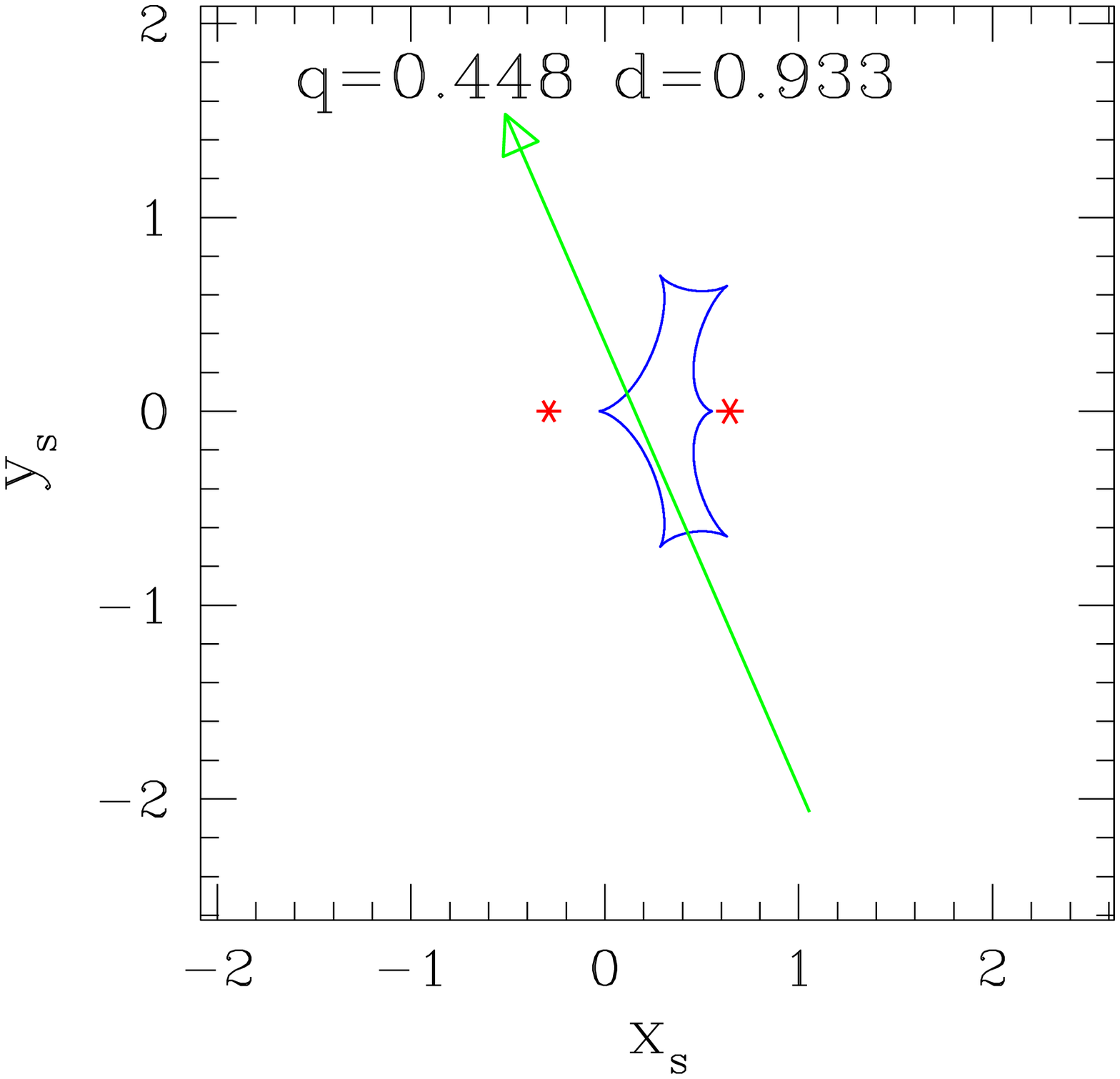} \hfill
\includegraphics[height=62mm,width=63mm]{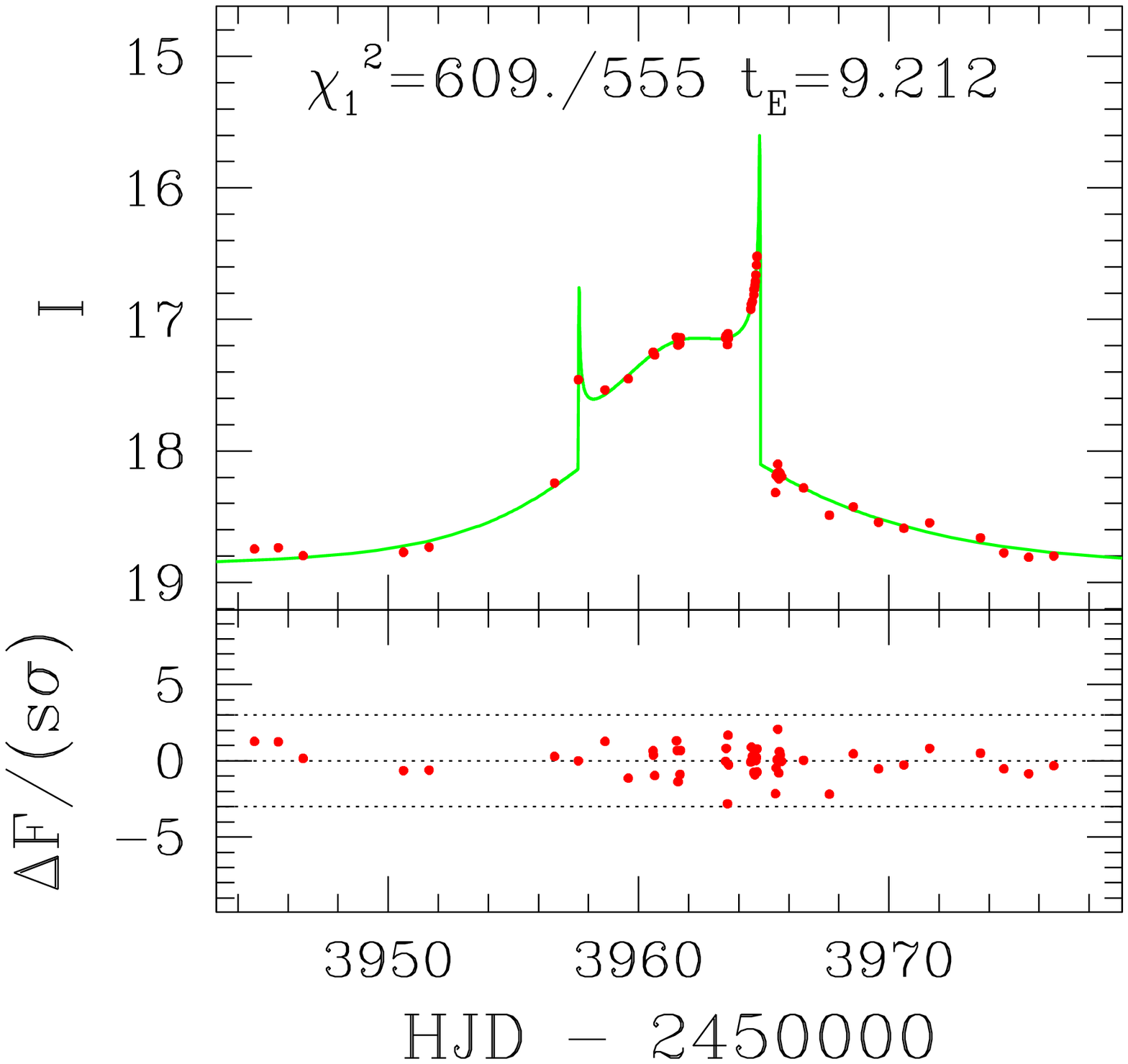}%

}

\noindent\parbox{12.75cm}{
\leftline {\bf OGLE 2006-BLG-460} 

\includegraphics[height=62mm,width=63mm]{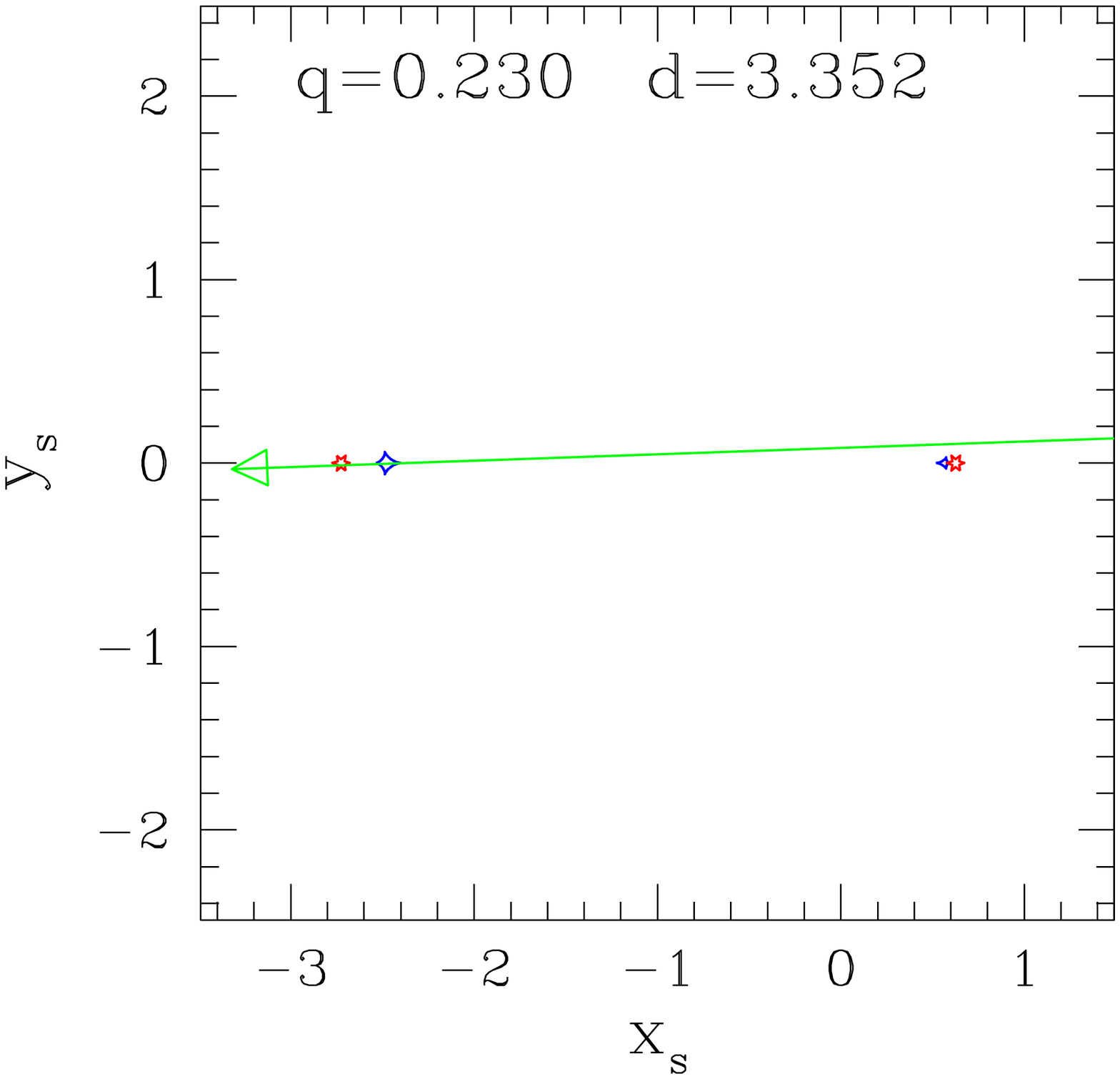} \hfill
\includegraphics[height=62mm,width=63mm]{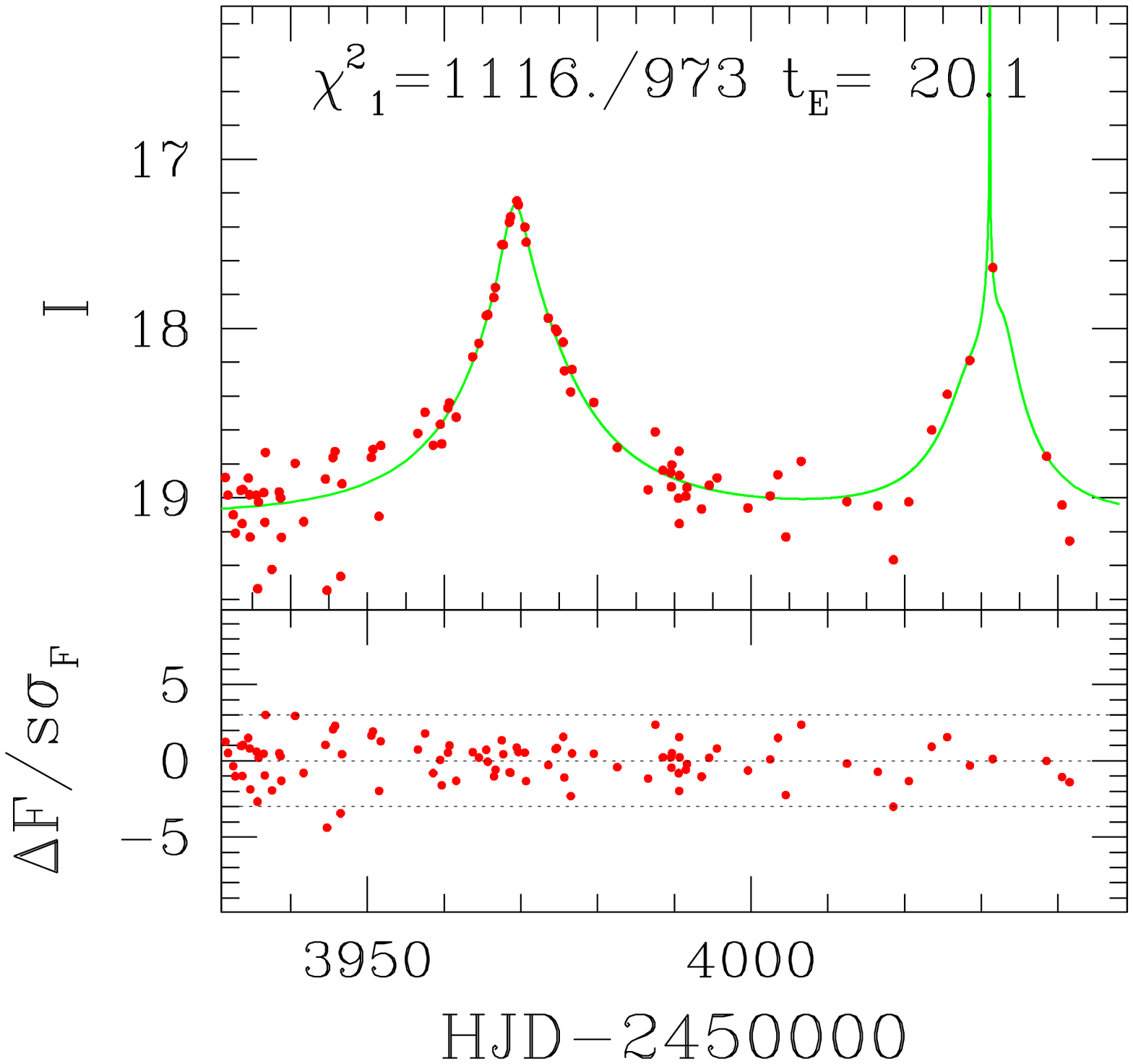}%

}

\noindent\parbox{12.75cm}{
\leftline {\bf OGLE 2007-BLG-006} 

\includegraphics[height=62mm,width=63mm]{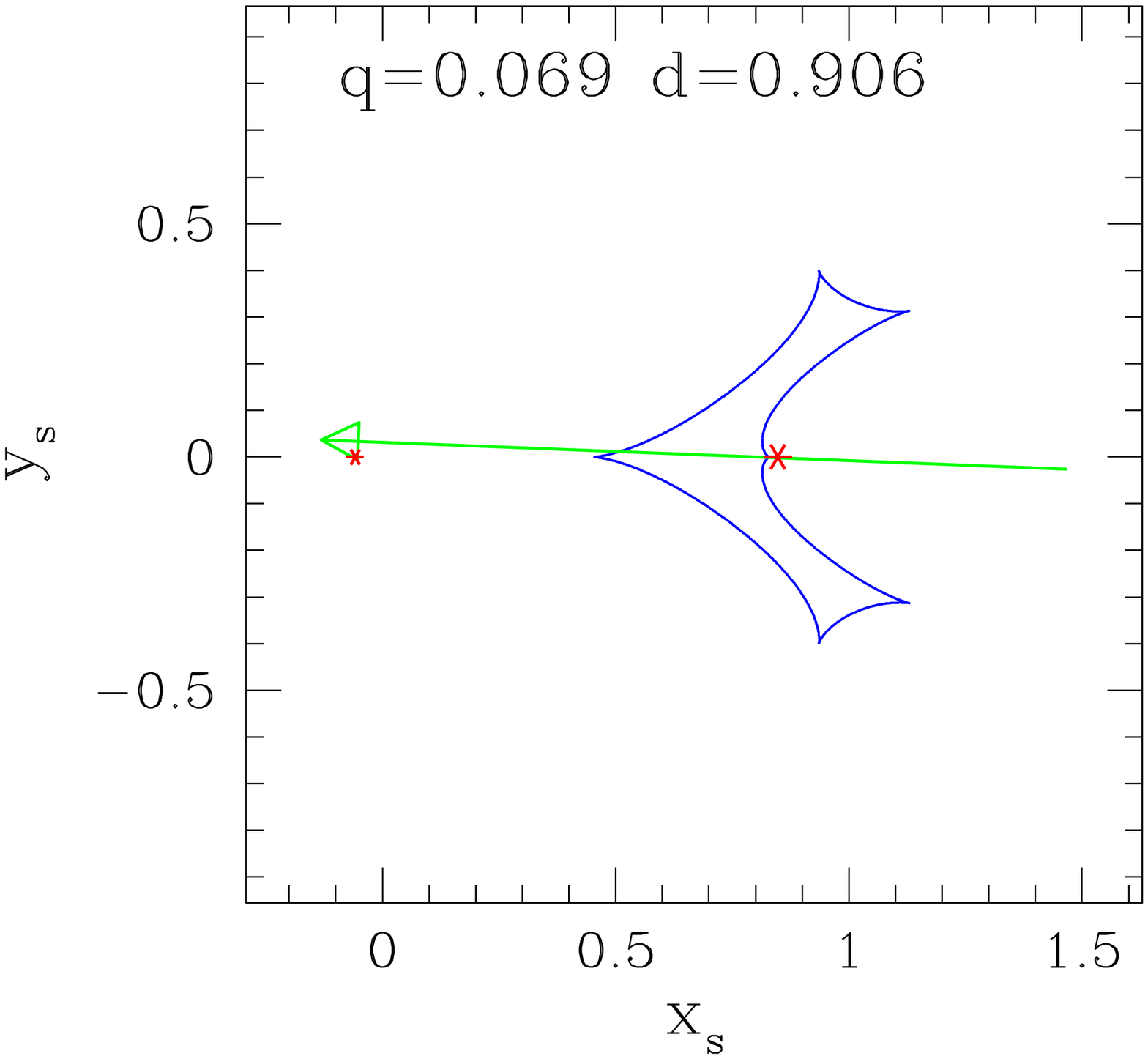} \hfill
\includegraphics[height=62mm,width=63mm]{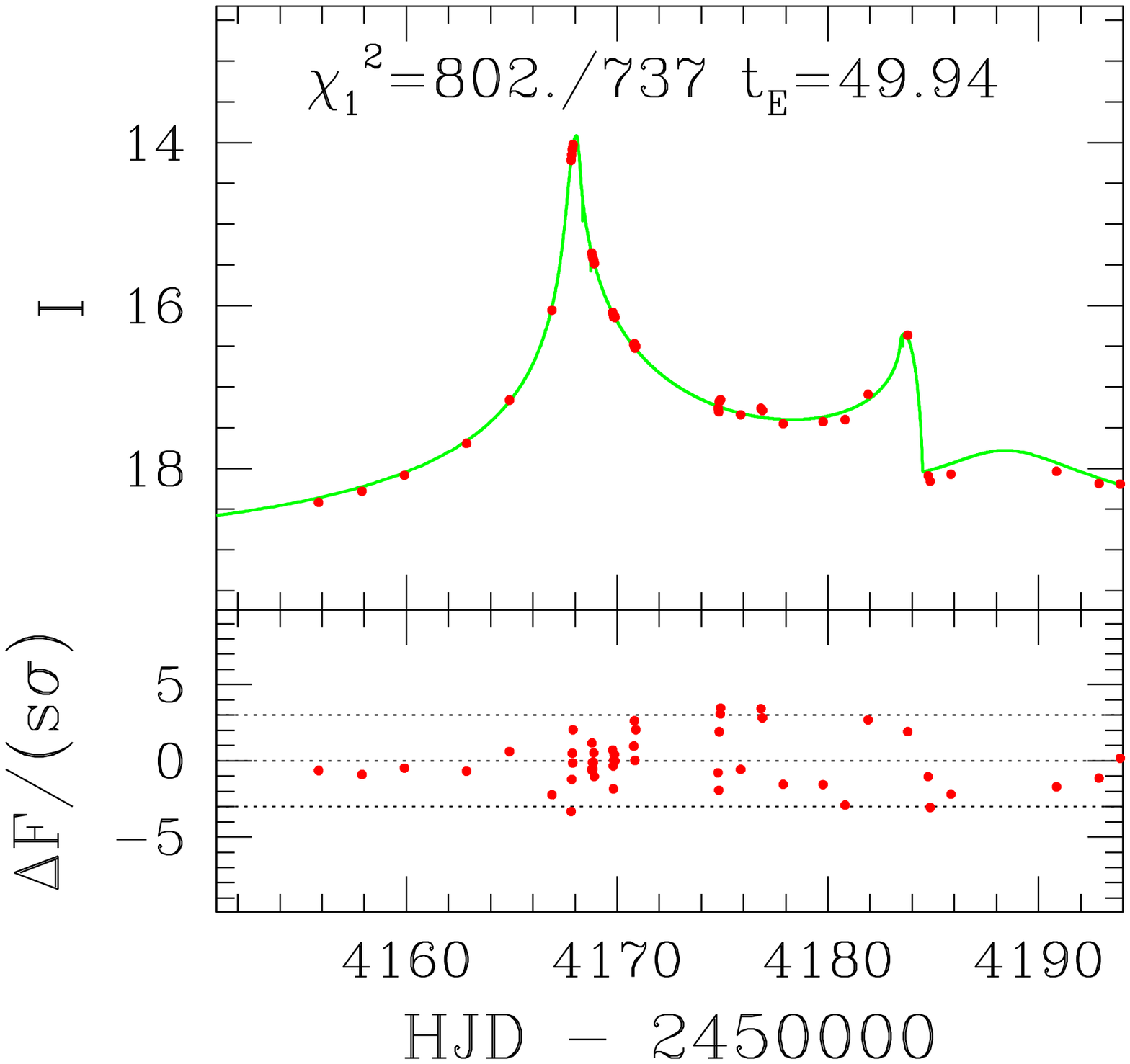}%

}

\noindent\parbox{12.75cm}{
\leftline {\bf OGLE 2007-BLG-069} 

\includegraphics[height=62mm,width=63mm]{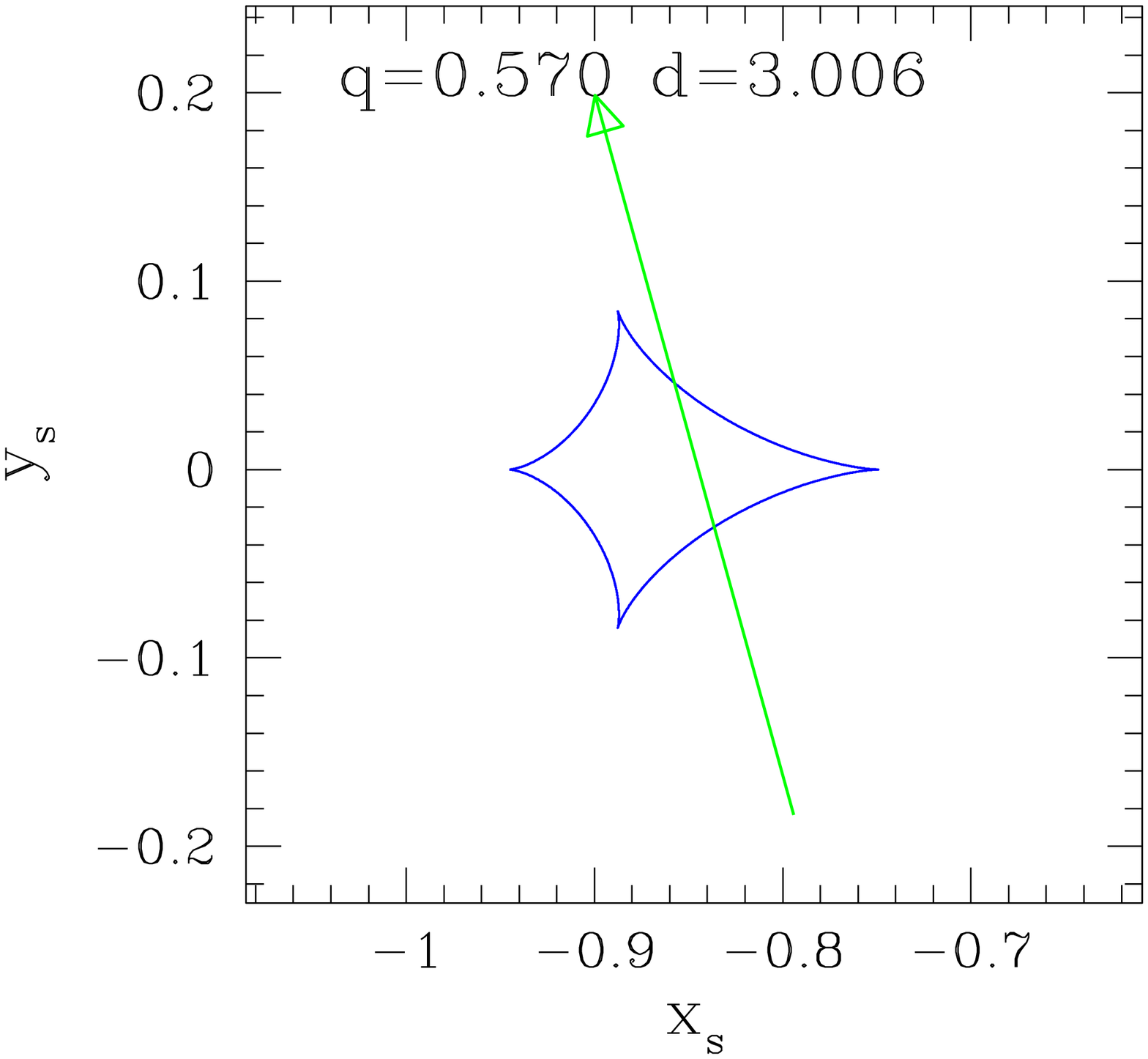} \hfill
\includegraphics[height=62mm,width=63mm]{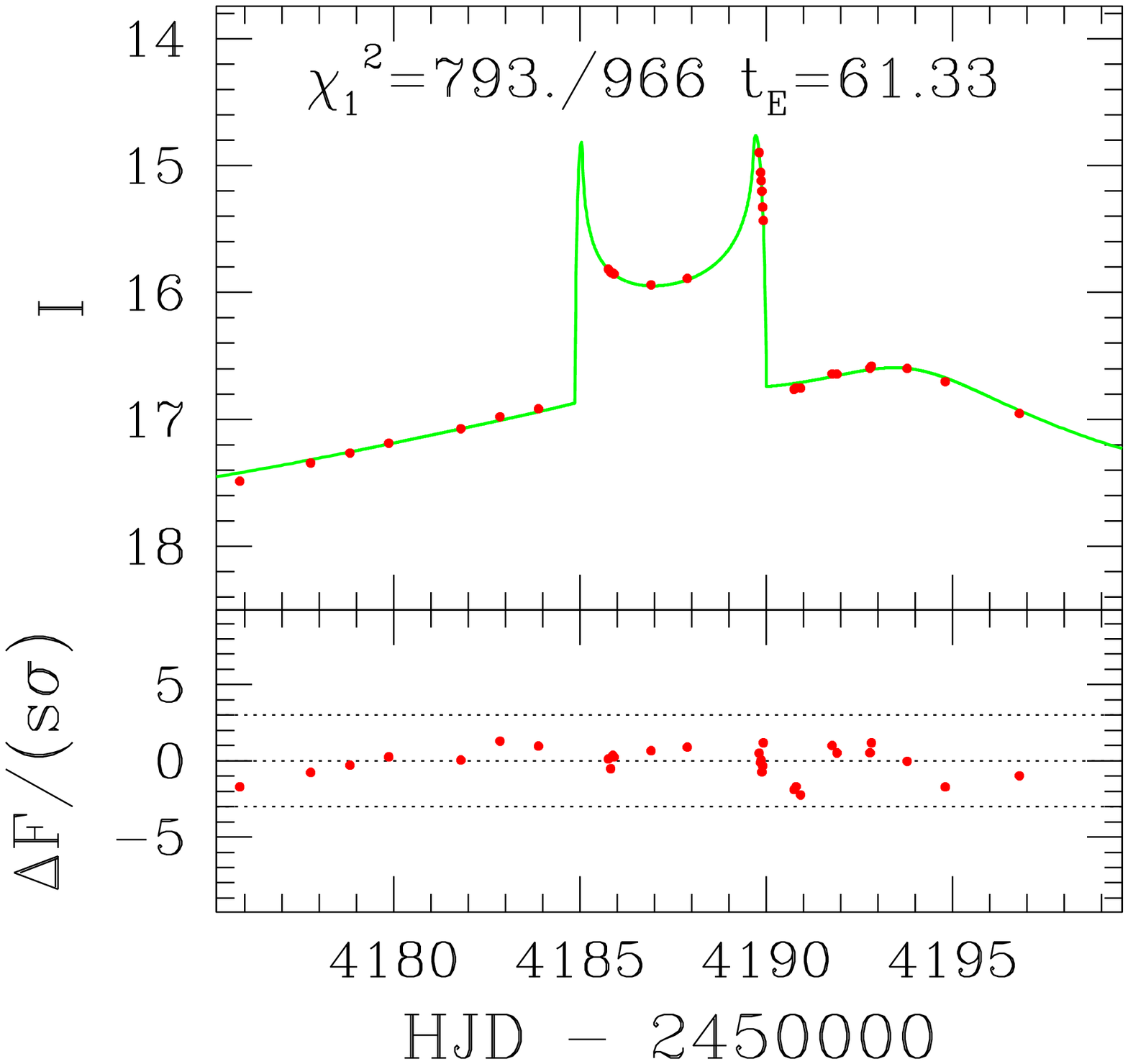}%

}

\noindent\parbox{12.75cm}{
\leftline {{\bf OGLE 2007-BLG-149} (1st model)}

\includegraphics[height=62mm,width=63mm]{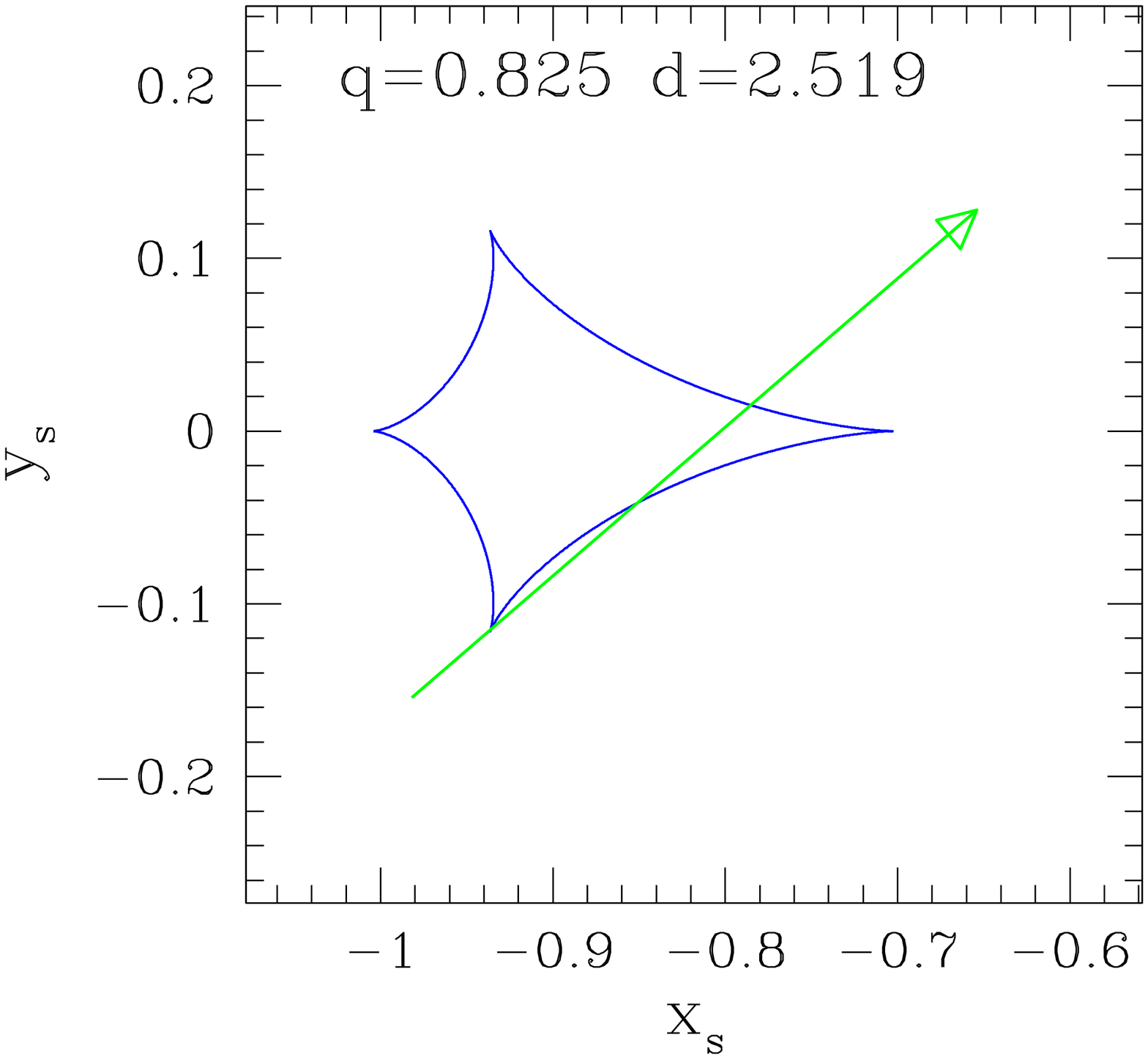} \hfill
\includegraphics[height=62mm,width=63mm]{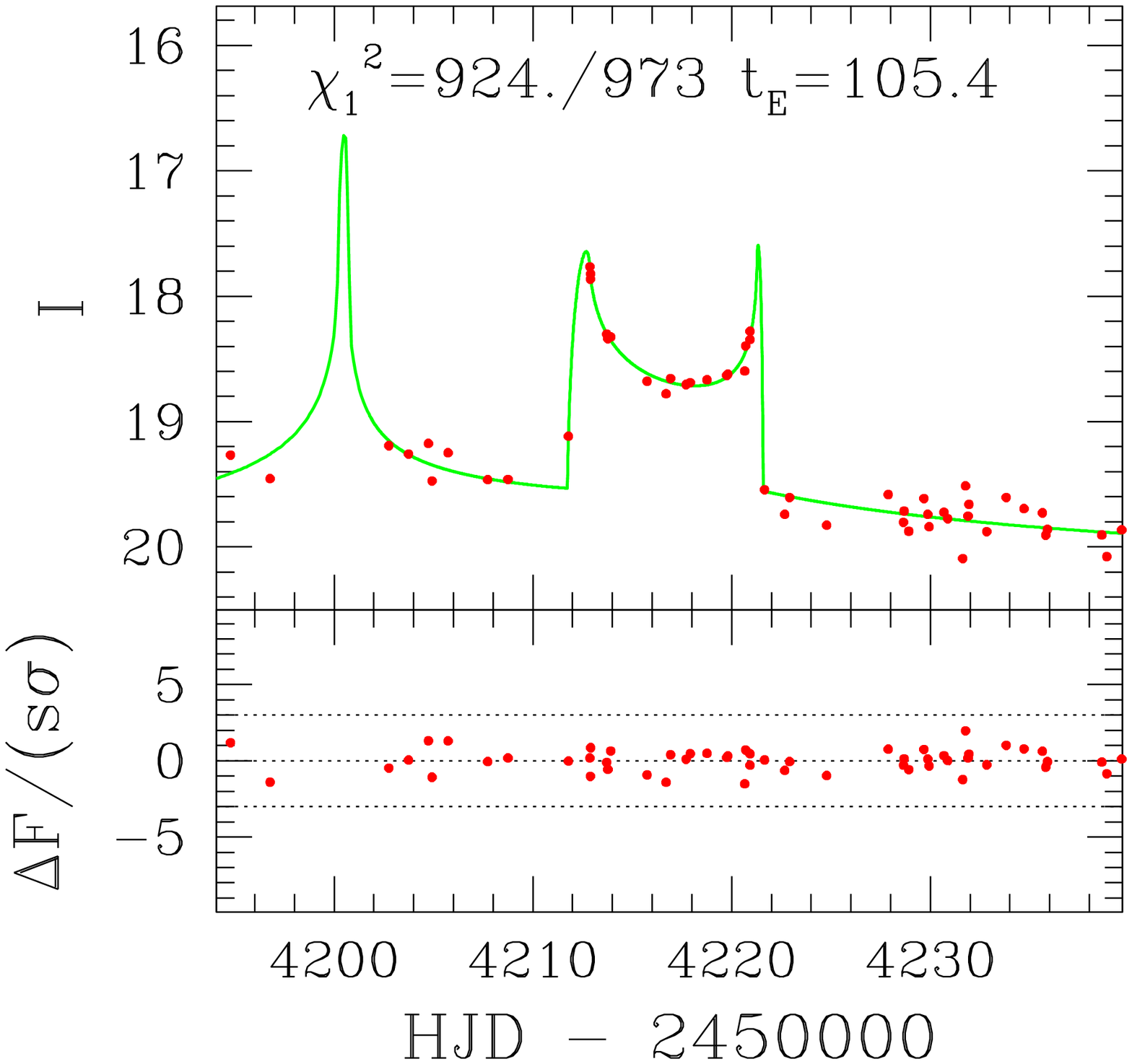}%

}
\noindent\parbox{12.75cm}{
\leftline {{\bf OGLE 2007-BLG-149} (2nd model)}

\includegraphics[height=62mm,width=63mm]{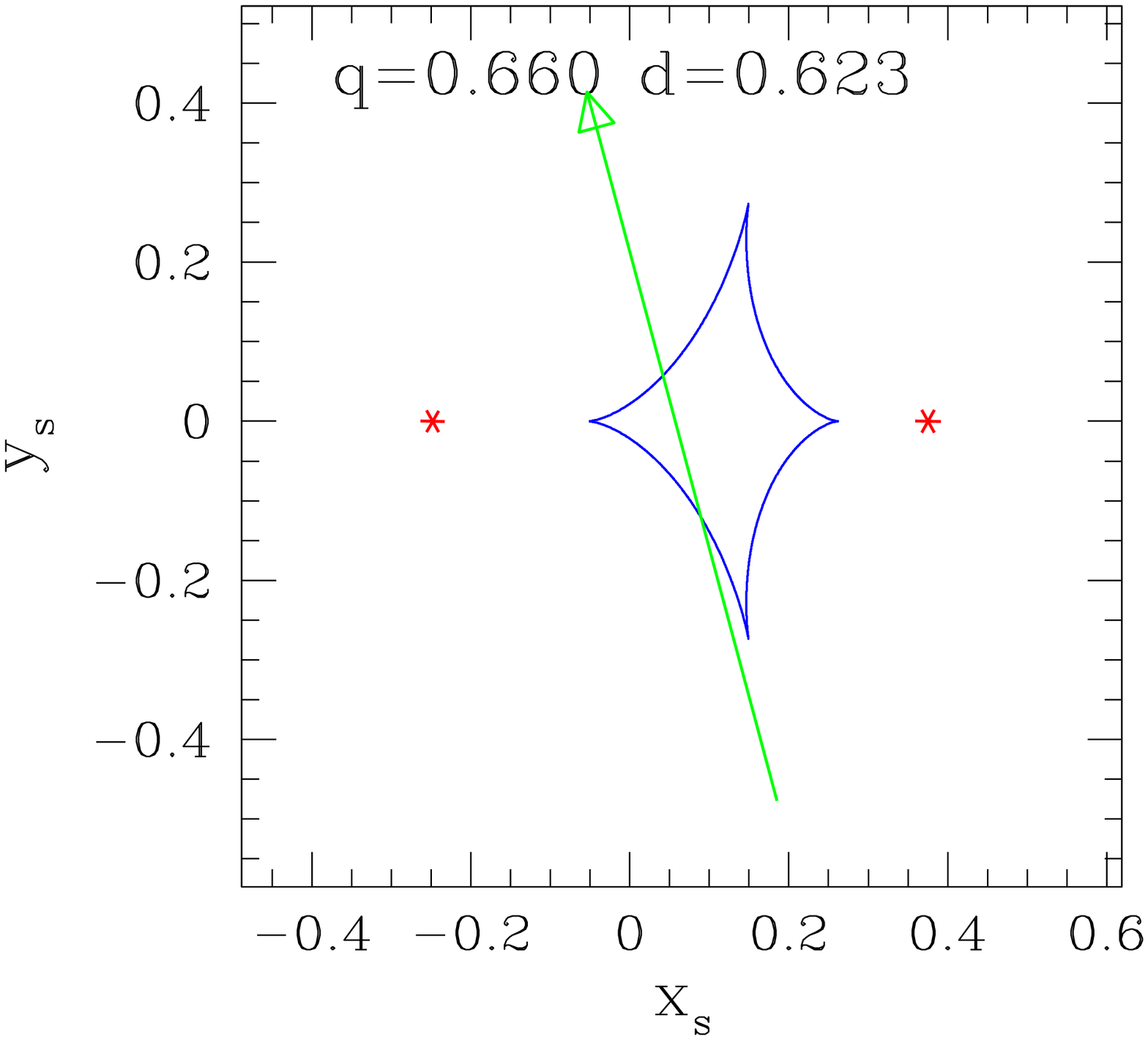} \hfill
\includegraphics[height=62mm,width=63mm]{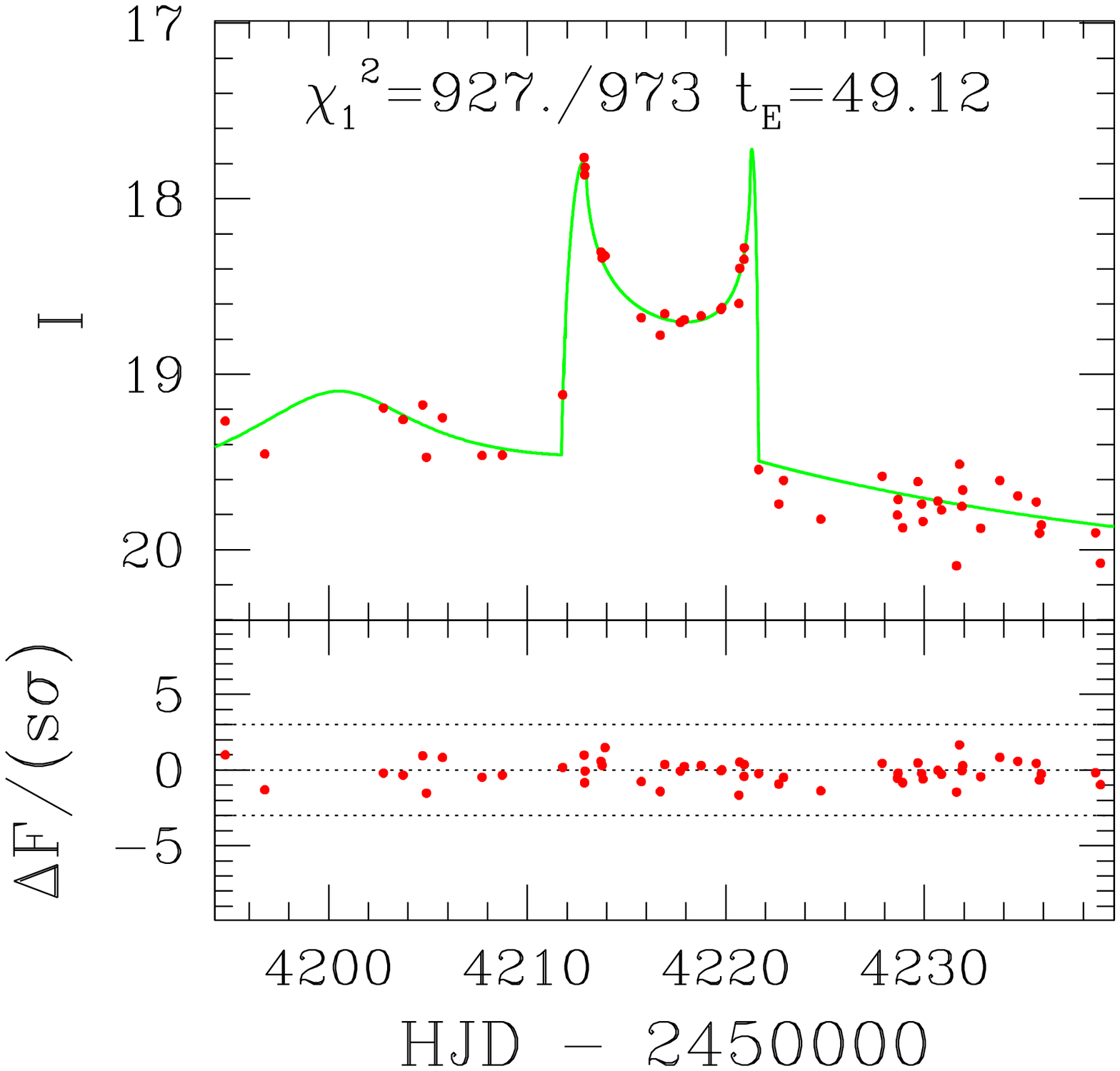}%

}
\noindent\parbox{12.75cm}{
\leftline {{\bf OGLE 2007-BLG-149} (3rd model)}

\includegraphics[height=62mm,width=63mm]{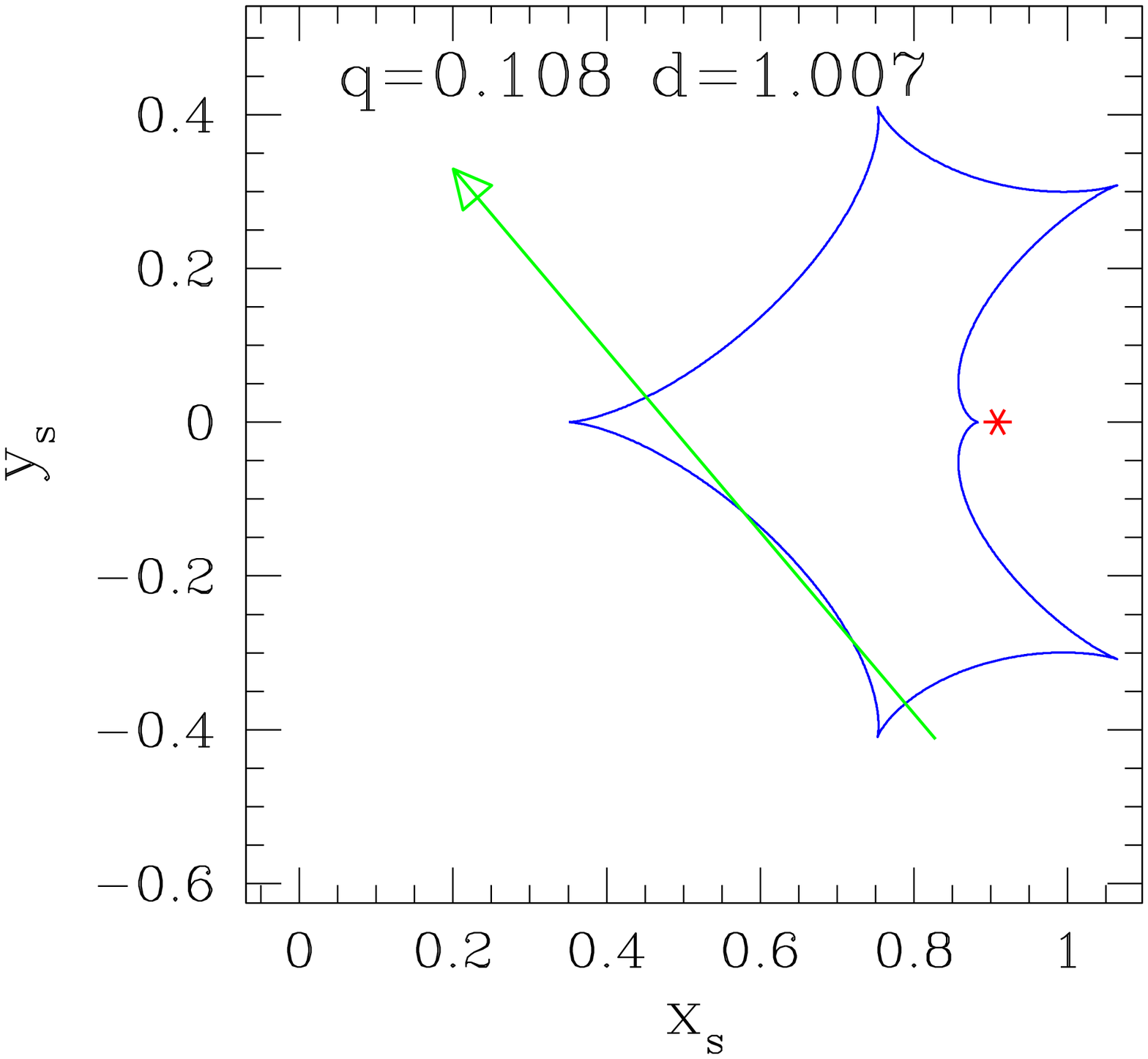} \hfill
\includegraphics[height=62mm,width=63mm]{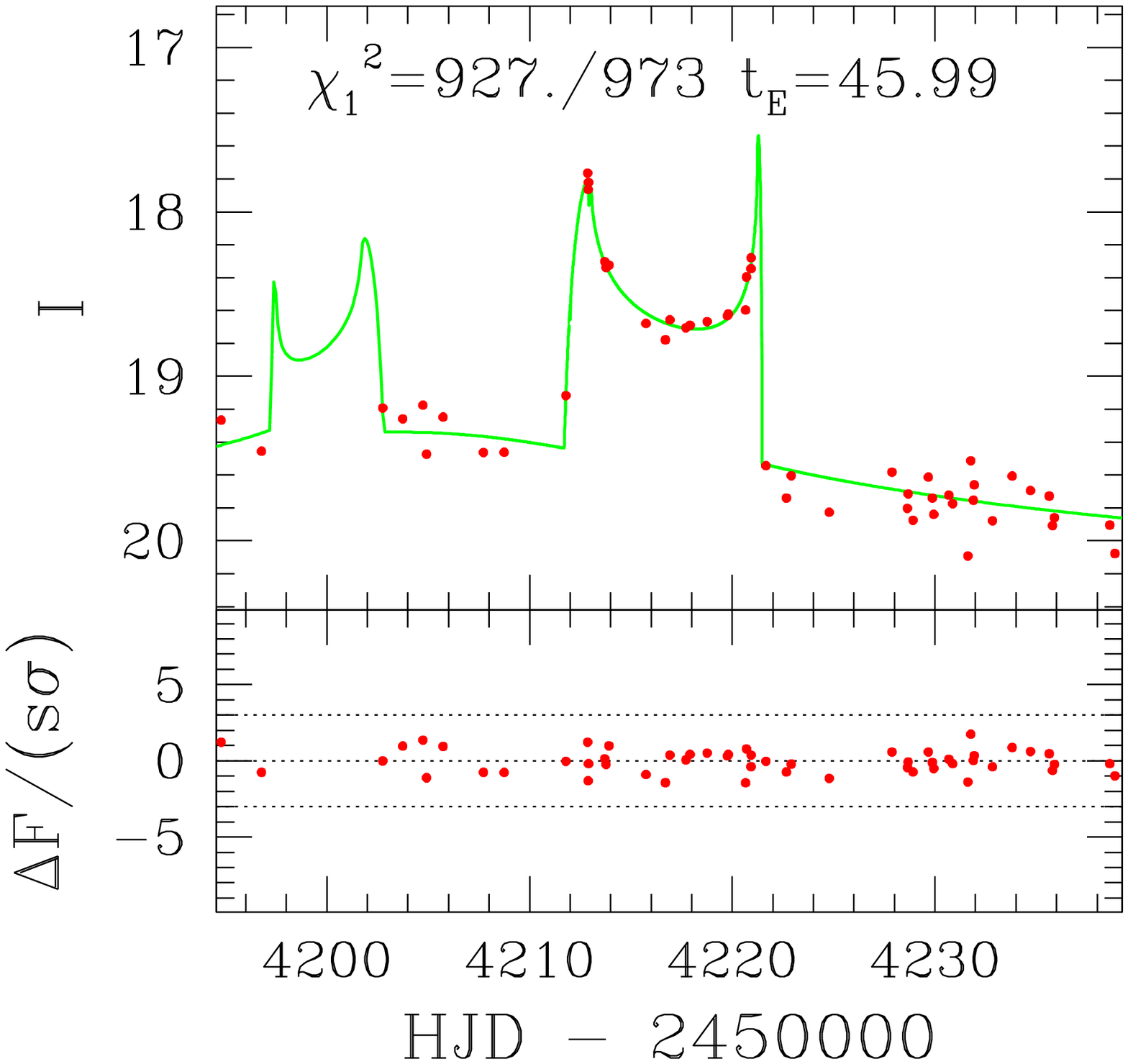}%

}

\noindent\parbox{12.75cm}{
\leftline {{\bf OGLE 2007-BLG-237} (1st model)}

\includegraphics[height=62mm,width=63mm]{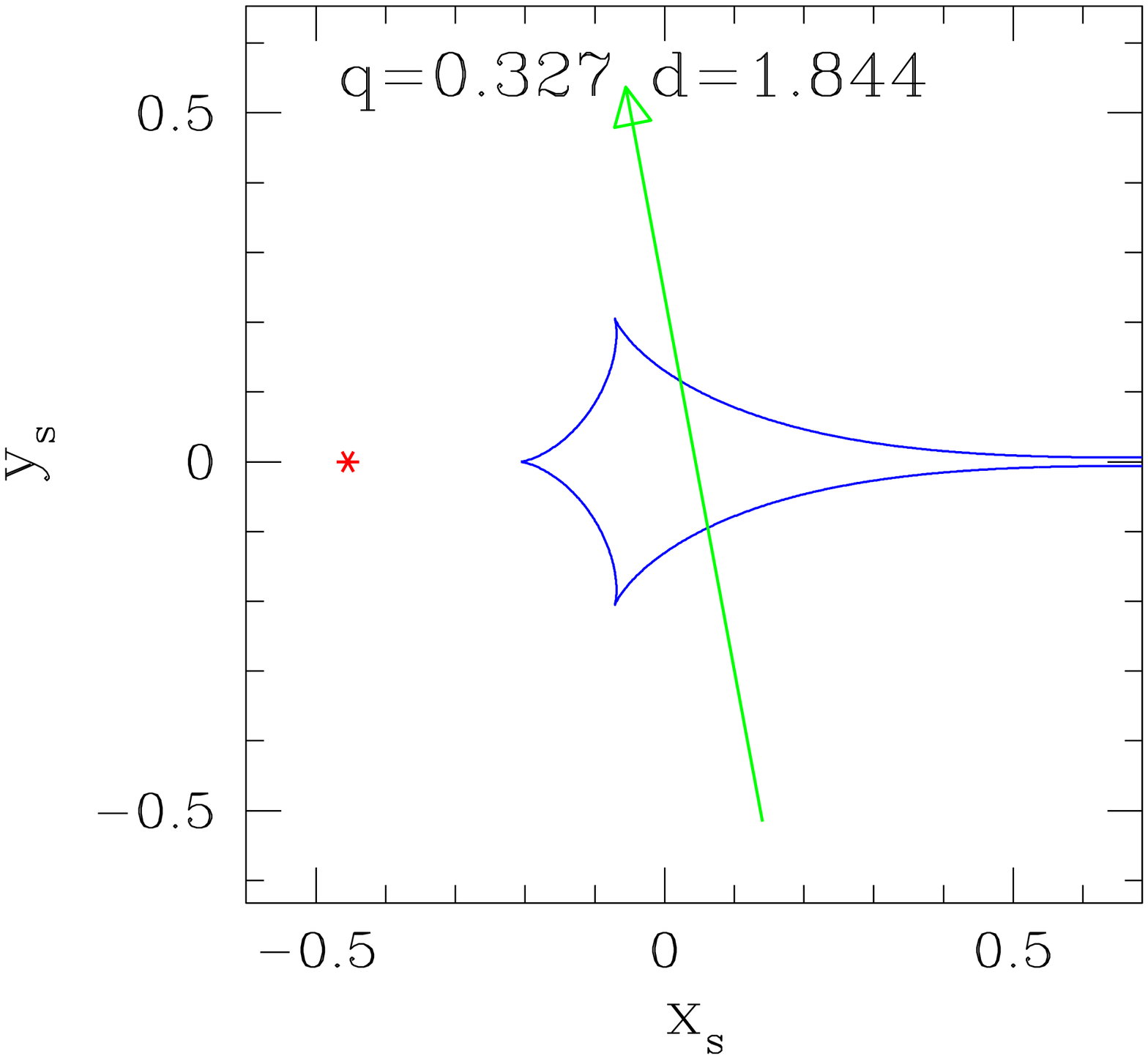} \hfill
\includegraphics[height=62mm,width=63mm]{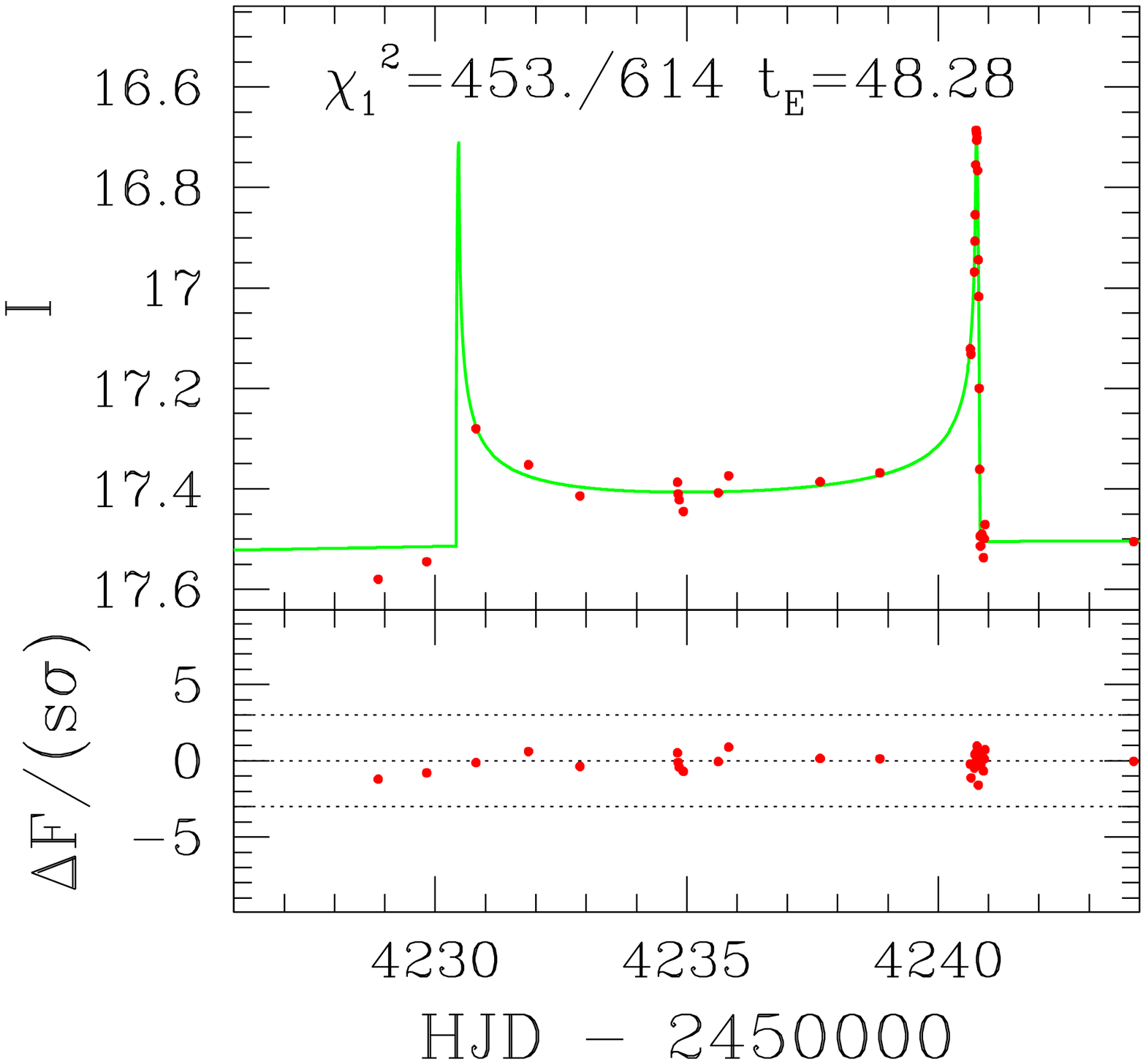}%

}

\noindent\parbox{12.75cm}{
\leftline {{\bf OGLE 2007-BLG-237} (2nd model)} 

\includegraphics[height=62mm,width=63mm]{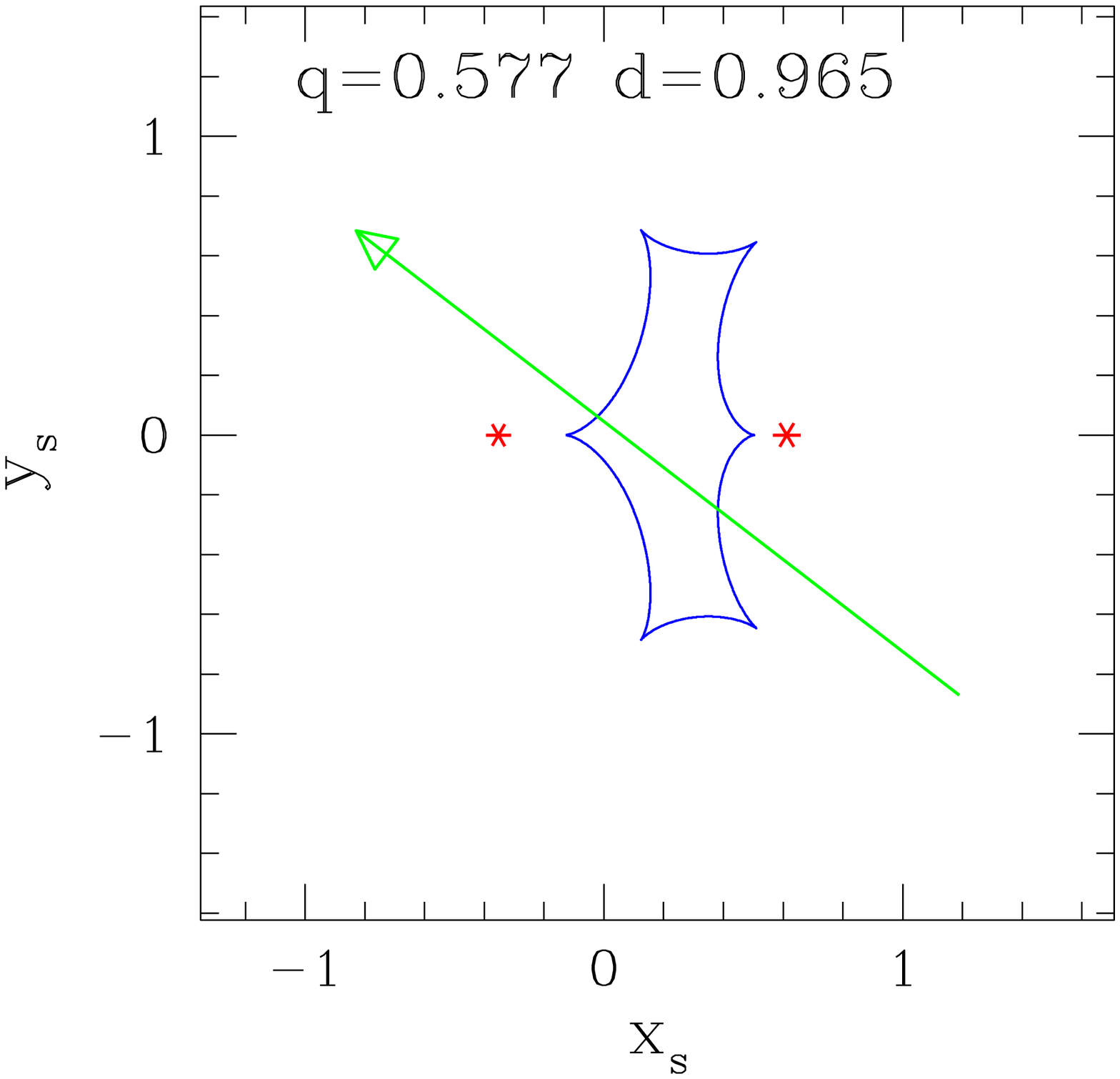} \hfill
\includegraphics[height=62mm,width=63mm]{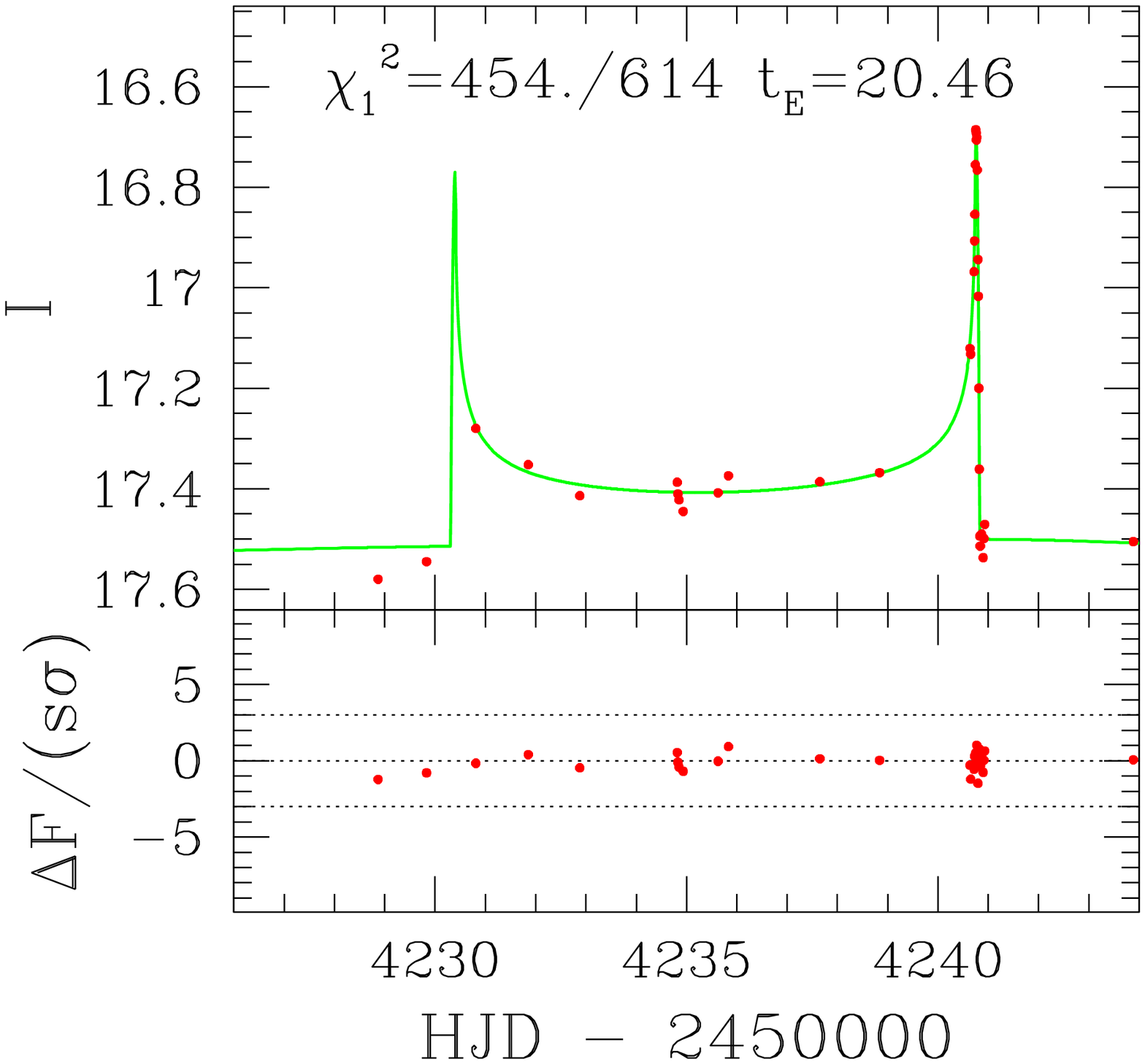}%

}

\noindent\parbox{12.75cm}{
\leftline {{\bf OGLE 2007-BLG-237} (3rd model)}

\includegraphics[height=62mm,width=63mm]{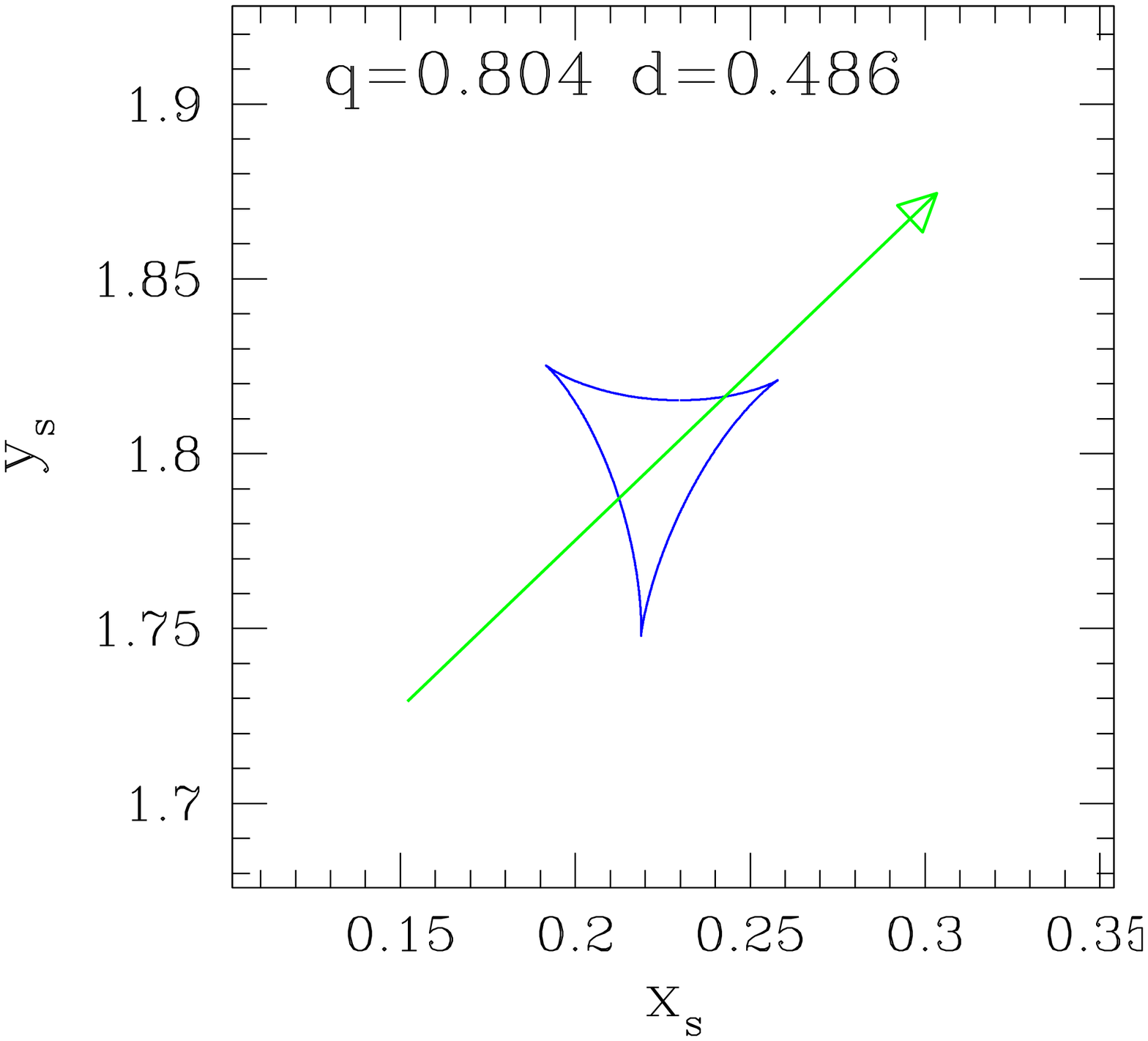} \hfill
\includegraphics[height=62mm,width=63mm]{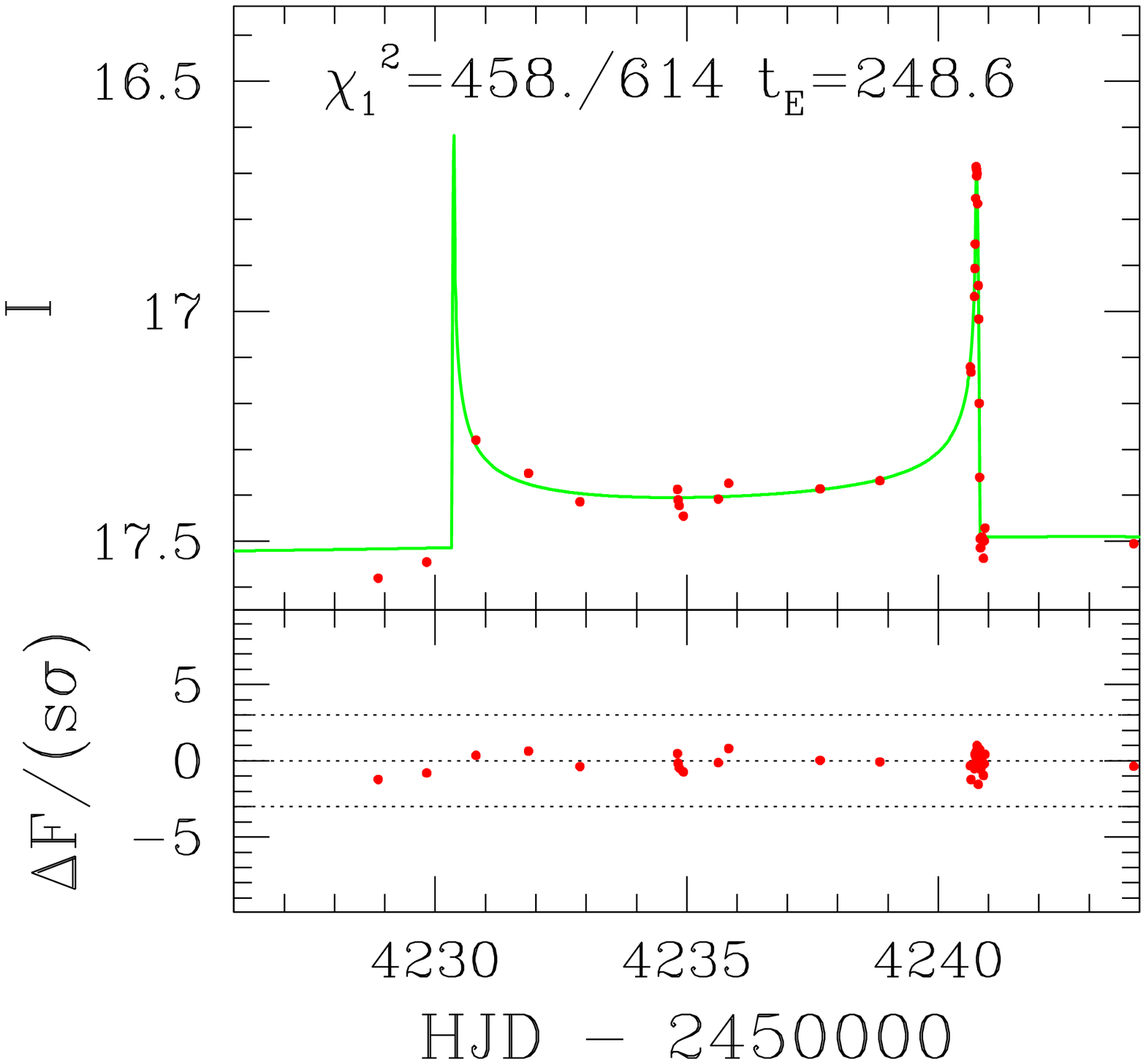}%

}

\noindent\parbox{12.75cm}{
\leftline {\bf OGLE 2007-BLG-327} 

\includegraphics[height=62mm,width=63mm]{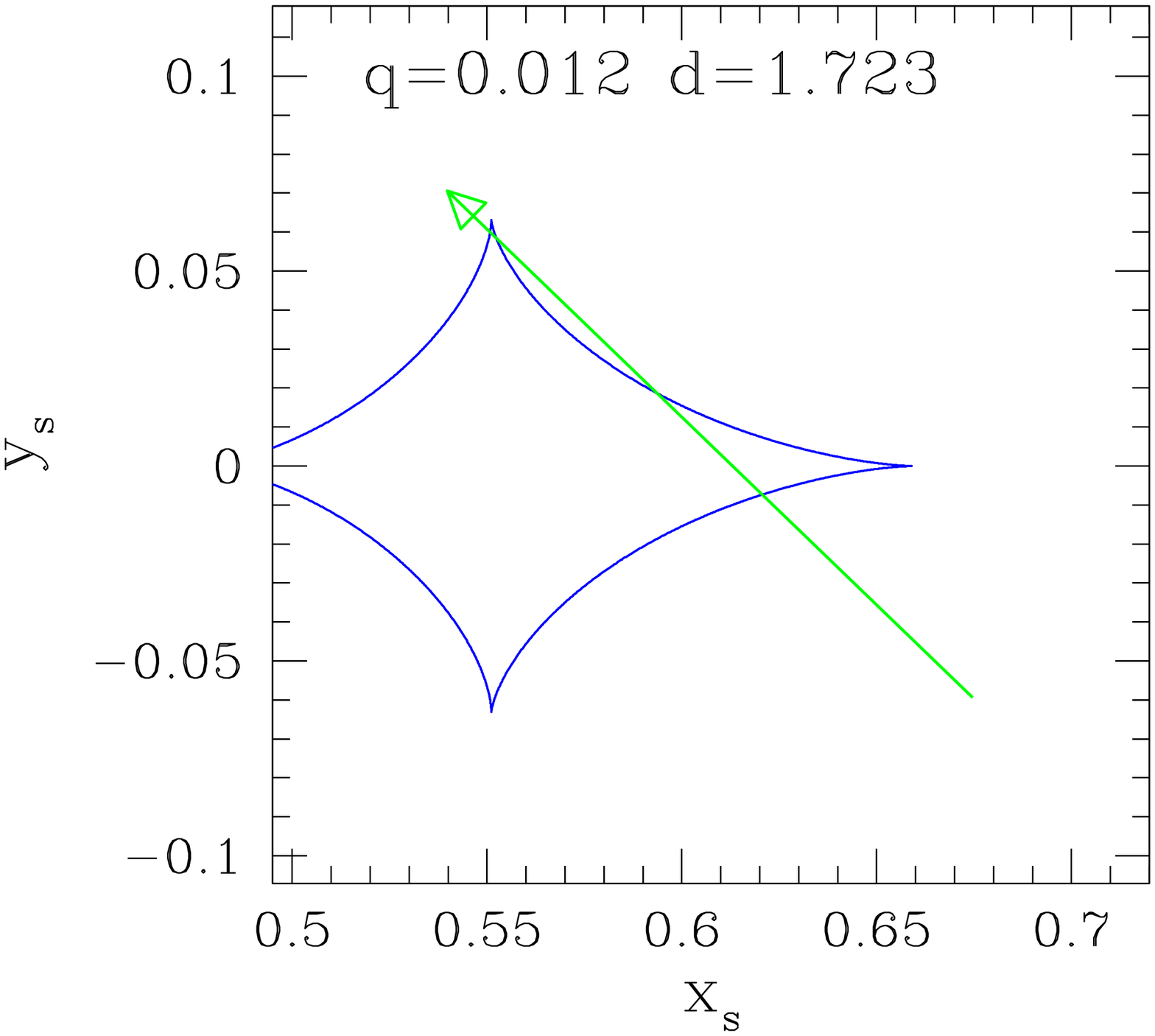} \hfill
\includegraphics[height=62mm,width=63mm]{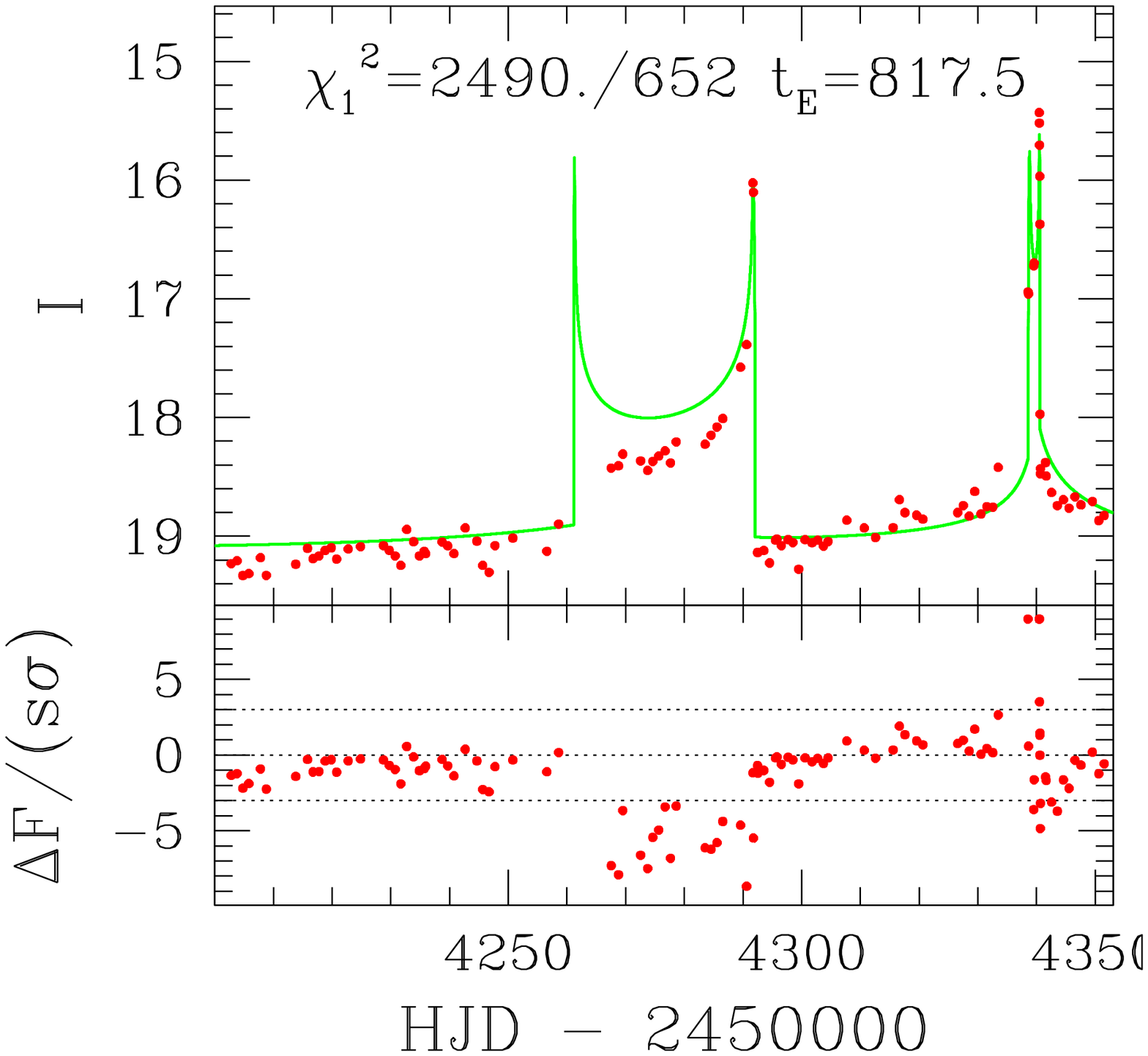}%

}

\noindent\parbox{12.75cm}{
\leftline {\bf OGLE 2007-BLG-363} 

\includegraphics[height=62mm,width=63mm]{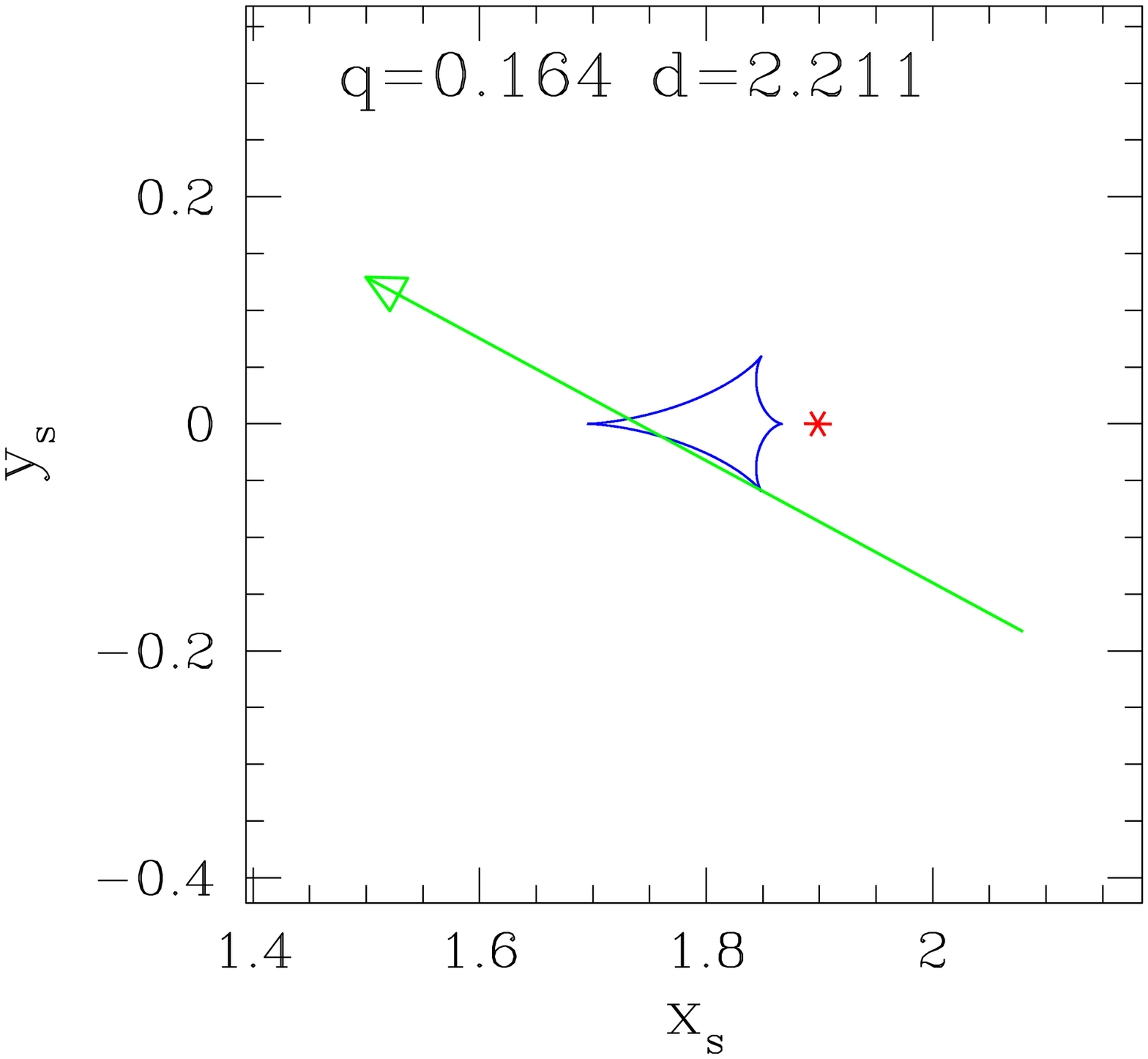} \hfill
\includegraphics[height=62mm,width=63mm]{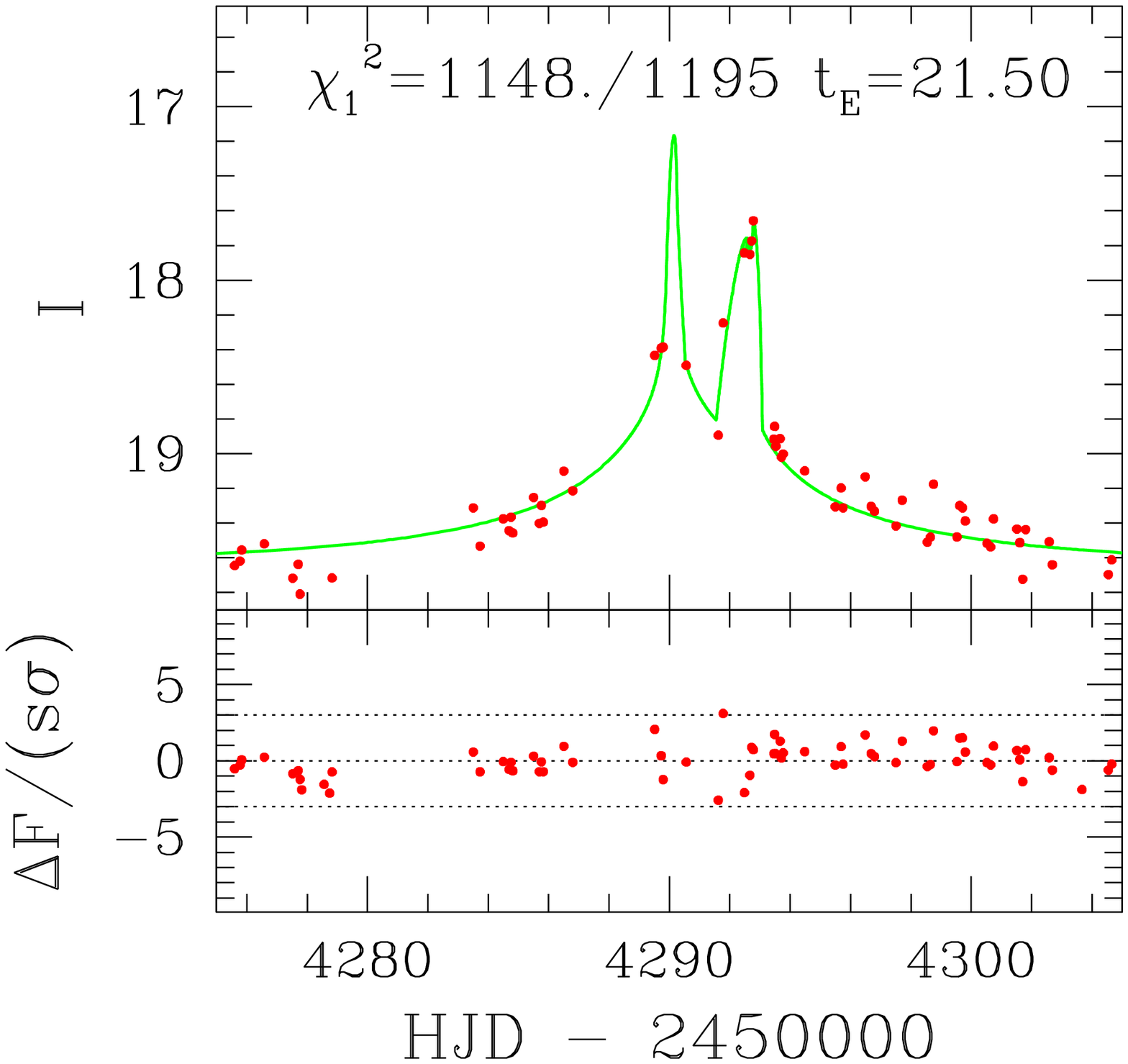}%

}

\noindent\parbox{12.75cm}{
\leftline {\bf OGLE 2007-BLG-373} 

\includegraphics[height=62mm,width=63mm]{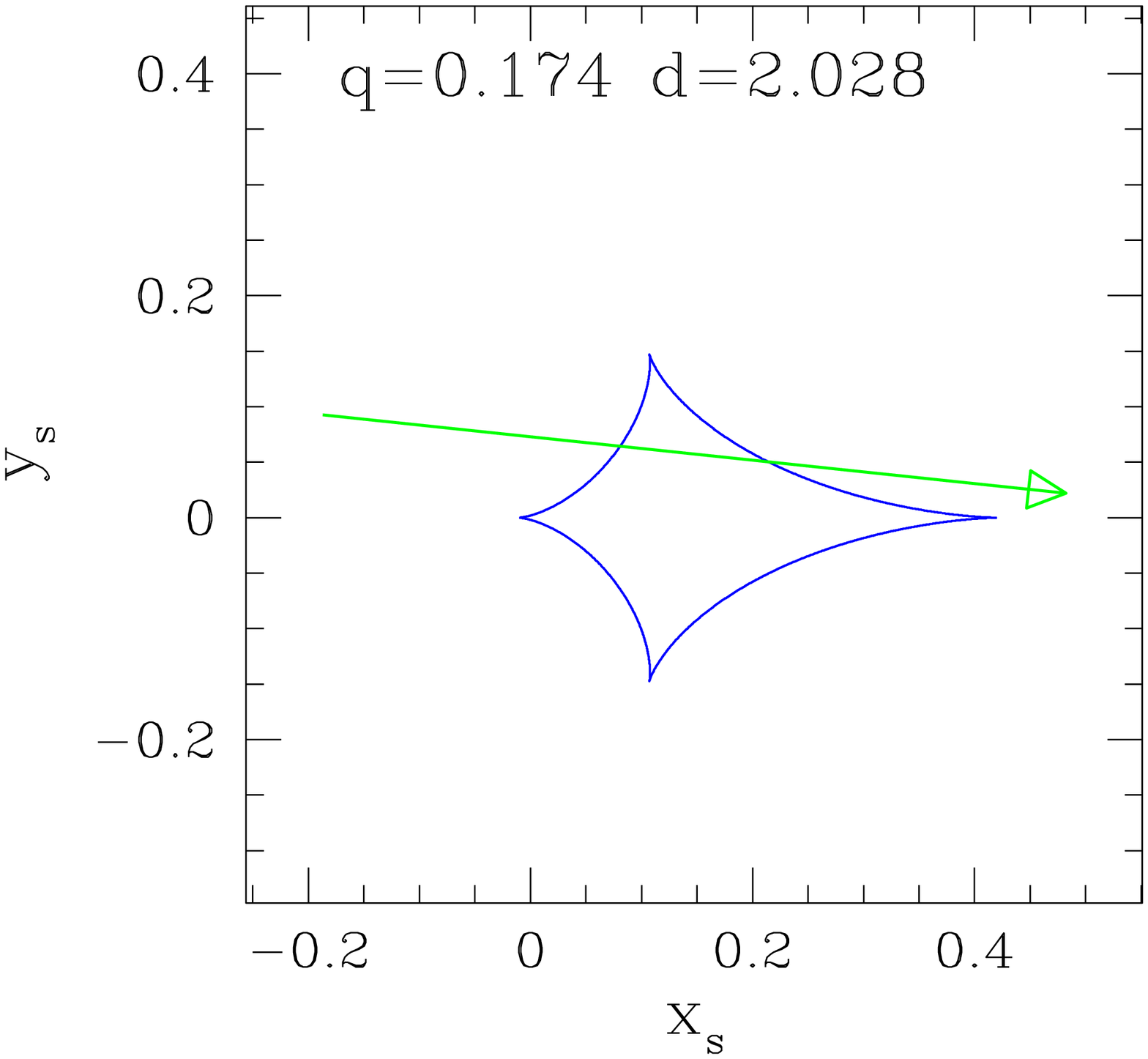} \hfill
\includegraphics[height=62mm,width=63mm]{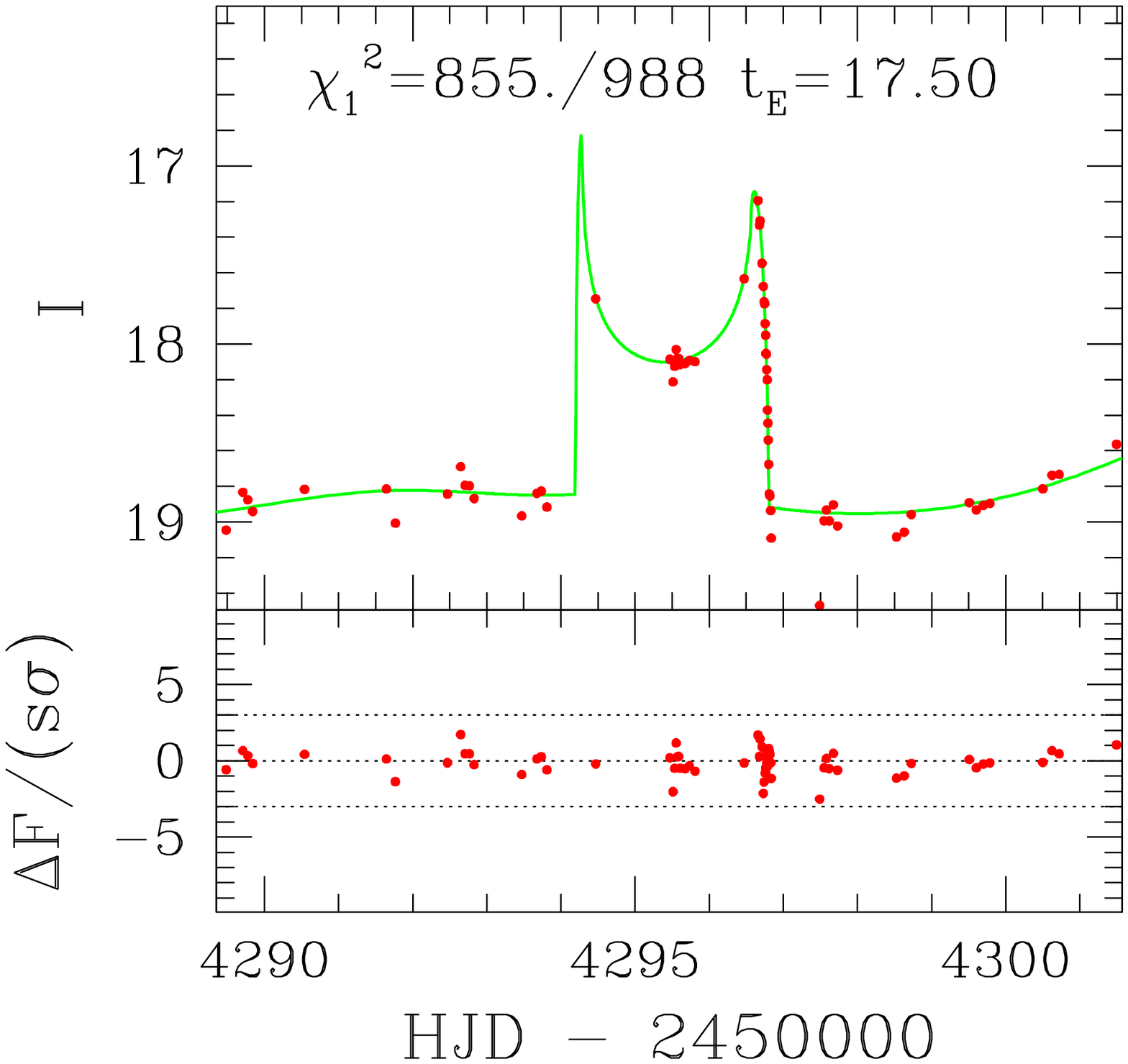}%

}

\noindent\parbox{12.75cm}{
\leftline {\bf OGLE 2007-BLG-399} 

\includegraphics[height=62mm,width=63mm]{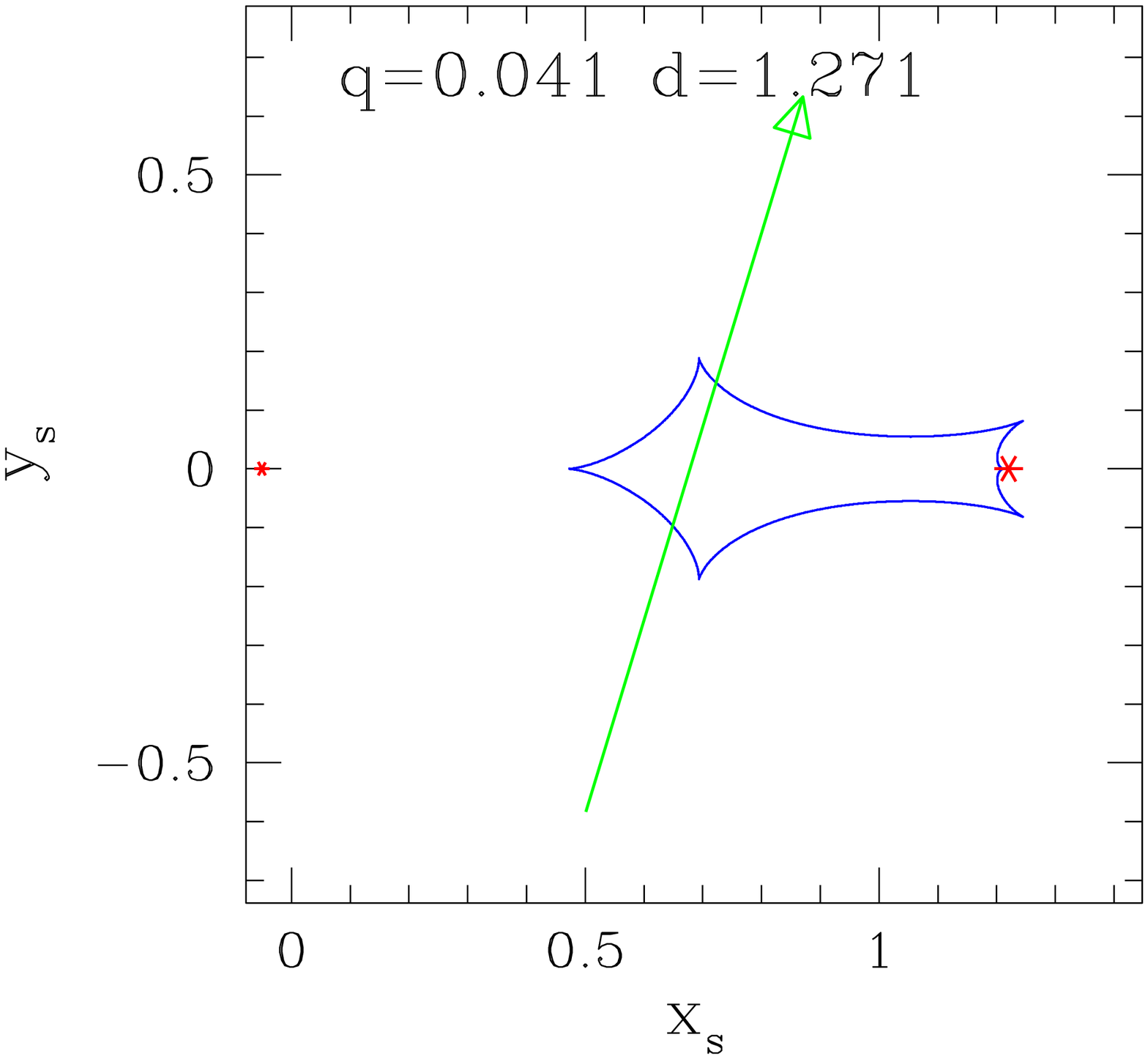} \hfill
\includegraphics[height=62mm,width=63mm]{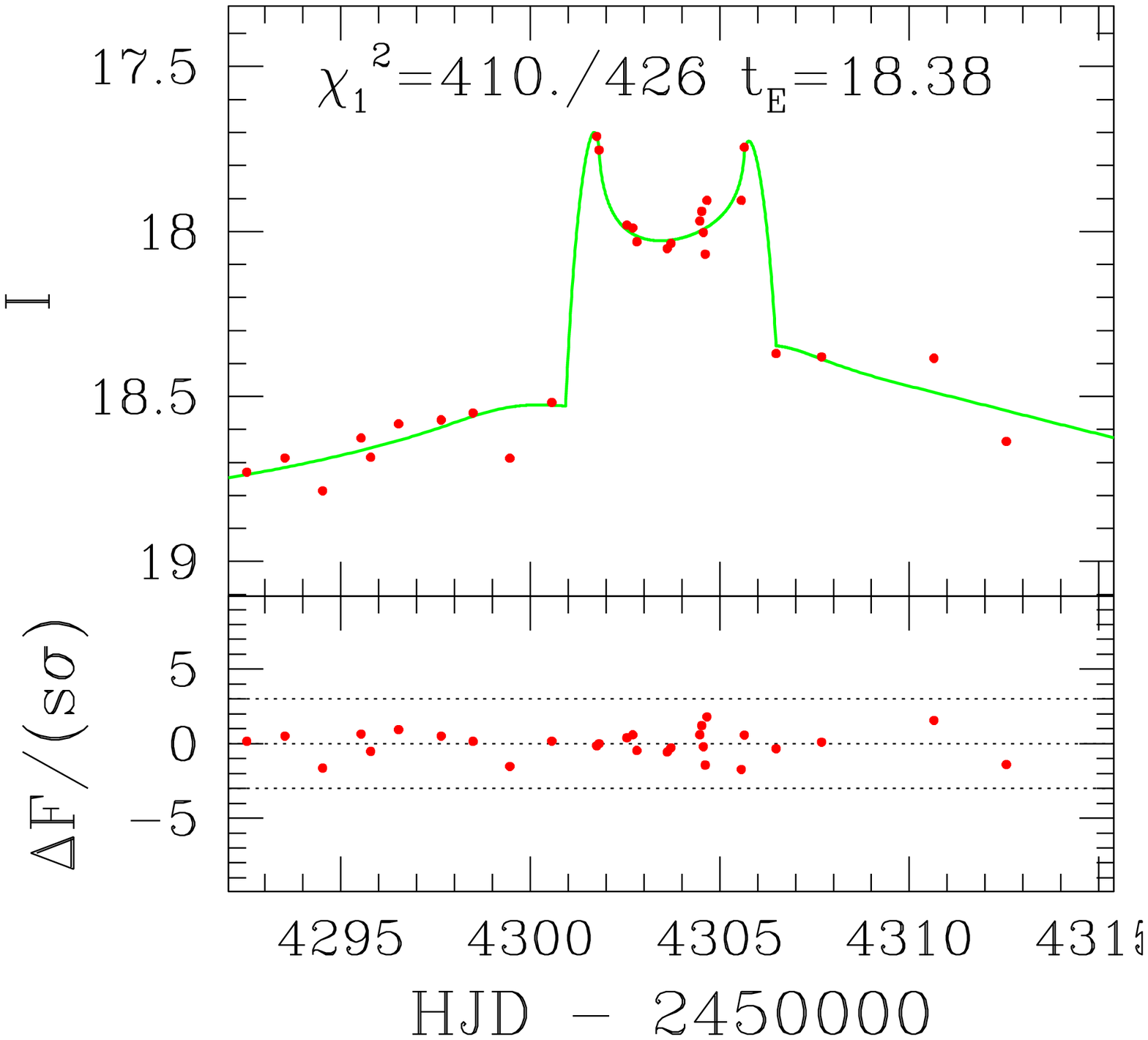}%

}

\noindent\parbox{12.75cm}{
\leftline {{\bf OGLE 2008-BLG-118} (1st model)}

\includegraphics[height=62mm,width=63mm]{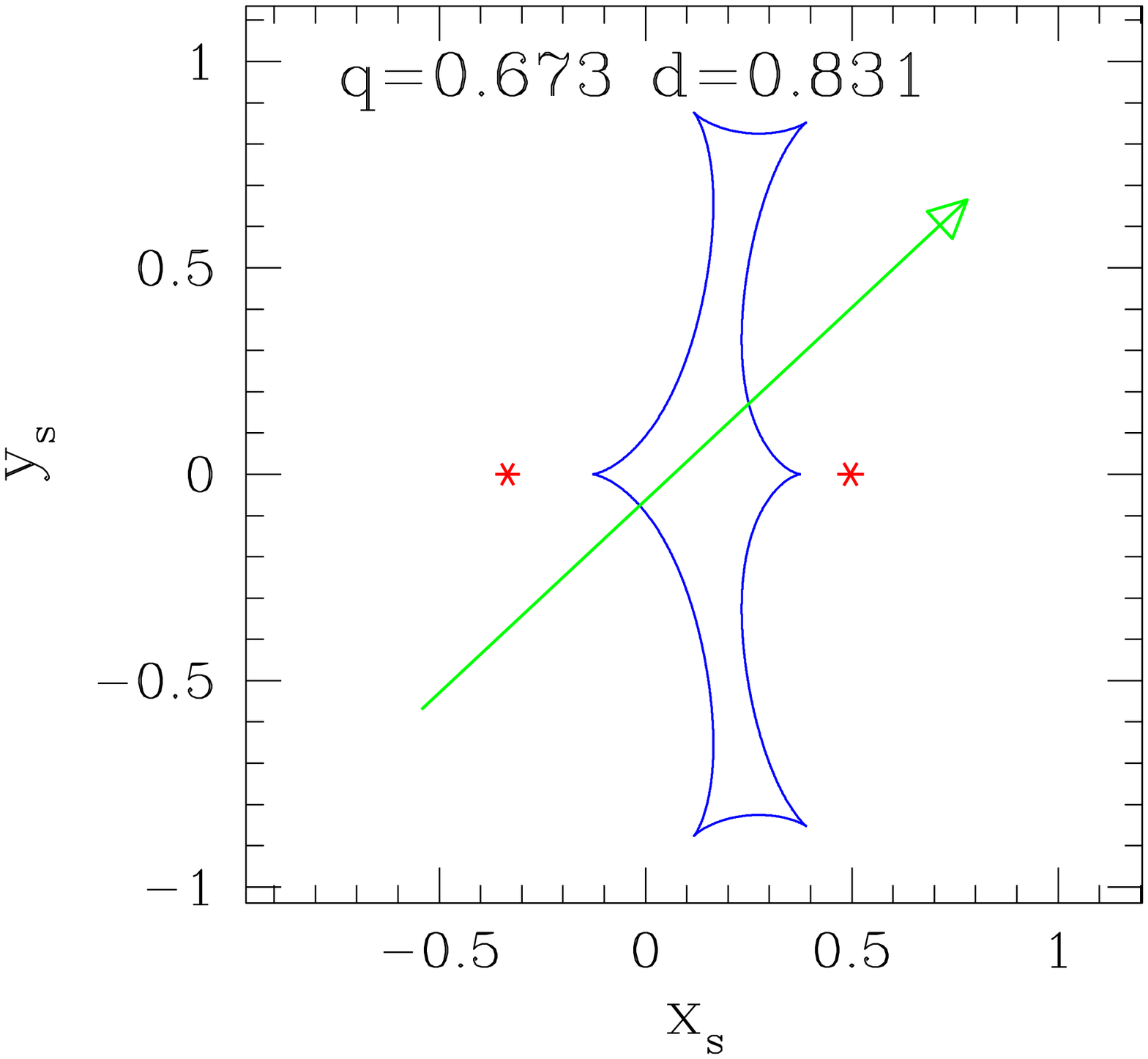} \hfill
\includegraphics[height=62mm,width=63mm]{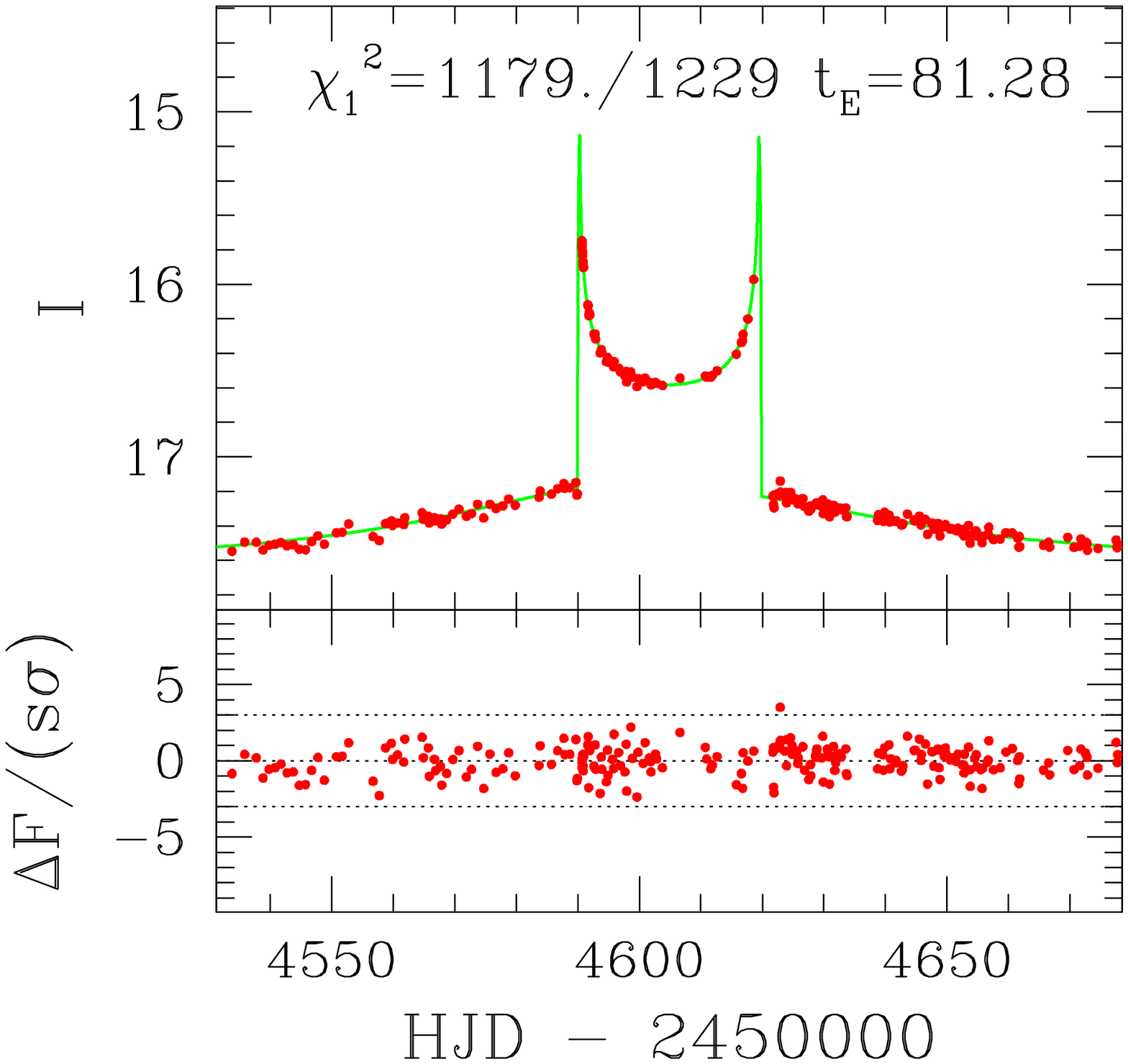}%

}

\noindent\parbox{12.75cm}{
\leftline {{\bf OGLE 2008-BLG-118} (2nd model)}

\includegraphics[height=62mm,width=63mm]{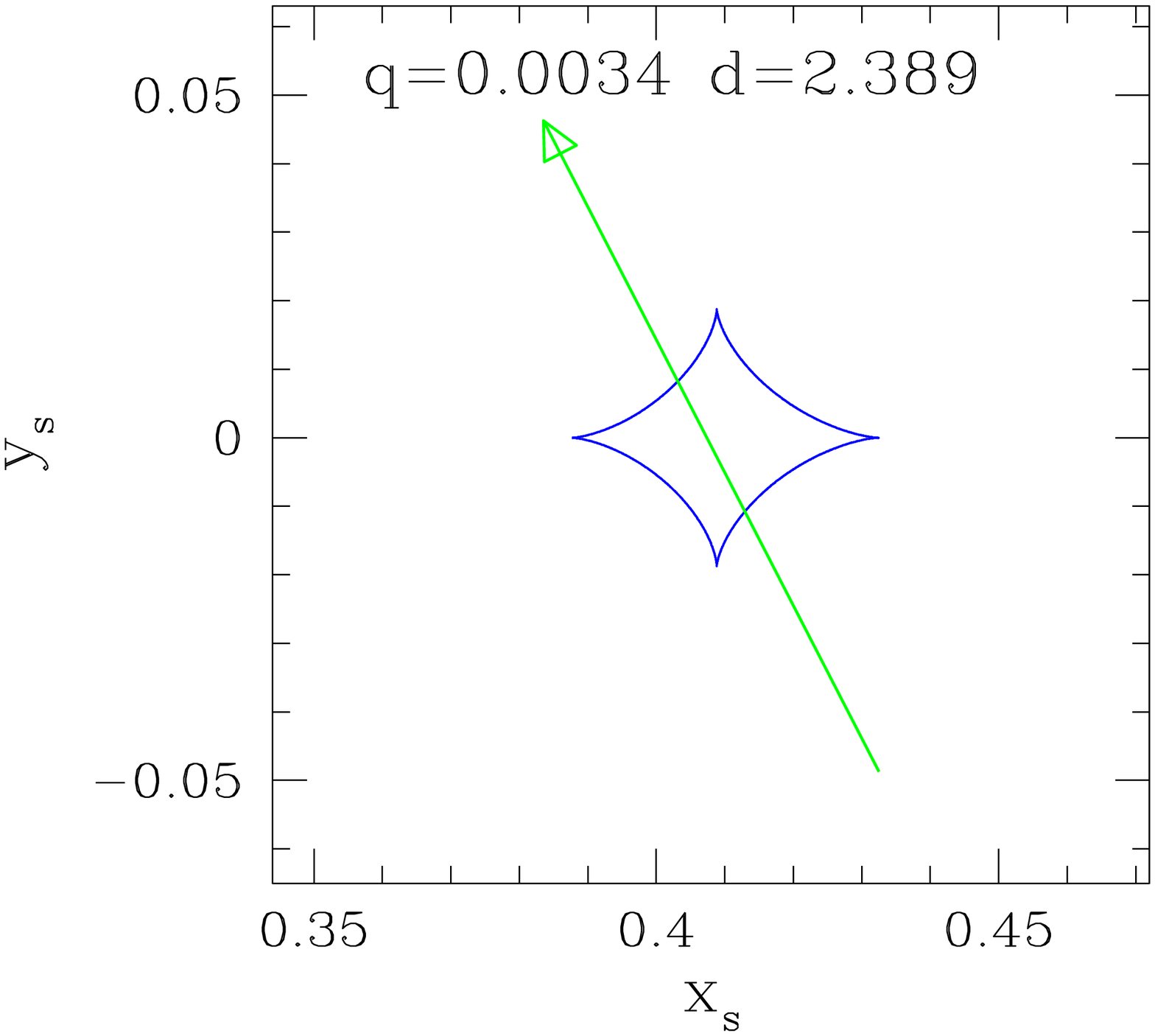} \hfill
\includegraphics[height=62mm,width=63mm]{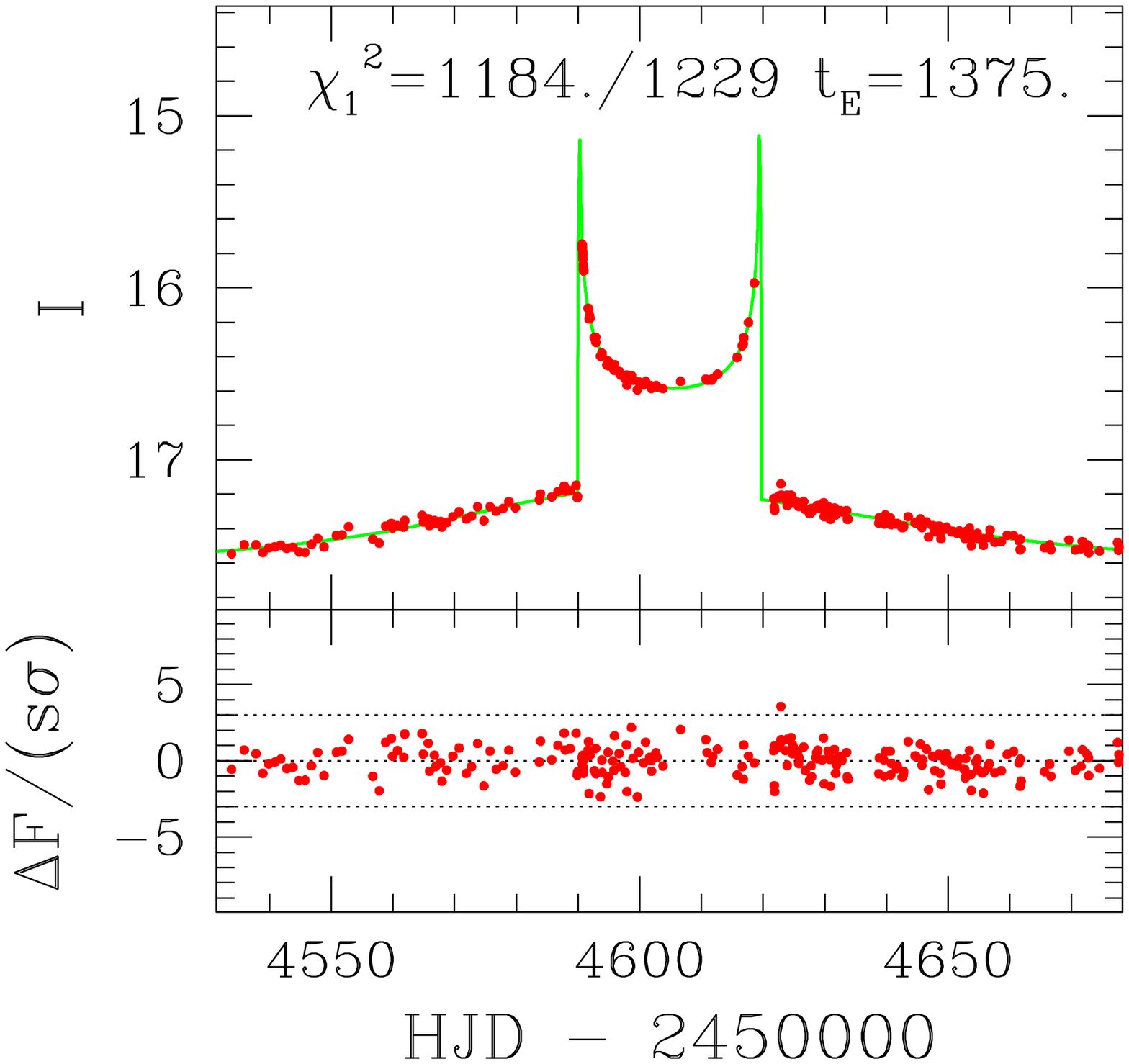}%

}

\noindent\parbox{12.75cm}{
\leftline {\bf OGLE 2008-BLG-125} 

\includegraphics[height=62mm,width=63mm]{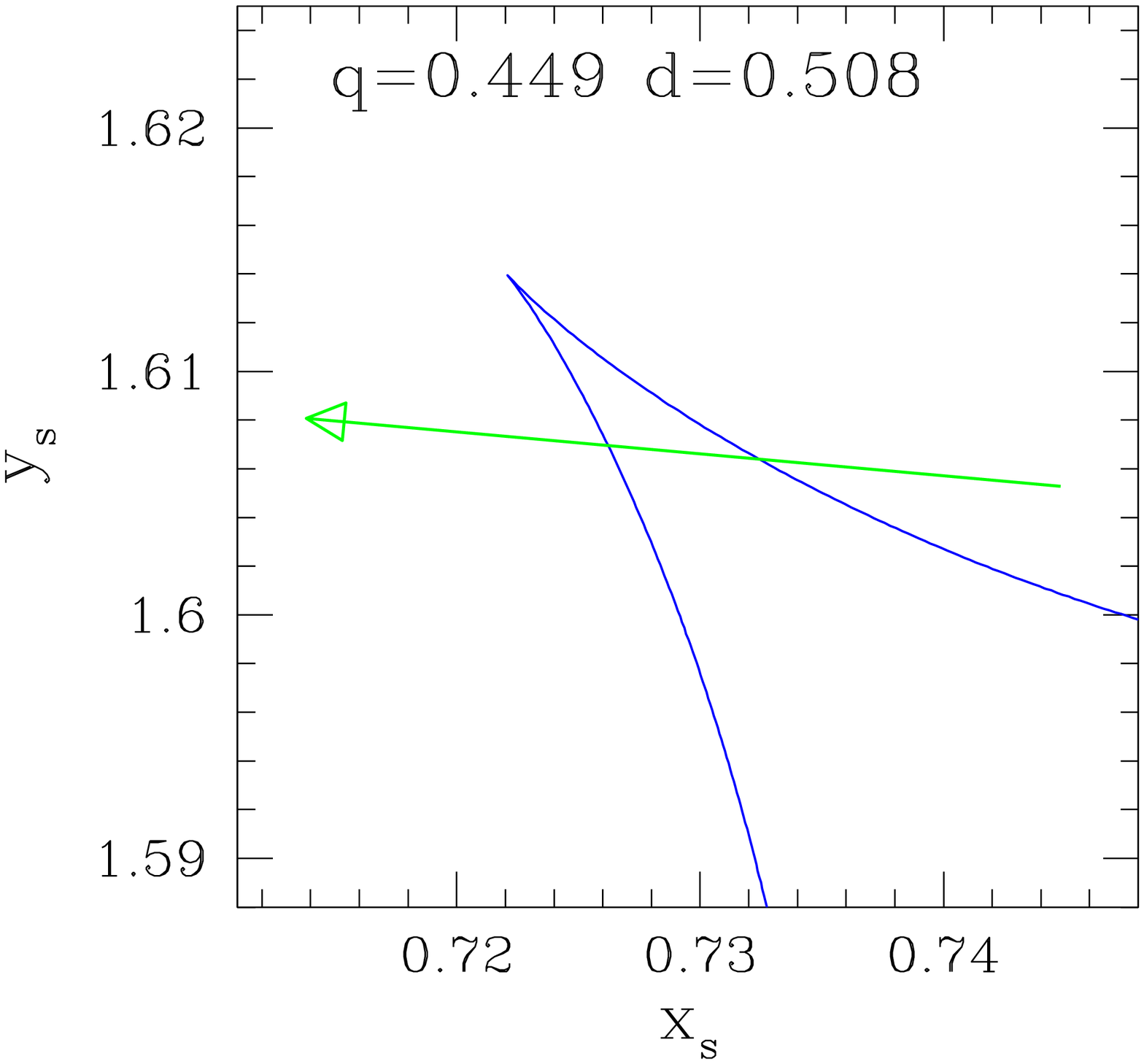} \hfill
\includegraphics[height=62mm,width=63mm]{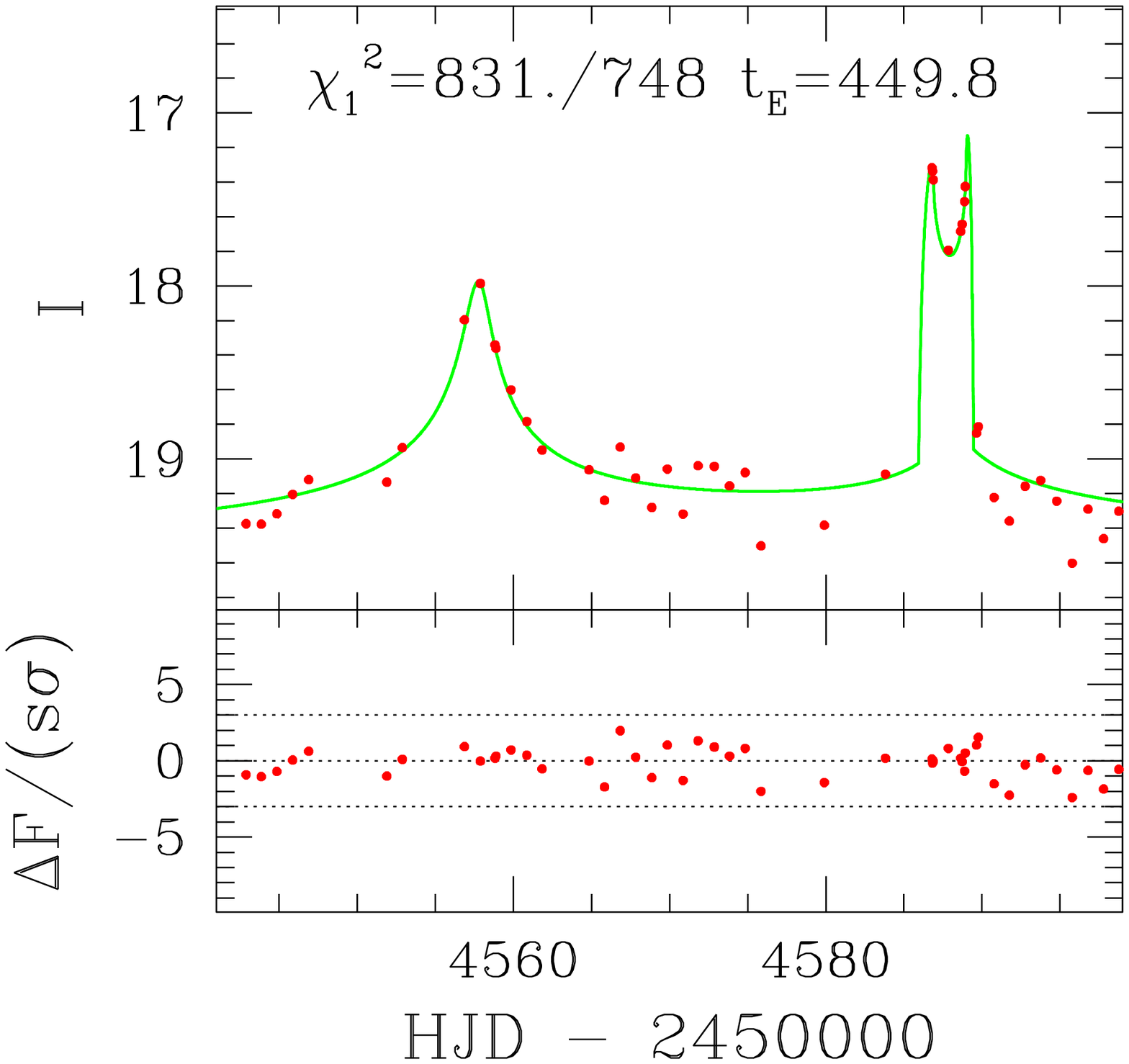}%

}

\noindent\parbox{12.75cm}{
\leftline {\bf OGLE 2008-BLG-243} 

\includegraphics[height=62mm,width=63mm]{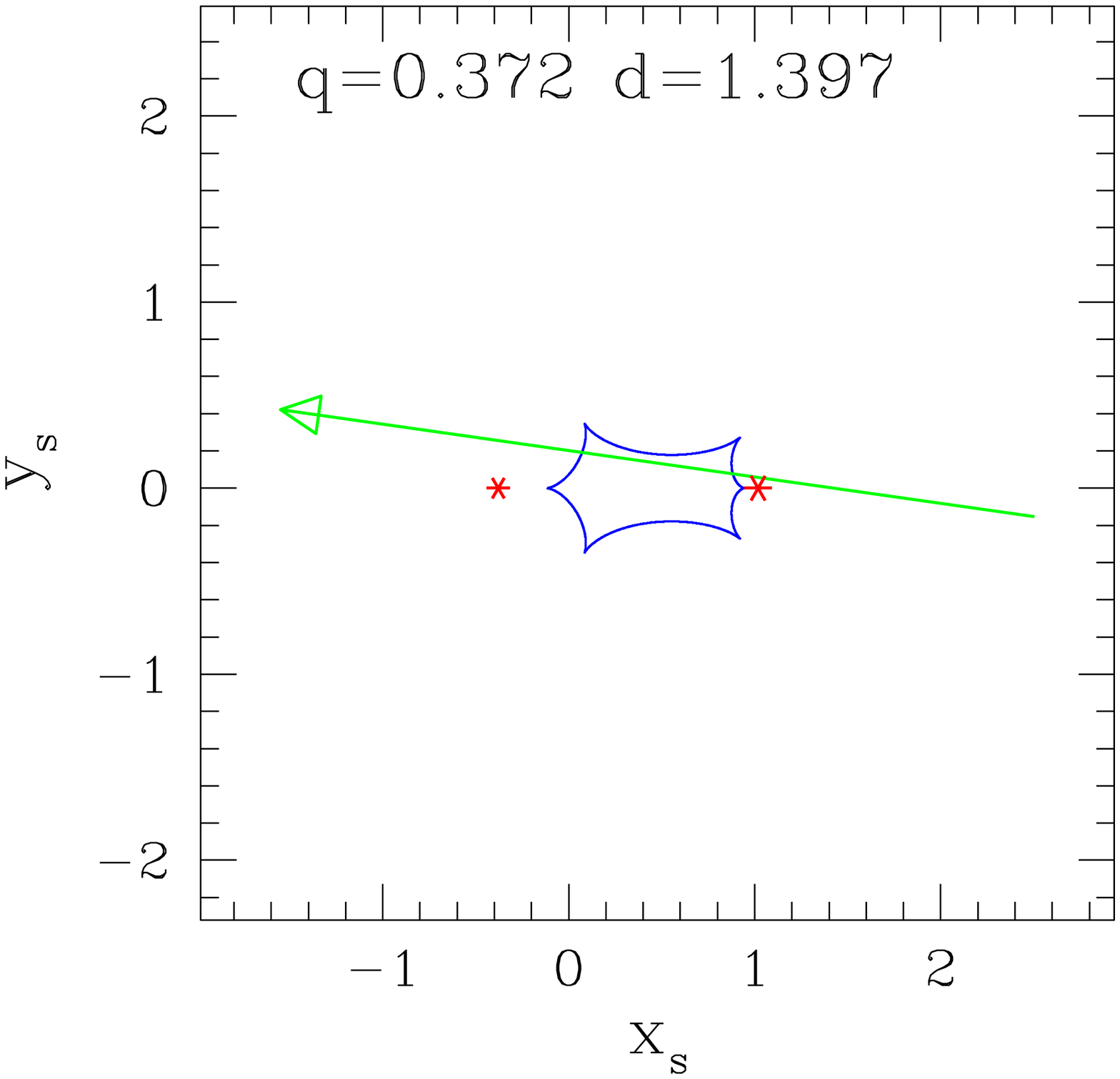} \hfill
\includegraphics[height=62mm,width=63mm]{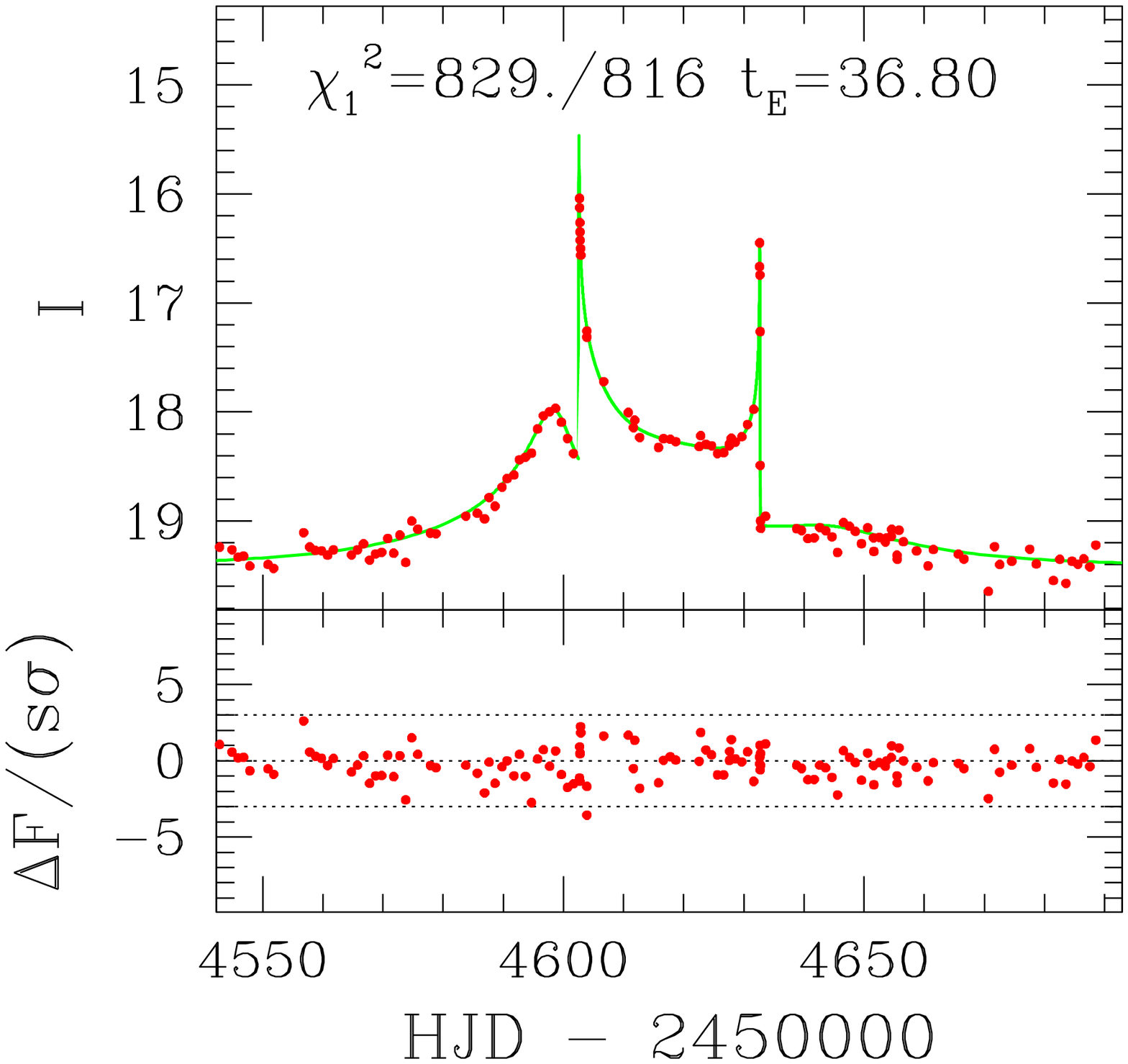}%

}

\noindent\parbox{12.75cm}{
\leftline {\bf OGLE 2008-BLG-263} 

\includegraphics[height=62mm,width=63mm]{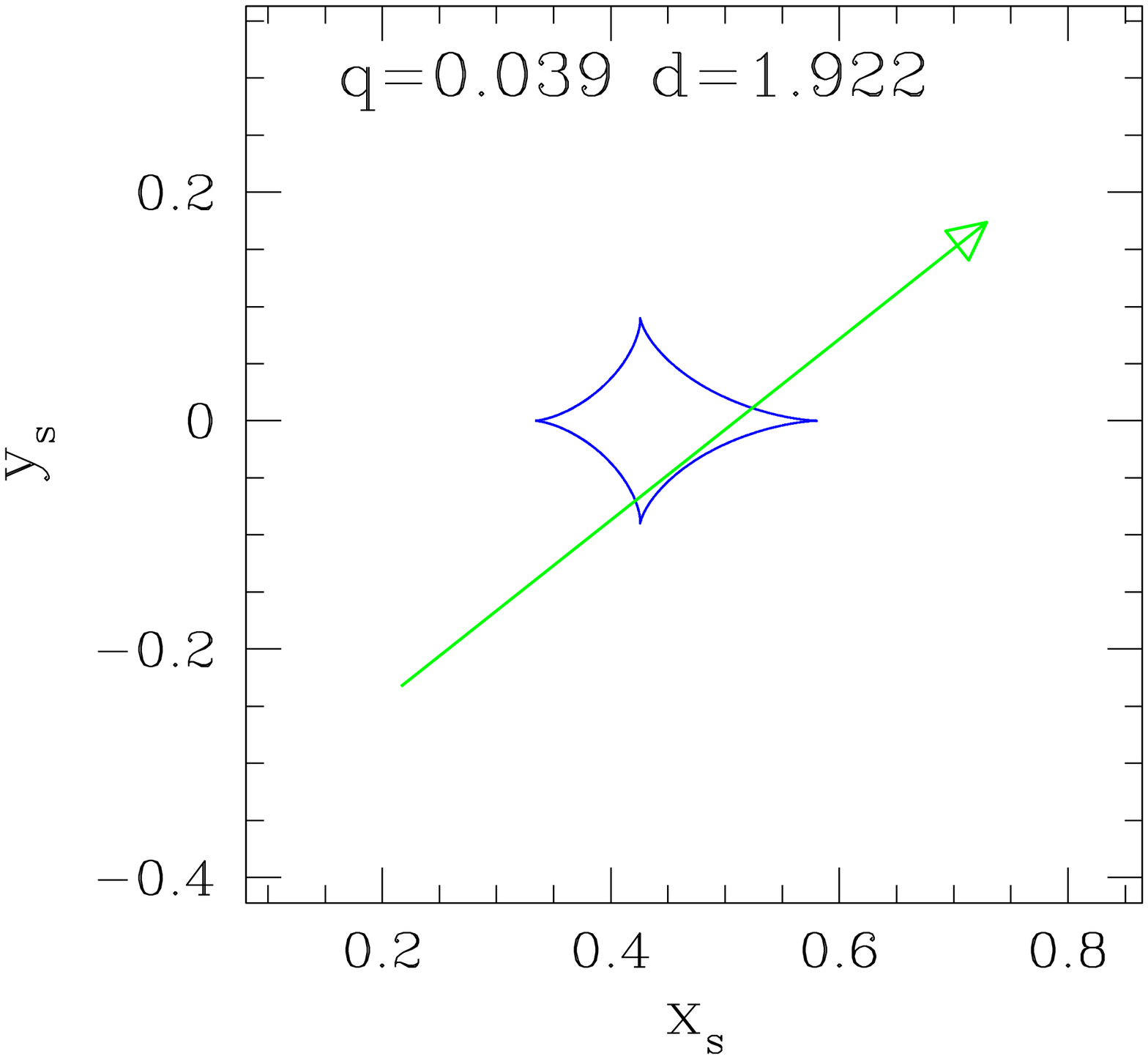} \hfill
\includegraphics[height=62mm,width=63mm]{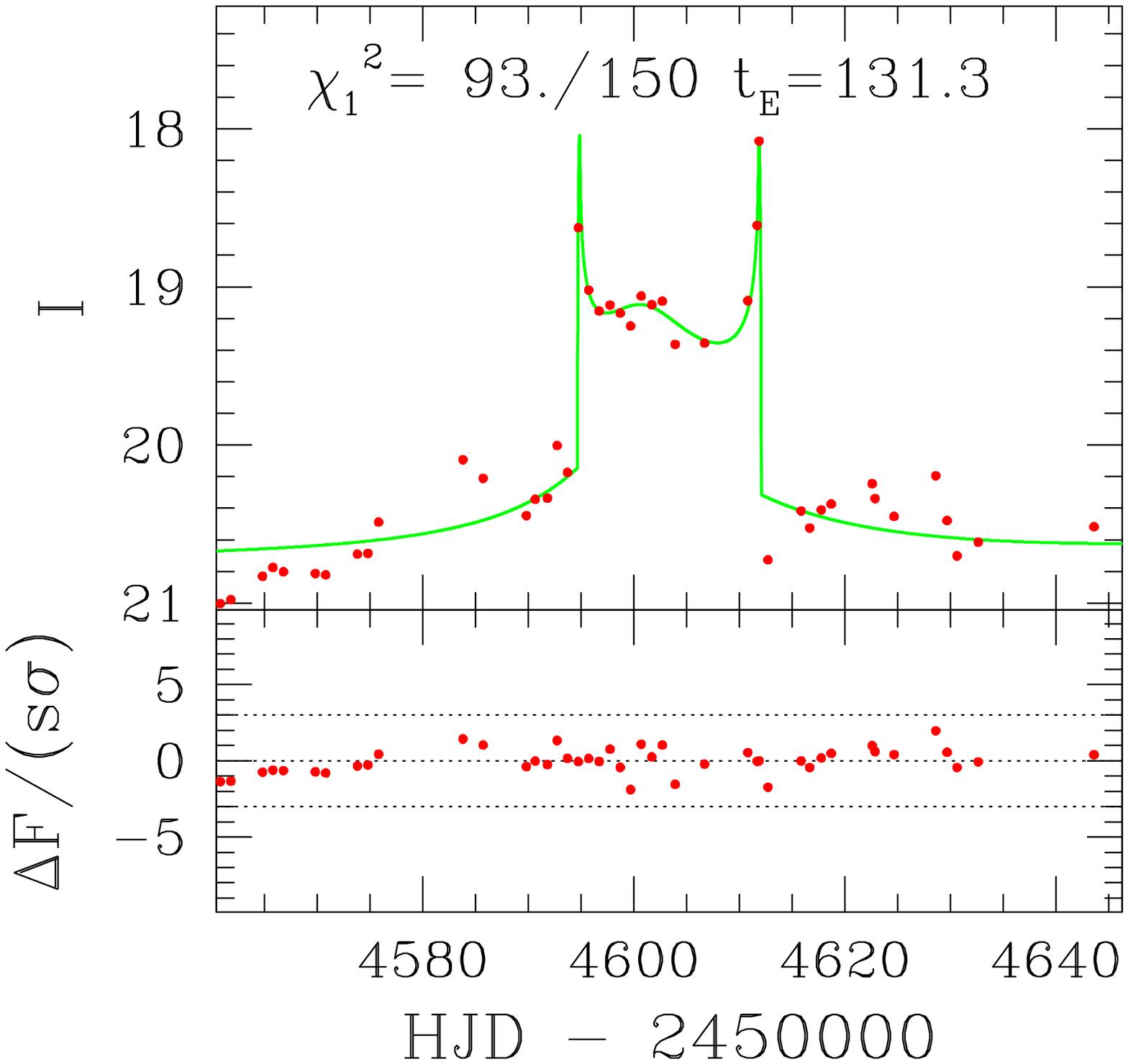}%

}

\noindent\parbox{12.75cm}{
\leftline {\bf OGLE 2008-BLG-330} 

\includegraphics[height=62mm,width=63mm]{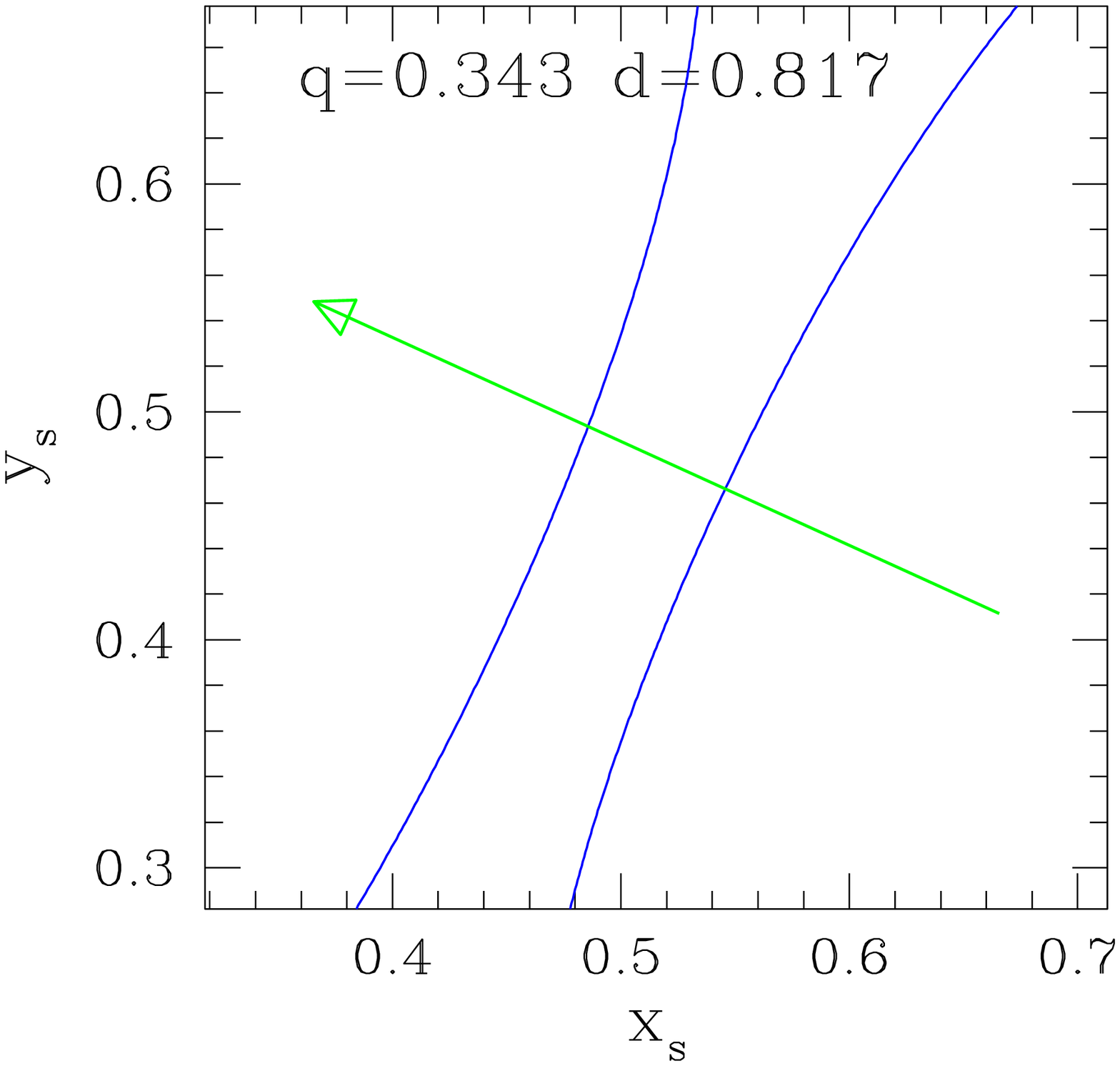} \hfill
\includegraphics[height=62mm,width=63mm]{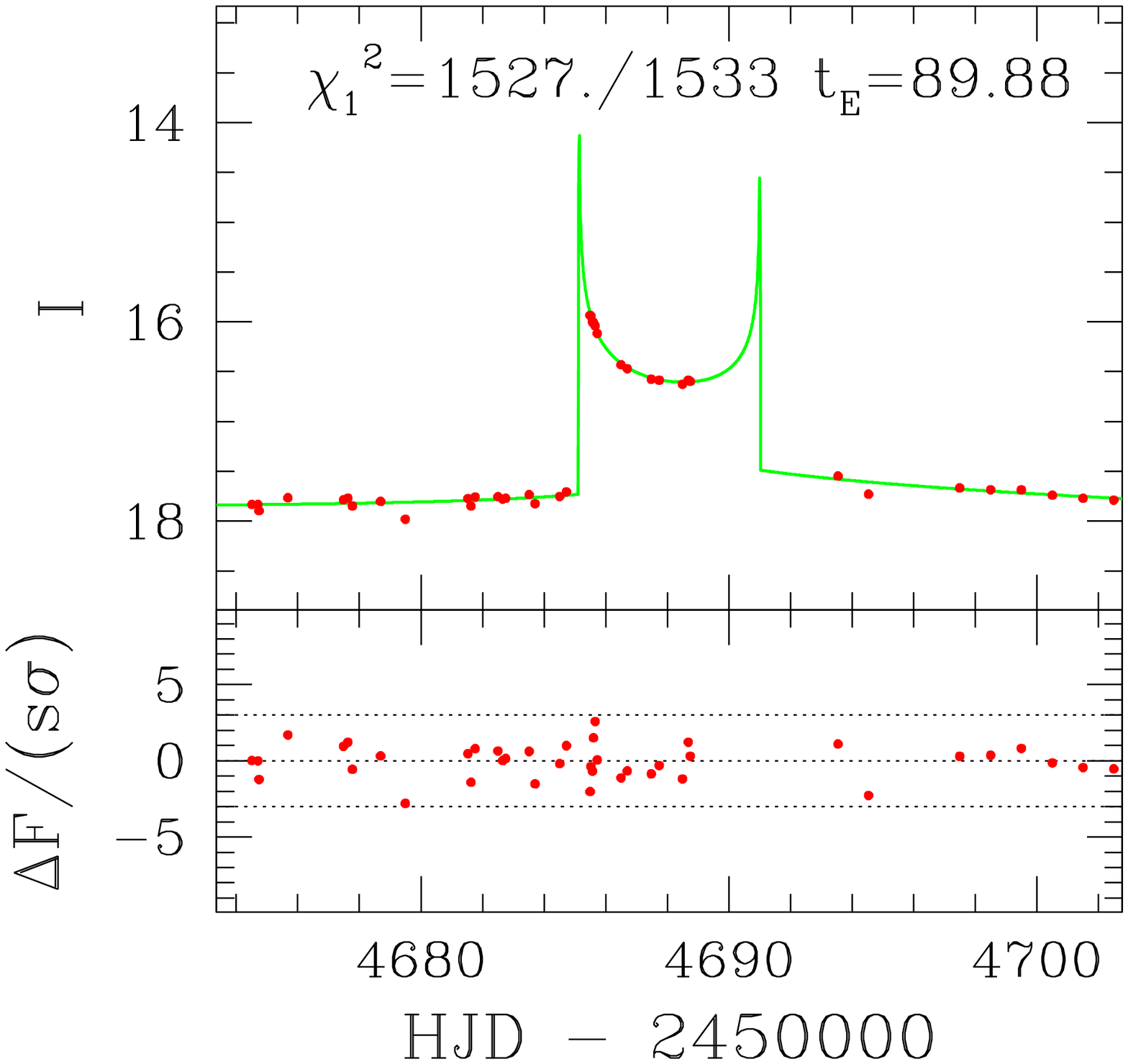}%

}

\noindent\parbox{12.75cm}{
\leftline {\bf OGLE 2008-BLG-355} 

\includegraphics[height=62mm,width=63mm]{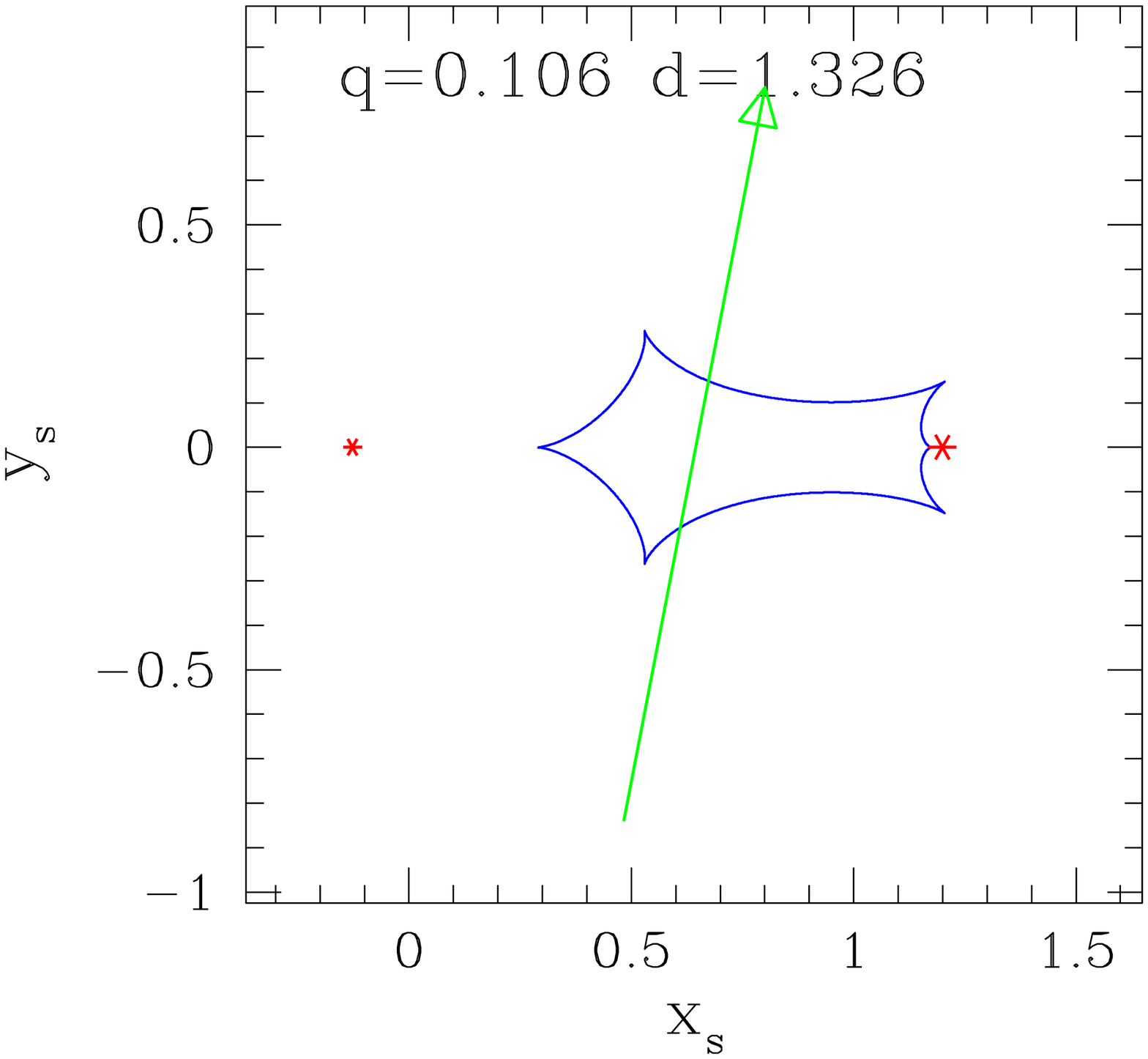} \hfill
\includegraphics[height=62mm,width=63mm]{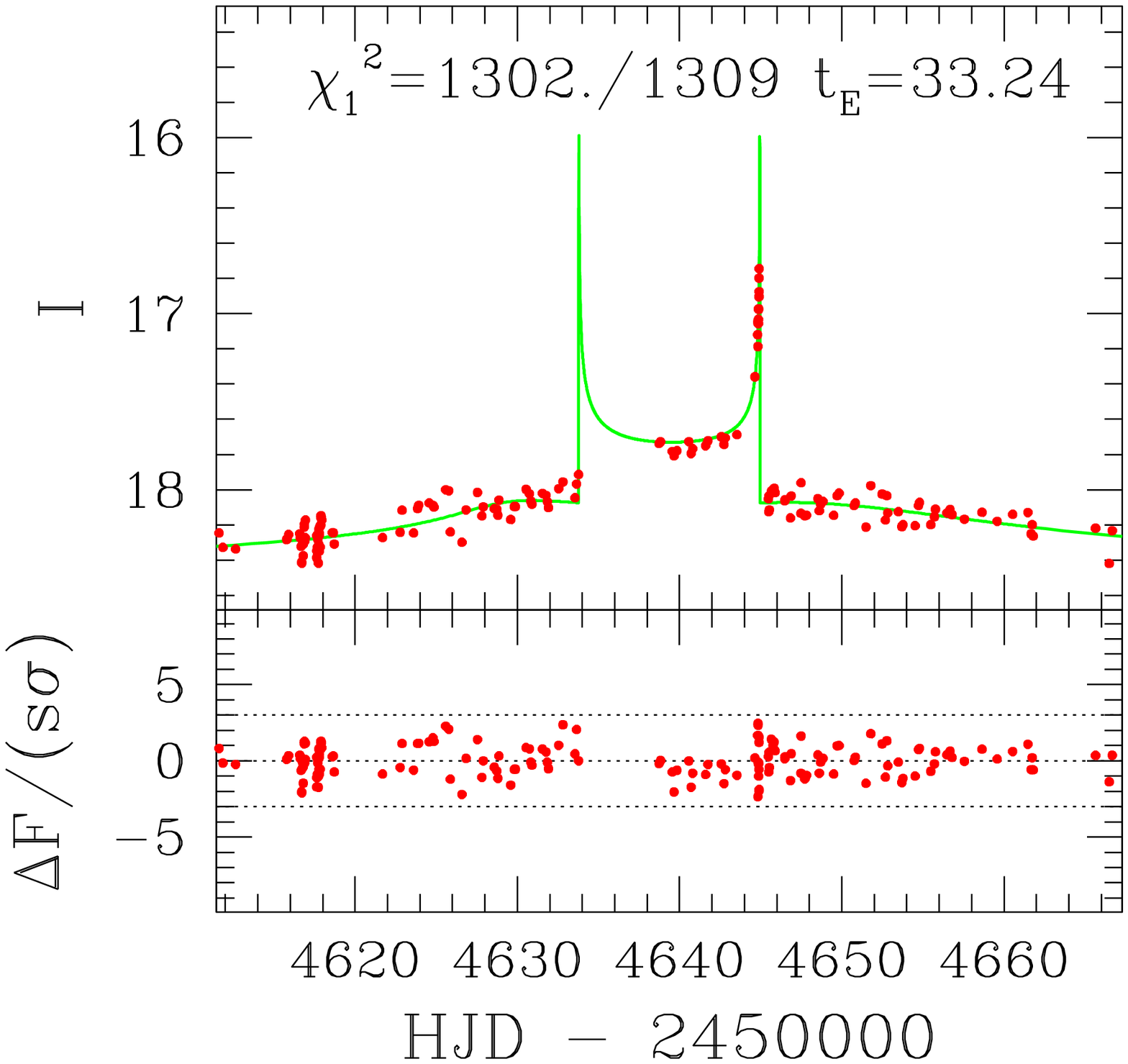}%

}

\noindent\parbox{12.75cm}{
\leftline {\bf OGLE 2008-BLG-493} 

\includegraphics[height=62mm,width=63mm]{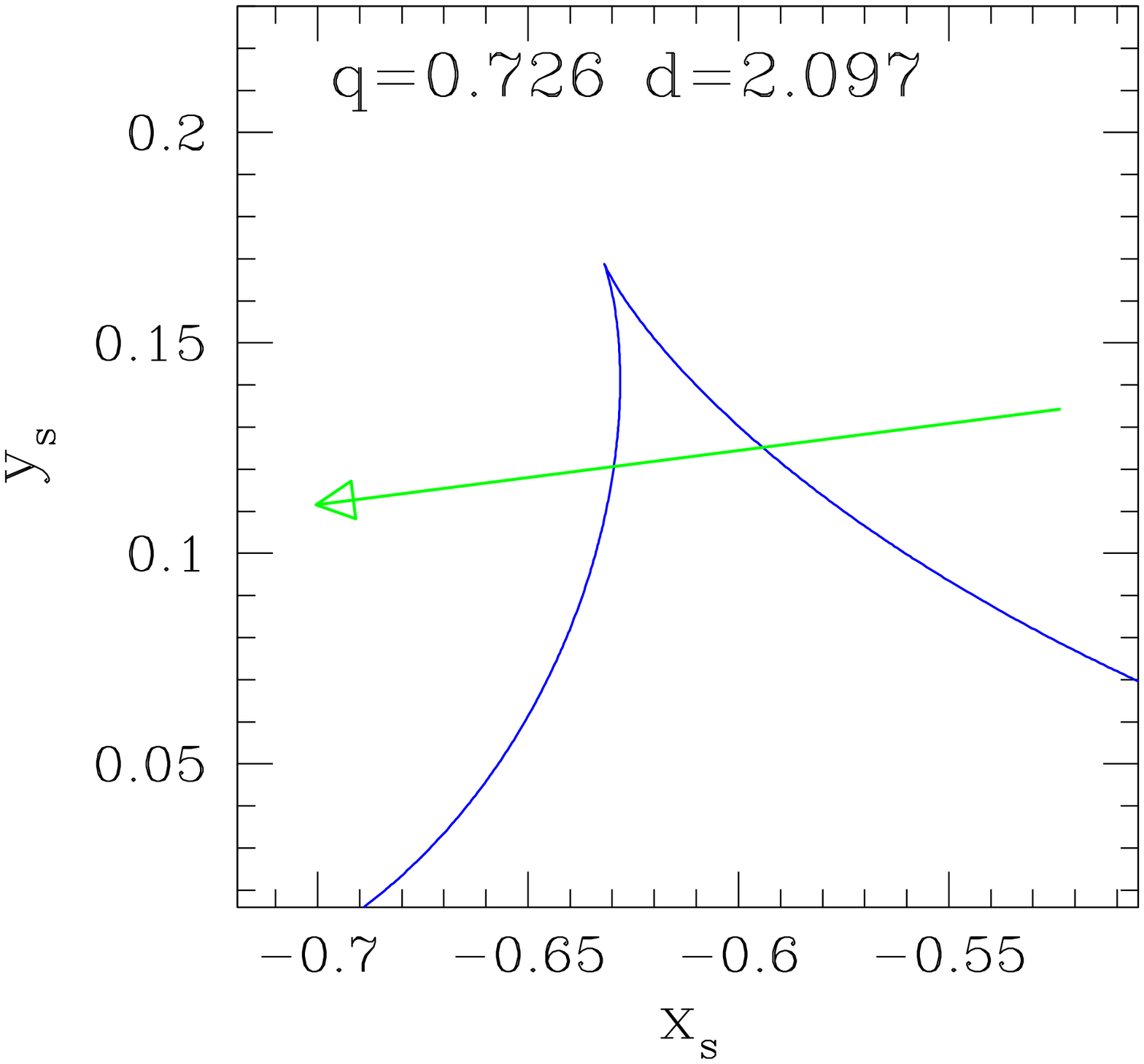} \hfill
\includegraphics[height=62mm,width=63mm]{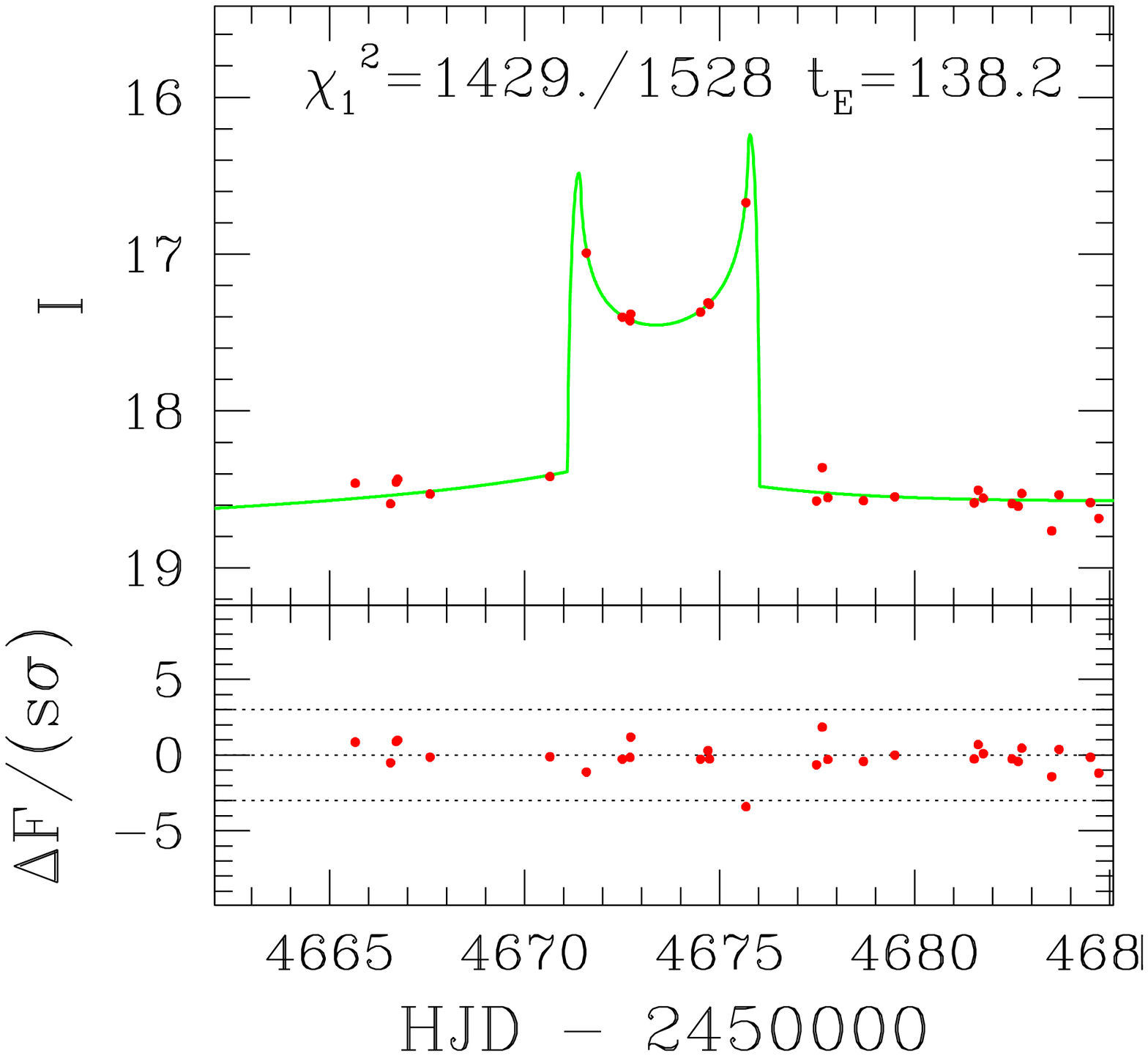}%

}

\noindent\parbox{12.75cm}{
\leftline {\bf OGLE 2008-BLG-513} 

\includegraphics[height=62mm,width=63mm]{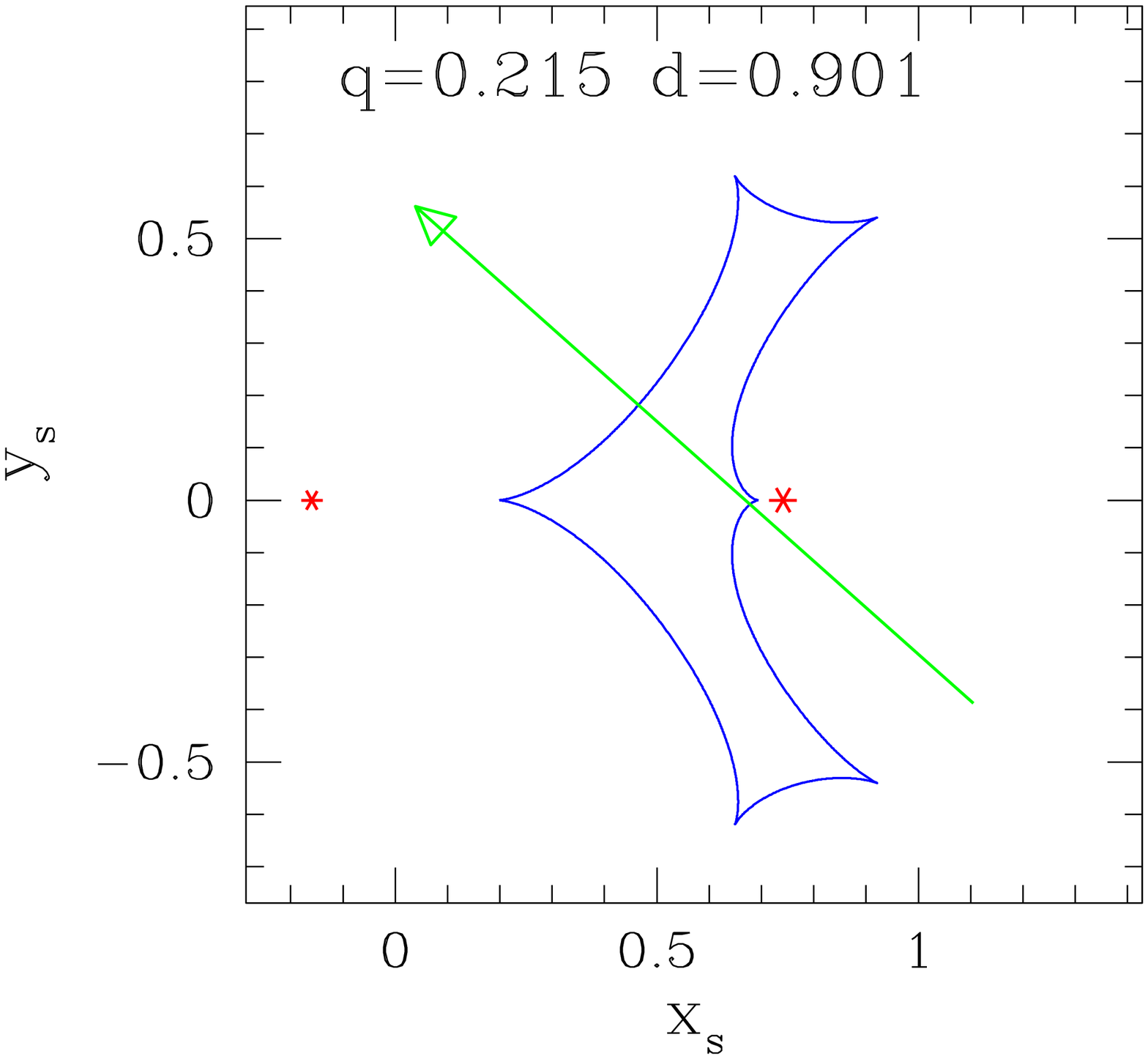} \hfill
\includegraphics[height=62mm,width=63mm]{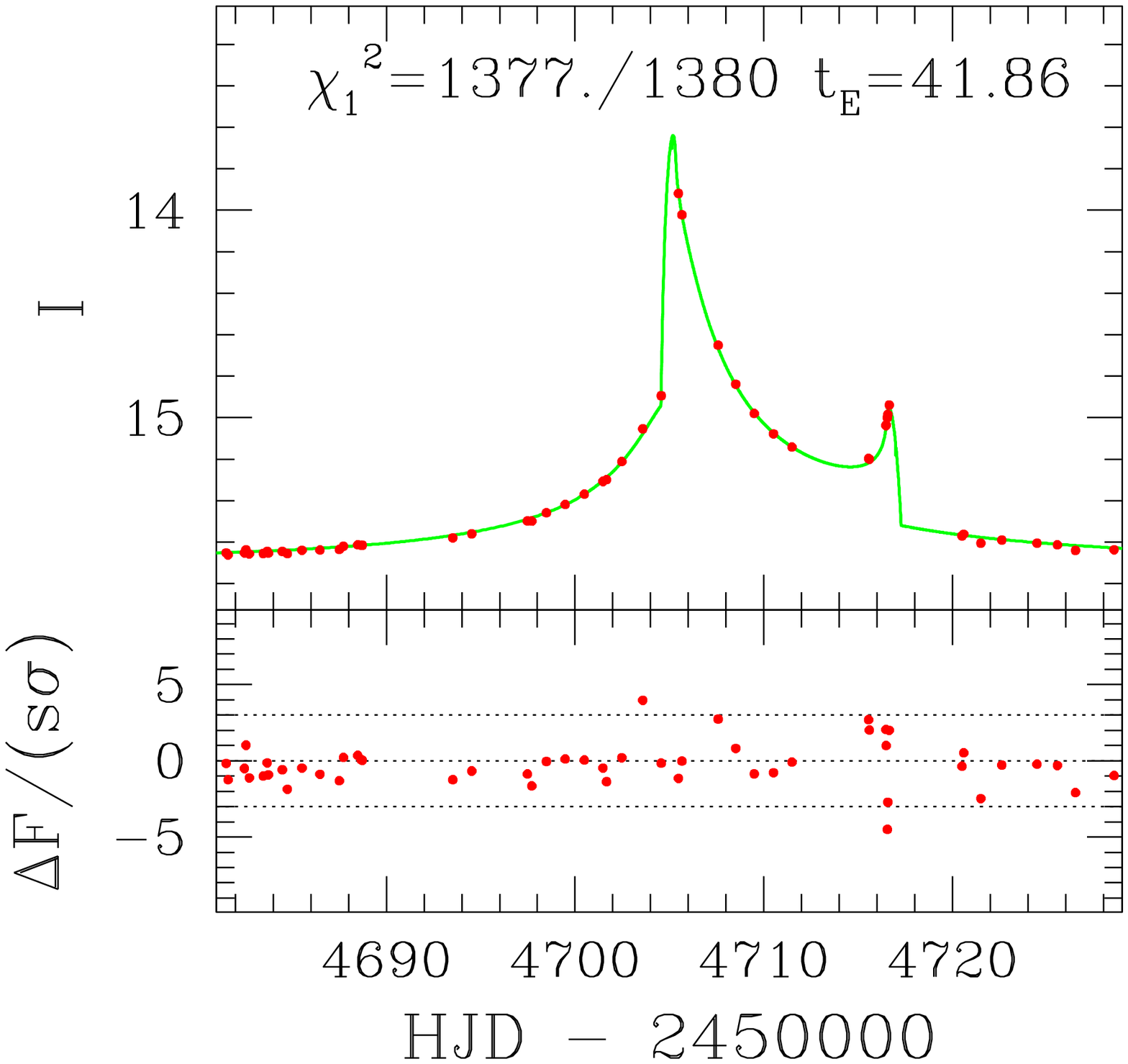}%

}

\noindent\parbox{12.75cm}{
\leftline {{\bf OGLE 2008-BLG-559} (1st model)}

\includegraphics[height=62mm,width=63mm]{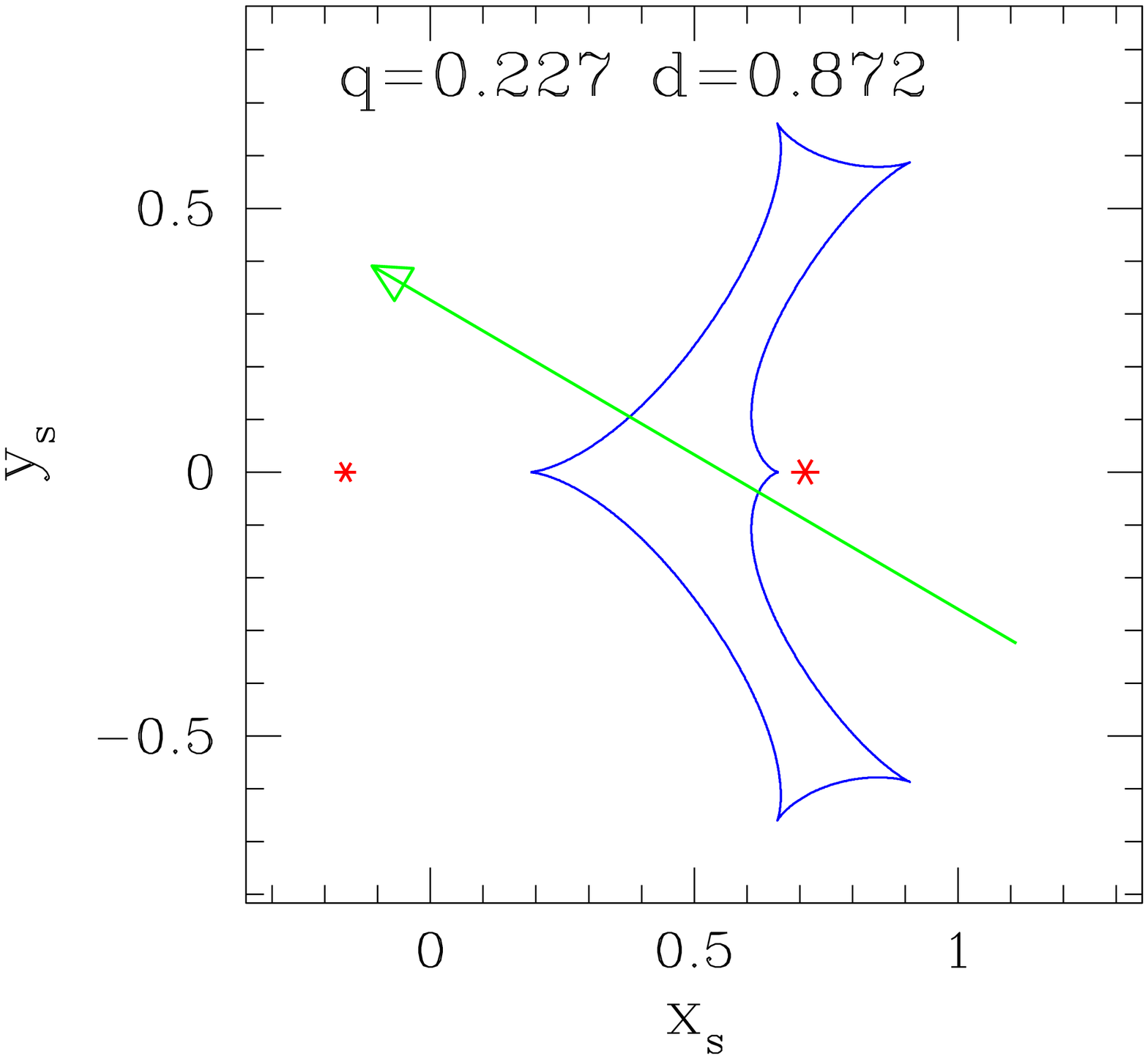} \hfill
\includegraphics[height=62mm,width=63mm]{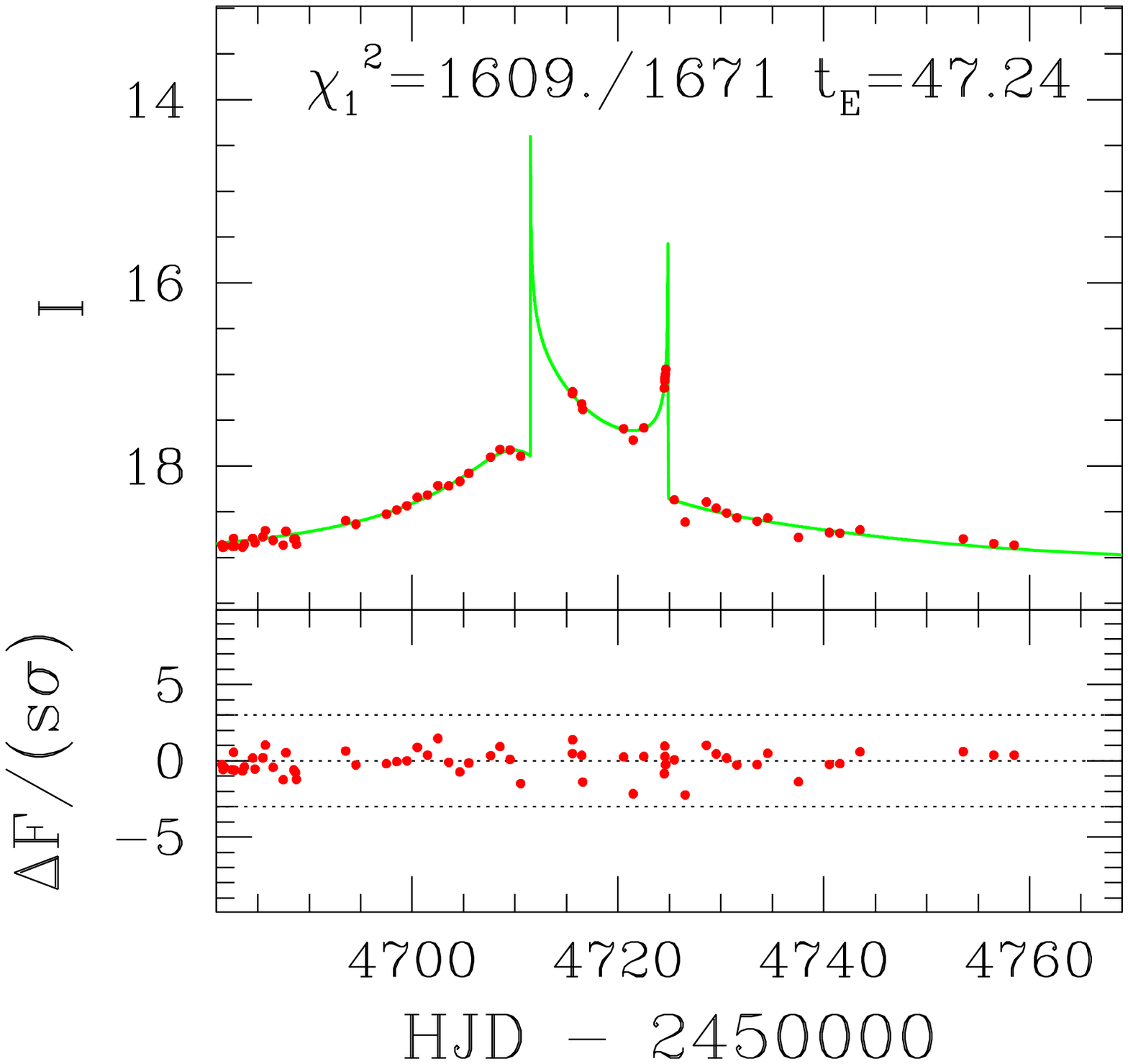}%

}

\noindent\parbox{12.75cm}{
\leftline {{\bf OGLE 2008-BLG-559} (2nd model)}

\includegraphics[height=62mm,width=63mm]{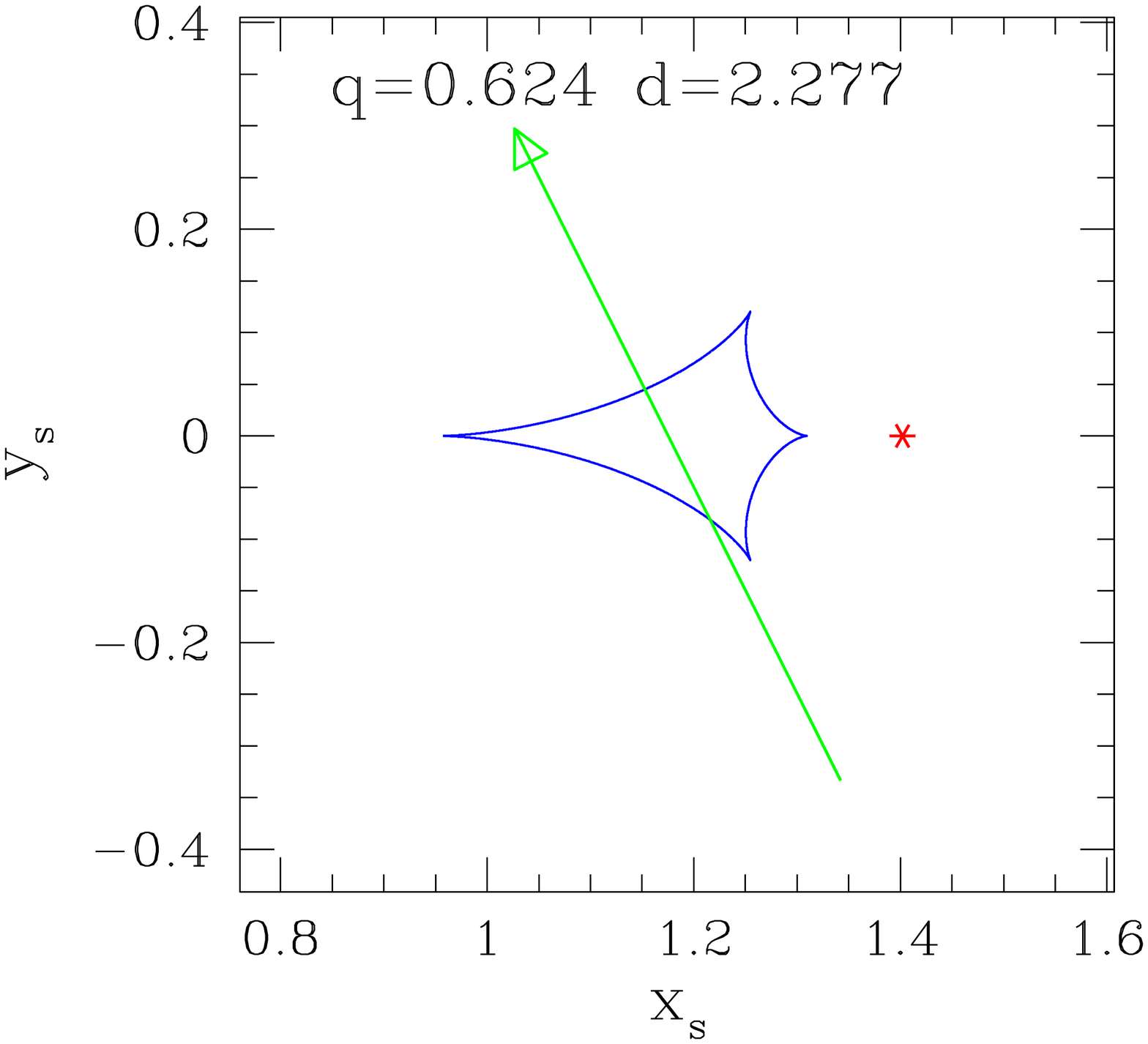} \hfill
\includegraphics[height=62mm,width=63mm]{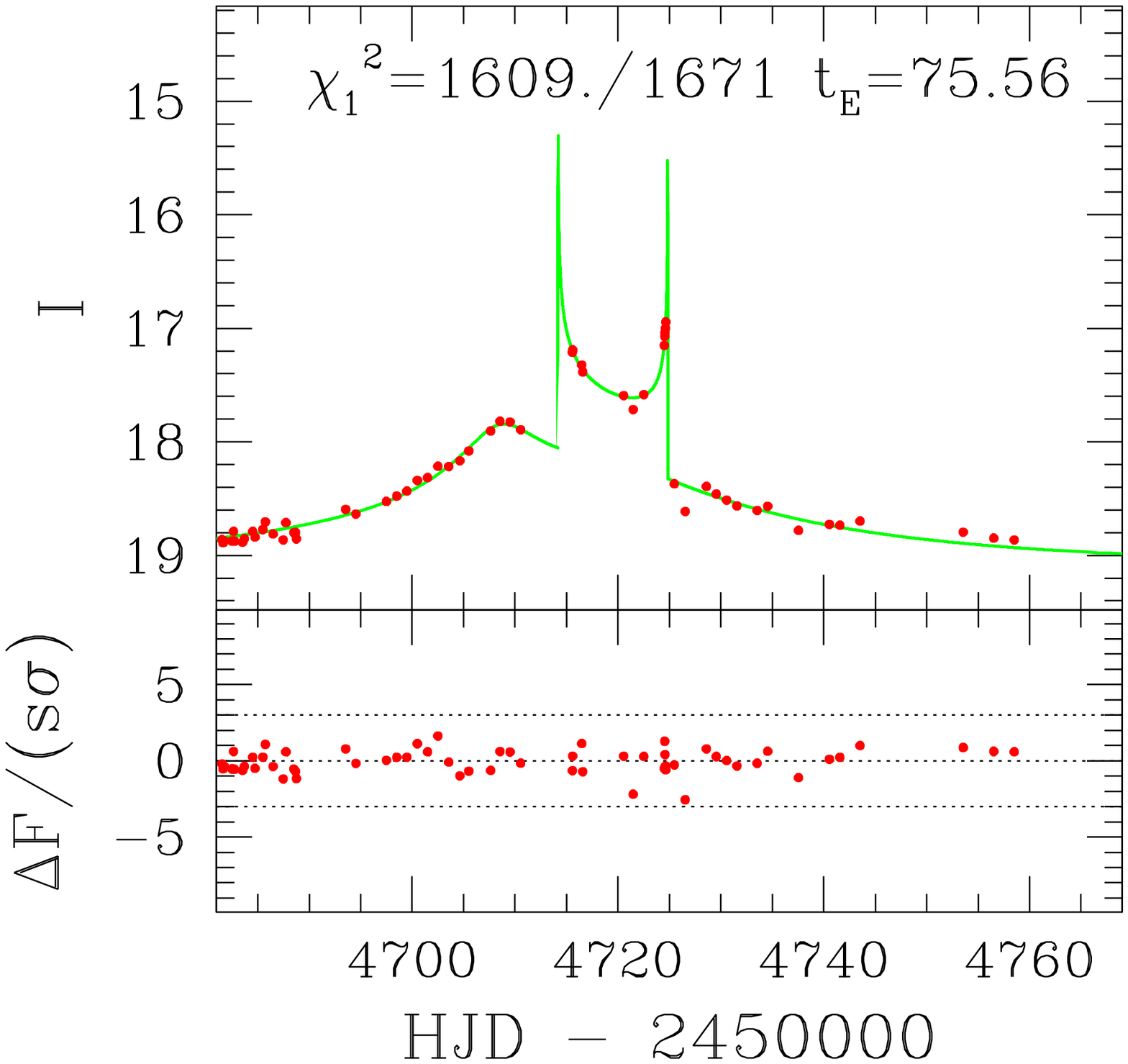}%

}

\noindent\parbox{12.75cm}{
\leftline {{\bf OGLE 2008-BLG-584} (1st model)}

\includegraphics[height=62mm,width=63mm]{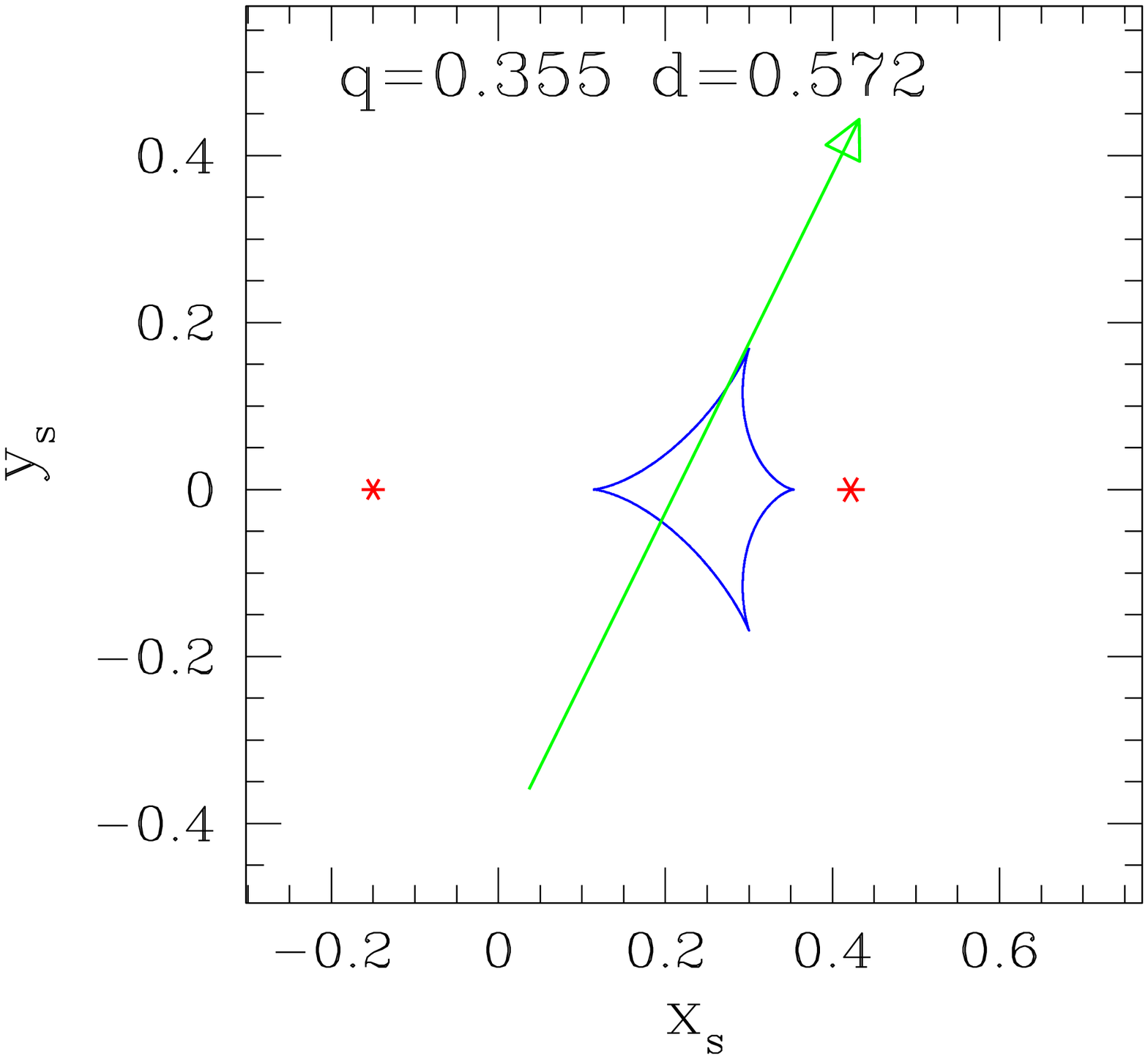} \hfill
\includegraphics[height=62mm,width=63mm]{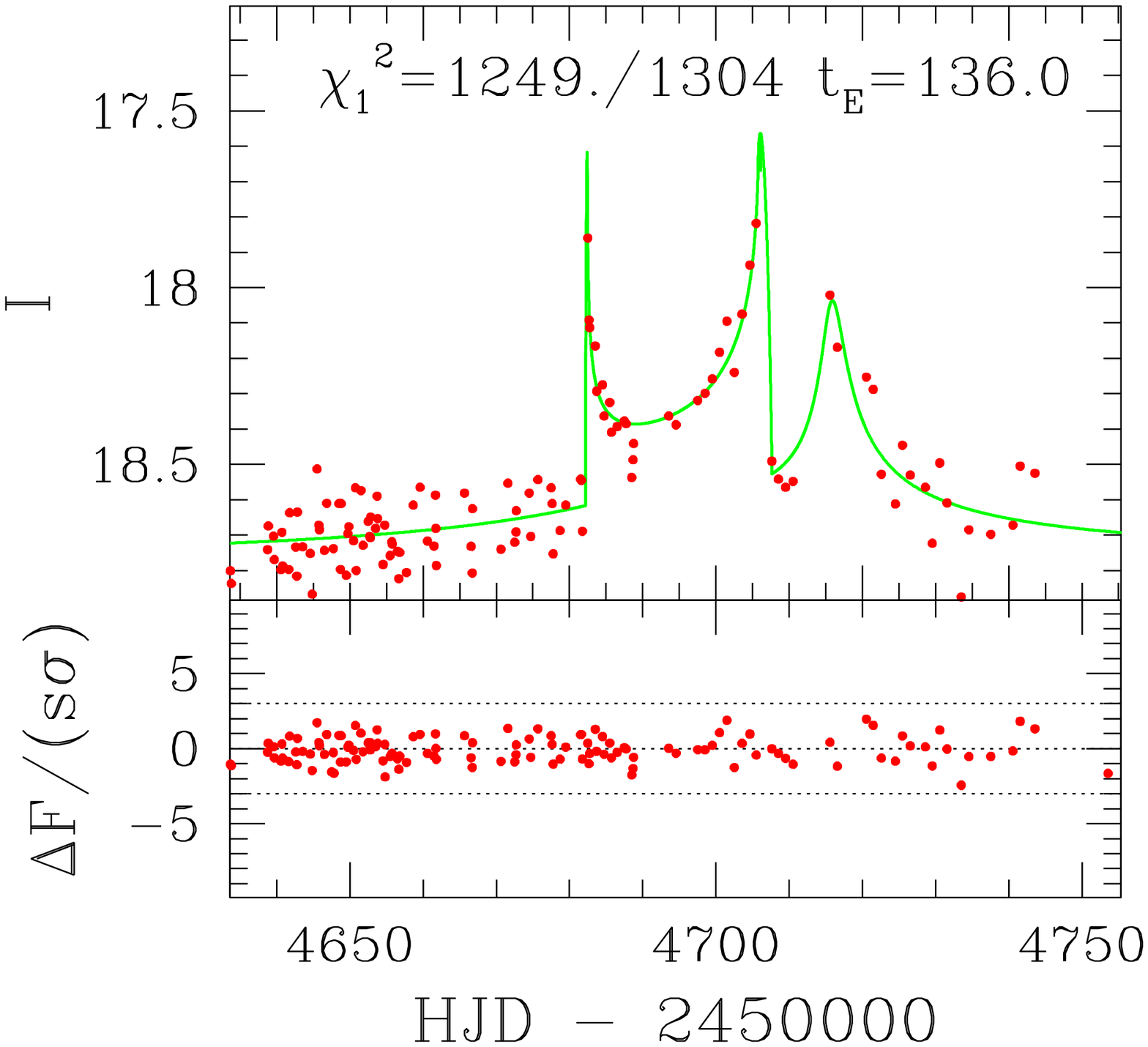}%

}

\noindent\parbox{12.75cm}{
\leftline {{\bf OGLE 2008-BLG-584} (2nd model)}

\includegraphics[height=62mm,width=63mm]{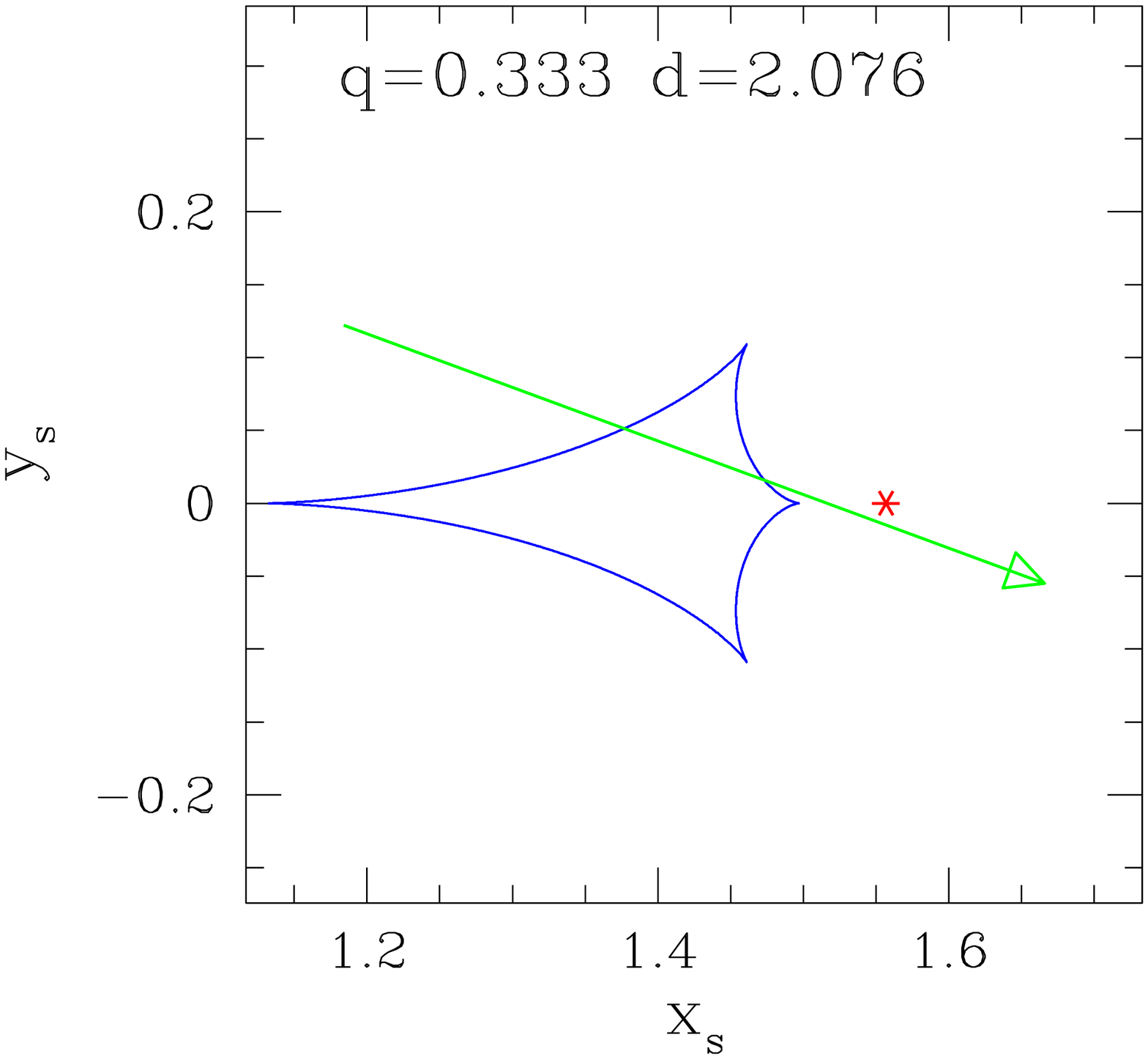} \hfill
\includegraphics[height=62mm,width=63mm]{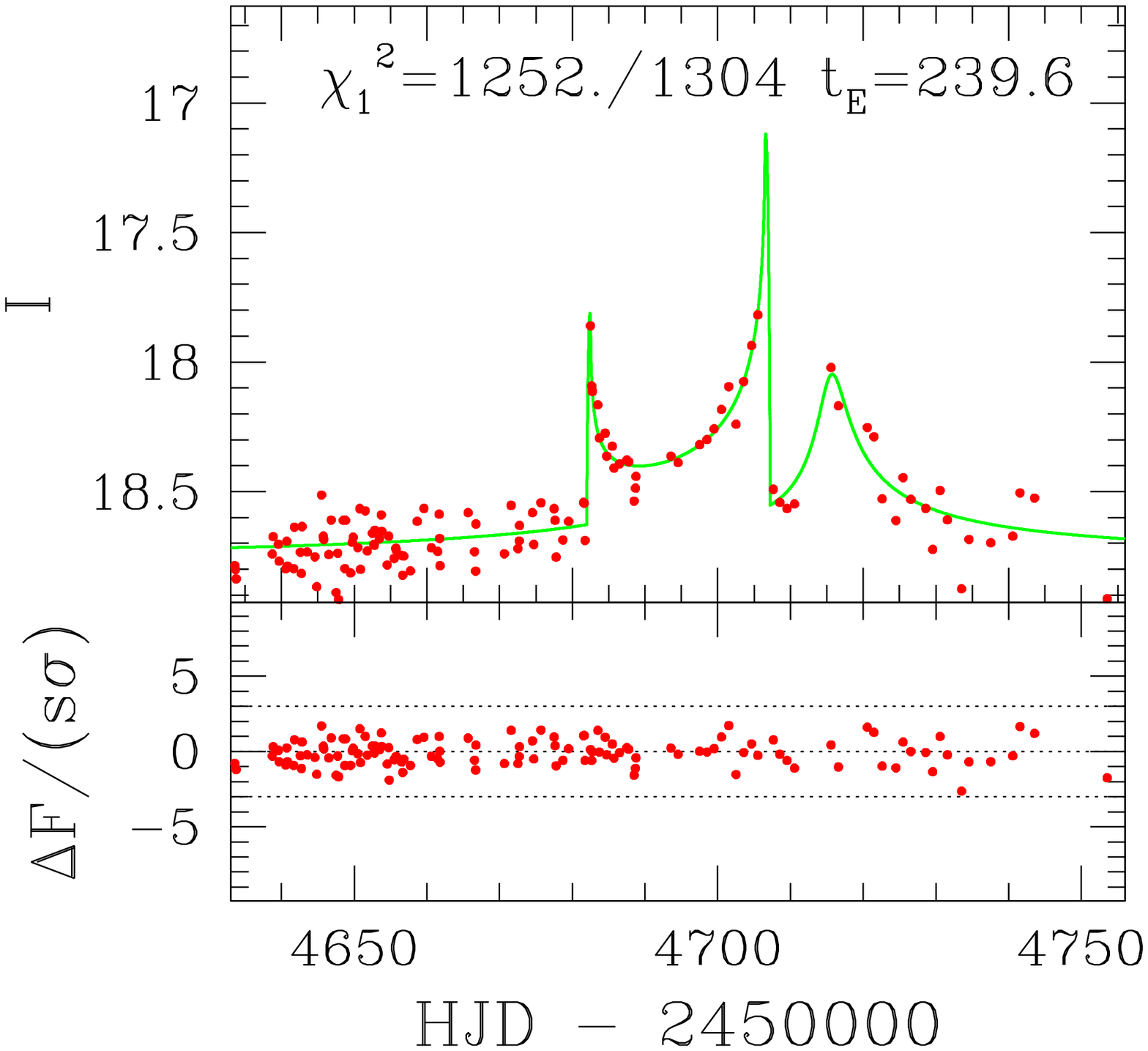}%

}

\noindent\parbox{12.75cm}{
\leftline {{\bf OGLE 2008-BLG-584} (3rd model)}

\includegraphics[height=62mm,width=63mm]{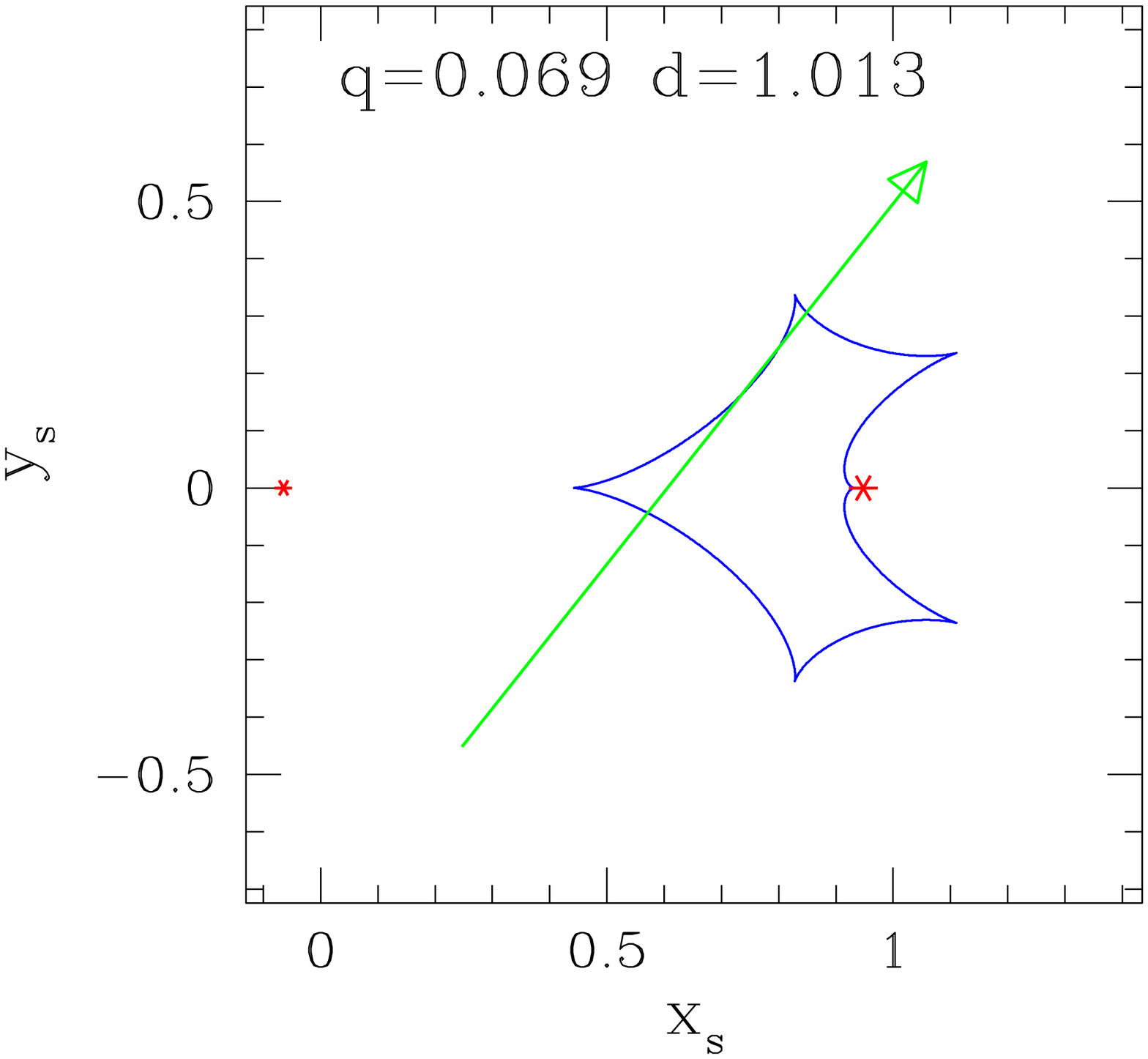} \hfill
\includegraphics[height=62mm,width=63mm]{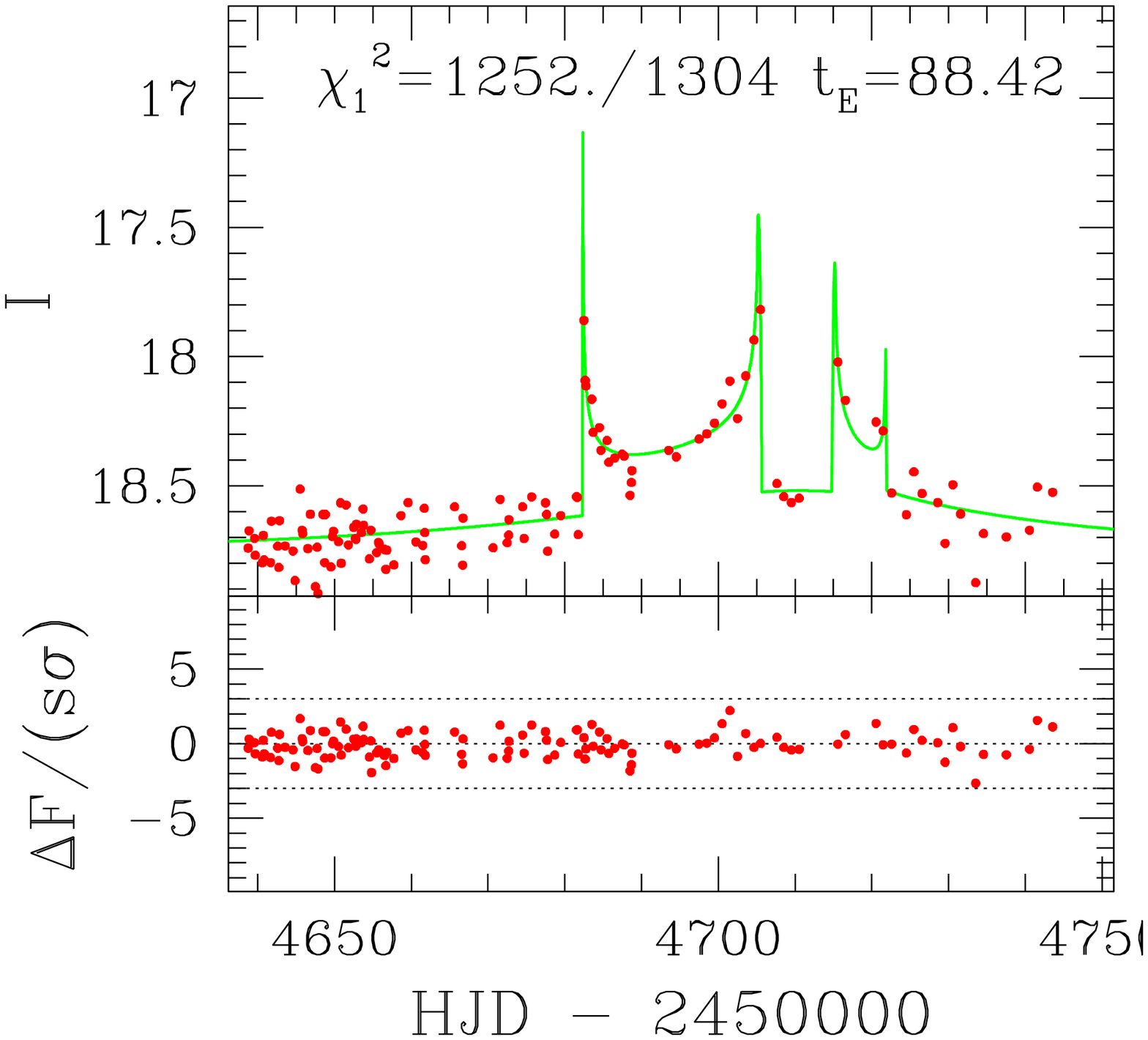}%

}

\noindent\parbox{12.75cm}{
\leftline {{\bf OGLE 2008-BLG-592} (1st model)}

\includegraphics[height=62mm,width=63mm]{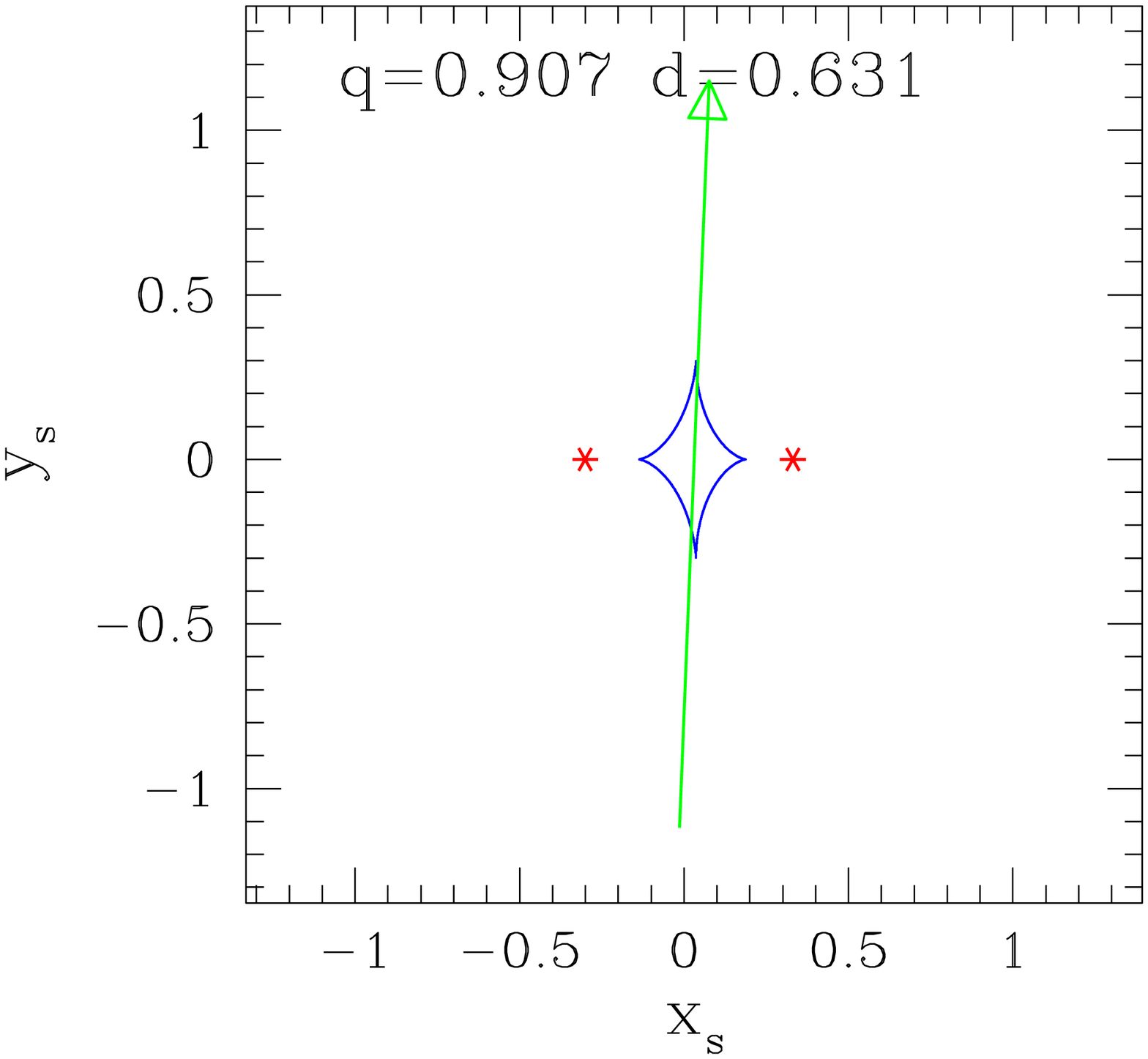} \hfill
\includegraphics[height=62mm,width=63mm]{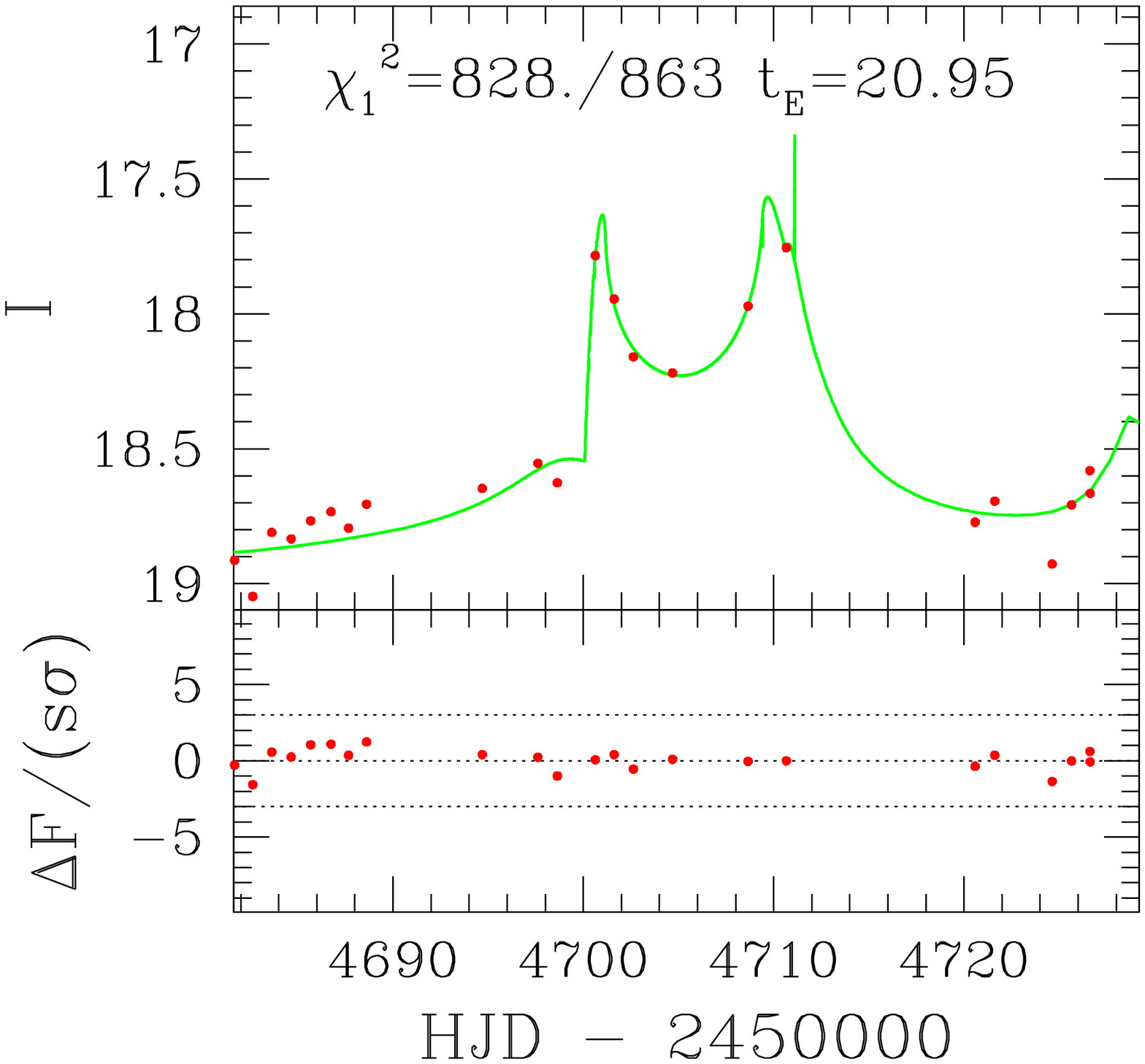}%

}

\noindent\parbox{12.75cm}{
\leftline {{\bf OGLE 2008-BLG-592} (2nd model)}

\includegraphics[height=62mm,width=63mm]{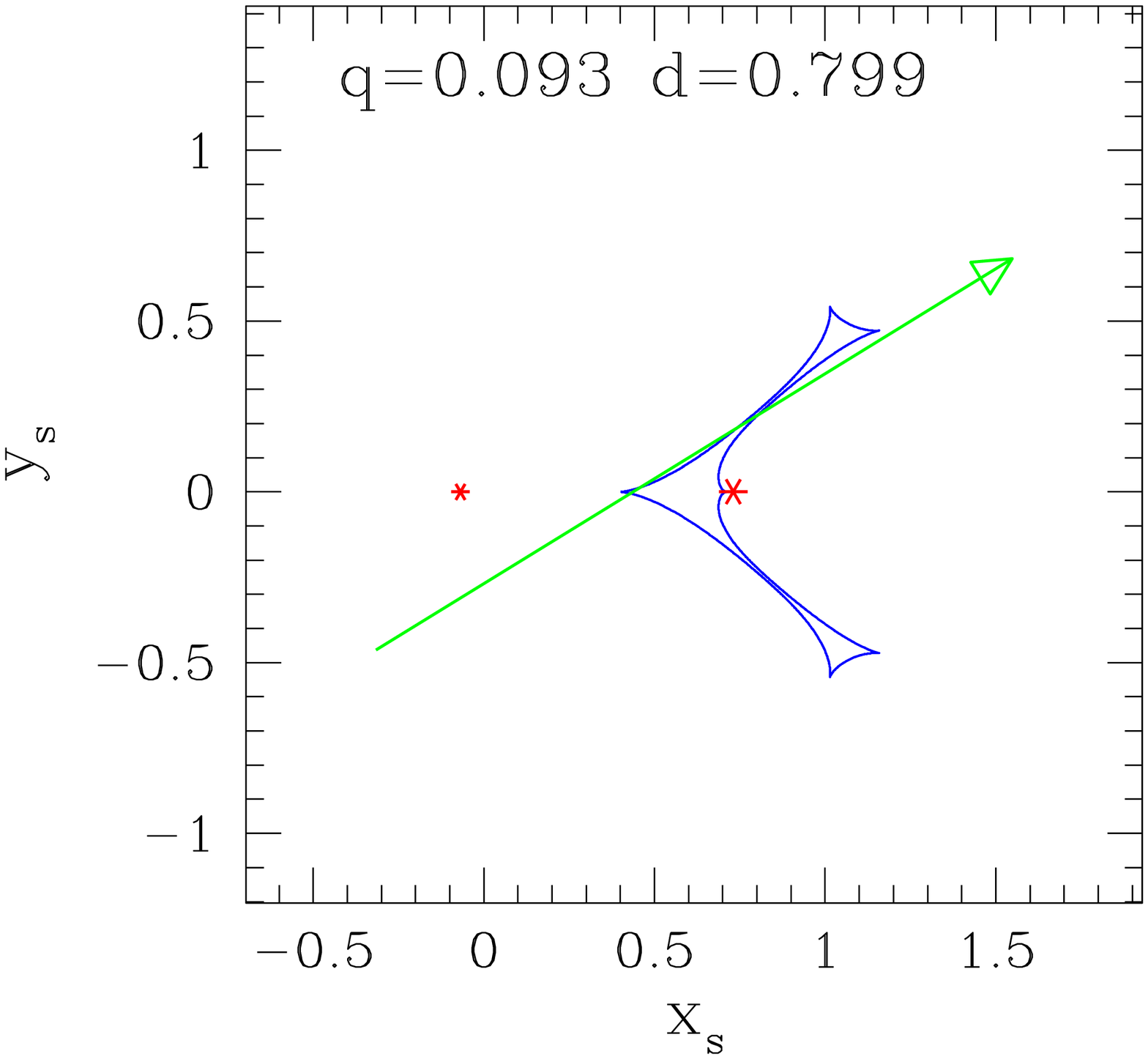} \hfill
\includegraphics[height=62mm,width=63mm]{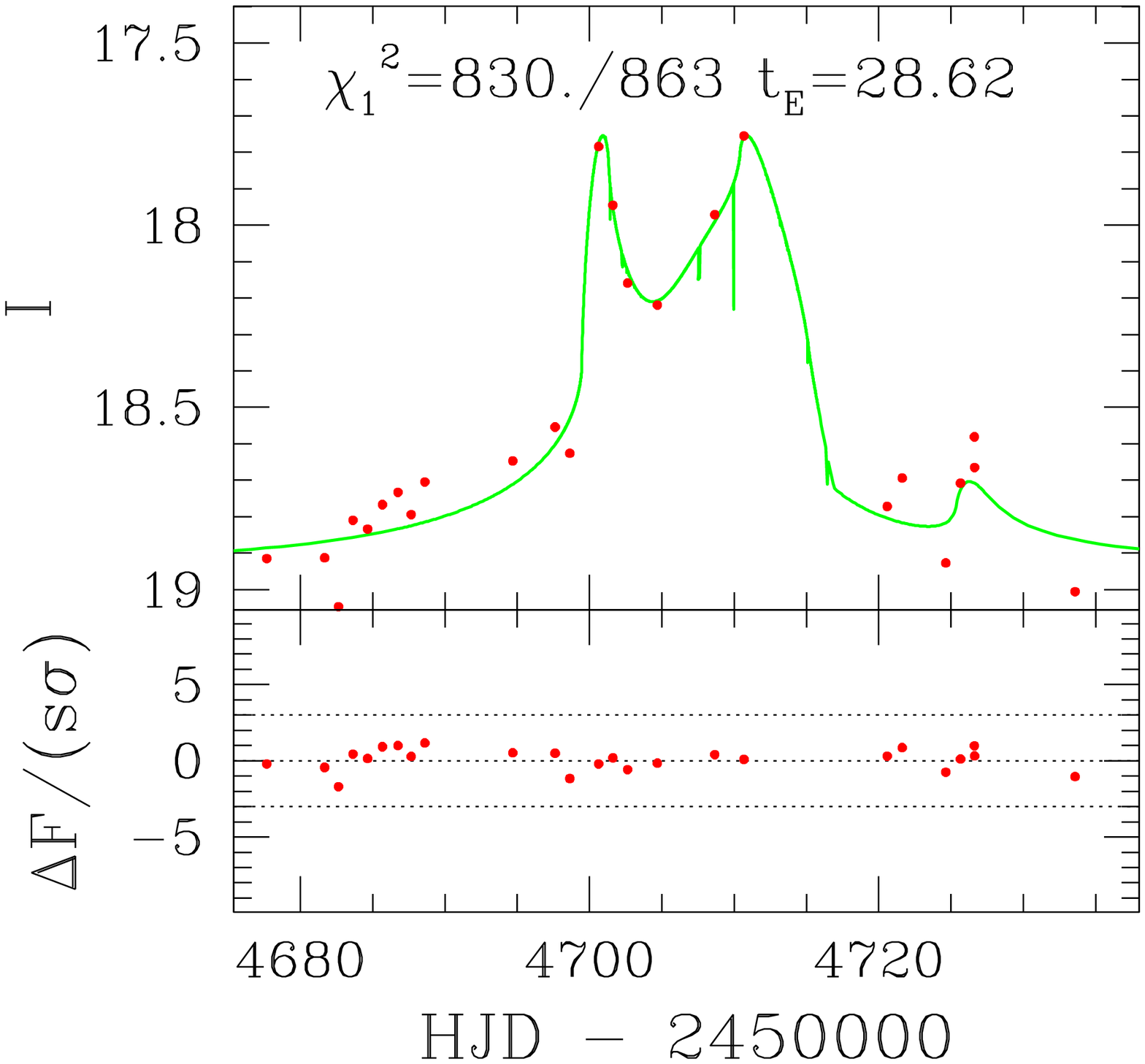}%

}

\noindent\parbox{12.75cm}{
\leftline {{\bf OGLE 2008-BLG-592} (3rd model)}

\includegraphics[height=62mm,width=63mm]{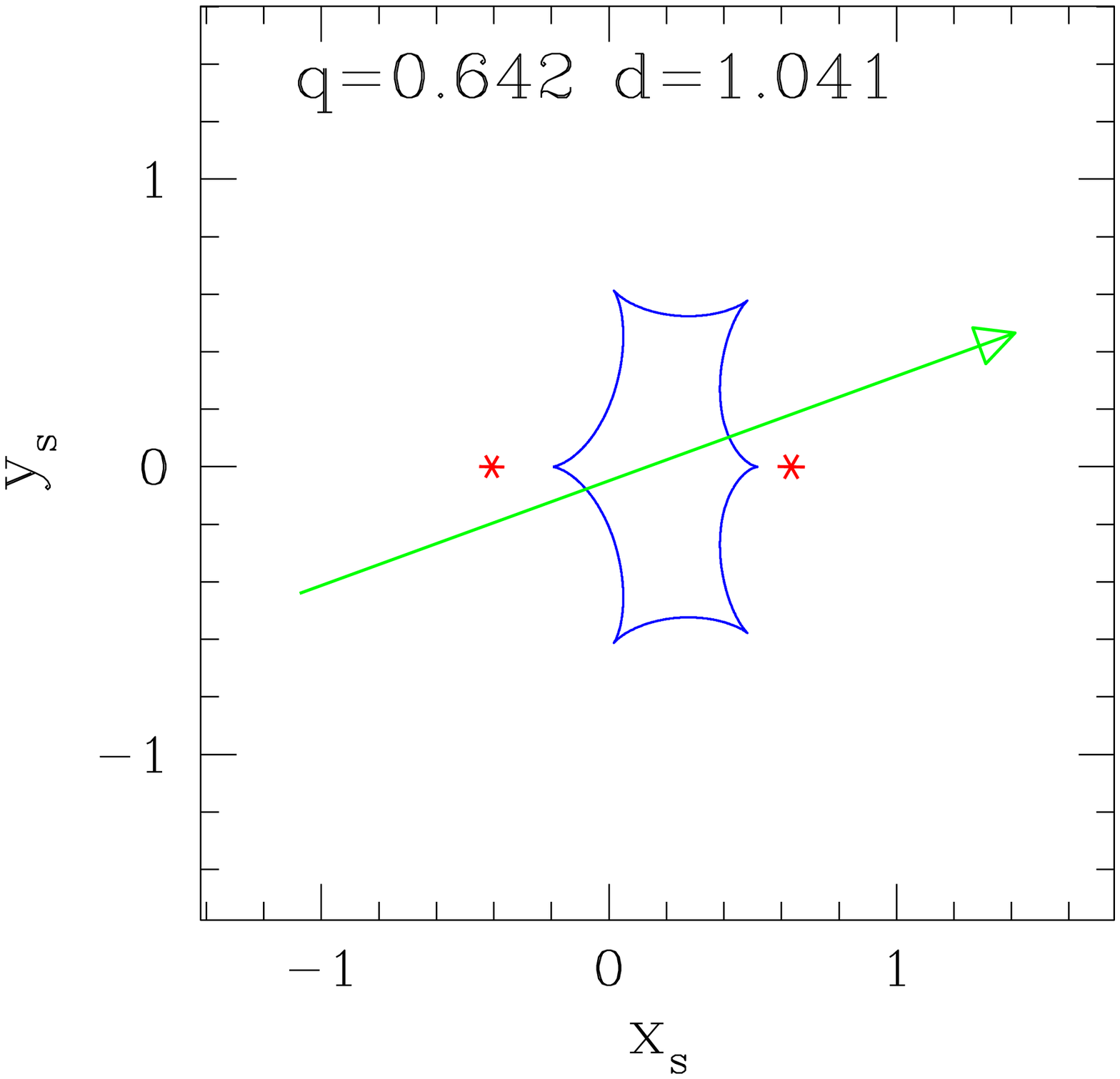} \hfill
\includegraphics[height=62mm,width=63mm]{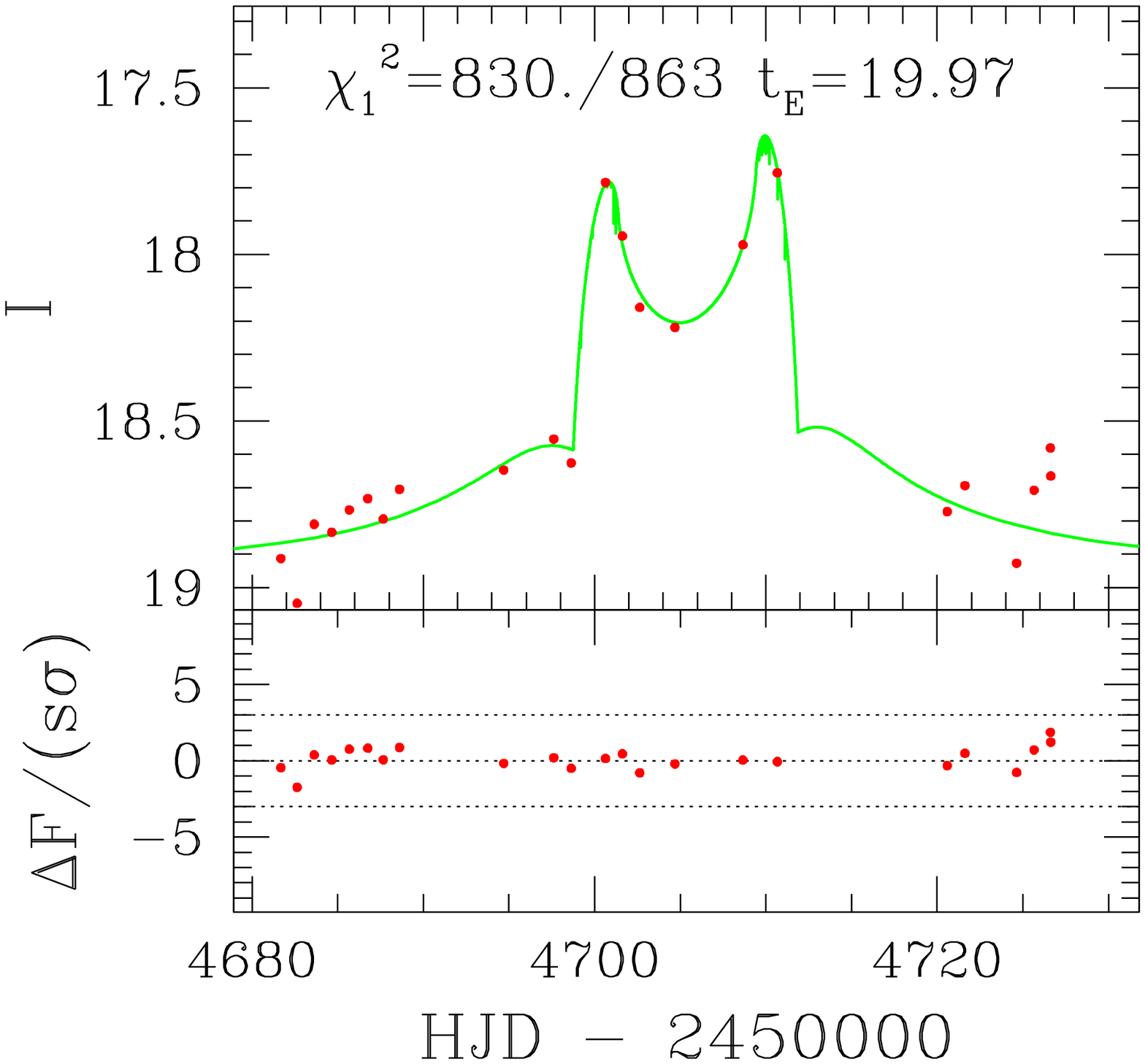}%

}

\vfill
\eject

%%END%%END%%END%%END%%END%%END%%END%%END%%END%%END%%END%%END%%END%%END%%END%%END%

\subsection{Comparison of binary lens and double source models}

Below we show the binary lens (on the left) and double source (on the right)
models of the light curves for some of the considered events. For all
binary lens models we have also calculated double source models. We do
not show events with evident jumps in the light curves, which can only
be modeled as caustic crossings. We include, however, few cases with
almost smooth observed light curves, despite the fact that their binary
lens models are formally far better.  
The light curve in a double source model is a sum of the constant
blended flux plus the two single lens light curves for the source
components, each shown with dotted lines. In the plots we use observed
fluxes (not magnitudes) since they are additive, which is important in
double source modeling. In majority of cases the binary
lens models give formally better fits as compared to the double source
models presented. On the other hand double source models, always
producing {\it simpler} light curves, look more natural in some cases.

\vskip 0.3cm

\noindent\parbox{12.75cm}{
\leftline {\bf OGLE 2006-BLG-038} 

\includegraphics[height=63mm,width=62mm]{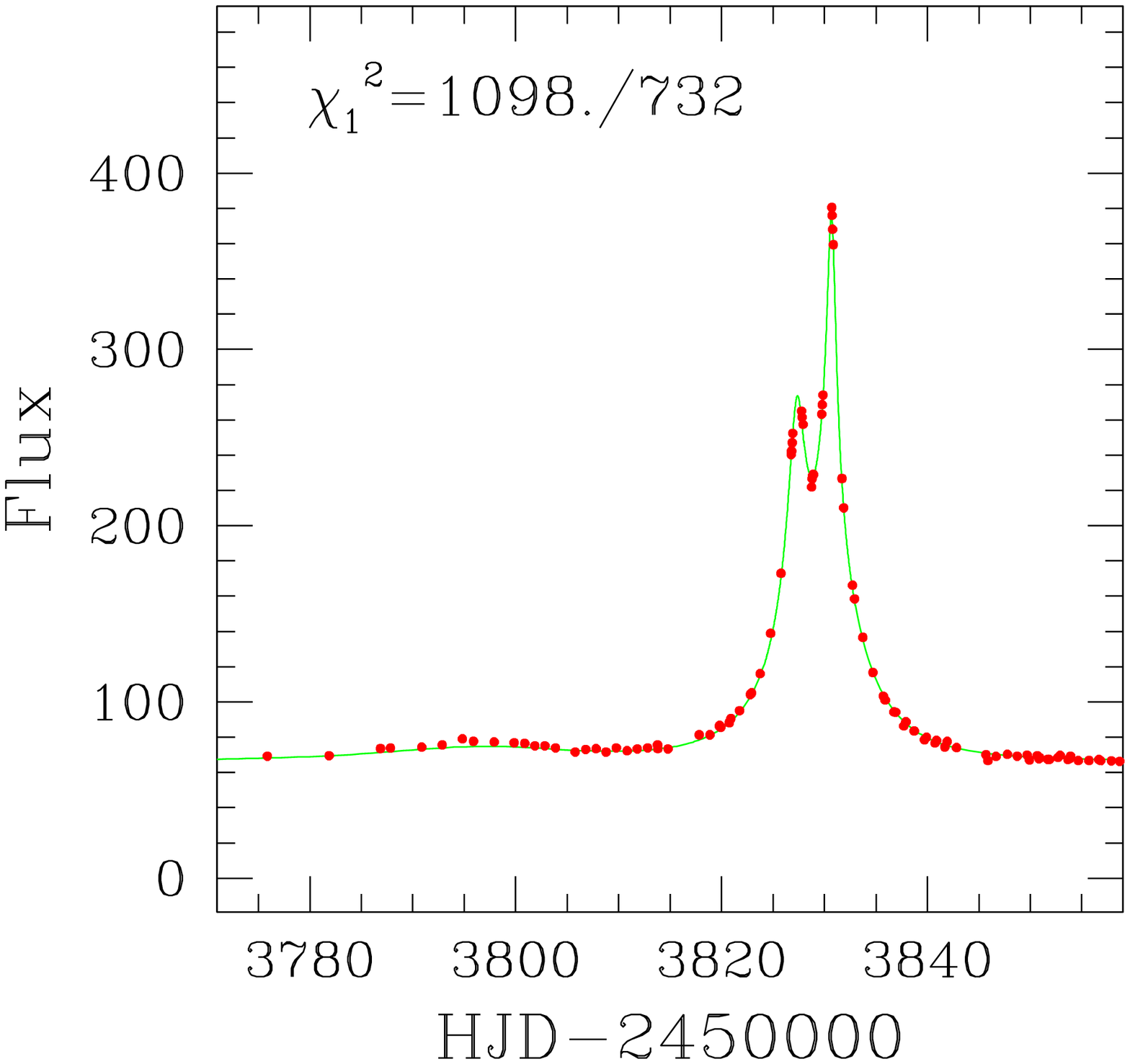}% 
\includegraphics[height=63mm,width=62mm]{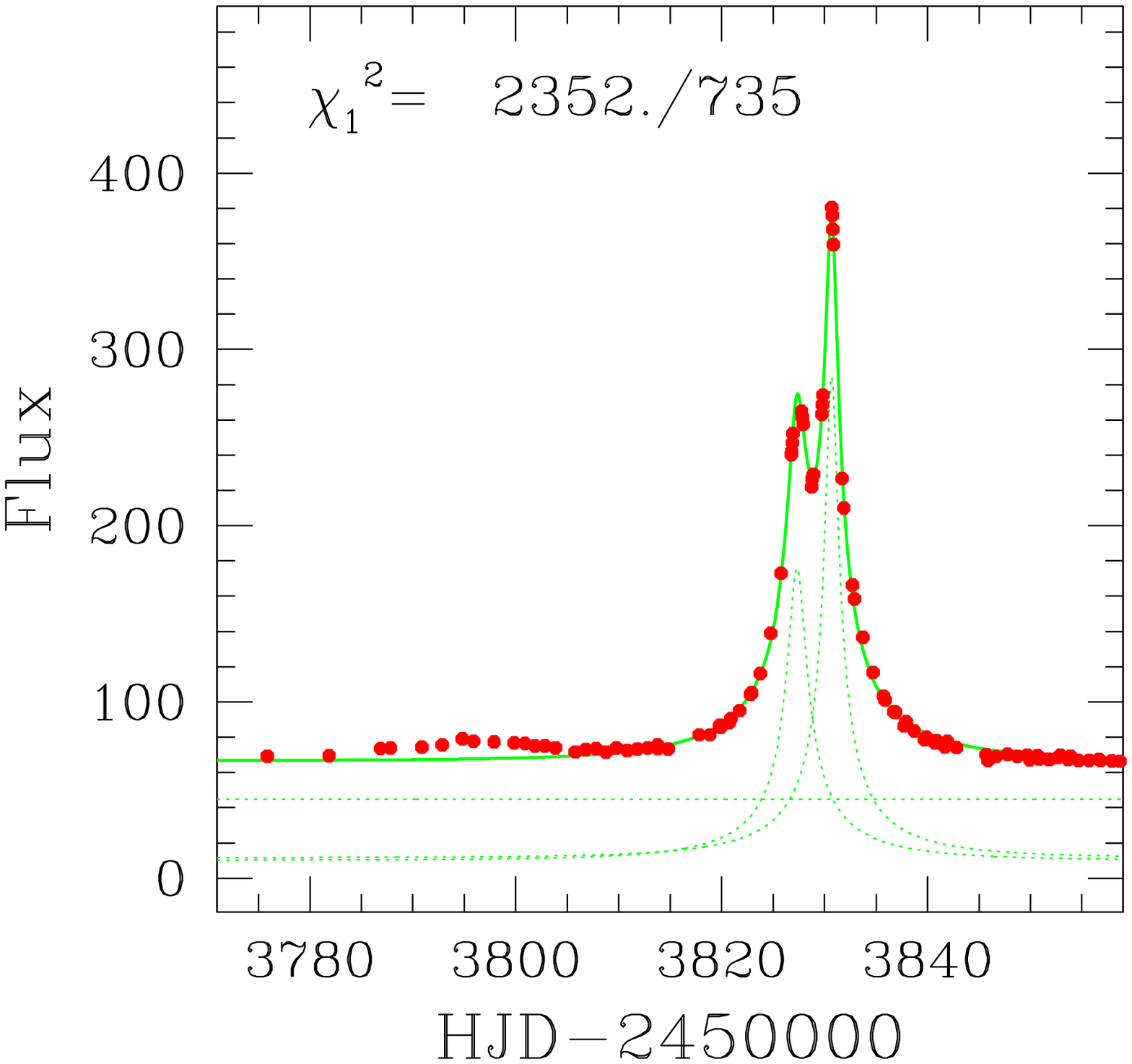}%

}

\noindent\parbox{12.75cm}{
\leftline {\bf OGLE 2006-BLG-450} 

\includegraphics[height=63mm,width=62mm]{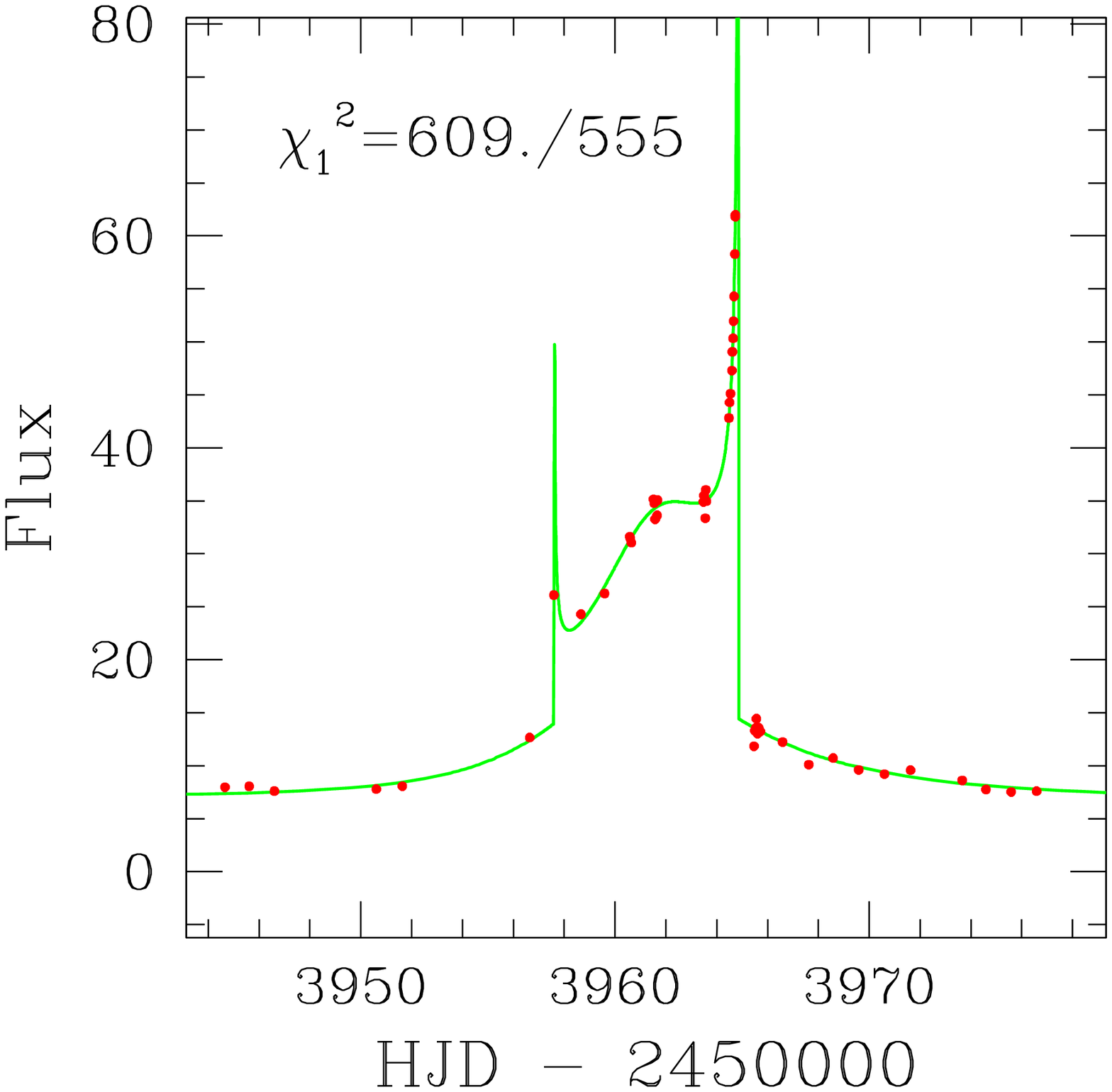}% 
\includegraphics[height=63mm,width=62mm]{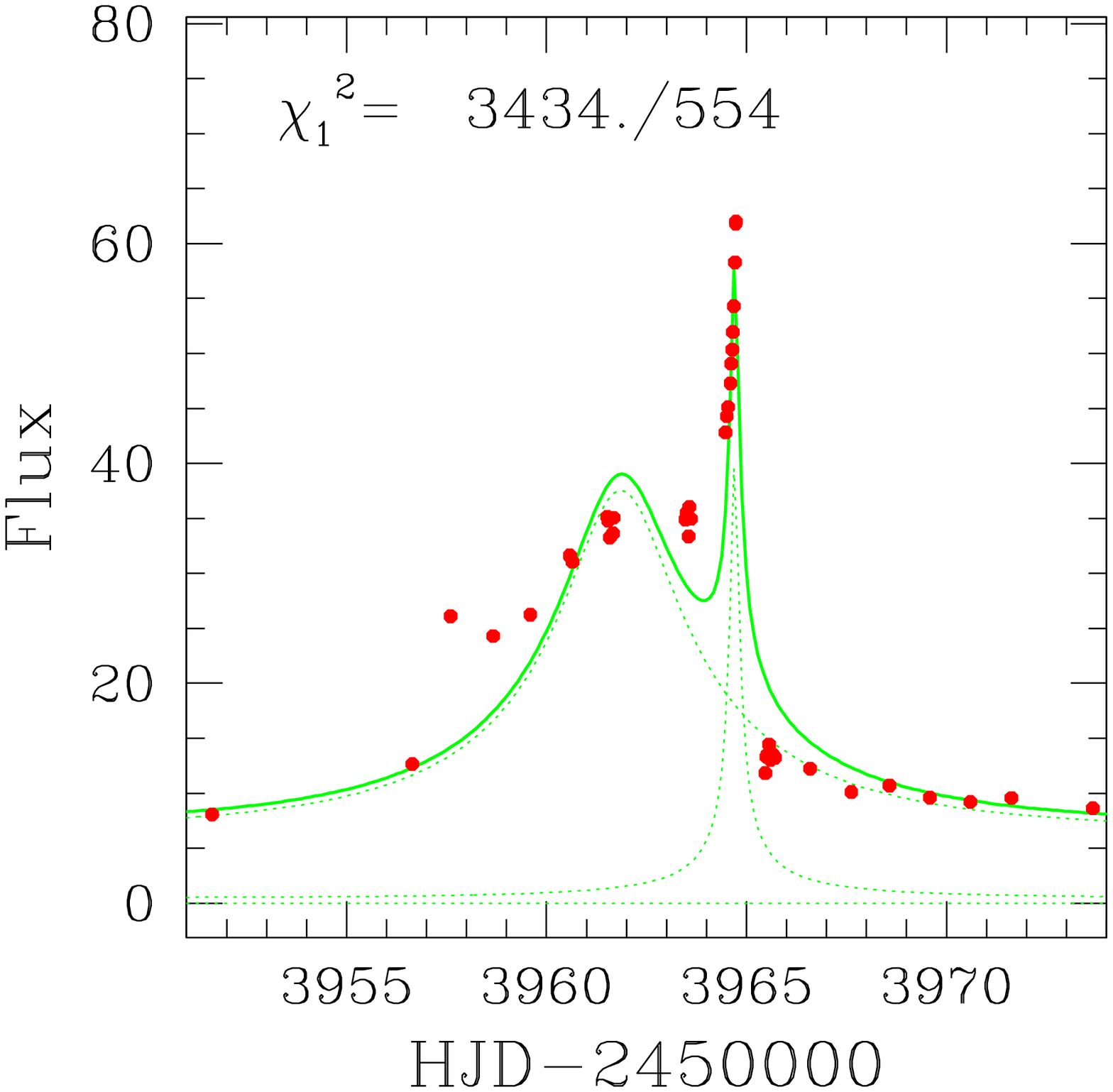}%

}

\noindent\parbox{12.75cm}{
\leftline {\bf OGLE 2006-BLG-460} 

\includegraphics[height=63mm,width=62mm]{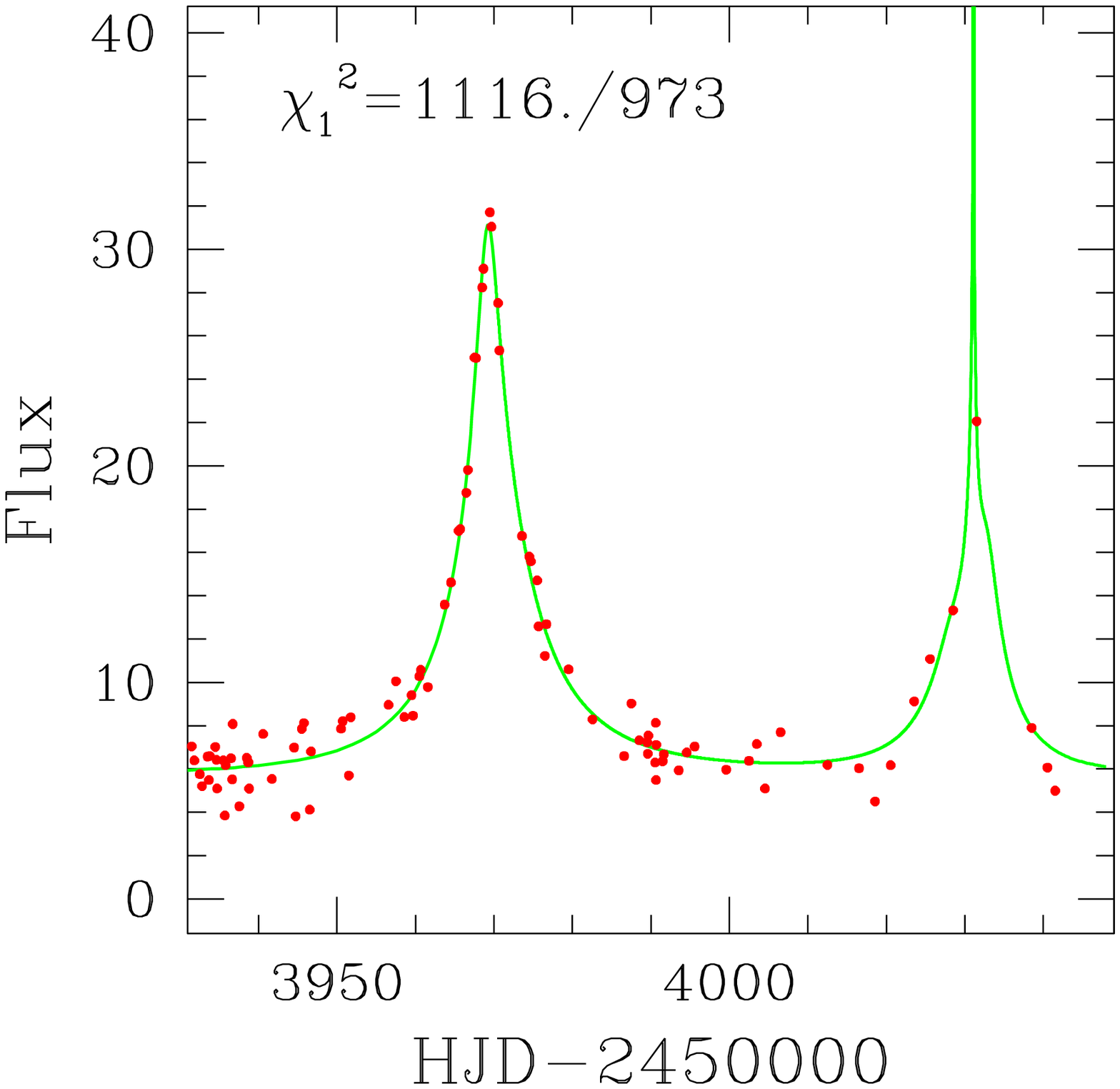}% 
\includegraphics[height=63mm,width=62mm]{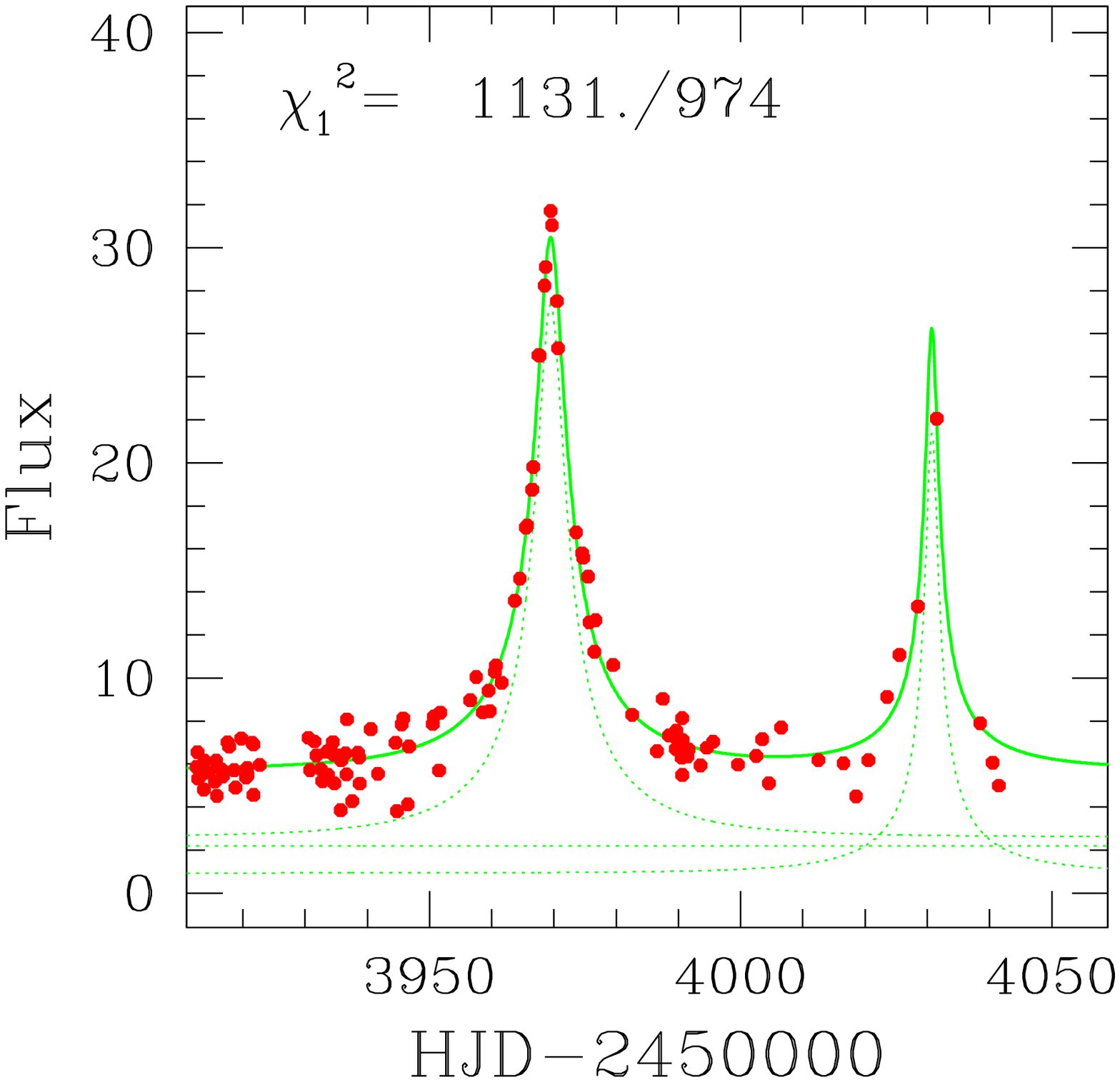}%

}

\noindent\parbox{12.75cm}{
\leftline {\bf OGLE 2007-BLG-006} 

\includegraphics[height=63mm,width=62mm]{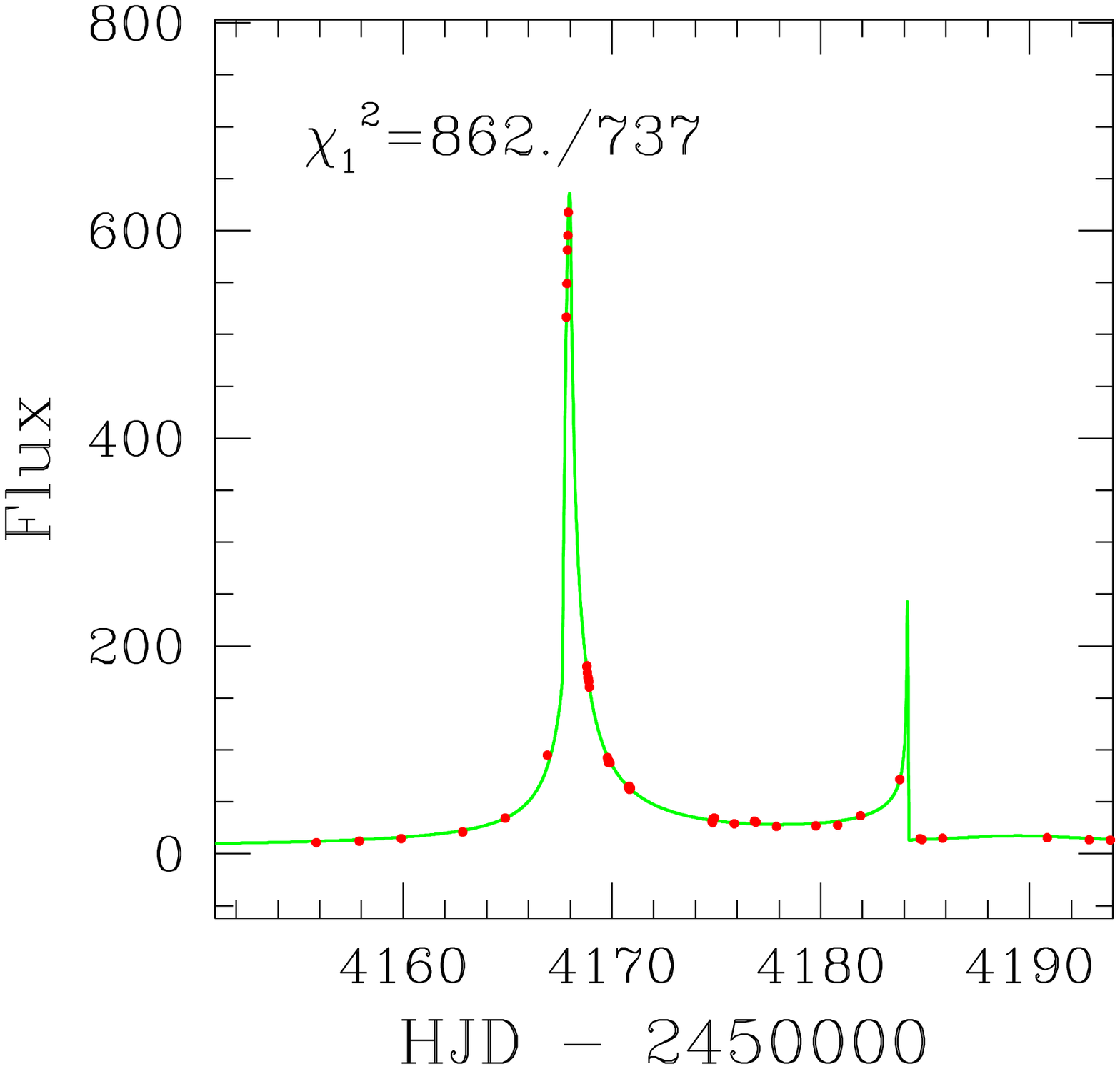}% 
\includegraphics[height=63mm,width=62mm]{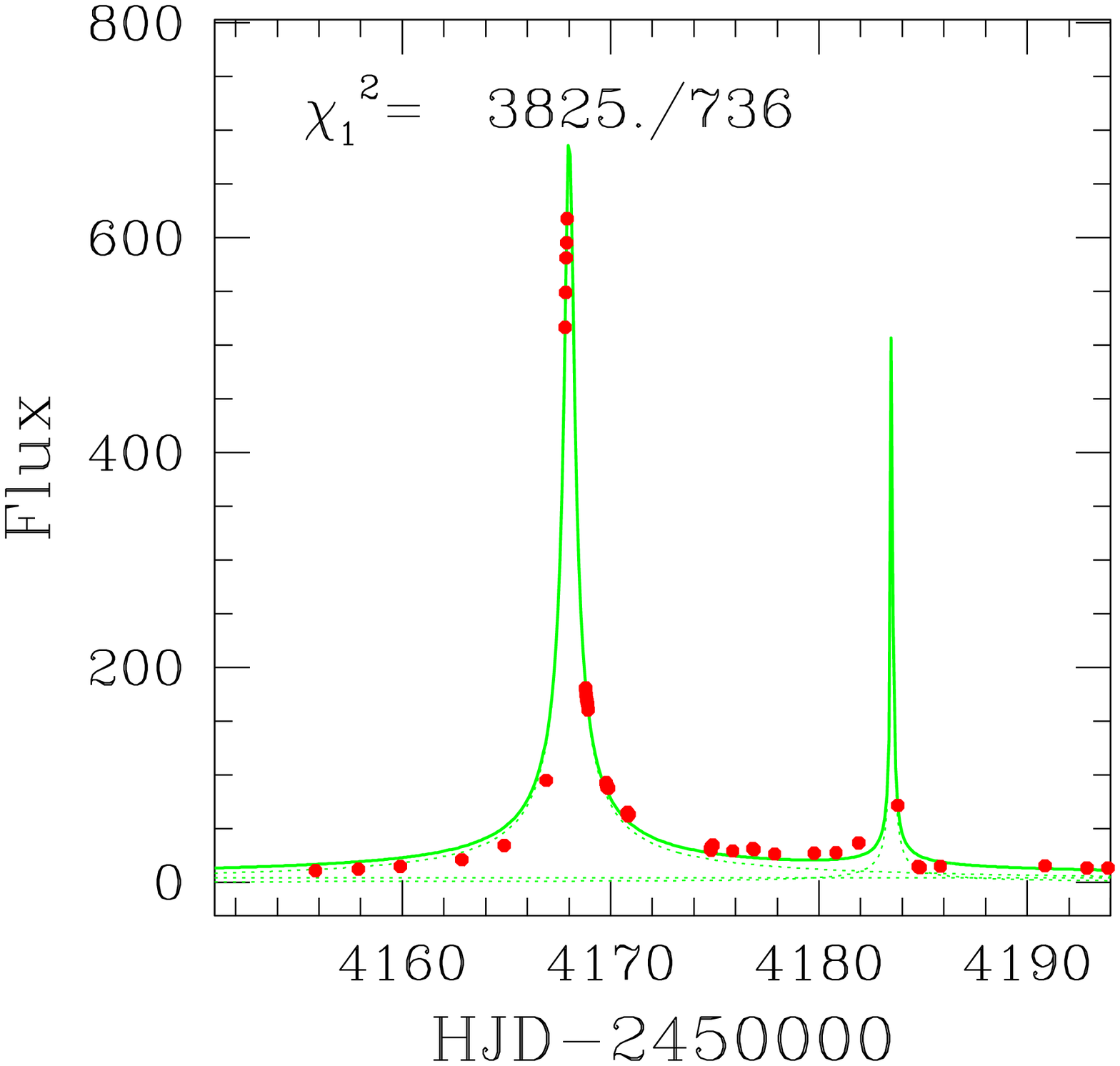}%

}

\noindent\parbox{12.75cm}{
\leftline {\bf OGLE 2007-BLG-237} 

\includegraphics[height=63mm,width=62mm]{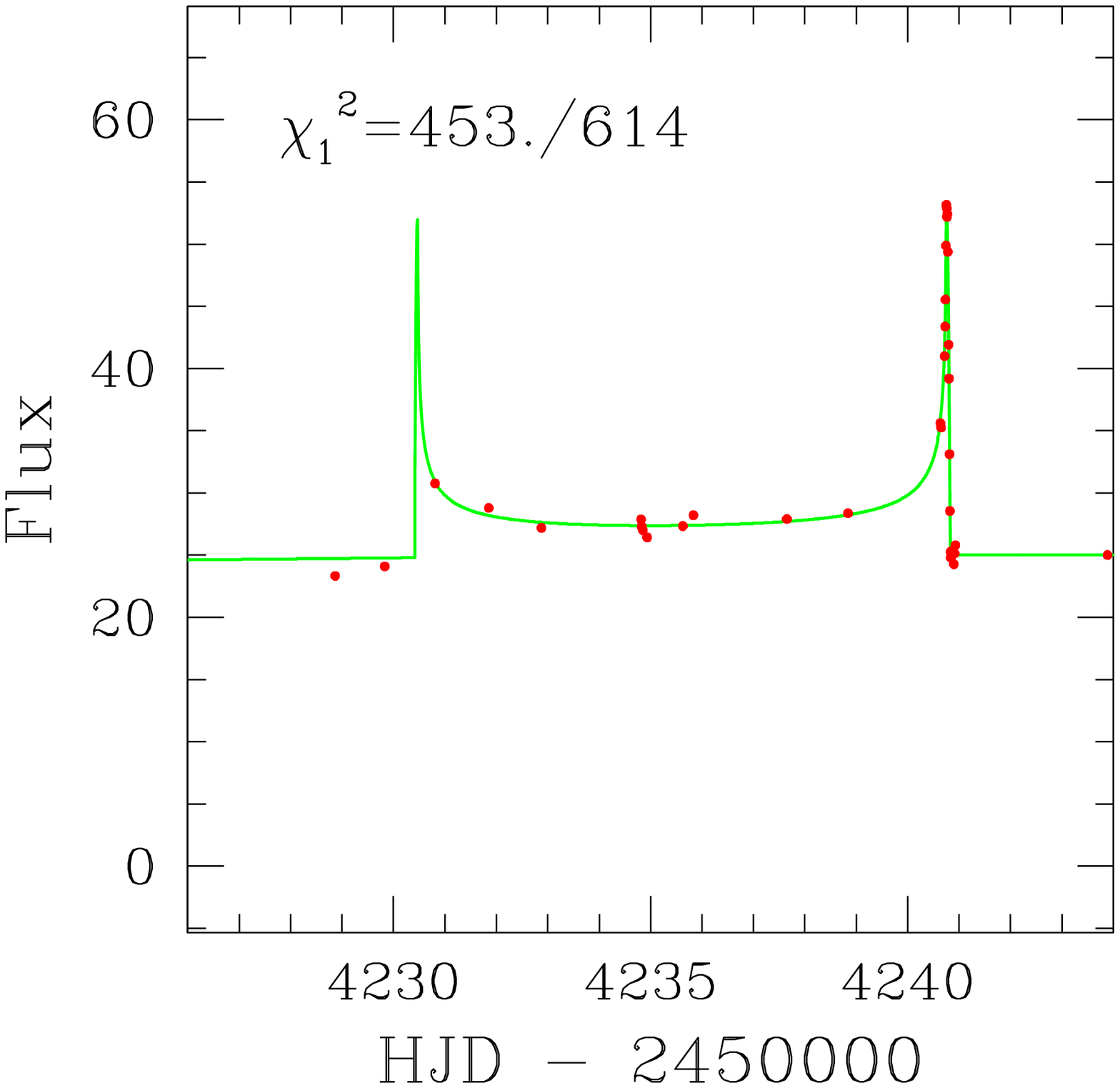}% 
\includegraphics[height=63mm,width=62mm]{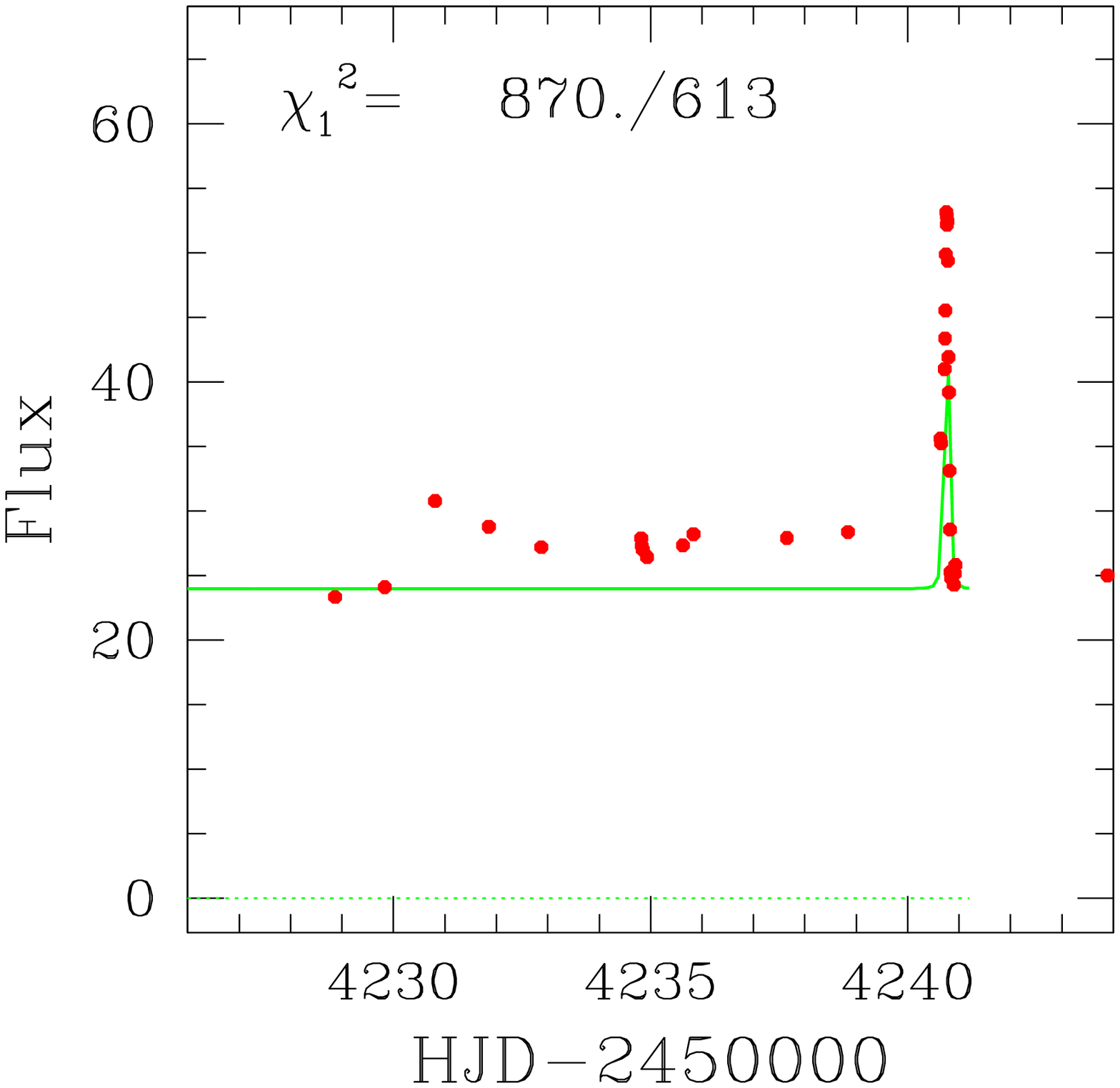}%

}

\noindent\parbox{12.75cm}{
\leftline {\bf OGLE 2007-BLG-363} 

\includegraphics[height=63mm,width=62mm]{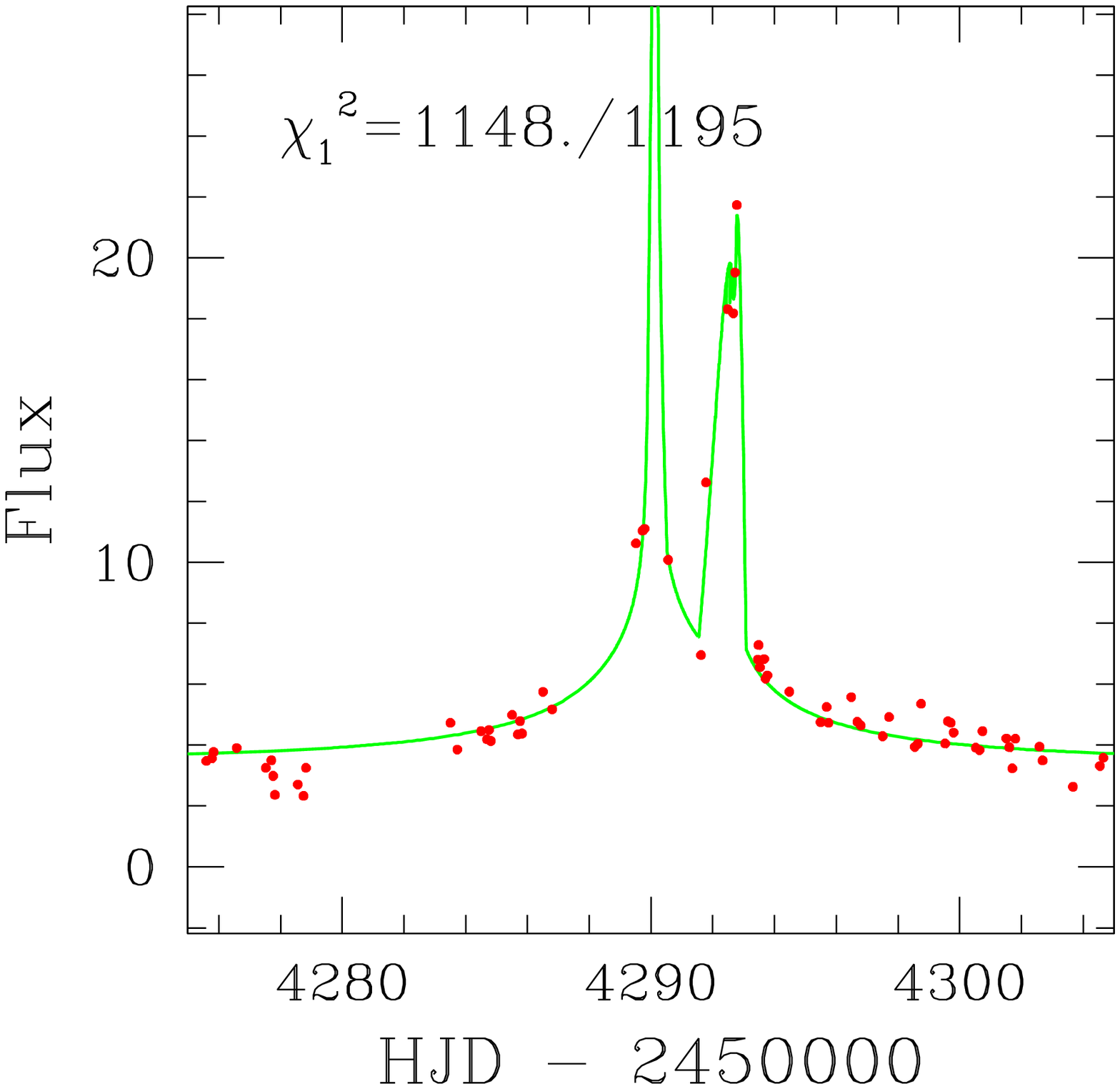}% 
\includegraphics[height=63mm,width=62mm]{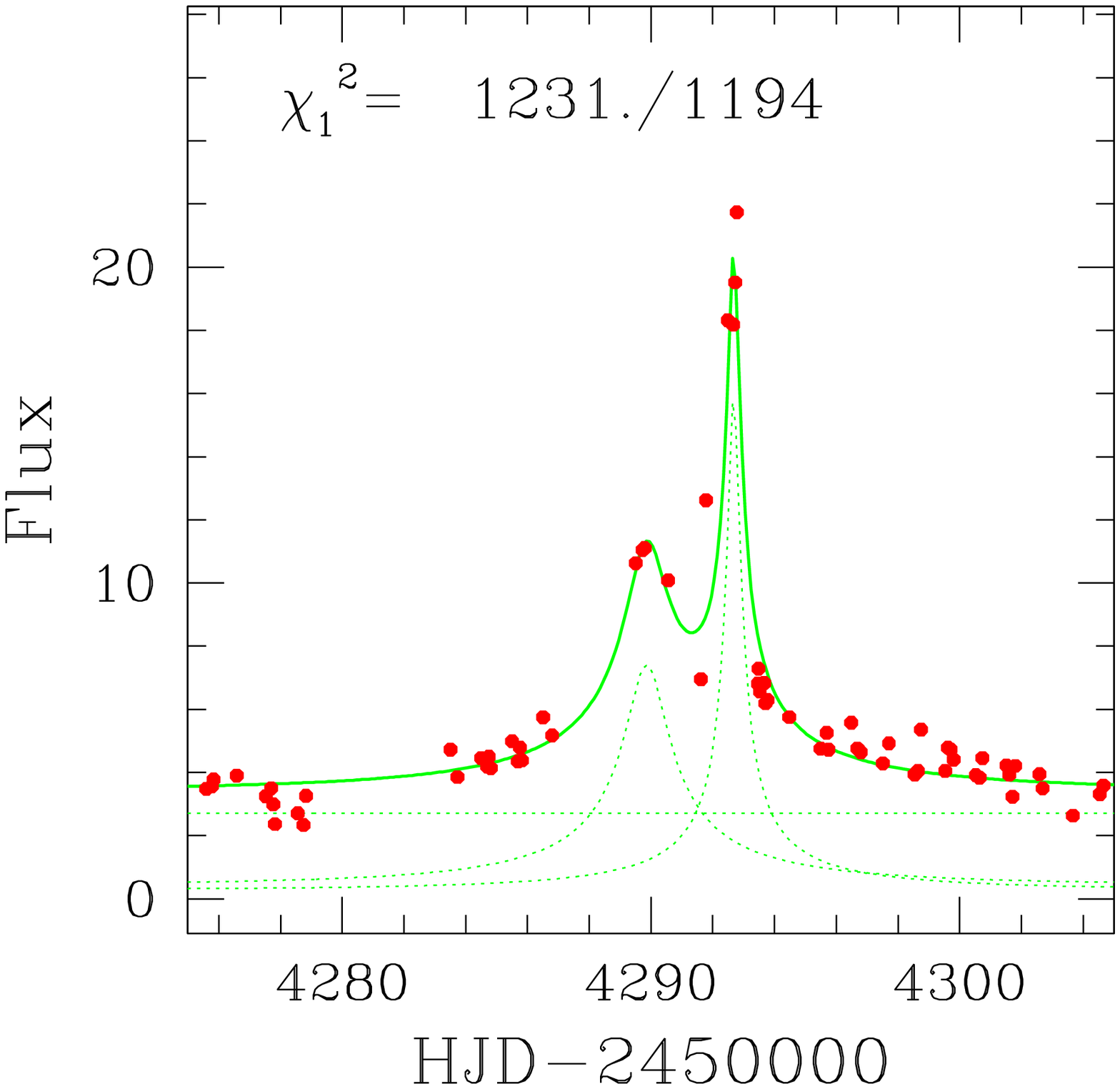}%

}

\noindent\parbox{12.75cm}{
\leftline {\bf OGLE 2007-BLG-399} 

\includegraphics[height=63mm,width=62mm]{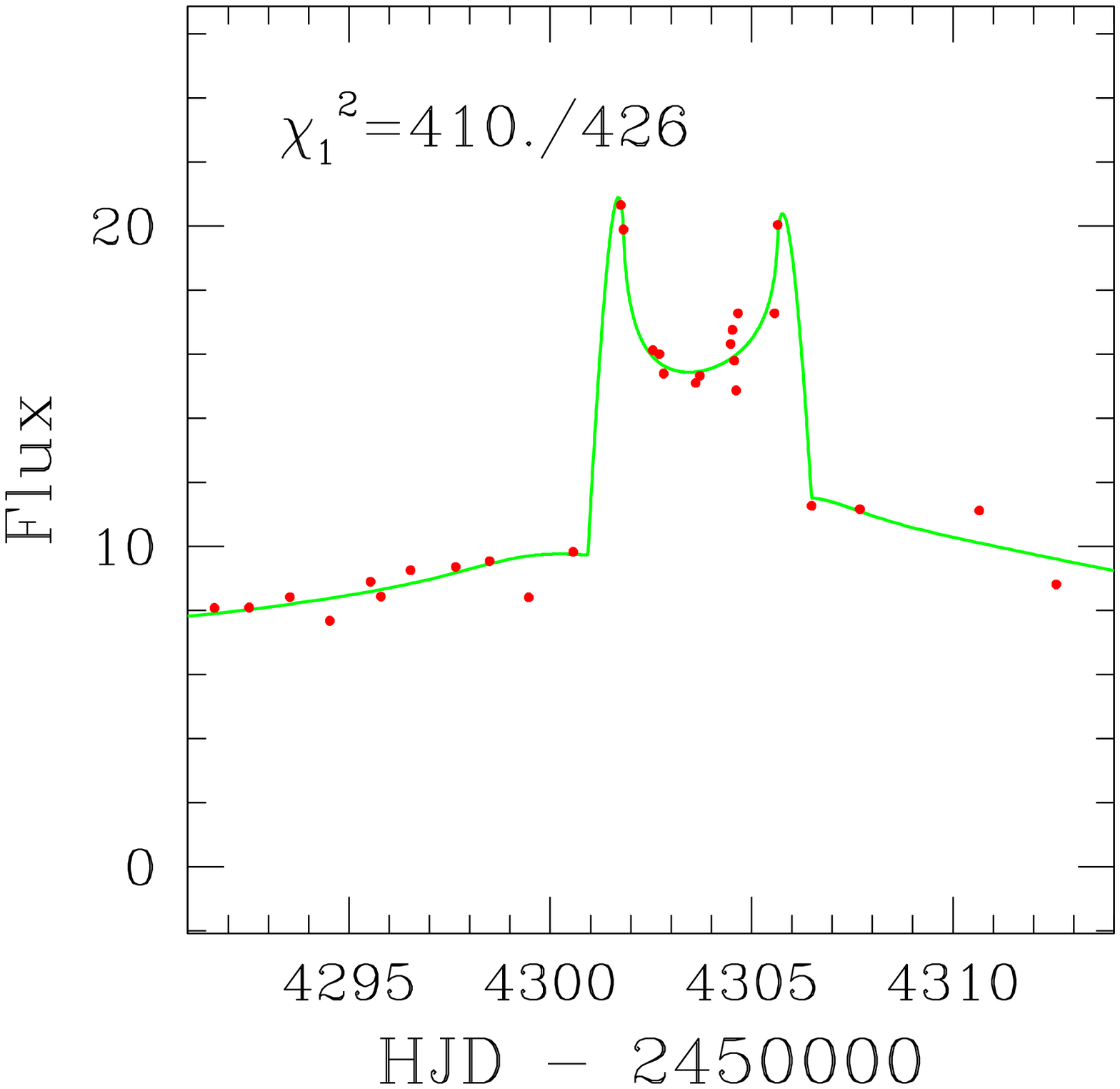}% 
\includegraphics[height=63mm,width=62mm]{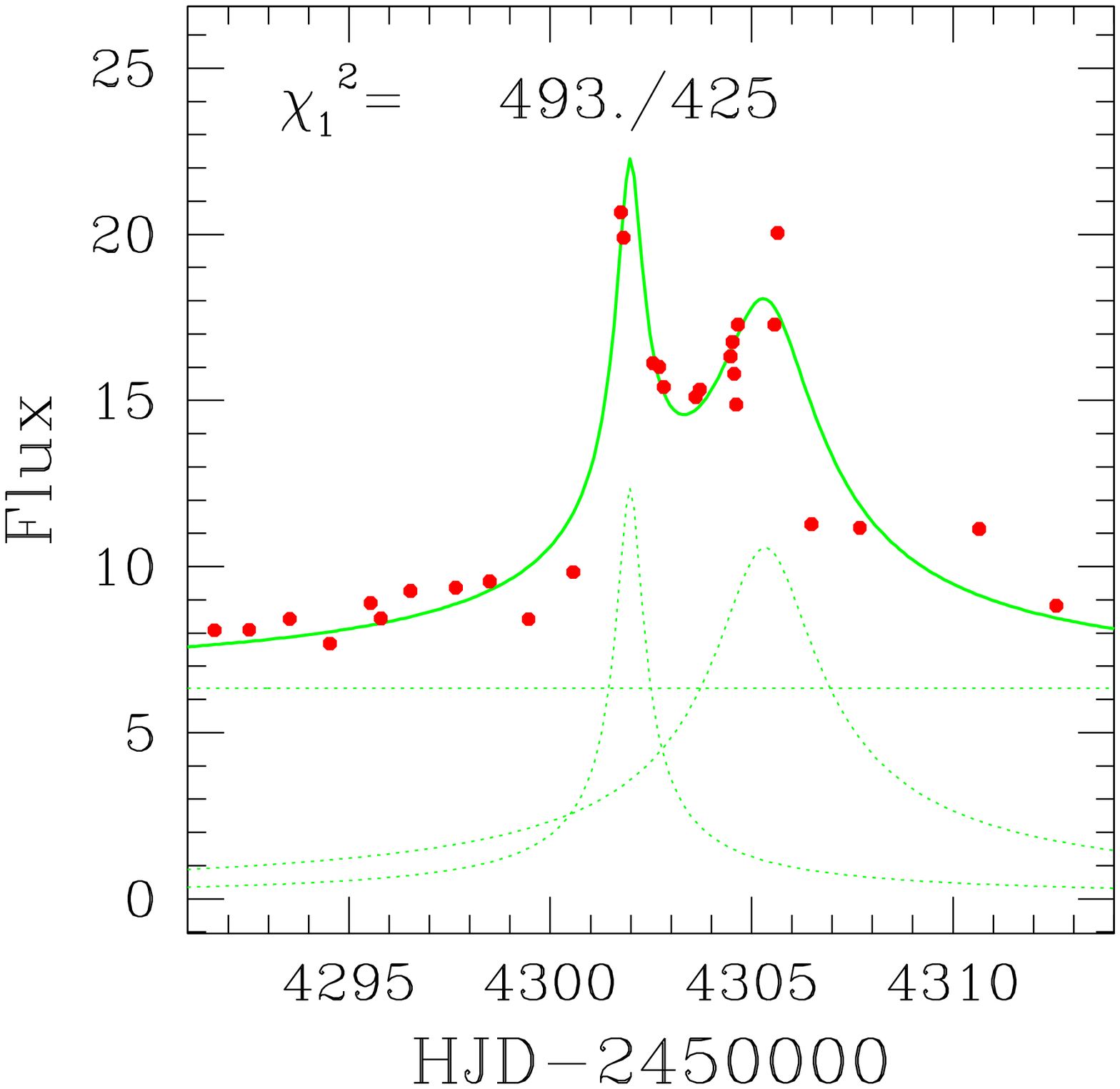}%

}

\noindent\parbox{12.75cm}{
\leftline {\bf OGLE 2008-BLG-513} 

\includegraphics[height=63mm,width=62mm]{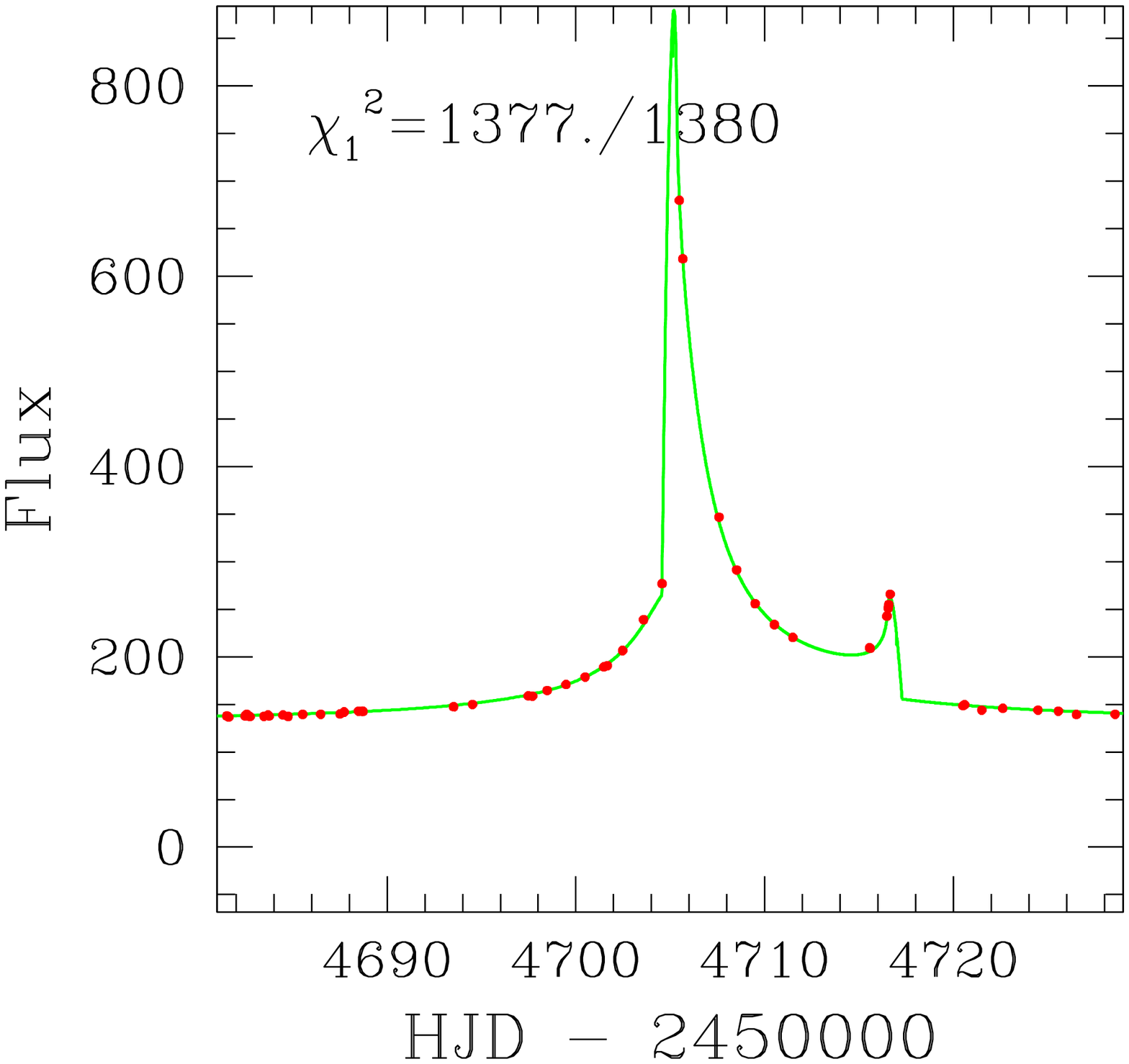}% 
\includegraphics[height=63mm,width=62mm]{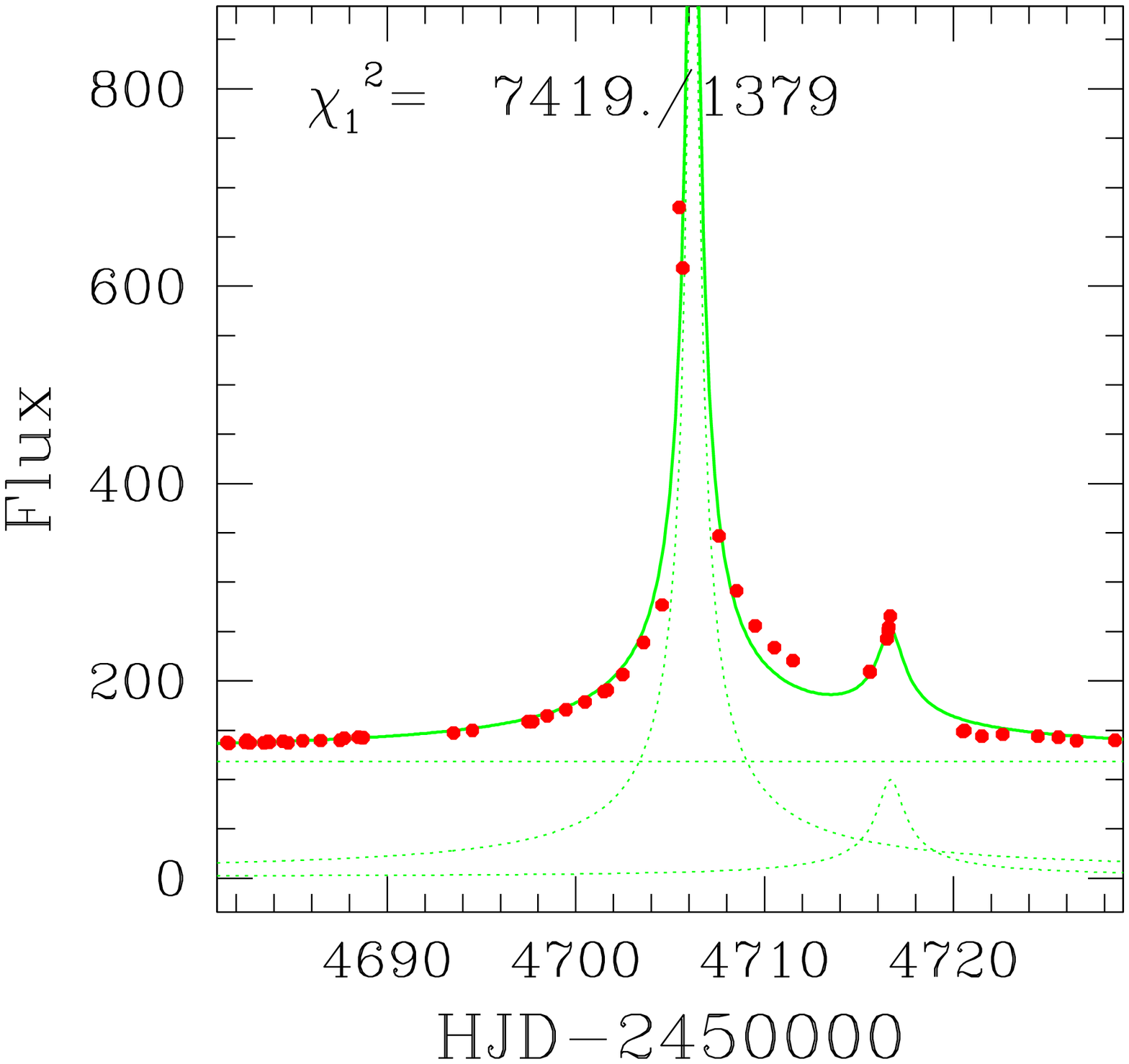}%

}

\noindent\parbox{12.75cm}{
\leftline {\bf OGLE 2008-BLG-559} 

\includegraphics[height=63mm,width=62mm]{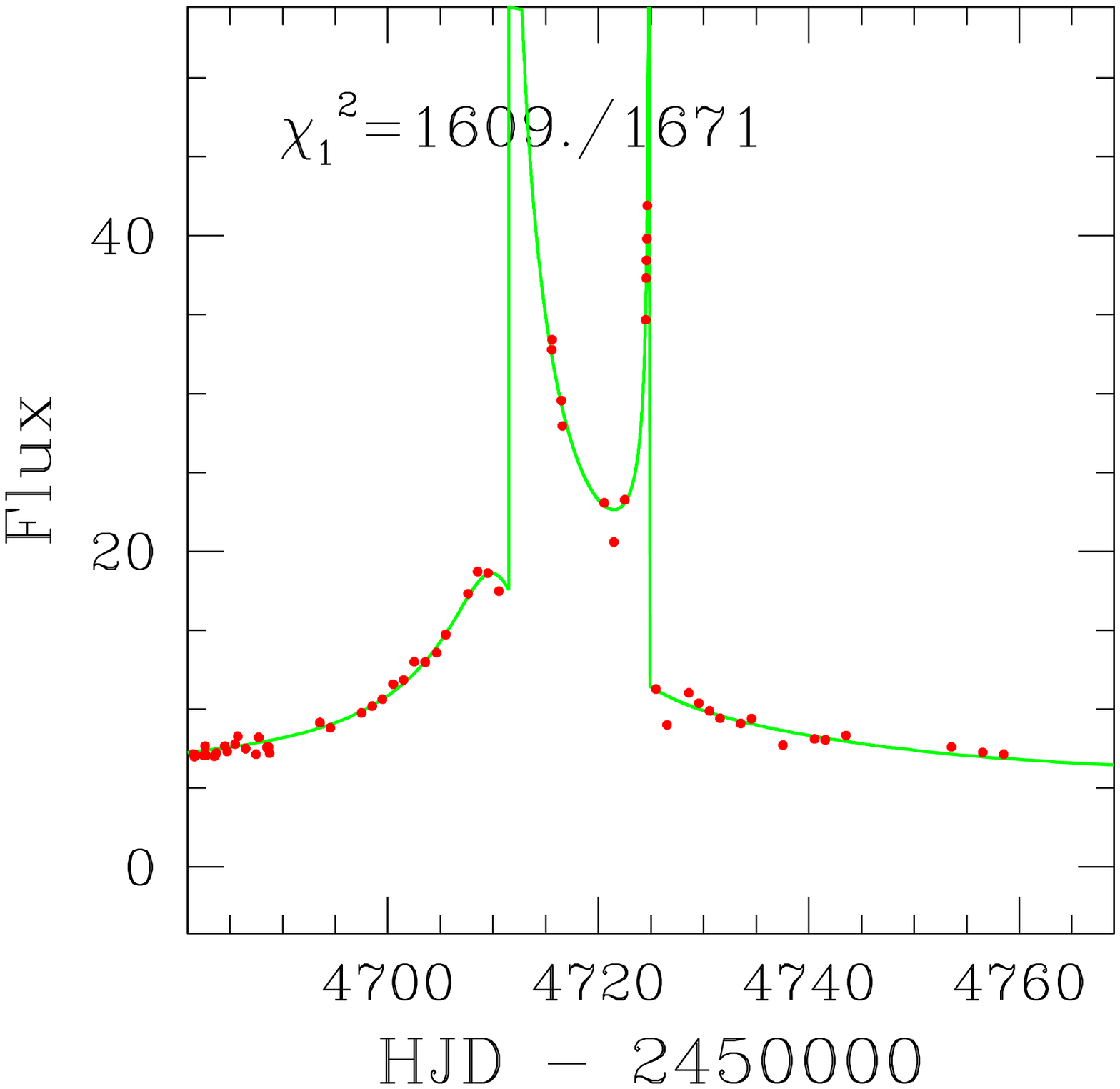}% 
\includegraphics[height=63mm,width=62mm]{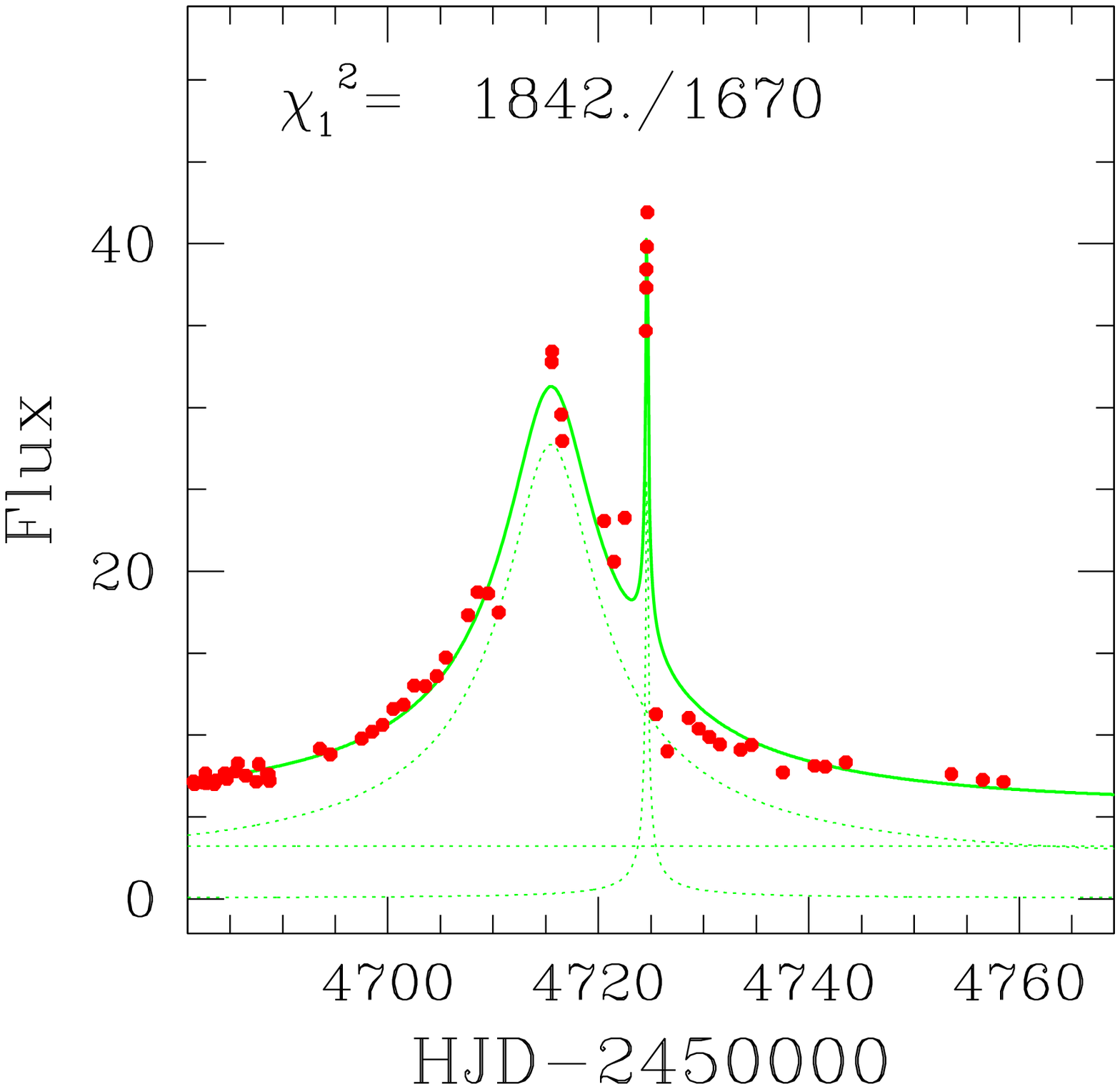}%

}

\noindent\parbox{12.75cm}{
\leftline {\bf OGLE 2008-BLG-592} 

\includegraphics[height=63mm,width=62mm]{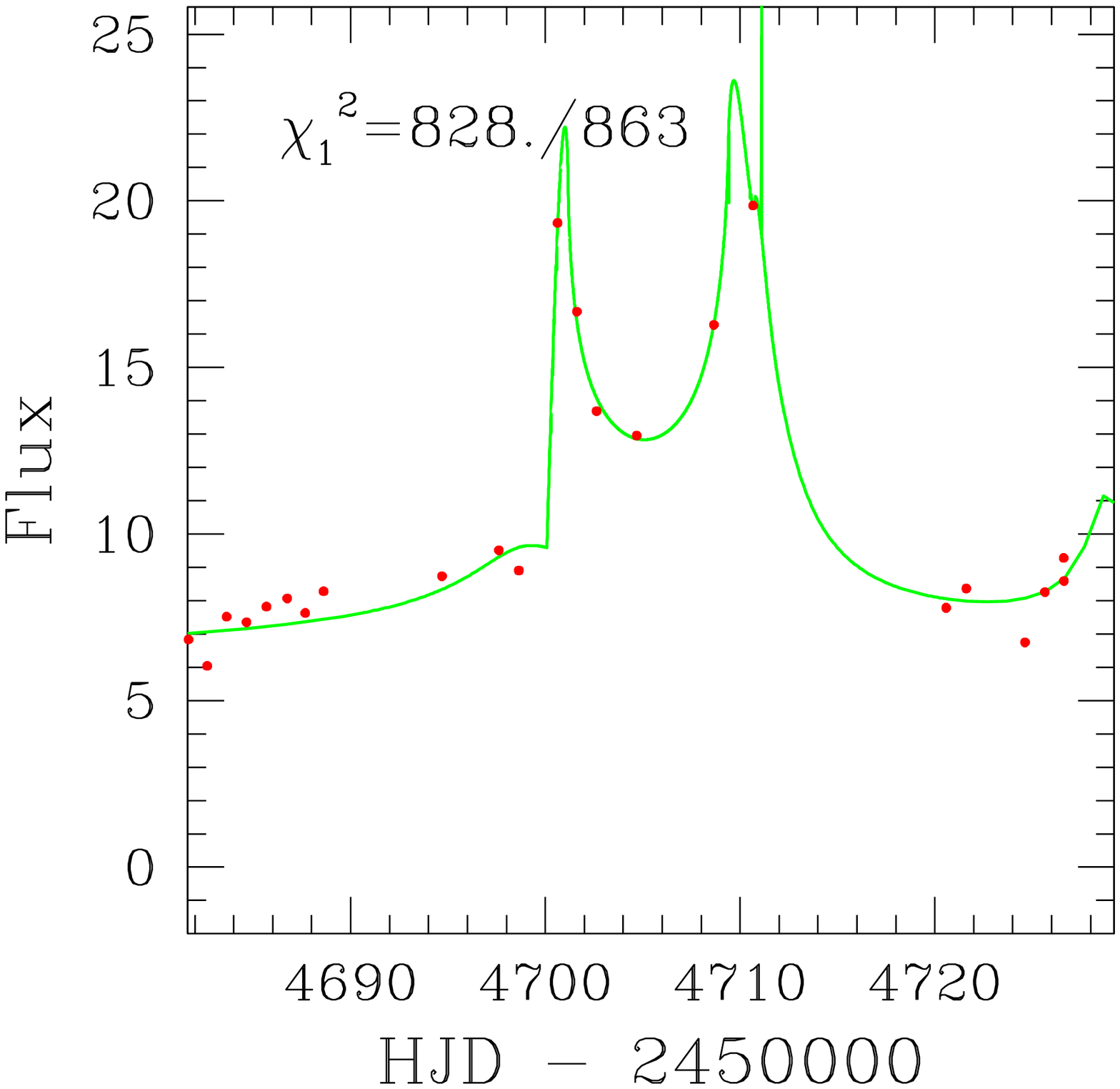}% 
\includegraphics[height=63mm,width=62mm]{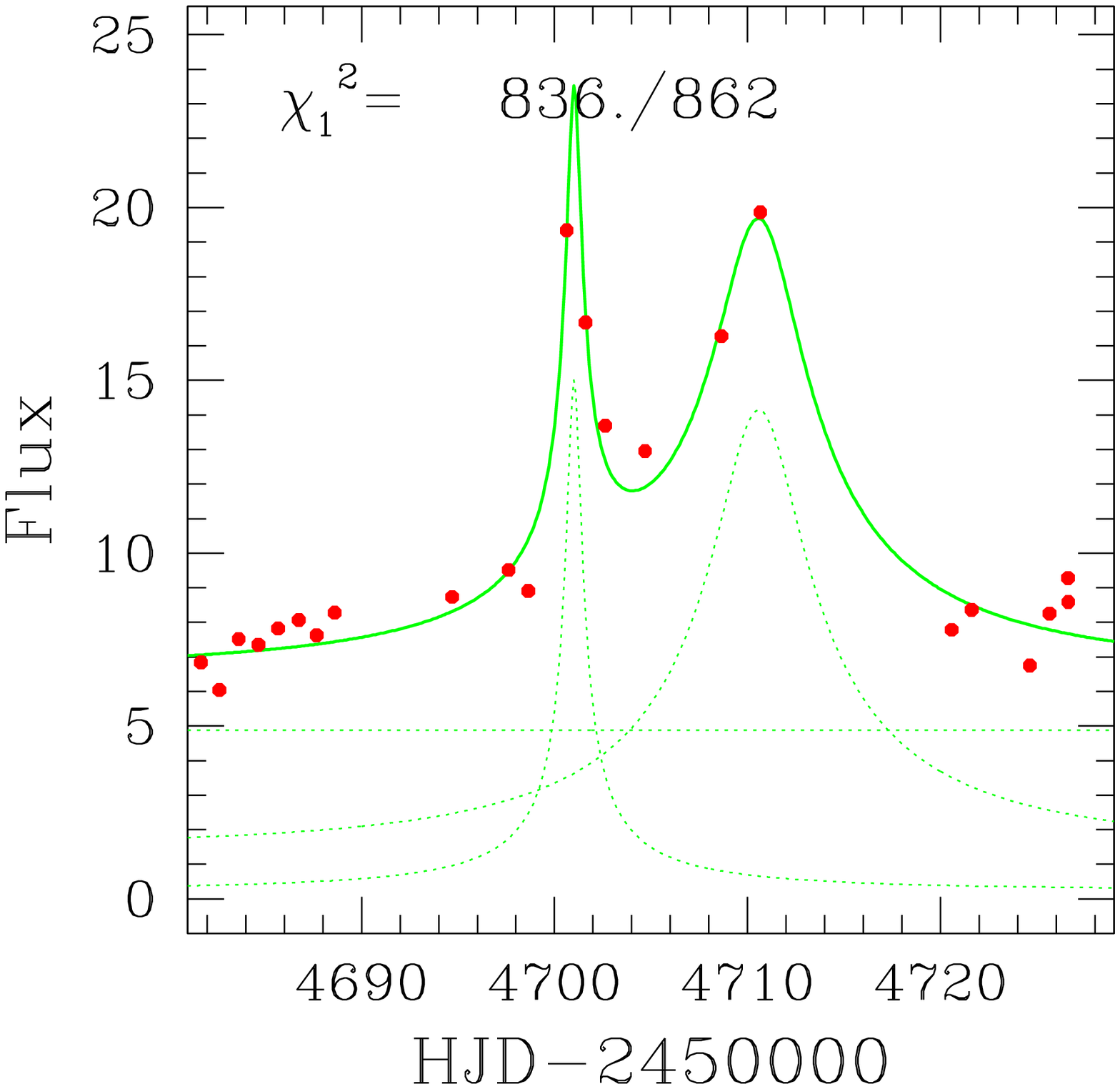}%

}

\subsection{Double source models}

\noindent\parbox{12.75cm}{
\centerline{{\bf OGLE 2006-BLG-023} \hfill {\bf OGLE-2006-BLG-061}\hfill}

\includegraphics[height=63mm,width=62mm]{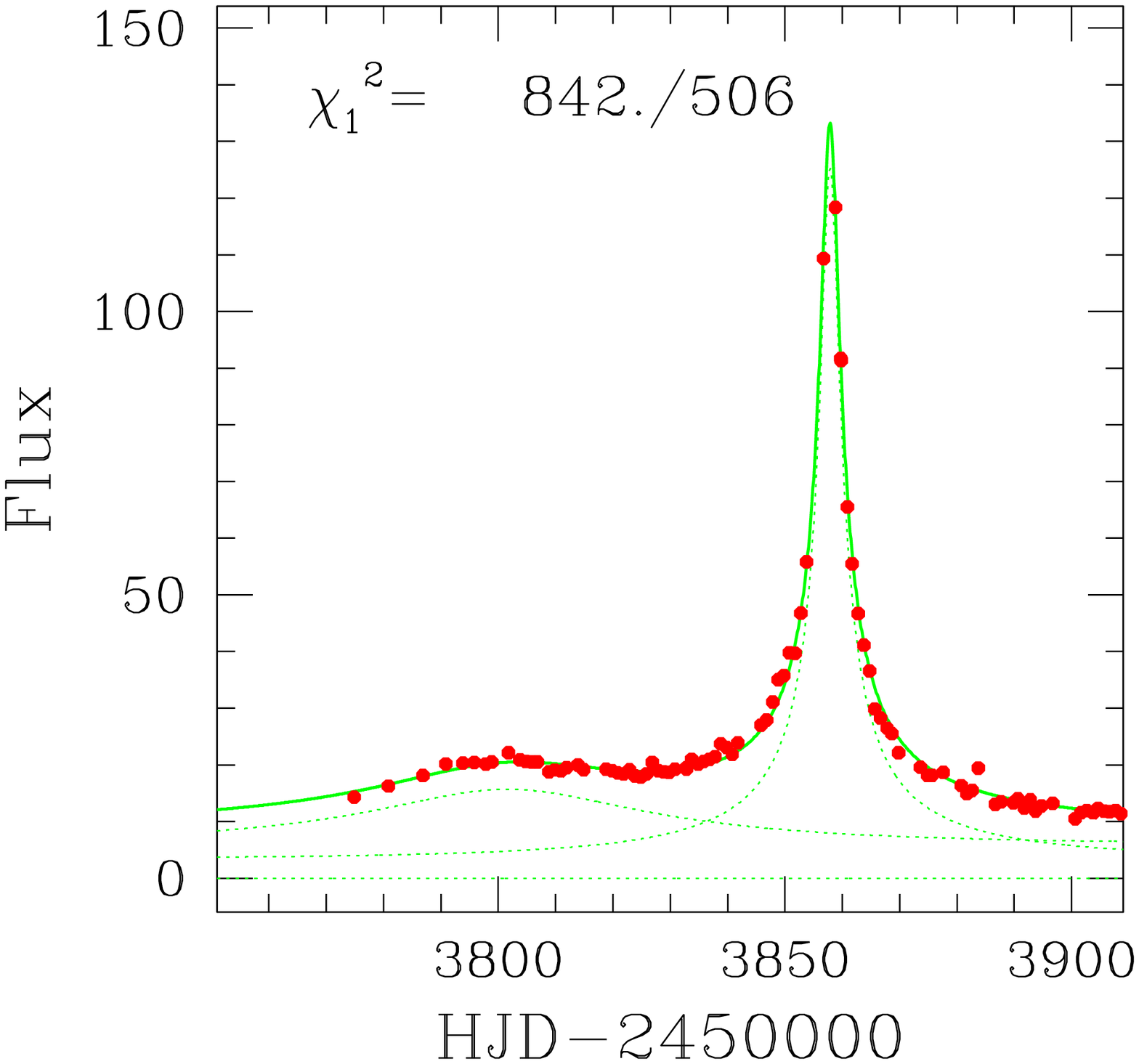} \hfill
\includegraphics[height=63mm,width=62mm]{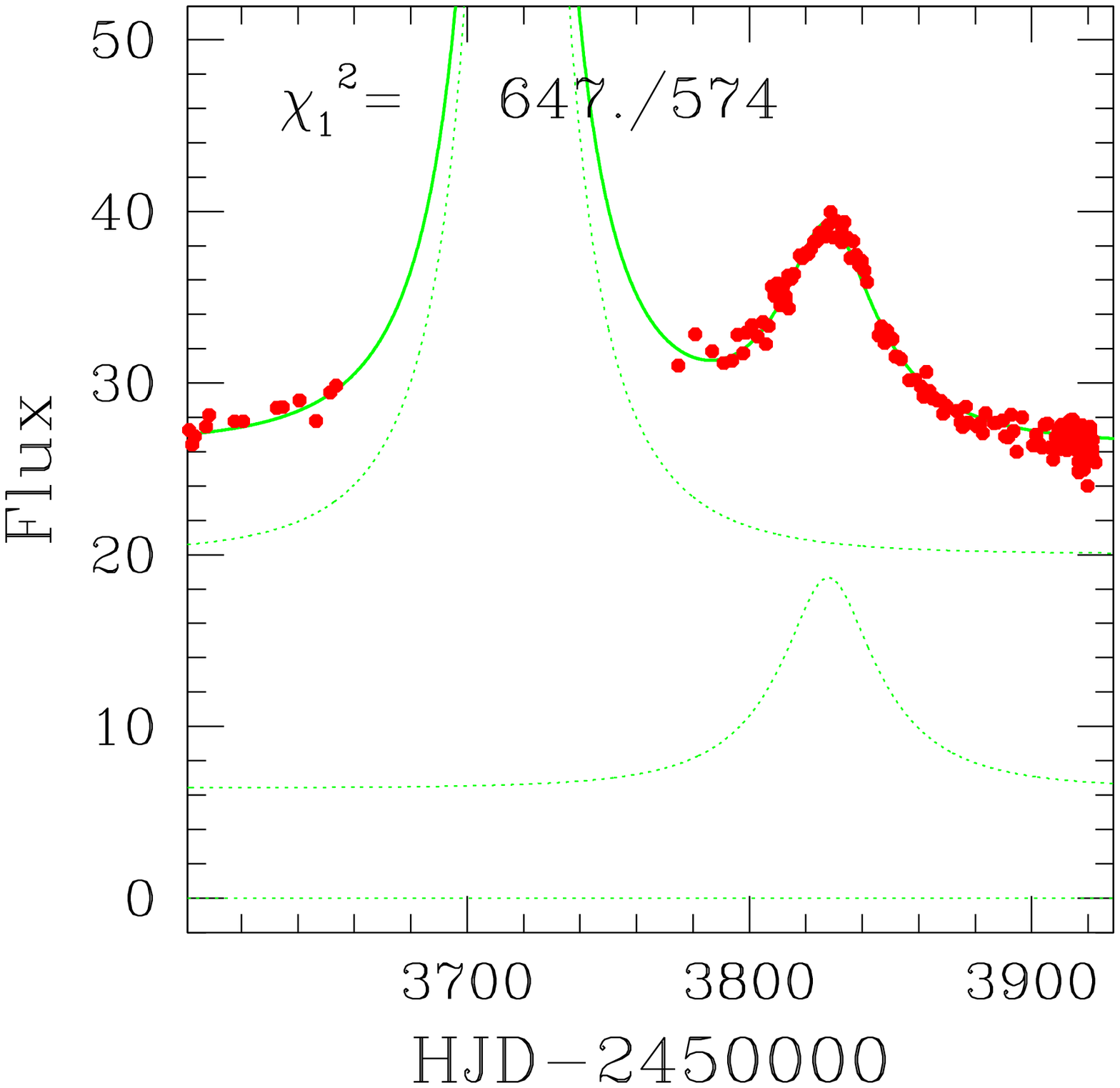} 

}

\noindent\parbox{12.75cm}{
\centerline{{\bf OGLE 2006-BLG-238} \hfill {\bf OGLE-2006-BLG-393}\hfill}

\includegraphics[height=63mm,width=62mm]{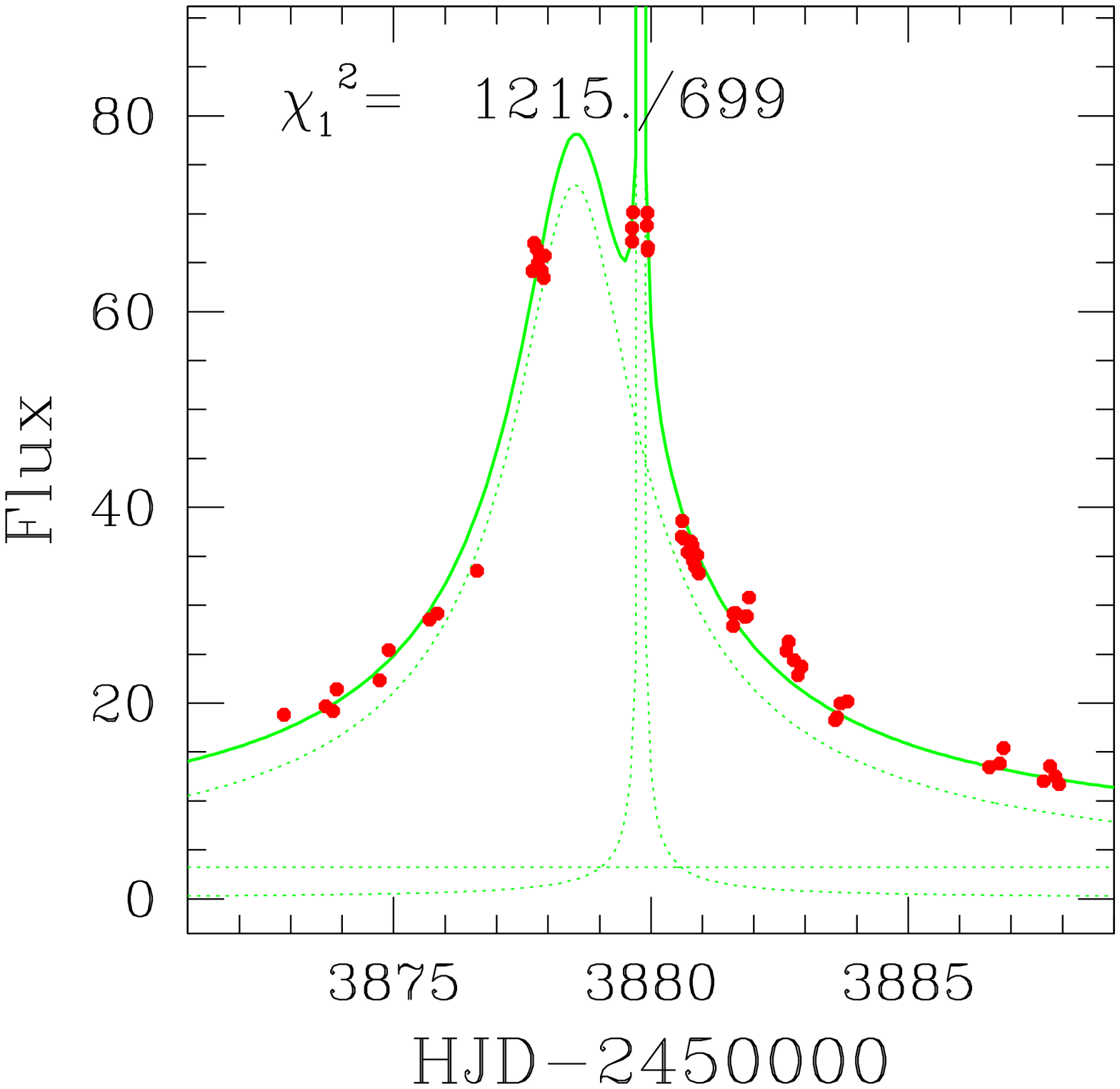} \hfill 
\includegraphics[height=63mm,width=62mm]{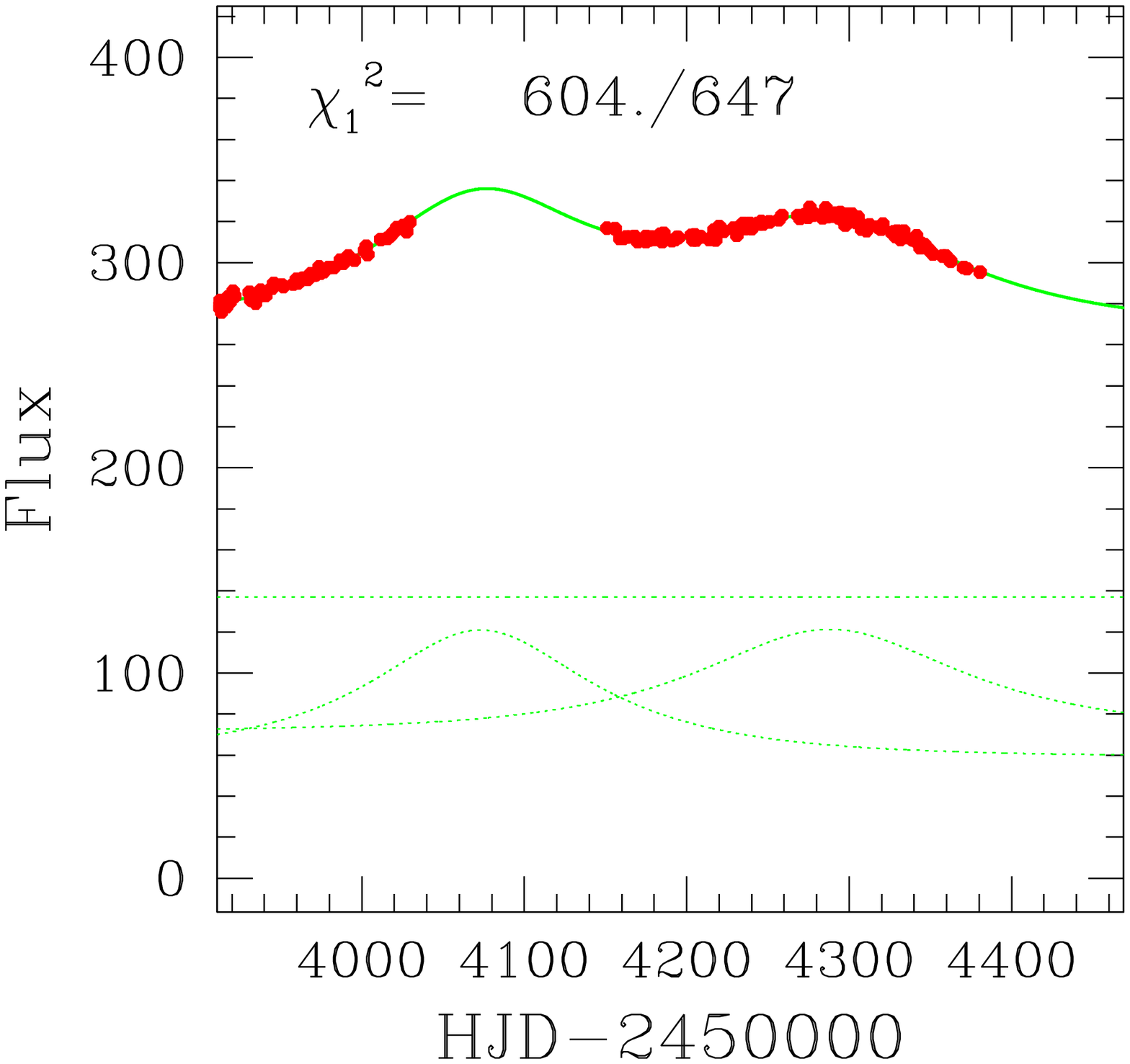} 

}
\noindent\parbox{12.75cm}{
\centerline{{\bf OGLE 2006-BLG-398} \hfill {\bf OGLE-2006-BLG-441}\hfill}

\includegraphics[height=63mm,width=62mm]{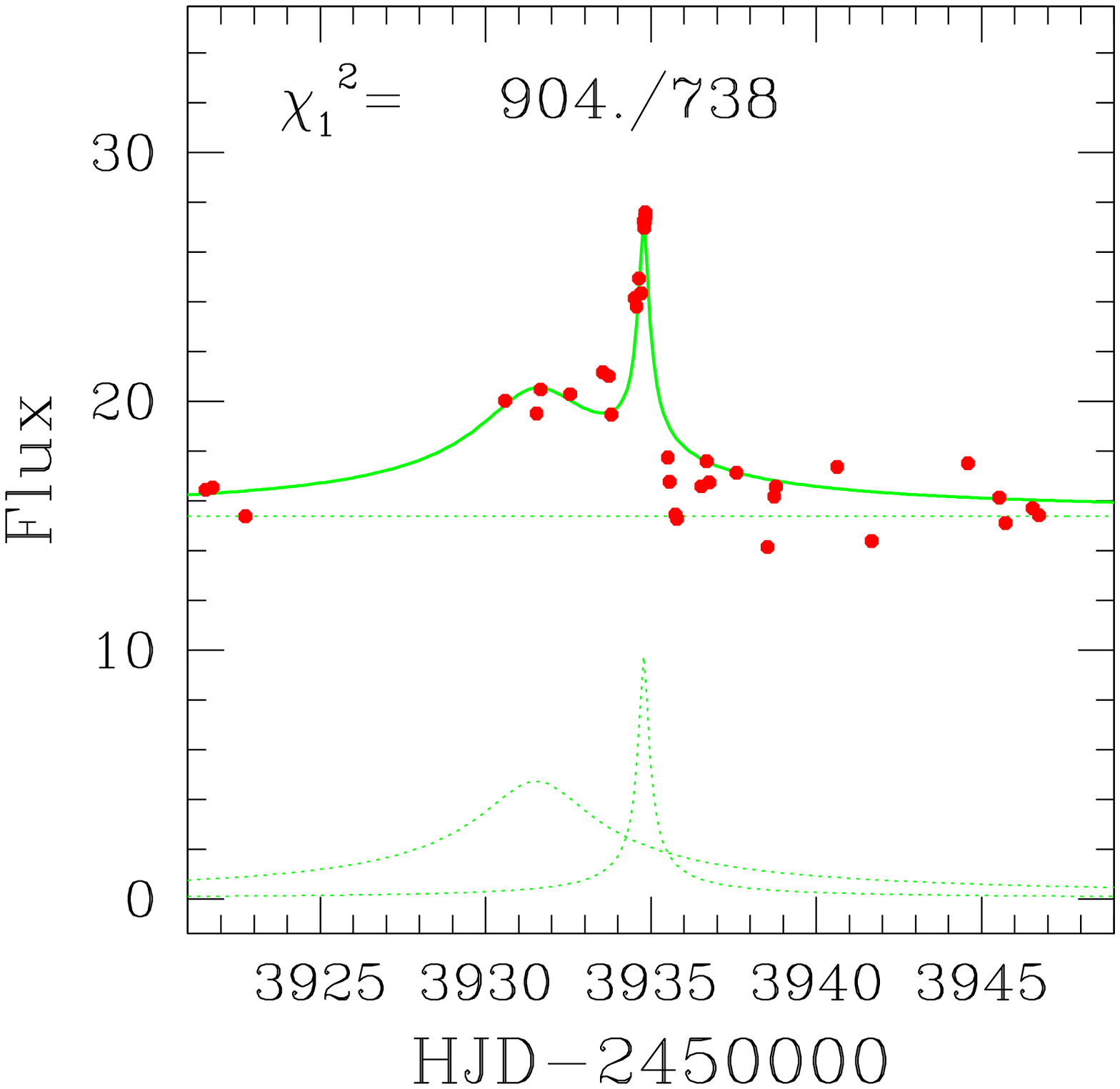} \hfill 
\includegraphics[height=63mm,width=62mm]{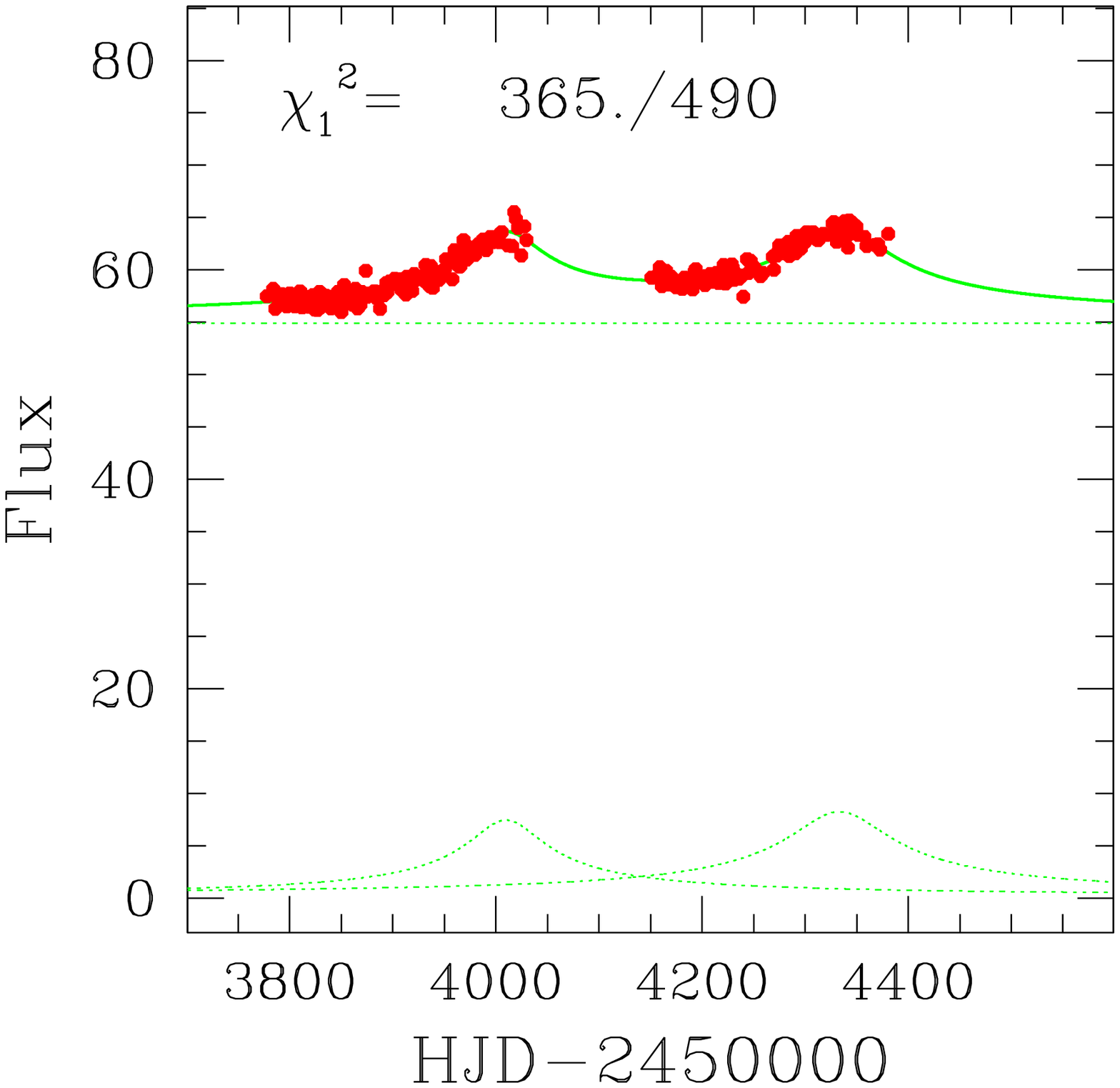} 

}
\noindent\parbox{12.75cm}{
\centerline{{\bf OGLE 2006-BLG-444} \hfill {\bf OGLE-2006-BLG-504}\hfill}

\includegraphics[height=63mm,width=62mm]{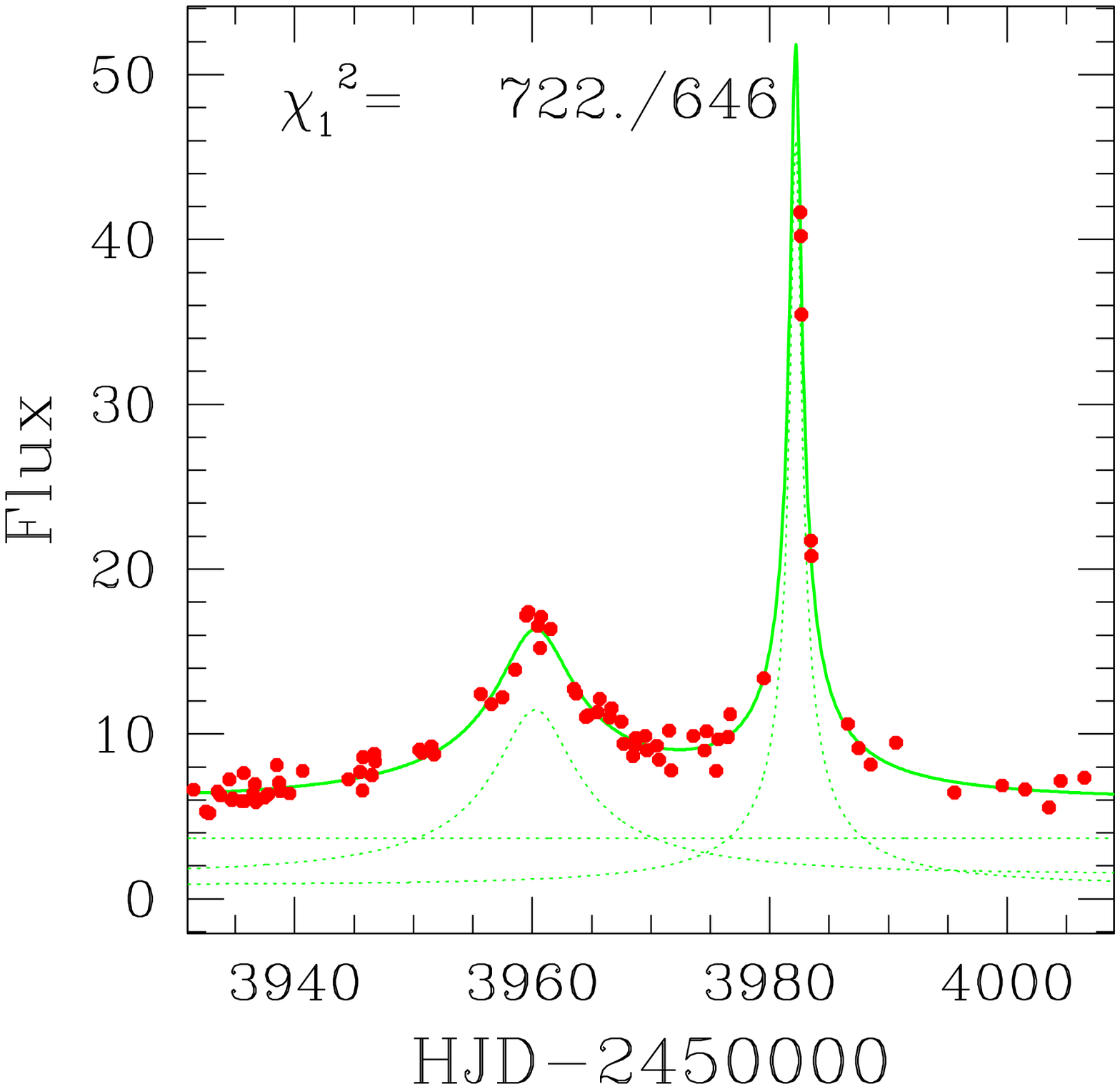} \hfill
\includegraphics[height=63mm,width=62mm]{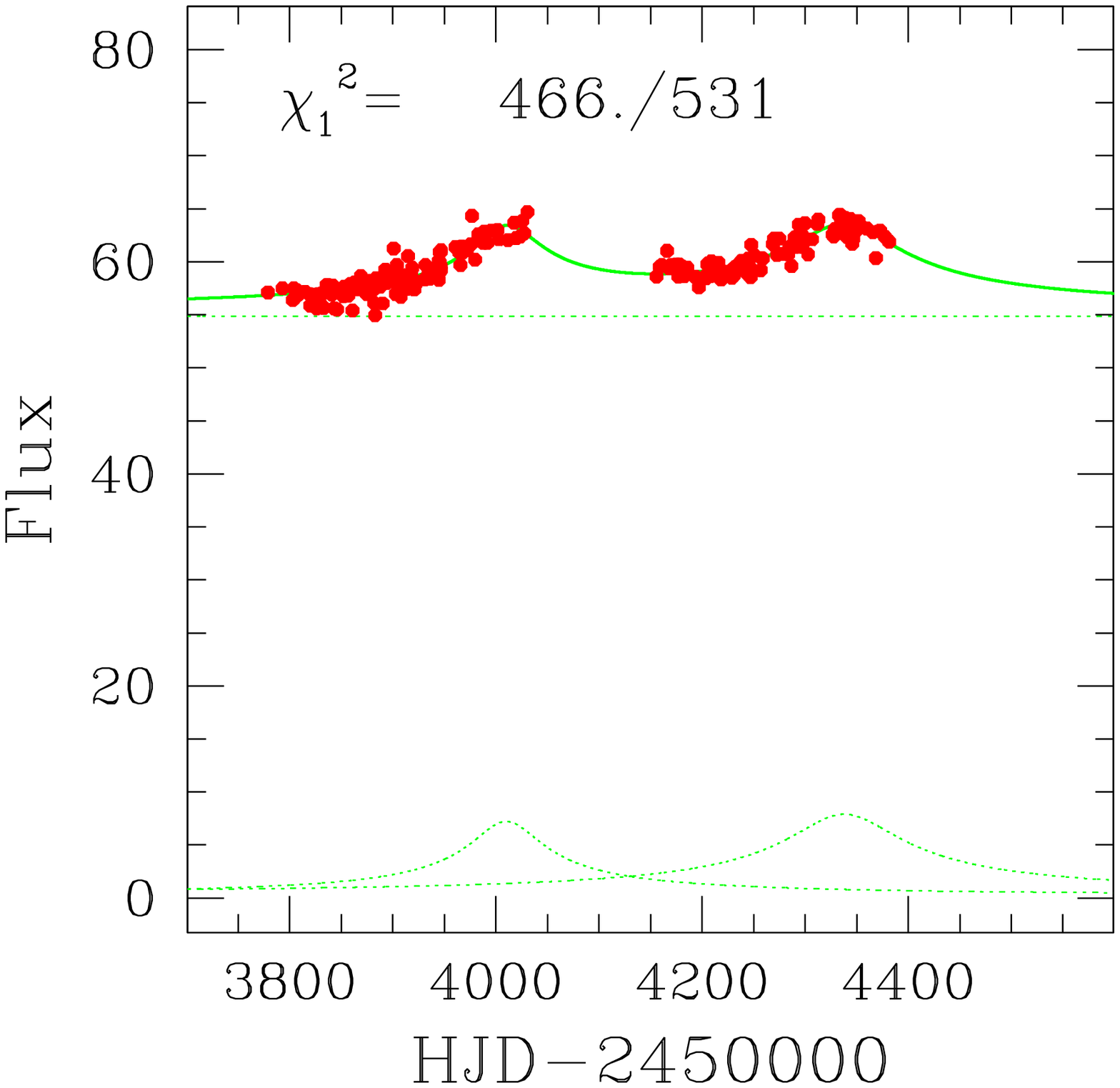}% 

}

\noindent\parbox{12.75cm}{
\centerline{{\bf OGLE 2007-BLG-040} \hfill {\bf OGLE-2007-BLG-080}\hfill}

\includegraphics[height=63mm,width=62mm]{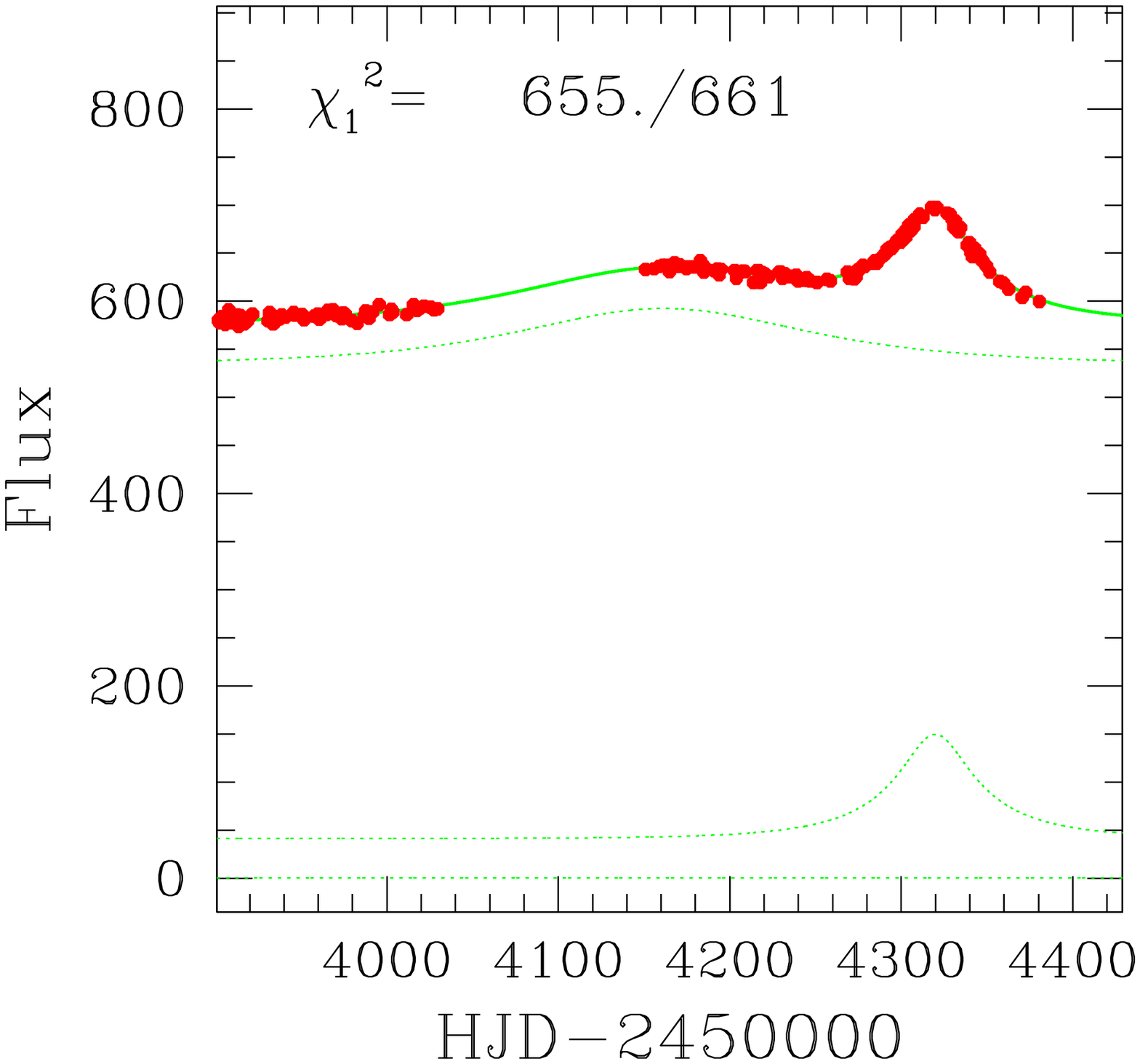} \hfill
\includegraphics[height=63mm,width=62mm]{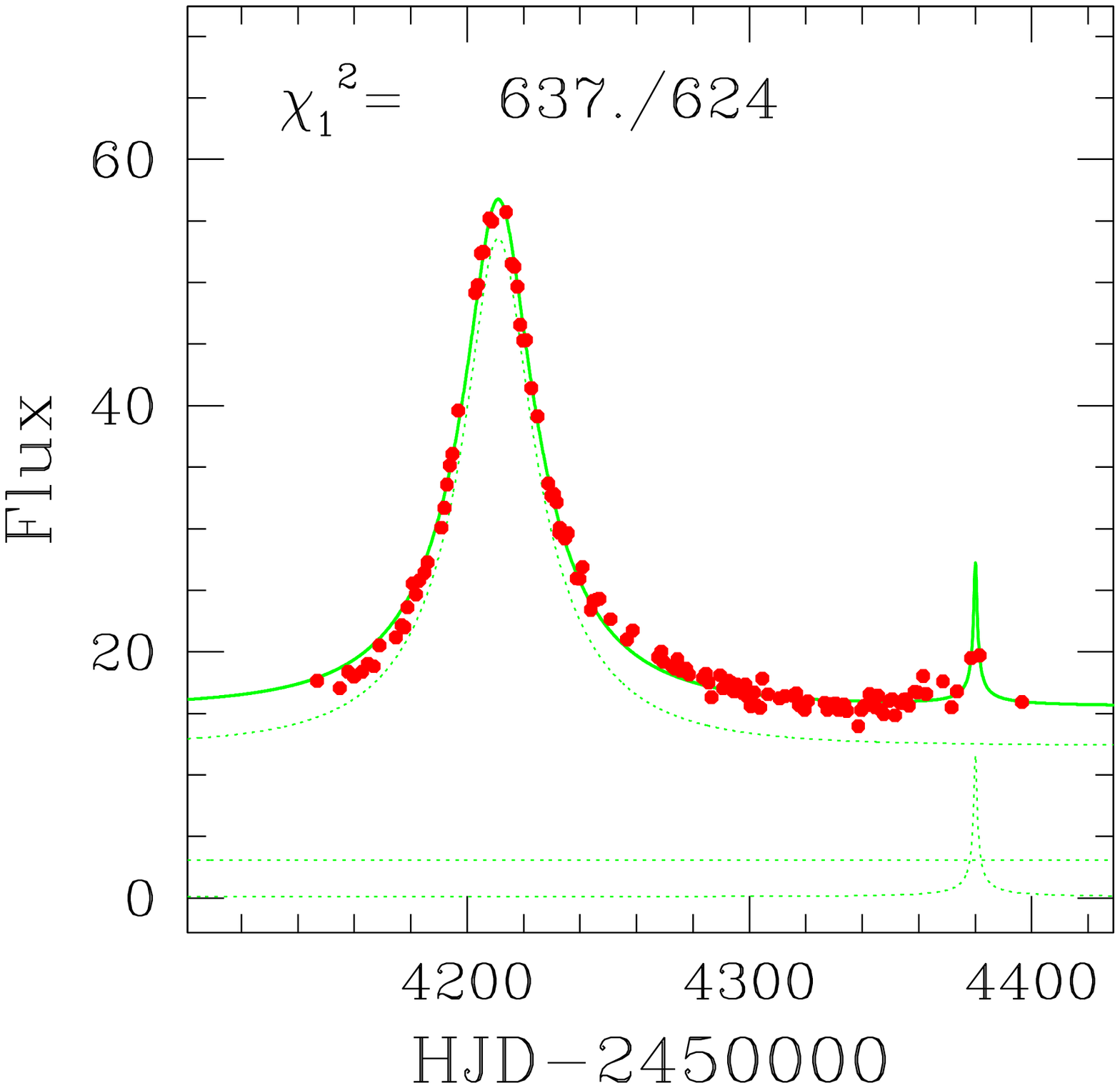}% 

}

\noindent\parbox{12.75cm}{
\centerline{{\bf OGLE 2007-BLG-091} \hfill {\bf OGLE-2007-BLG-159}\hfill}

\includegraphics[height=63mm,width=62mm]{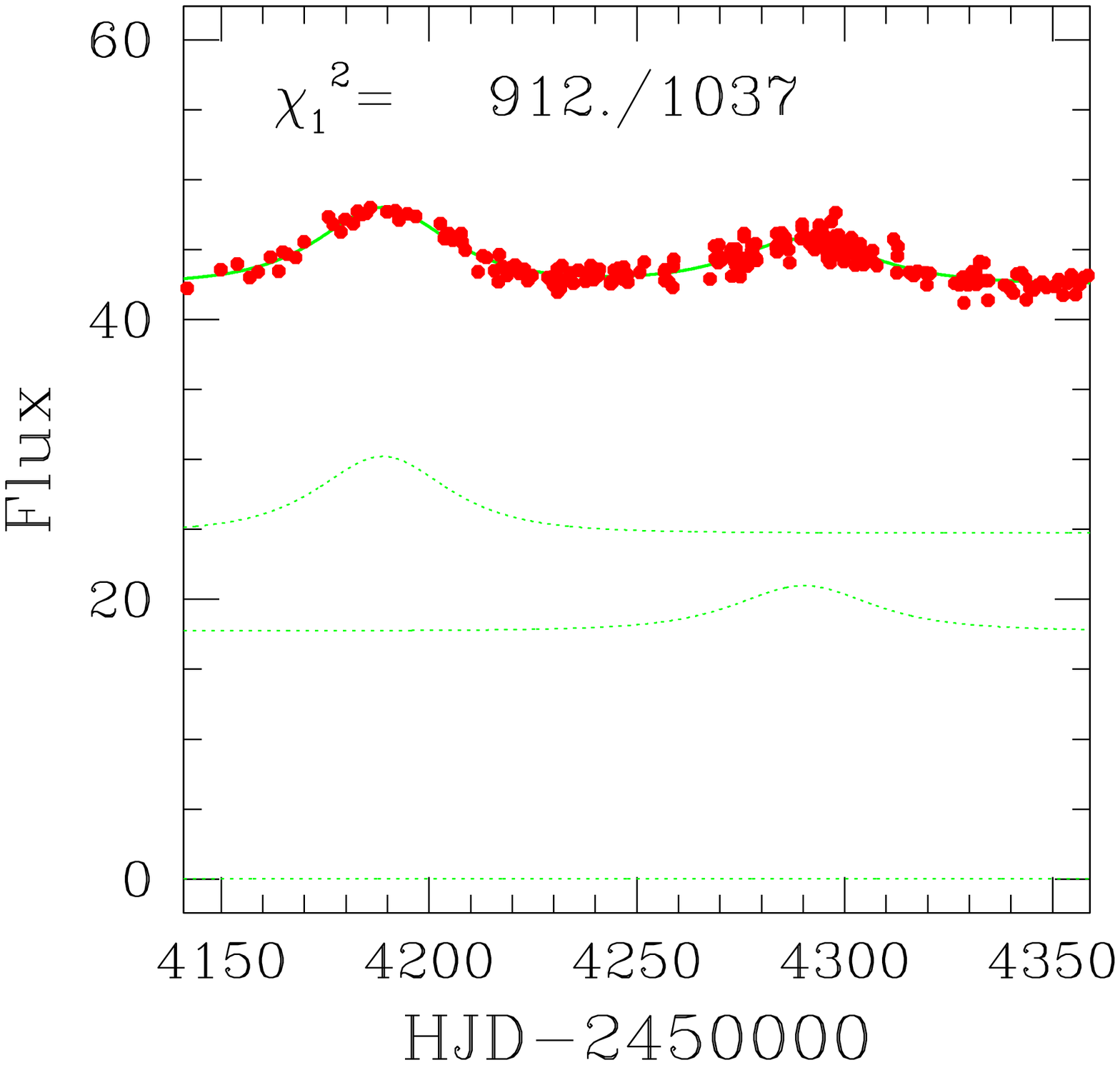} \hfill
\includegraphics[height=63mm,width=62mm]{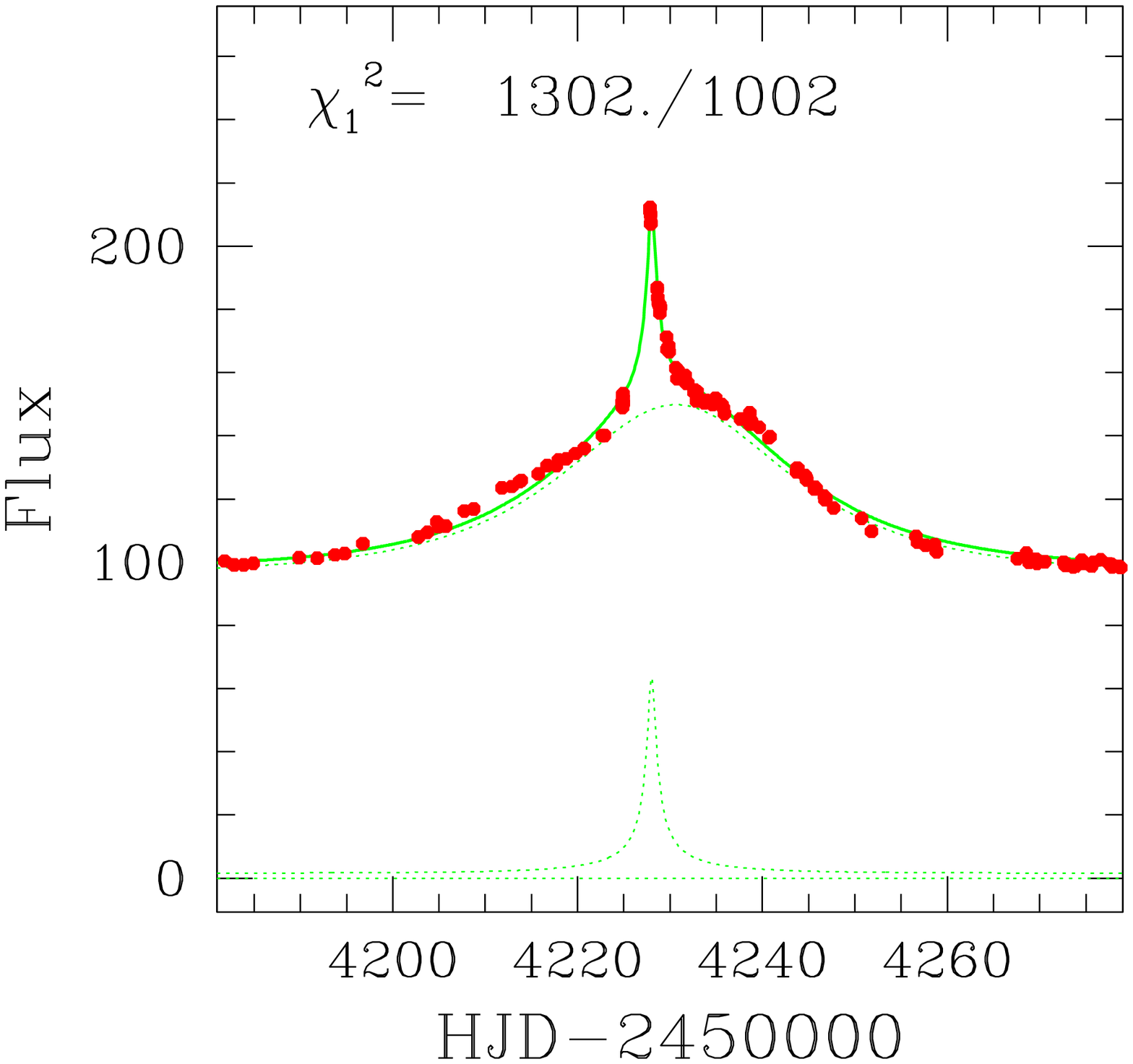}% 

}

\noindent\parbox{12.75cm}{
\centerline{{\bf OGLE 2007-BLG-236} \hfill {\bf OGLE-2007-BLG-355}\hfill}

\includegraphics[height=63mm,width=62mm]{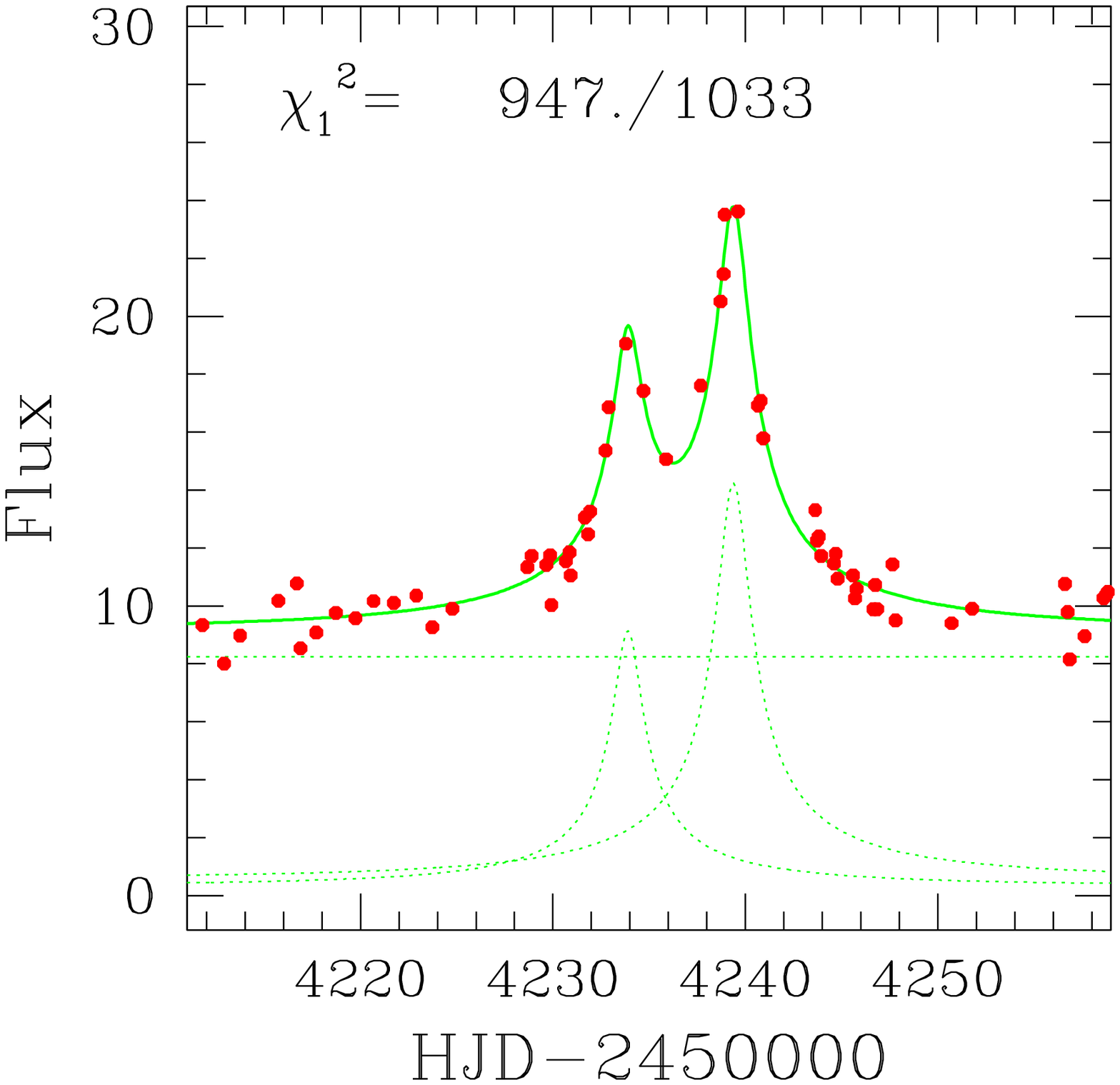} \hfill
\includegraphics[height=63mm,width=62mm]{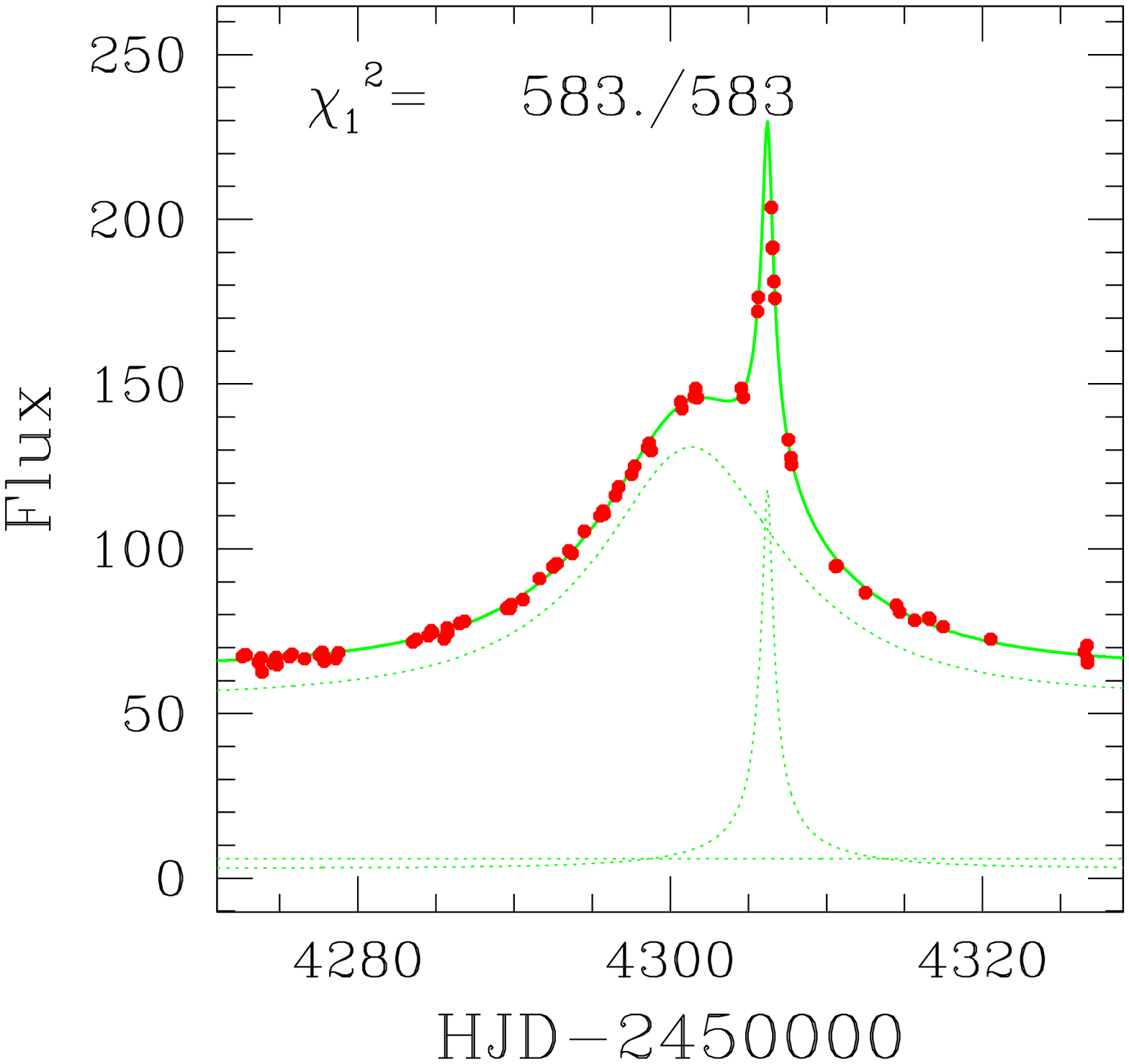}% 

}

\noindent\parbox{12.75cm}{
\centerline{{\bf OGLE 2007-BLG-491} \hfill {\bf OGLE-2007-BLG-594}\hfill}

\includegraphics[height=63mm,width=62mm]{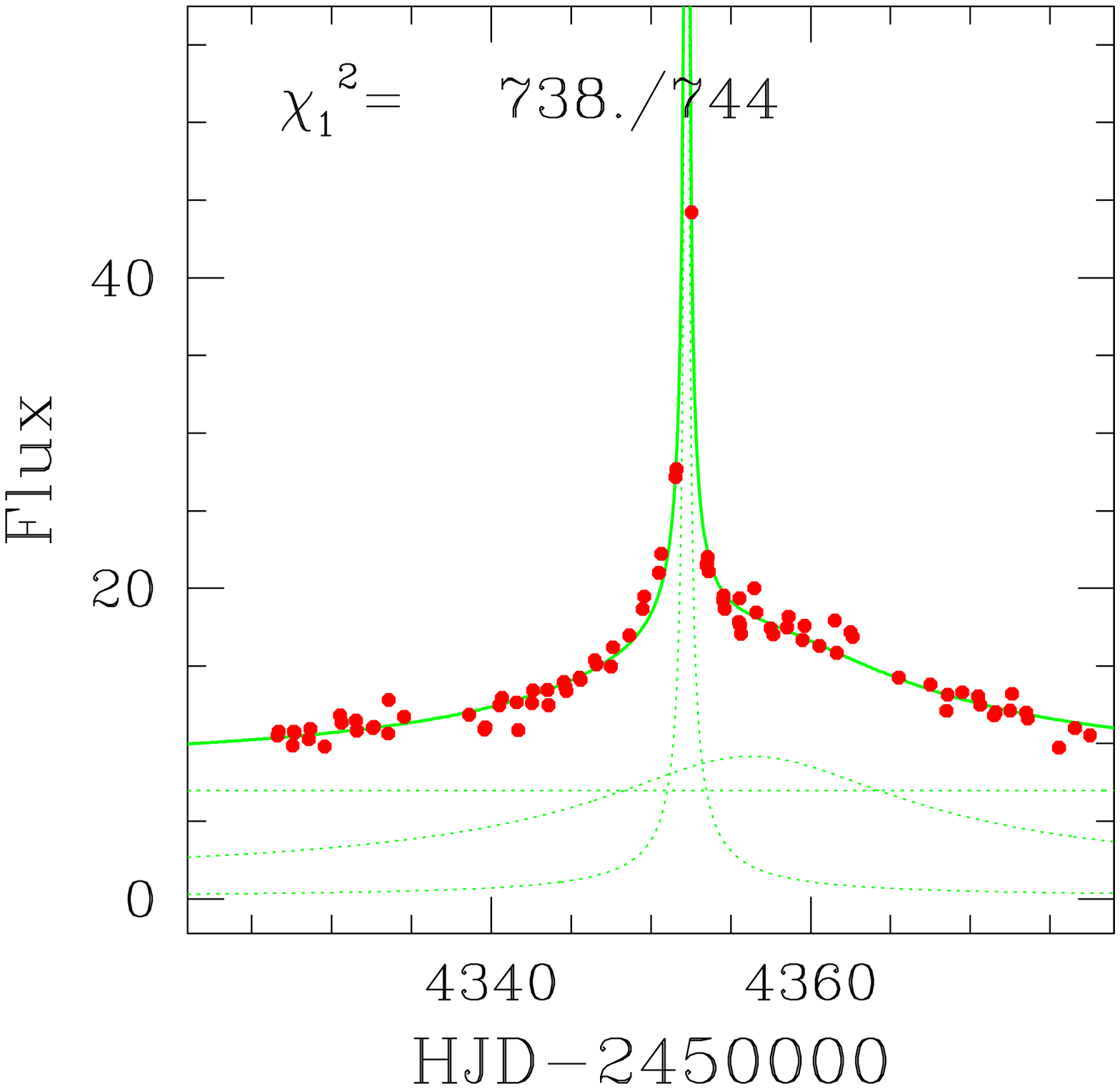} \hfill
\includegraphics[height=63mm,width=62mm]{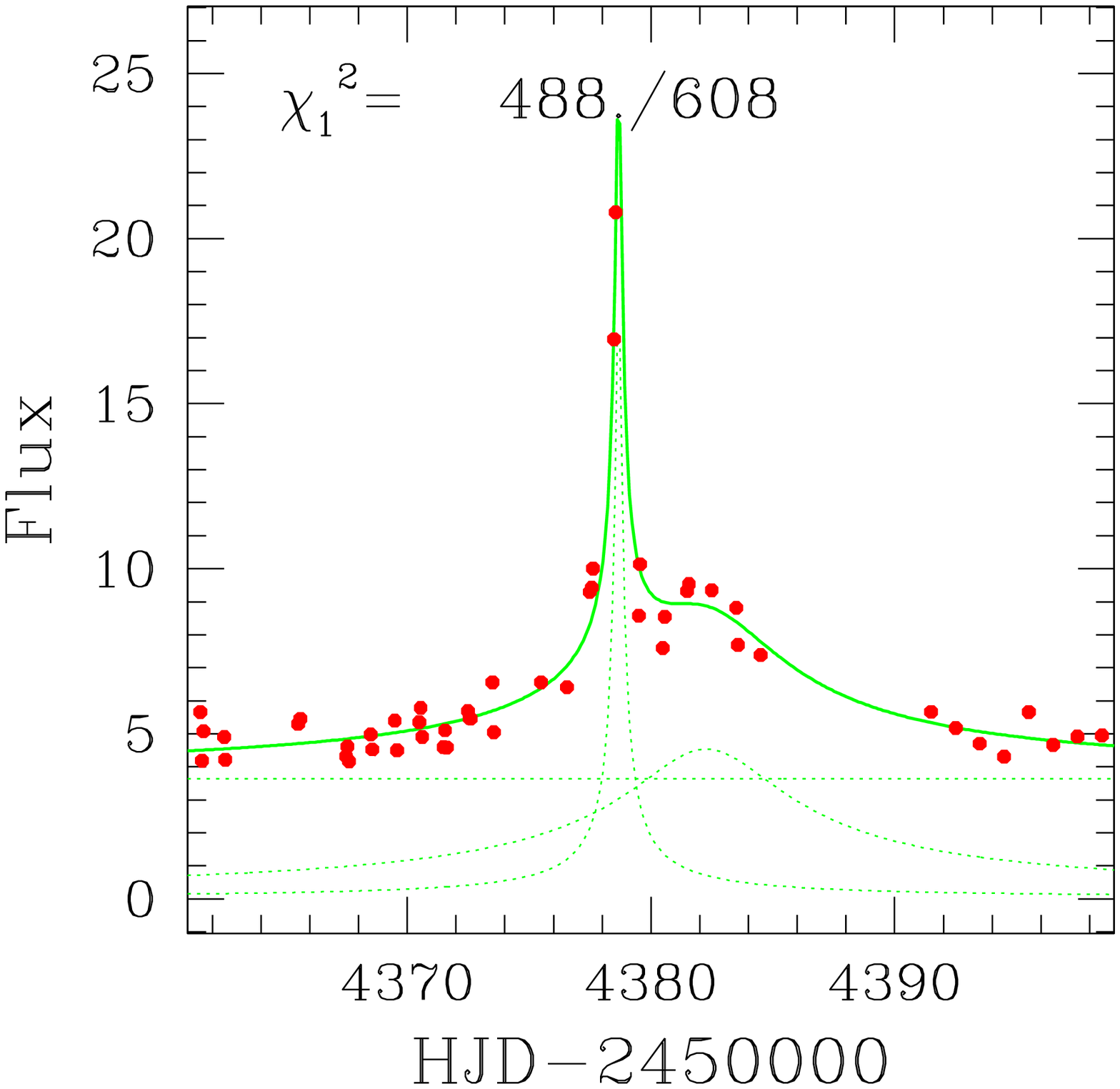}% 

}

\noindent\parbox{12.75cm}{
\centerline{{\bf OGLE 2008-BLG-078} \hfill {\bf OGLE-2008-BLG-084}\hfill}

\includegraphics[height=63mm,width=62mm]{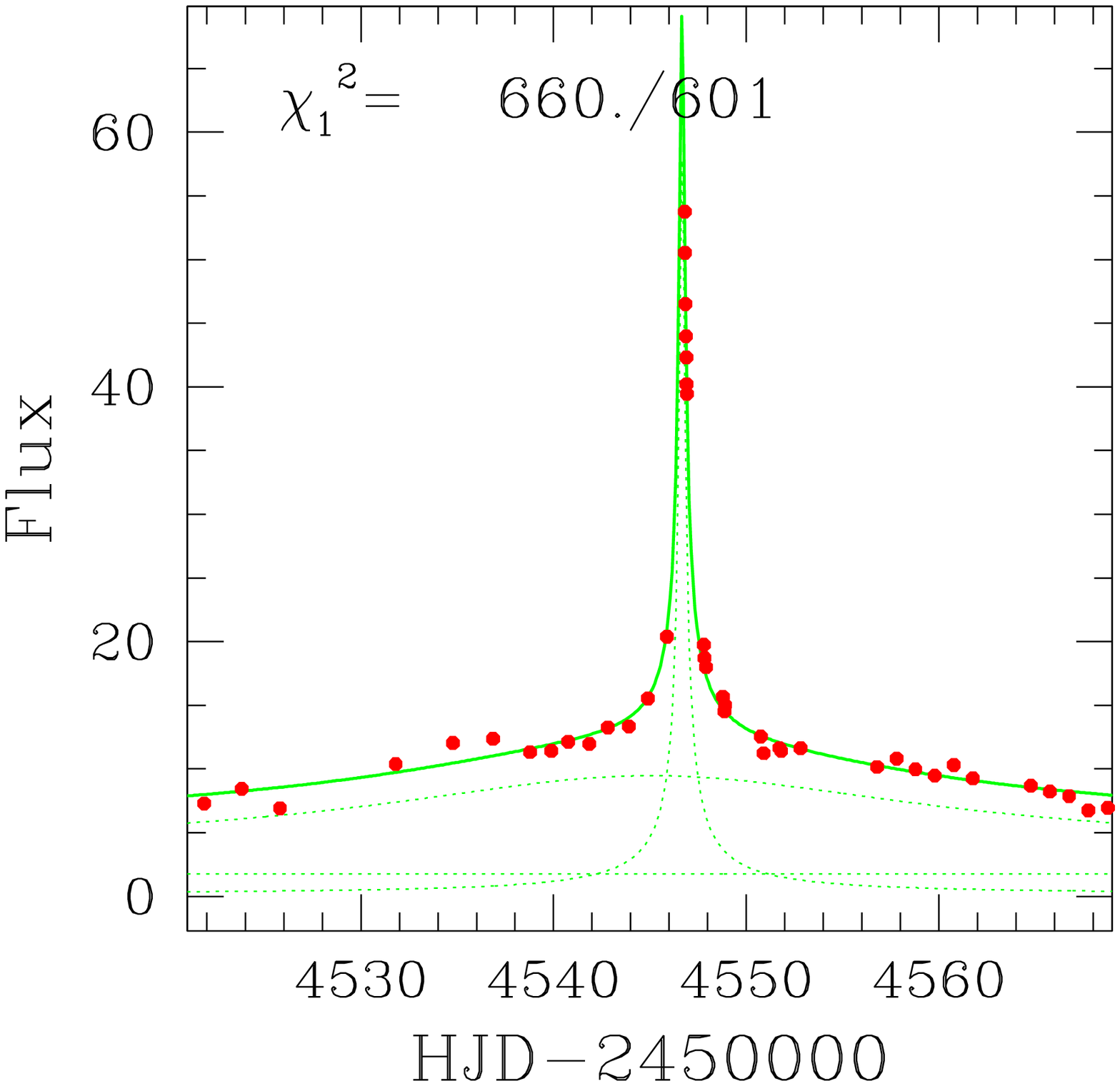} \hfill
\includegraphics[height=63mm,width=62mm]{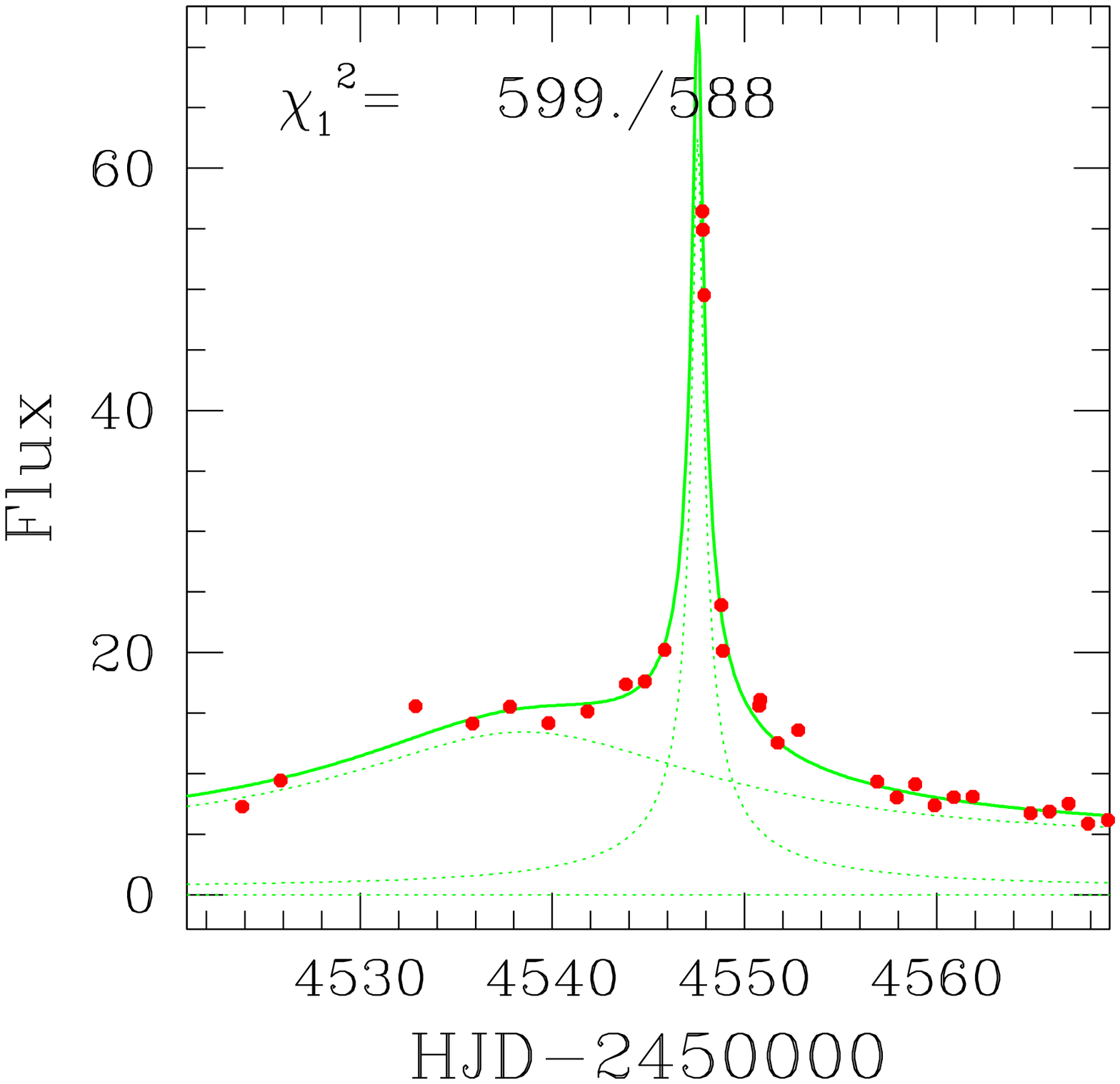}% 

}

\noindent\parbox{12.75cm}{
\centerline{{\bf OGLE 2008-BLG-092} \hfill {\bf OGLE-2008-BLG-098}\hfill}

\includegraphics[height=63mm,width=62mm]{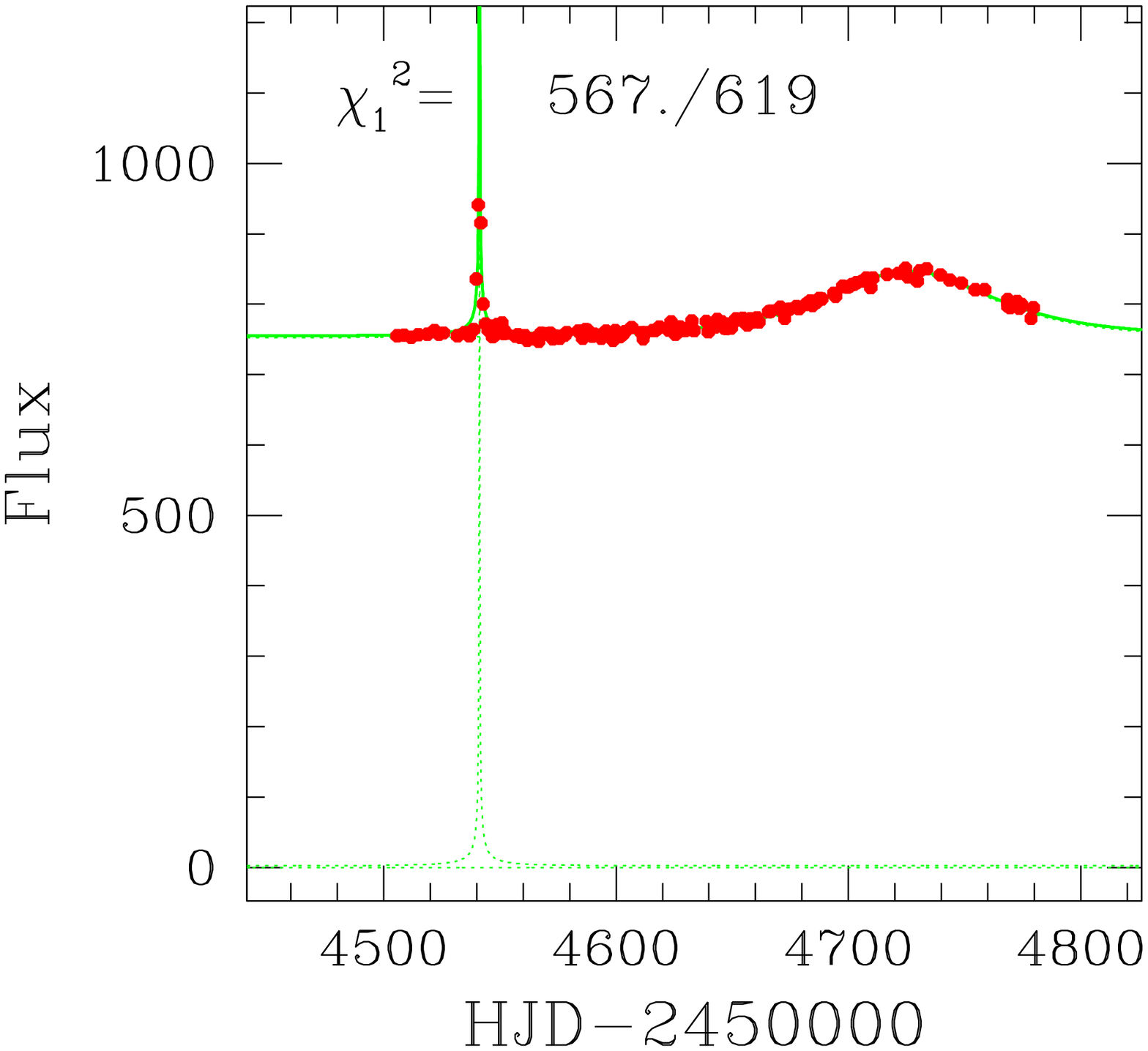} \hfill
\includegraphics[height=63mm,width=62mm]{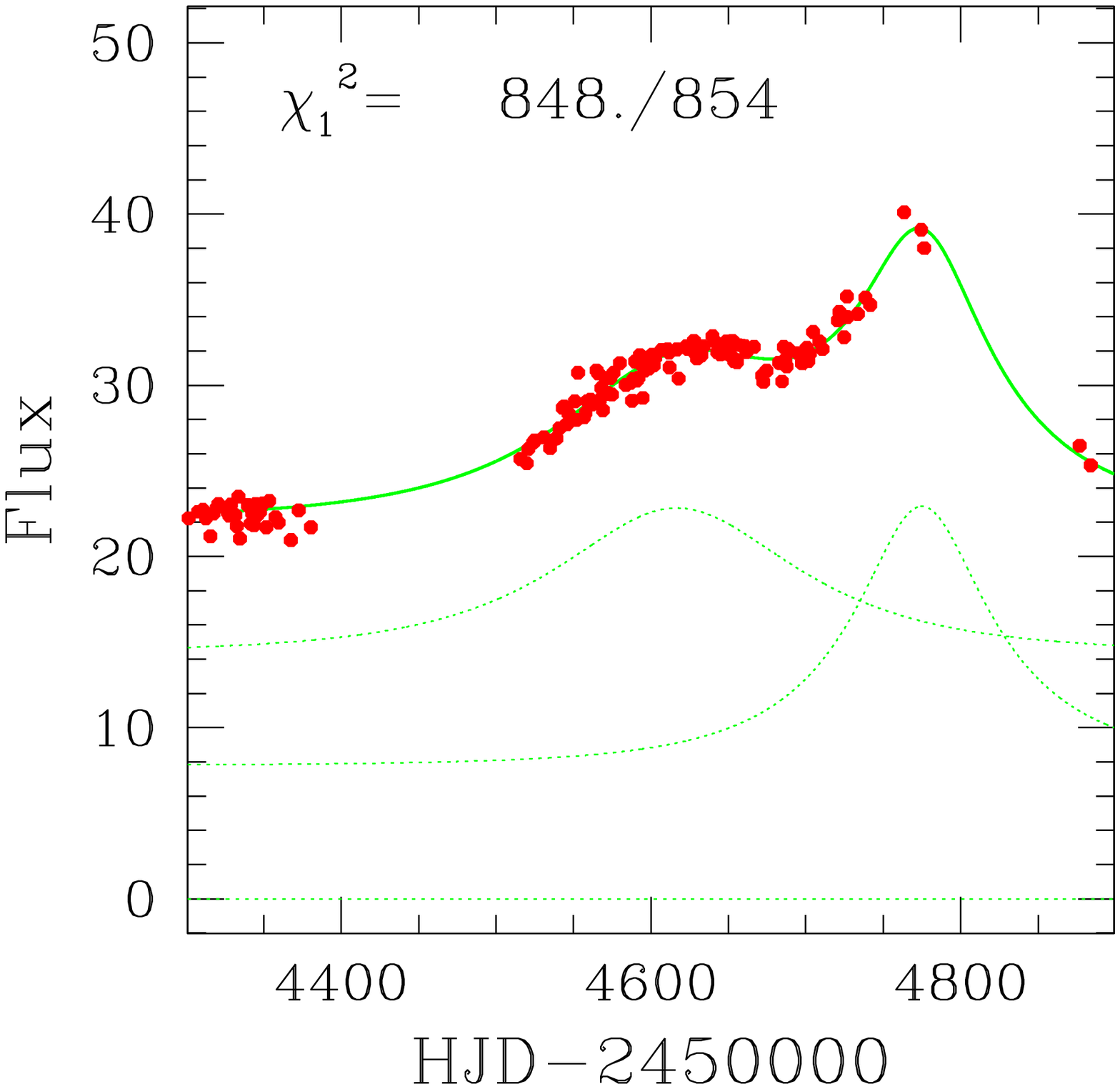}% 

}

\noindent\parbox{12.75cm}{
\centerline{{\bf OGLE 2008-BLG-110} \hfill {\bf OGLE-2008-BLG-143}\hfill}

\includegraphics[height=63mm,width=62mm]{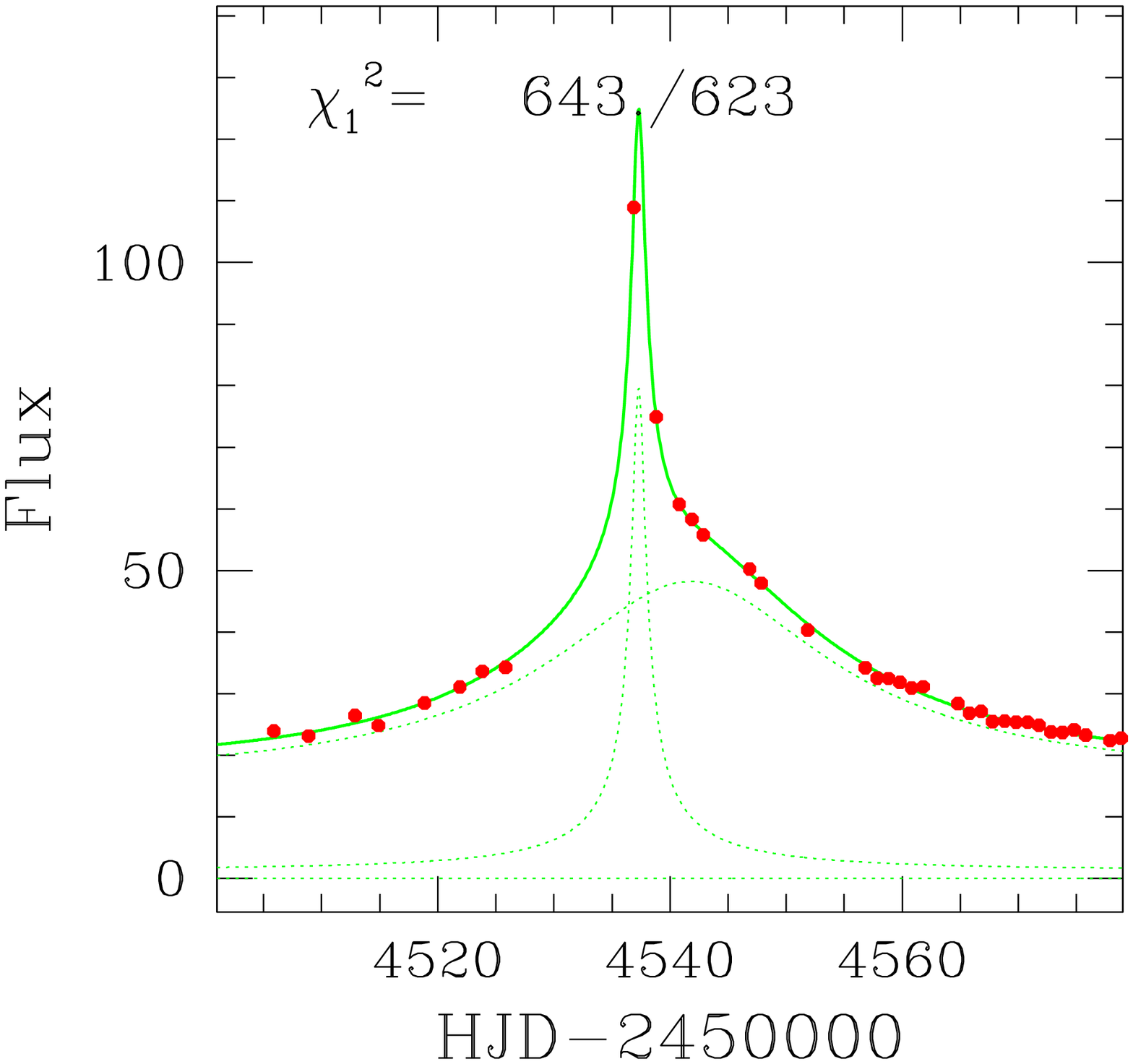} \hfill
\includegraphics[height=63mm,width=62mm]{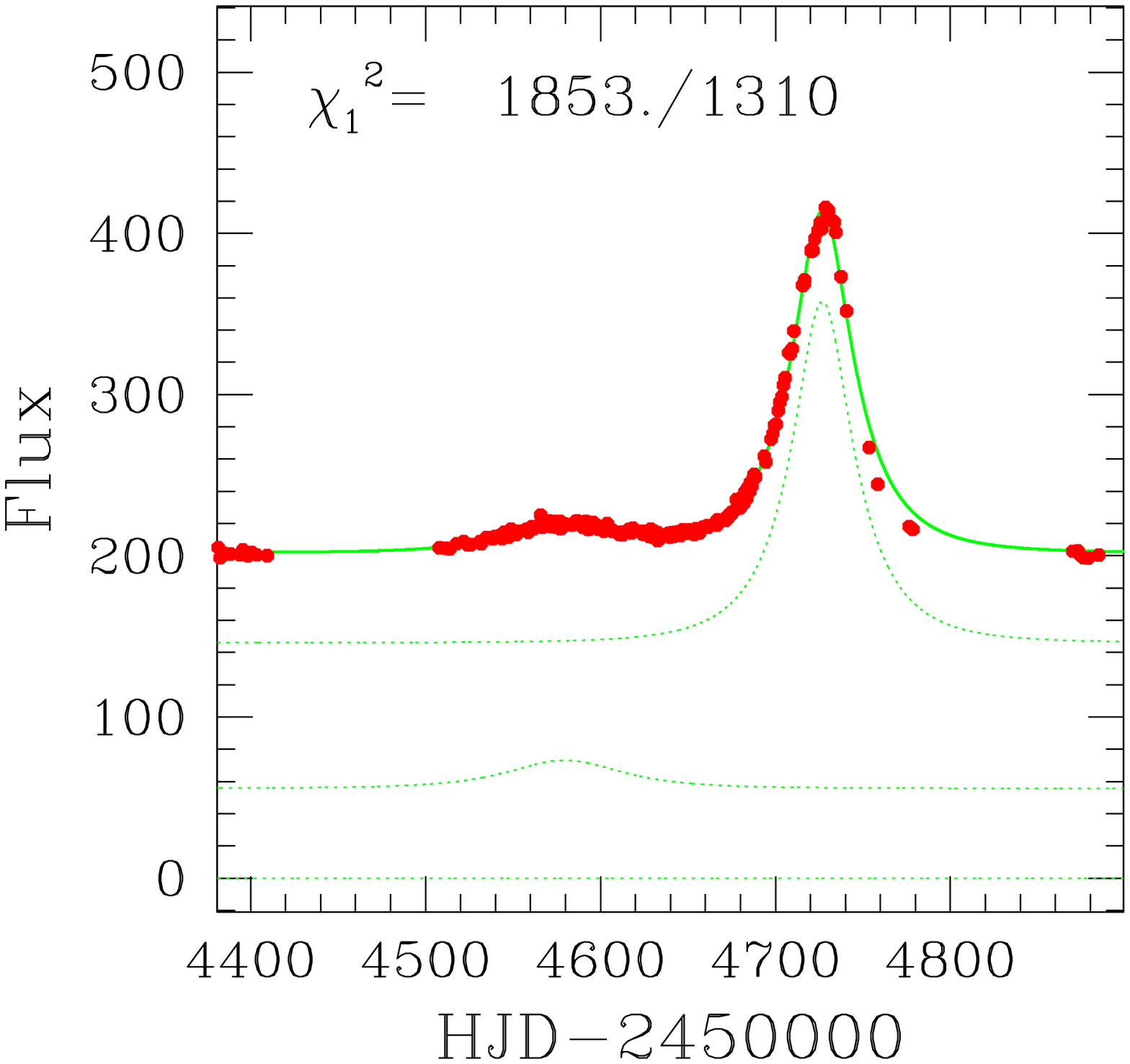}% 

}

\noindent\parbox{12.75cm}{
\centerline{{\bf OGLE 2008-BLG-146} \hfill {\bf OGLE-2008-BLG-210}\hfill}

\includegraphics[height=63mm,width=62mm]{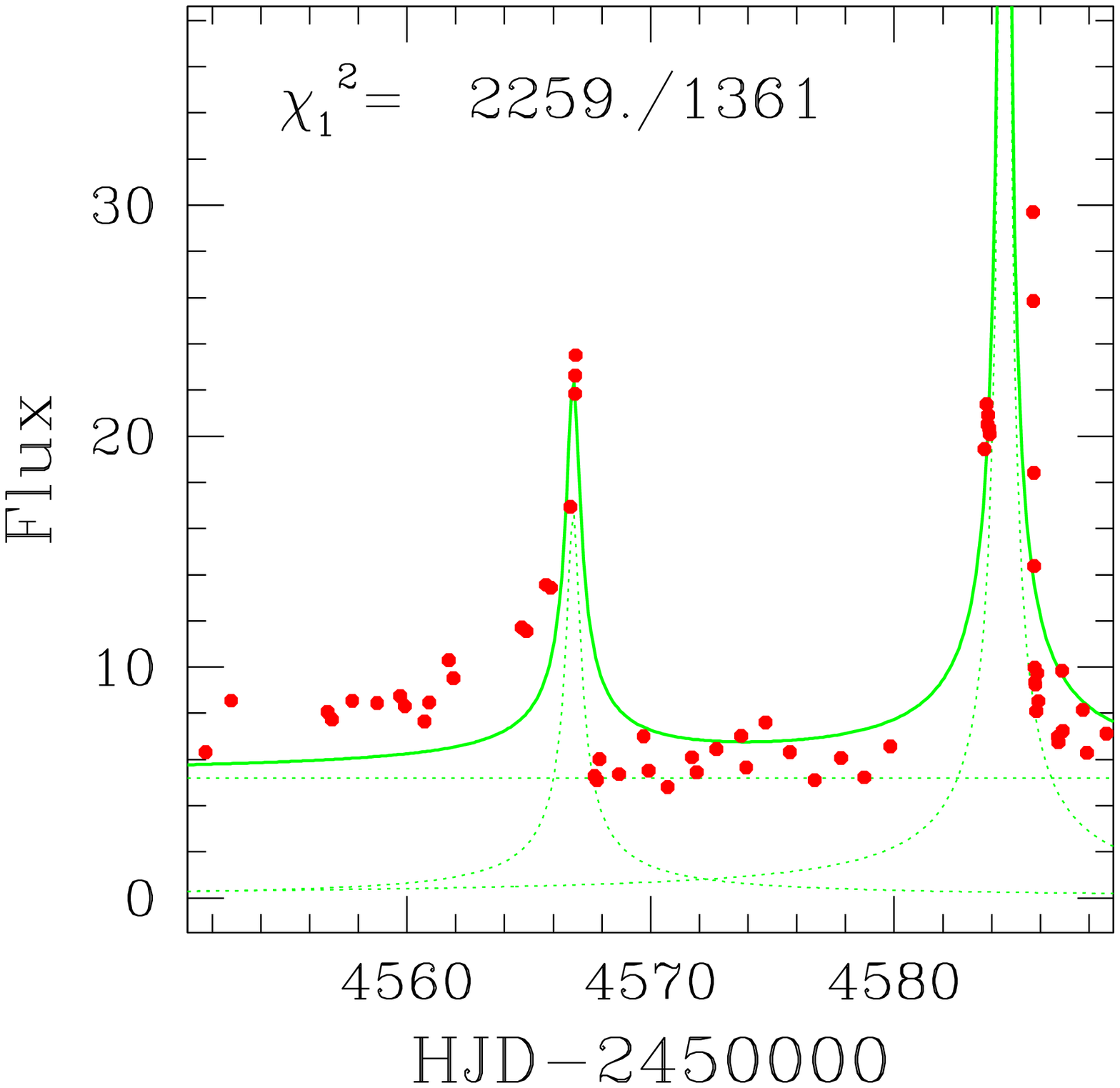} \hfill
\includegraphics[height=63mm,width=62mm]{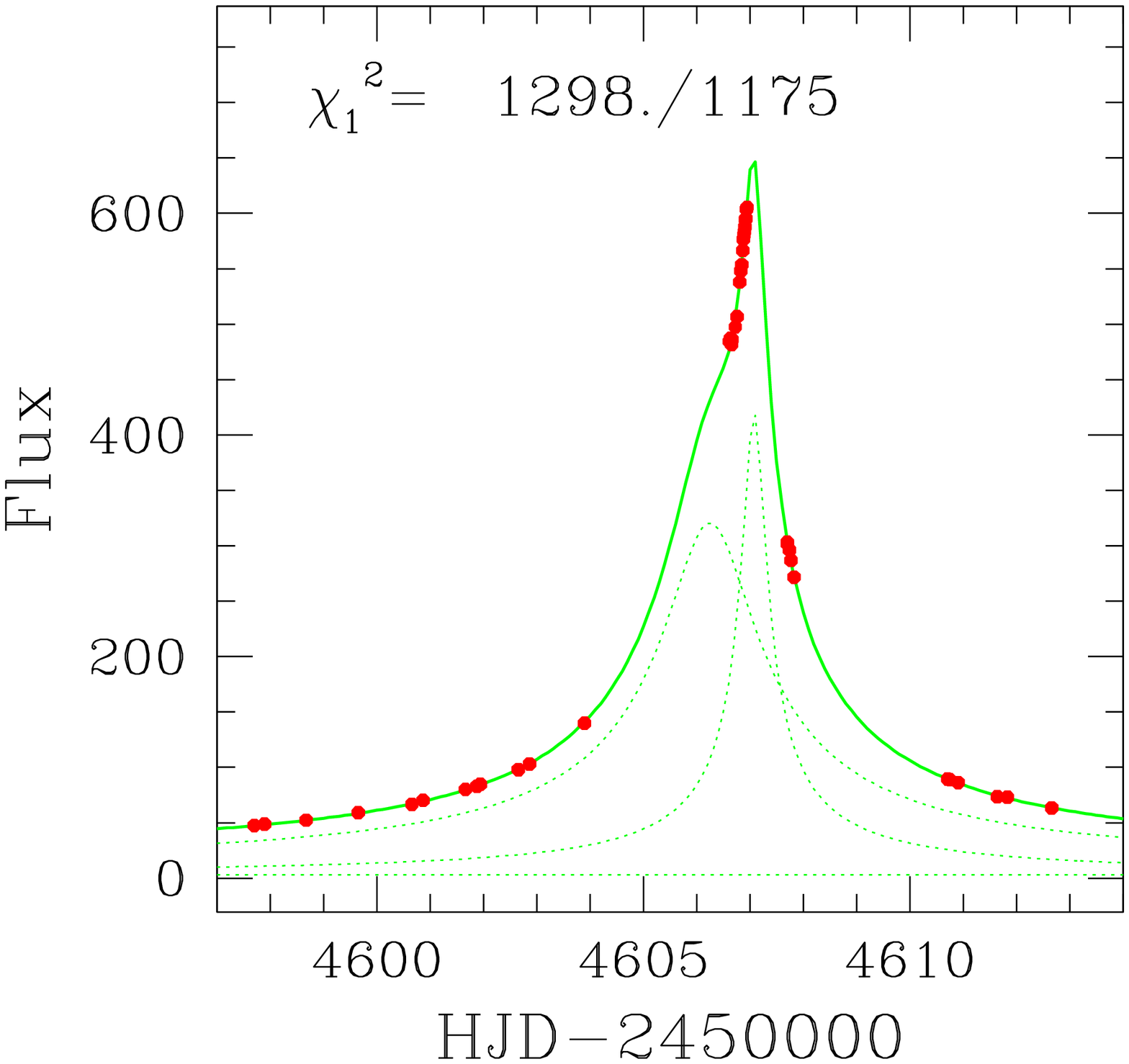}% 

}

\noindent\parbox{12.75cm}{
\centerline{{\bf OGLE 2008-BLG-353} \hfill} 

\includegraphics[height=63mm,width=62mm]{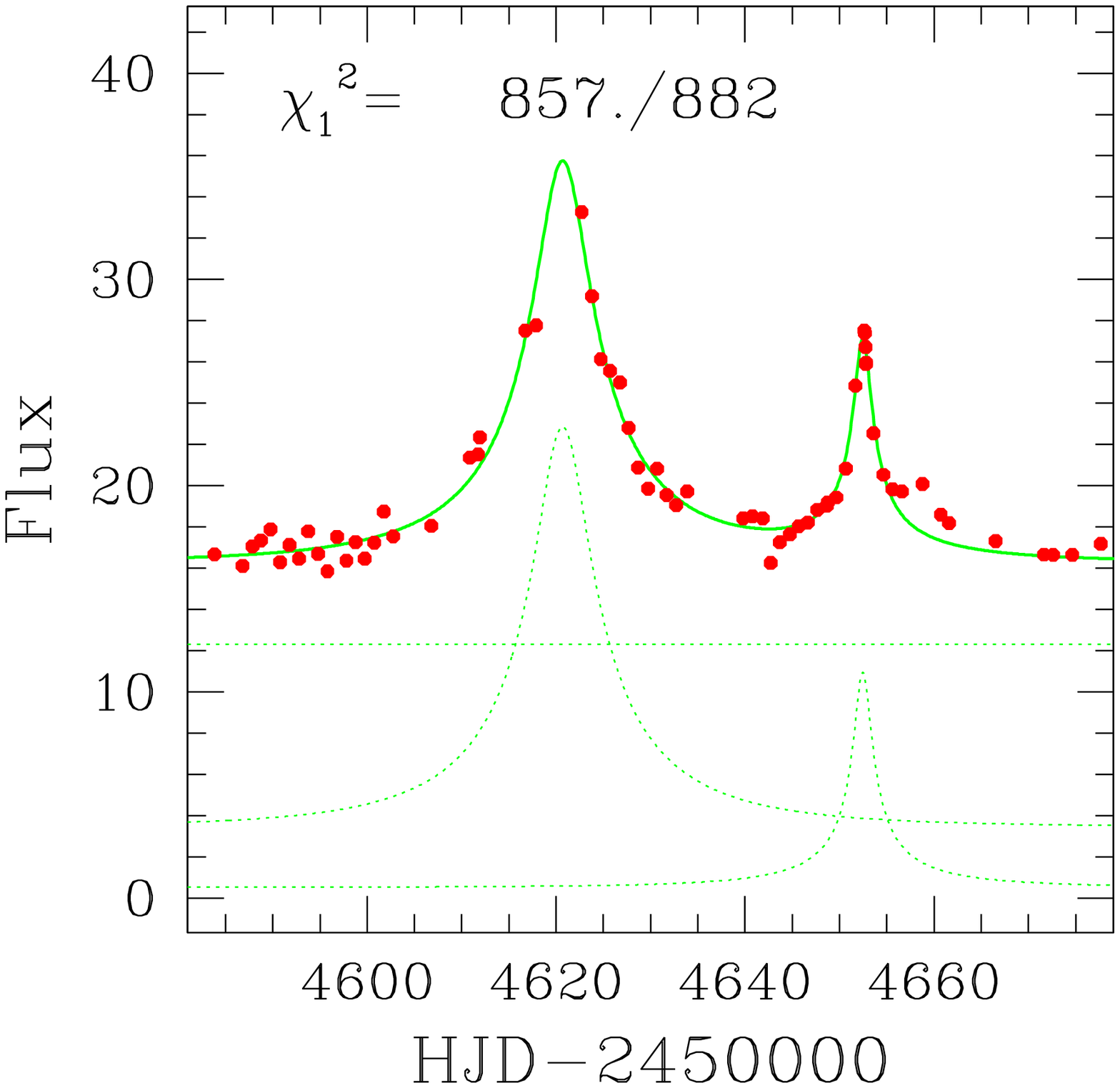} \hfill

}


\begin{references}
\refitem{Alard, C.}{2000}{\AAS}{144}{363}
\refitem{Alard, C., and Lupton, R.H.}{1998}{\ApJ}{503}{325}
\refitem{Alcock, C., \etal}{2000}{\ApJ}{541}{270} 
\refitem{Beaulieu, J.-P. \etal}{2006}{Nature}{439}{437}
\refitem{Bond, I.A. \etal}{2004}{\ApJ}{606}{L155} 
\refitem{Cassan, A.}{2008}{\AA}{491}{587}
\refitem{DiStefano, R., and Mao, S.}{1996}{\ApJ}{457}{93}
\refitem{Dominik, M.}{1998}{\AA}{333}{L79}
\refitem{Gaudi, B.S., and Han, Ch.}{2004}{\ApJ}{611}{528}
\refitem{Gaudi, B.S., \etal}{2008}{Science}{319}{927} 
\refitem{Gould, A. \etal}{2006}{\ApJ}{644}{L37}
\refitem{Griest, K.}{1991}{\ApJ}{366}{412}
\refitem{Jaroszyński, M.}{2002}{\Acta}{52}{39 (Paper I)}
\refitem{Jaroszyński, M., and Skowron, J.}{2008}{\Acta}{58}{345}
\refitem{Jaroszyński, M., Udalski, A., Kubiak, M., Szymański, M., Pietrzyński, G., Soszyński, I., Żebruń, K., Szewczyk, O., and Wyrzykowski, Ł.}{2004}{\Acta}{54}{103 (Paper~II)}
\refitem{Jaroszyński, M., Skowron, J., Udalski, A., Kubiak, M., Szymański, M., Pietrzyński, G., Soszyński, I., Żebruń, K., Szewczyk, O., and Wyrzykowski, Ł.}{2006}{\Acta}{56}{307 (Paper~III)}
\refitem{Kiraga, M., and Paczyński, B.}{1994}{\ApJ}{430}{L101}
\refitem{Mao, S., and DiStefano, R.}{1995}{\ApJ}{440}{22}
\refitem{Mao, S., and Loeb, A.}{2001}{\ApJ}{547}{L97}
\refitem{Mao, S., and Paczyński, B.}{1991}{\ApJ}{374}{L37}
\refitem{Paczyński, B.}{1996}{Ann. Rev. Astron. Astrophys.}{34}{419}%{``Gravitational Lenses'', Springer, Berlin}
\refitem{Schneider, P., and Weiss, A.}{1986}{\AA}{164}{237}
\refitem{Skowron, J.}{2009}{\it PhD Thesis}{}{University of Warsaw Astronomical Observatory}
\refitem{Skowron, J., Jaroszyński, M., Udalski, A., Kubiak, M., Szymanski, M.K., Pietrzynski, G., Soszynski, I., Szewczyk, O., Wyrzykowski, Ł., and Ulaczyk, K.}{2007}{\Acta}{57}{281 (Papaer IV)}
\refitem{Skowron, J., Wyrzykowski, Ł., Mao, S., and Jaroszyński, M.}{2009}{\MNRAS}{393}{999}
\refitem{Smith, M.C., Woźniak, P., Mao, S., and Sumi, T.}{2007}{\MNRAS}{380}{805}
\refitem{Sumi, T., \etal}{2010}{\ApJ}{710}{1641}
\refitem{Trimble, V.}{1990}{\MNRAS}{242}{79}
\refitem{Udalski, A.}{2003}{\Acta}{53}{291}
\refitem{Udalski, A., \etal}{2005}{\ApJ}{628}{L109}
\refitem{Udalski, A., Szymański, M., Kaluzny, J., Kubiak, M., Mateo, M., Krzemiński, W., and Paczyński, B.}{1994}{\Acta}{44}{227}
\refitem{Wood, A., and Mao, S.}{2005}{\MNRAS}{362}{945}
\refitem{Wyrzykowski, Ł.}{2005}{\it PhD Thesis}{}{Warsaw University Astronomical Observatory}
\end{references}
\end{document}